\documentclass[usenatbib]{mnras}
\usepackage{amsmath}
\usepackage{amssymb}
\usepackage{morefloats}
\usepackage{graphicx}

\title[The first 10 yr of dust evolution in SN~1987A]{Modelling supernova line profile 
asymmetries to determine ejecta dust masses:  SN~1987A from days 714 to 
3604}
\author[Antonia Bevan and M. J. Barlow]{Antonia Bevan$^{1}$ and M. J. 
Barlow$^{1}$\\
$^{1}$Department of Physics and Astronomy, University College London, 
Gower Street, London WC1E 6BT, UK}

\begin{document}

\date{Submitted on 27 August 2015}

\pagerange{\pageref{firstpage}--\pageref{lastpage}} \pubyear{2015}

\maketitle

\label{firstpage}

\hyphenation{CCSNe}
\hyphenation{CCSN}
\hyphenation{CTIO}
\hyphenation{VLT}
\hyphenation{AAT}
\hyphenation{HST}

\begin{abstract}

The late time optical and near-IR line profiles of many core-collapse supernovae exhibit a red-blue asymmetry as a result of greater extinction by internal dust of radiation emitted from the receding parts of the supernova ejecta.  We present here a new code, {\sc DAMOCLES}, that models the effects of dust on the line profiles of core-collapse supernovae in order to determine newly formed dust masses.
We find that late-time dust-affected line profiles may exhibit an extended red scattering wing (as noted by \citet{Lucy1989}) and that they need not be flux-biased towards the blue, although the profile peak will always be blueshifted. We have collated optical spectra of SN~1987A from a variety of archival sources and have modelled  
the H$\alpha$ line from days 714 to 3604 and the [O~{\sc i}]~6300,6363~\AA\ doublet between days 714 and 1478. 
Our line profile fits rule out day~714 dust masses $>3\times10^{-3}$ M$_{\odot}$ for all grain types apart from 
pure magnesium silicates, for which no more than 0.07~M$_{\odot}$ can be accommodated. Large grain radii ($\geqslant 0.6~\mu$m) are generally required to fit the line profiles even at the earlier epochs.  We find that a large dust mass ($\ge 0.1$~M$_{\odot}$) had formed by day 3604 and infer that the majority of the present dust mass must have formed after this epoch. Our findings agree with recent estimates from spectral energy distribution fits for the dust mass evolution of SN~1987A and support the inference that the majority of SN~1987A's dust formed many years after the initial explosion.

\end{abstract}

\begin{keywords}
radiative transfer --
supernovae: general --
supernovae: individual: SN 1987A  --  
ISM: supernova remnants.
\end{keywords}

\section{Introduction}

Core-collapse supernovae (CCSNe) have long been thought to be potential 
dust factories \citep{Hoyle1970, Kozasa1991, Todini2001}. However 
over the previous decade observations at mid-infrared (mid-IR) wavelengths of 
warm dust emission from CCSNe had suggested that the quantities of dust 
produced, typically $\leq$ 10$^{-3}$~M$_\odot$ during the first 1000 d 
\citep{Sugerman2006, Meikle2007, Kotak2009, Andrews2010, Fabbri2011} were 
much less than the 0.1-1.0~M$_\odot$ of dust per CCSN  
estimated to be needed \citep{Morgan2003, Dwek2007} in order to account 
for the very large dust masses measured in some high redshift galaxies 
\citep{Omont2001, Bertoldi2003, Watson2015}. However, recent {\em 
Herschel} far-IR and sub-mm observations of cold dust masses as high as 
0.2-0.8$M_{\odot}$ in several young supernova remnants have resulted in a 
re-evaluation of the rate of dust production by CCSNe \citep{Barlow2010, 
Matsuura2011, Gomez2012}. The {\em Hershel} dust mass estimates were based 
on fitting dust spectral energy distributions (SEDs) that peaked at far-IR 
wavelengths. Following the end of the {\em Herschel} mission in 
2013 there is likely to be a long wait for far-IR facilities with 
comparable or better sensitivities than {\em Herschel} to become 
available, providing an incentive to make use of alternative methods to 
estimate the dust masses that form in supernova ejecta.

The absorption and scattering of optical or near-IR radiation by 
newly-formed dust within the ejecta of supernovae can result in an 
asymmetry between the red and blueshifted components, with redwards 
emission from the far side of the ejecta undergoing greater absorption. 
\citet{Lucy1989} identified a progressive blueshifting of the [O~{\sc 
i}]~$\lambda\lambda$6300,6363~\AA\ doublet from SN~1987A between days 529 and 739 
after outburst, with the doublet in the later spectrum being blueshifted 
by $\sim 600 $~km~s$^{-1}$. Since then, such red-blue asymmetries have 
been frequently observed in the late-time ($ > 400$ d) spectra of 
supernova ejecta and there is now a growing data base of such observations (e.g.
\citet{Lucy1989,Fabbri2011,Mauerhan2012,Milisavljevic2012}).

SN~1987A as an archetypal object is critical to our growing understanding 
of the formation and evolution of dust in CCSNe.  Since its outburst there 
have been numerous observations at many wavelengths and many epochs. 
Mid-infrared emission from warm dust (T$\sim$400~K) was observed by day 
$\sim$450 \citep{Roche1989, Bouchet1991, Wooden1993} and by day 775 the 
emitting dust mass was estimated to have been between
$\sim5-20\times10^{-4}$~M$_\odot$ \citep{Wooden1993, Ercolano2007, 
Wesson2015}. Beginning from 23 yr after outburst, the {\em Herschel 
Space Observatory} detected much larger quantities 
(0.4--0.8~M$_\odot$ of T$\sim$20~K cold dust emitting at far-IR 
and submillimetre wavelengths \citep{Matsuura2011, Matsuura2015}. This 
emission has been confirmed by ALMA observations to originate from the 
ejecta of SN~1987A \citep{Indebetouw2014}.

We here seek to model the effects of dust on line profiles with a view to 
providing both an alternative way of determining dust masses formed in the 
ejecta of CCSNe and in order to investigate the effects 
of dust on the shapes of line profiles emitted from these objects.  We 
present a new code, {\sc DAMOCLES} (Dust Affected Models Of Characteristic Line 
Emission in Supernovae), that utilizes a Monte Carlo methodology in order 
to model line profiles in expanding atmospheres.  The code can treat dust 
composed of multiple species and grain sizes with variable ejecta density 
and velocity distributions.  Both clumped and smooth geometries may be 
modelled.

\begin{figure}
\includegraphics[trim =20 5 30 0,clip=true,scale=0.47]{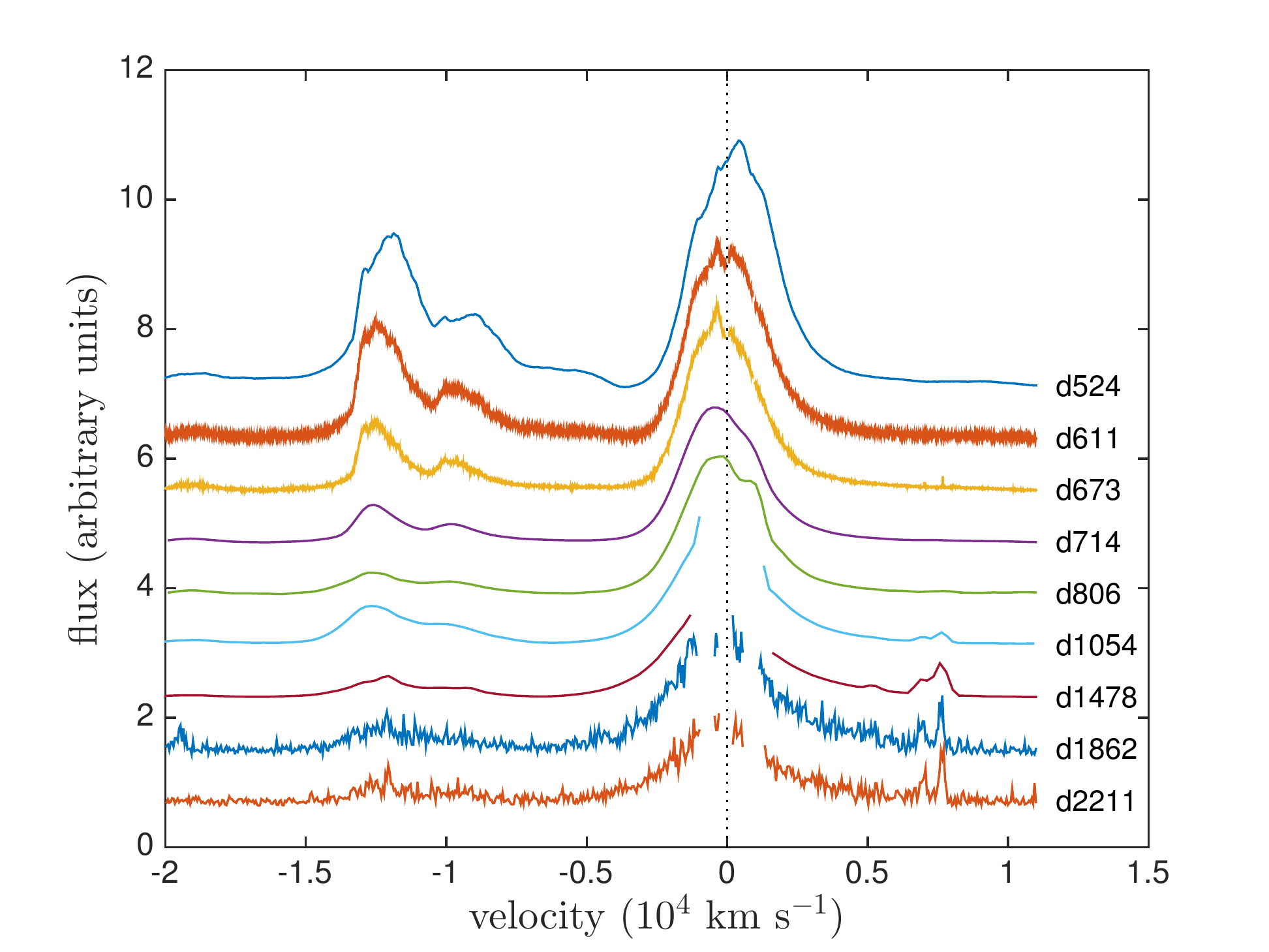}
\caption{Archival data showing the evolution of the H$\alpha$ and
[O~{\sc i}] line profiles from SN~1987A at the earlier of the epochs considered. The 
spectral gaps at the last two epochs correspond to where narrow line 
emission from the equatorial ring has been removed. The spectra have been
continuum-subtracted and offsets have ben applied for display purposes.}
\label{Ha_evol_early}
\end{figure}

In this paper, we collate optical spectra from the archives of four 
different telescopes in order to study the effects of dust formation on 
the H$\alpha$ line and on the [O~{\sc i}]~$\lambda$6300,6363~\AA\ doublet.  
We model epochs spanning a range of approximately 8 yrs from the first 
indications of blueshifting in the H$\alpha$ line between days 600--700, using 
both smooth and clumped geometries.  We compare our derived dust masses to 
those obtained by \citet[hereafter 2015]{Wesson2015} and \citet[hereafter DA15]{Dwek2015} and consider the implied dust formation rate.  We 
present our testing of the new code against analytical cases and 
previously published optically thick models \citep{Lucy1989}. We also 
investigate the sensitivity of line profiles to each of the variables and
 note the range of signatures that observed line profiles may exhibit 
in the presence of dust.

In Section \ref{spectra}, we detail the observed spectra that we used for 
our modelling.  In Section \ref{code}, we discuss the details of the 
{\sc DAMOCLES} code and in Section \ref{params} we present our testing of the 
code and our parameter sensitivity analyses.  Our modelling of the 
H$\alpha$ and [O~{\sc i}]~$\lambda$6300,6363~\AA\ lines is presented in 
Section \ref{results} and we discuss our findings in Section 
\ref{discuss}.


\begin{figure}
\includegraphics[trim =25 0 10 0,clip=true,scale=0.47]{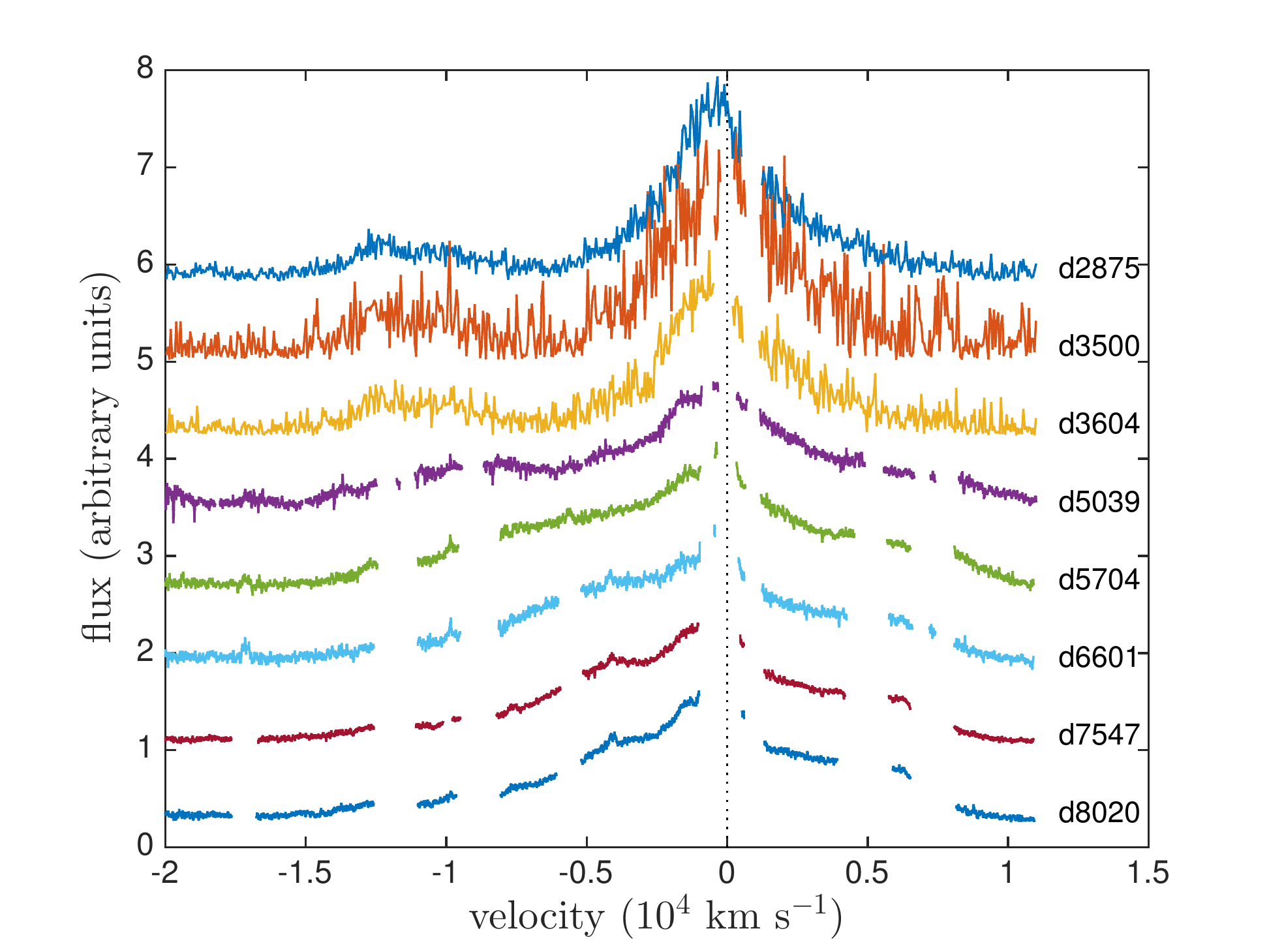}
\caption{Archival data showing the evolution of the H$\alpha$
line profile from SN~1987A at the later epochs. The spectral gaps 
correspond to where narrow line emission from the equatorial ring has been 
removed. The spectra have been continuum-subtracted and offsets applied 
for display purposes.}
\label{Ha_evol_late}
\end{figure}


\begin{table*}
	\begin{minipage}{180mm}
	\caption{Details of the archival data for SN 1987A.}
	\label{tb:data}
  	\begin{tabular}{@{} ccccccccl @{}}
    	\hline
	Date & Age & Telescope  & Inst & $\lambda_{\rm min}$ & $\lambda_{\rm max}$ & Res. & Res. Power & Reference \\
	& (d) & & &(\AA) & (\AA)& (\AA)\\
	\hline
31 Jul 1988 & 524 & AAT & FORS & 5500 & 10190 & 20 & & \citet{Spyromilio1991} \\
26 Oct 1988 & 611 & AAT & UCLES & 6011 & 7336 &  & 30000 & \citet{Hanuschik1993, Spyromilio1993}\\
27 Dec 1988 & 673 & AAT & UCLES & 5702 & 10190 &  & 30000 & \citet{Hanuschik1993, Spyromilio1993}\\
06 Feb 1989 & 714 & CTIO-1.5m & Cass. & 6420 & 10380 & 16 & & \citet{Phillips1990}\\
09 May 1989 & 806 & CTIO-1.5m & Cass. & 6430 & 10330 & 16 & & \citet{Phillips1990}\\
12 Jan 1990 & 1054 & CTIO-4m & RC & 3565 & 10000 & 11 & & \cite{Suntzeff1991} \\
12 Mar 1991 & 1478 & CTIO-4m & RC & 3245 & 9175 & 11 & & \\
30 Mar 1992 & 1862 & {\em HST} & STIS & 4569 & 6818 & 4.4 &  & \citet{Wang1996}\\
14 Mar 1993 & 2211 & {\em HST} & STIS & 4569 & 6818 & 4.4 &  & \citet{Wang1996}\\
07 Jan 1995 & 2875 & {\em HST} & STIS & 4569 & 6818 & 4.4 &  & \citet{Chugai1997}\\
23 Sep 1996 & 3500 & {\em HST} & STIS & 4569 & 6818 & 4.4 &  \\ 
05 Jan 1997 & 3604 & {\em HST} & STIS & 4569 & 6818 & 4.4 &  \\
10 Dec 2000 & 5039 & VLT & UVES & 4760 & 6840 &  & 50000 & \citet{Groeningsson2006, Groeningsson2007}\\
06 Oct 2002 & 5704 & VLT & UVES & 4760 & 6840 &  & 50000 & \citet{Groeningsson2006, Groeningsson2007, Groningsson2008}\\
21 Mar 2005 & 6601 & VLT & UVES & 4760 & 6840 &  & 50000 &\citet{Groeningsson2006, Groeningsson2007}\\
23 Oct 2007 & 7547 & VLT & UVES & 4760 & 6840 &  & 50000 & \citet{Groeningsson2007}\\
07 Feb 2009 & 8020 & VLT & UVES & 4800 & 6800 &  & 50000 & \citet{Tziamtzis2010}\\
    \hline
  \end{tabular}
\end{minipage}
\end{table*}

\section{Archival spectra of SN 1987A}
\label{spectra}

SN~1987A has been the most intensively observed supernova in history, with 
a wealth of both spectral and photometric data available to model.  From 
the archives of a number of different telescopes we have collated optical 
spectra acquired over a wide range of epochs.  At the earlier epochs we 
use spectra obtained by the Anglo-Australian Telescope (AAT) and the Cerro 
Tololo Inter-American Observatory (CTIO) and at later epochs we 
use spectra from the archives of the {\em Hubble Space Telescope} ({\em HST}) and the Very 
Large Telescope (VLT).  An explosion date of 1987 February 23 is adopted 
throughout and epochs are measured relative to this date.  Full details of 
all observations may be found in Table \ref{tb:data}. The spectral 
resolutions of the grating spectrograph observations are listed in 
column~7, while column~8 lists the spectral resolving powers of the 
echelle spectrograph observations.

Wavelength ranges encompassing the H$\alpha$ line and [O~{\sc 
i}]~$\lambda\lambda$6300,6363~\AA\ doublet were selected in order to trace their 
evolution from day 524, near the time of the first indications of dust 
formation \citep{Wooden1993}, to day 8020, near the current era. Optical 
spectroscopy obtained at the AAT using the Faint Object Red Spectrogaph 
(FORS) during the first 2 yr after outburst was kindly supplied by Dr 
Raylee Stathakis \citep{Spyromilio1991, Hanuschik1993, Spyromilio1993} and 
optical spectra from the CTIO were donated by Dr Mark Phillips 
\citep{Suntzeff1991}.

The evolution of the H$\alpha$ and [O~{\sc i}] line profiles is presented 
in Figs \ref{Ha_evol_early} and \ref{Ha_evol_late}.  At later epochs, 
the broad H$\alpha$ profile emitted by the ejecta becomes contaminated by 
narrow line emission from the equatorial ring.  These lines have been 
removed for the purposes of modelling the broad line. A continuum fit has 
been subtracted from each spectrum and a velocity correction has been 
applied for a recession velocity of 287 km~s$^{-1}$ 
\citep{Groningsson2008}.

\subsection{Contamination of the H$\alpha$ profiles}

The H$\alpha$ profile at day 714 exhibits a very slight inflection visible 
at $V \approx +900$ km~s$^{-1}$.  By day 806, this slight inflection has 
developed into a noticeable shoulder in the line profile of H$\alpha$ (see 
Fig. \ref{Ha}).

Although these features are similar in nature to features produced by dust 
absorption in the flat-topped region (as discussed in Section \ref{beta}), 
we conclude that this shoulder is an early appearance of the unresolved 
[N {\sc ii}] $\lambda$6583~\AA\ line from the equatorial ring \citep{Kozma1998b}.  Unresolved nebular [N~{\sc ii}] lines at $\lambda=$ 6583~\AA\ and 
$\lambda=$ 6548~\AA\ either side of the H$\alpha$ rest-frame velocity at 
6563~\AA\ are certainly seen by day 1054 
and have to be removed in order to consider the evolution of the broad 
H$\alpha$ profile (see Fig. \ref{Ha_evol_early}). We do not remove this 
potential contaminant at earlier epochs but try to fit the broad line 
profiles around it.

\begin{table}
\centering
\caption{H$\alpha$ FWHM and the HWZI determined by the zero intensity velocity on the 
blue side of the line.  The tabulated line widths have been corrected for the relevant instrumental resolution.}
\begin{tabular}{c cc}
day & FWHM (\AA) & HWZI (\AA) \\
\hline
524 & 3200 & 3600 \\
611 & 2700 & 3400 \\
673 & 1600 & 3700 \\
714 & 3100 & 4500 \\
806 & 3200 & 5500 \\
1054 & 2100 & 5600 \\
1478 & 1400 & 6600 \\
1862 & 1600 & 6800 \\
2211 & 1400 & 6700 \\
2875 & 2700 & 6700 \\
3500 & 3500 & 7000 \\
3604 & 2100 & 7000

\end{tabular}

\label{FWHM}
\end{table}%

By day 1054, all three of the narrow nebular lines are strong.  They 
remain unresolved in the low spectral resolution CTIO data at days 1054 
and 1478 and therefore contaminate the entire central region of the 
H$\alpha$ line profile.  Their presence renders two CTIO H$\alpha$ 
profiles from days 1054 and 1478 unusable for modelling purposes.  The HST 
and VLT H$\alpha$ profiles at later epochs ($\ge$ 1862 d) have a higher 
spectral resolution and it was therefore easier to remove the narrower 
[N~{\sc ii}] and H$\alpha$ lines from the broad H$\alpha$ profiles (for 
example Figs \ref{Ha_evol_early} and \ref{Ha_evol_late}). Although this 
does remove a potentially informative section of the profile ($+500$ 
km~s$^{-1}<v<+1500$ km~s$^{-1}$), we achieve good fits to the overall line 
profiles at these epochs.

\subsection{The evolution of the maximum and minimum velocities}

For a freely expanding medium, the velocity of any fractional radial 
element should not change with time.  The maximum velocity of any 
line-emitting region is therefore expected to be constant.  However, at 
the epochs we consider here, it appears that the maximum velocities of the 
H$\alpha$ line, as determined by the velocity at zero intensity on the 
blue side, generally increase over time (see Table \ref{FWHM}).  We 
attribute this to the start of the freeze-out phase in the outer regions 
of the ejecta, while the hydrogen neutral fraction is still increasing in 
the denser inner regions \citep{Danziger1991,Fransson1993}.

The onset of a fixed ionization structure in the ejecta causes the rate of 
H$\alpha$ flux decline to slow.  Since the outer, faster moving regions 
reach this state at earlier times than the inner, slower moving regions, 
the relative flux contribution of the outer regions is increased.  At 
early epochs ($t<900$ d) the flux contribution from hydrogen in the 
core dominates the overall H$\alpha$ flux, whereas at later epochs ($t > 
900$ d) the flux from the envelope dominates \citep{Fransson1993, 
Kozma1998a}.  This shift likely explains apparent broadening of the line 
with the higher velocity material becoming increasingly noticeable in the 
line profiles.  This may also explain the increase in half-width zero intensity (HWZI) velocities at 
these epochs with the relative flux from the very densest regions dropping 
more rapidly relative to the outer line-emitting region. The full width at
half-maximum (FWHM) remains relatively steady (see Table 
\ref{FWHM}). However, the FWHM values presented in Table \ref{FWHM} were difficult 
to determine accurately since the peak of the broad line profile is 
contaminated by narrow line emission from the equatorial ring.


\section{The {\sc DAMOCLES} code}
\label{code}

Monte Carlo methods have long been used to model radiative transfer 
problems in diverse environments and there are several examples of codes 
which apply the technique in application to supernovae (for example 
\citet{Maeda2003, Lucy2005c, Jerkstrand2012,Owen2015}).  Whilst there are 
numerous codes that treat dust or gas or both in order to produce an 
overall SED, there is a dearth of codes 
designed to focus on the shapes of individual line profiles.  Although a 
velocity field is naturally considered in codes that seek to reproduce the 
spectra of supernovae, absorption and scattering by dust is not and thus 
the resulting shapes of line profiles are potentially unrepresentative of 
those emerging from dusty ejecta at late times.

In this work we aim to model single or doublet line profiles produced by a 
moving atmosphere in a dusty medium.  Since a comparatively small 
wavelength range is considered, a fully self-consistent radiative transfer 
model is unnecessarily expensive.  Instead any energy packet that is 
absorbed during the simulation may simply be removed on the grounds that 
it would be reemitted outside the wavelength range of interest. This approach is clearly not applicable to SED radiative transfer models that treat continuum emission from dust.  The 
extinction due to dust is assumed to be temperature-independent and it is 
therefore unnecessary to iteratively calculate the temperature of the 
ejecta as in a fully self-consistent calculation of the SED.  Though 
clearly the total energy transferred through the medium is not conserved 
in the wavelength range of interest, the signature of the normalized line 
profile is preserved.

The {\sc DAMOCLES} code builds on the work of \citet{Lucy1989} who employed a 
similar approach to model the broad [O~{\sc i}]~$\lambda$6300,6363~\AA\ 
doublet seen in SN~1987A at early epochs (up to $\sim$ day 775).  It 
models the transport of initially monochromatic energy packets through a 
smooth or clumped dusty medium having a smooth velocity field. The 
velocity field and the inner and outer ejecta radii are free parameters. 
The late-time ($> 400$ d) line emission is assumed to be optically 
thin, with an emissivity distribution proportional to the square of the 
local gas density, i.e. proportional to the product of the recombining 
proton and electron densities in the case of H$\alpha$ or to the product 
of the neutral oxygen and electron densities in the case of collisionally 
excited [O~{\sc i}] emission.

\subsection{The energy packet formalism}
\label{packets}

The initial radiation field is inherently tied to the distribution of gas 
throughout the supernova ejecta which is declared as a power law $\rho(r) 
\propto r^{-\beta}$ between $R_{\rm in}$ and $R_{\rm out}$. $R_{\rm out}$ is 
calculated directly from the epoch of the line to be modelled and the 
declared maximum line velocity.  The emissivity distribution is also 
specified as a power law with $i(\rho) \propto \rho^{k}$.  However this is 
generally taken to be $i(r) \propto r ^{-2\beta}$ since the majority of 
lines modelled are optically thin recombination lines or collisionally 
excited lines and therefore $i(\rho) \propto \rho^2$.  The radiation is 
quantised into monochromatic packets with equal energy $E_{0}=nh\nu_{0}$.  
In Monte Carlo simulations (that model non-moving media) packets are 
usually taken to be of constant energy.  When the frequency of a packet is 
altered after an event, the energy of that packet is kept constant and the 
number of real photons contained within it is assumed to change.  
However, in the case of dust scattering, the number of real photons is 
conserved and thus the energy of the packet is altered.  This is most 
easily achieved by weighting each packet over all scattering events as

\begin{equation}
w=\prod_{\rm scat} \frac{\nu'}{\nu}
\end{equation} 

\noindent where $w$ is the weight of the packet and $\nu$ and $\nu'$ are 
the frequencies of the packet before and after the scattering event 
respectively.  The final energy of each packet is then $E=wE_0$, where 
$E_0$ is the initial energy of the packet.

The emissivity distribution is calculated by dividing the ejecta into a specifiable number of shells between $R_{\rm in}$ and $R_{\rm out}$ overlaid on the Cartesian grid
and the number of packets to be emitted isotropically in each shell 
calculated according to the specified emissivity and density power laws.  For each packet a 
location within that shell and an initial trajectory is randomly sampled 
from an isotropic distribution such that

\begin{align}
\phi&=2\pi\eta \\
 \cos \theta&=2\xi -1
\end{align}

\noindent where $0<\eta<1$ and $0<\xi<1$ are random numbers, $\phi$ is the 
azimuthal angle and $\cos \theta$ is the radial direction cosine.  At 
emission and at each scattering event the frequency of the packet is 
recalculated according to the specified radial velocity field $v(r) 
\propto V_{\rm max}r^{\alpha}$ (see Section \ref{transport}).

\begin{figure*}
\includegraphics[trim =10 25 45 15,clip=true,scale=0.34]{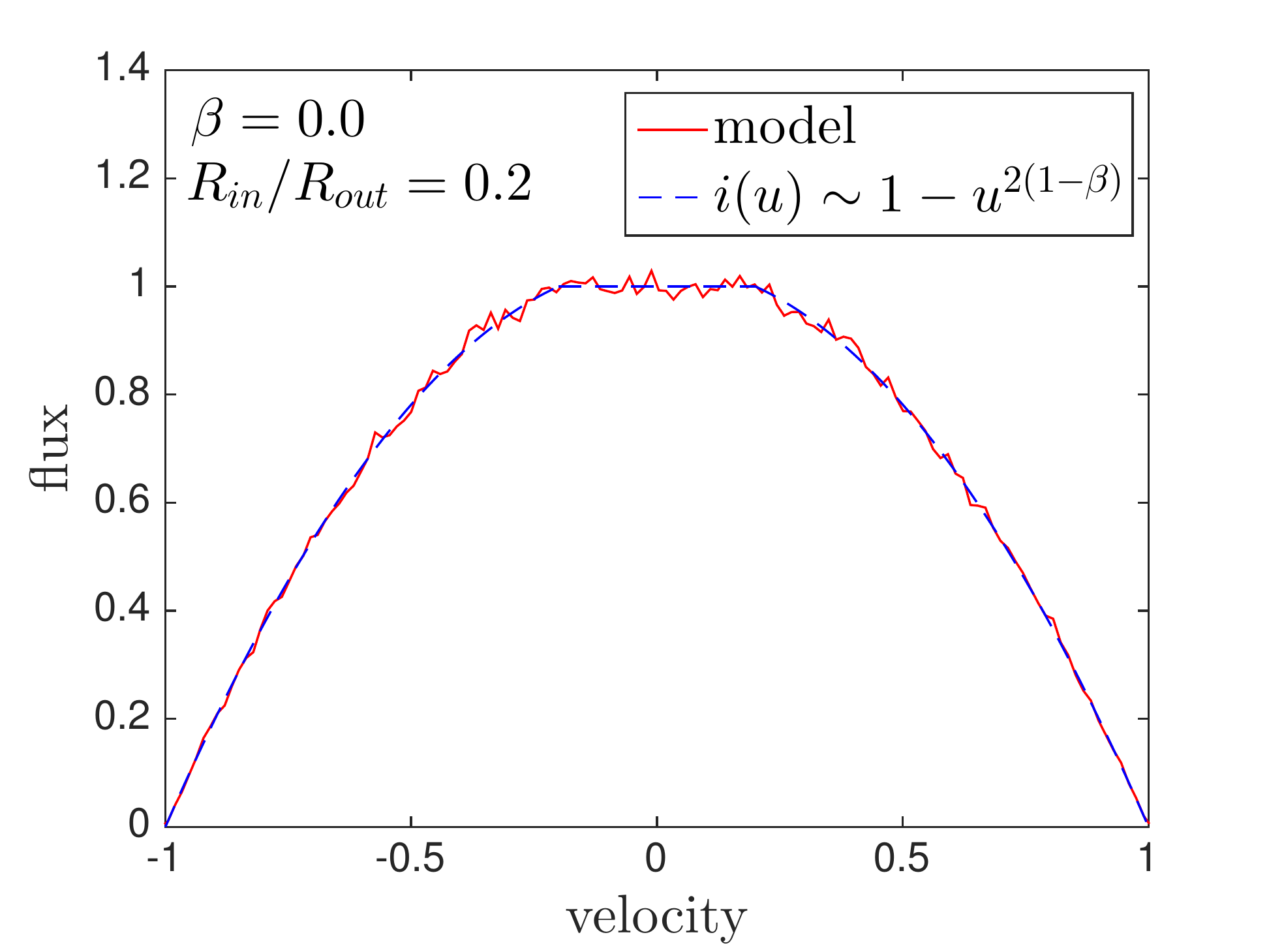} 
\includegraphics[trim =37 25 45 15,clip=true,scale=0.34]{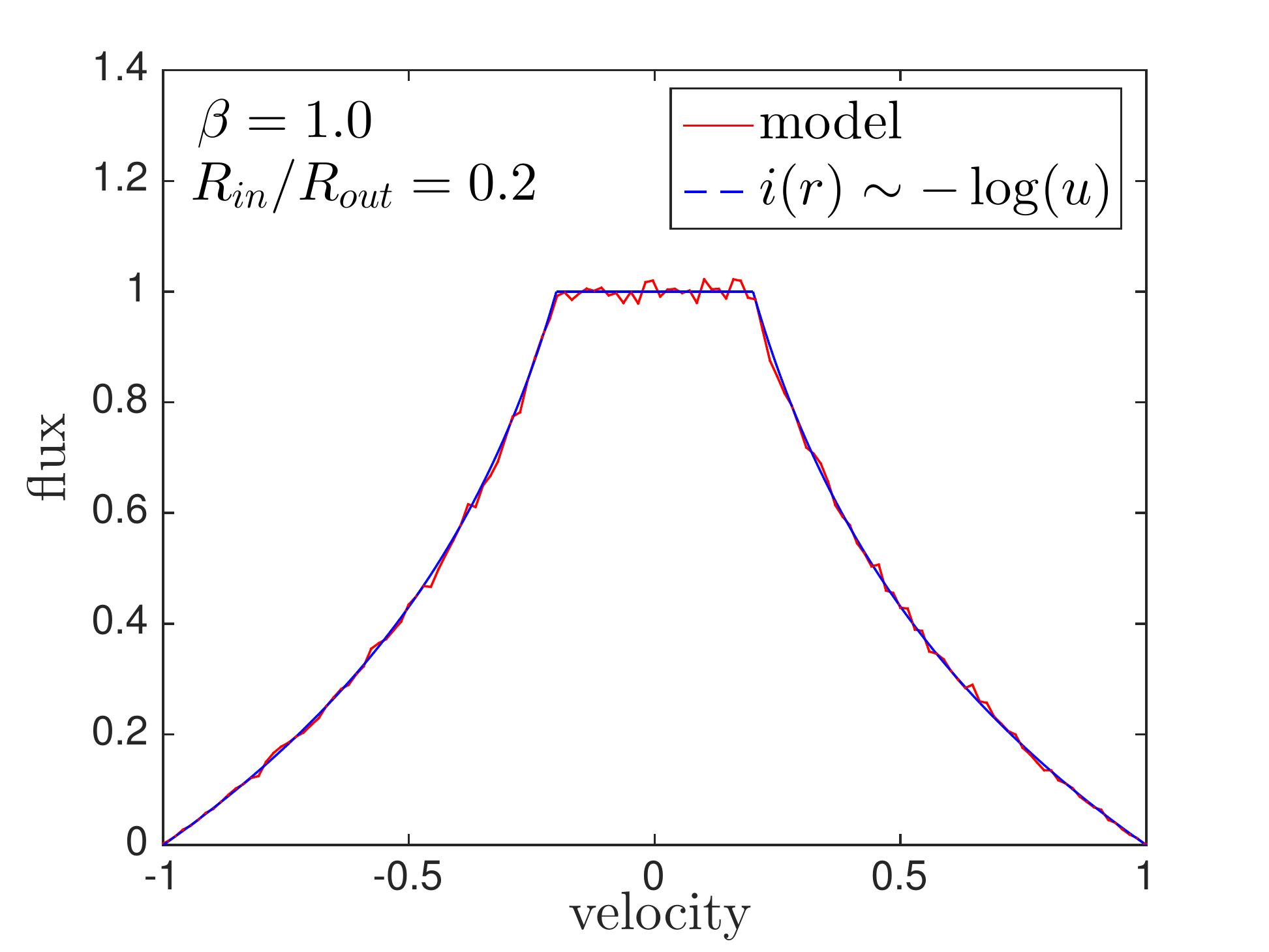}
\includegraphics[trim =37 25 45 15,clip=true,scale=0.34]{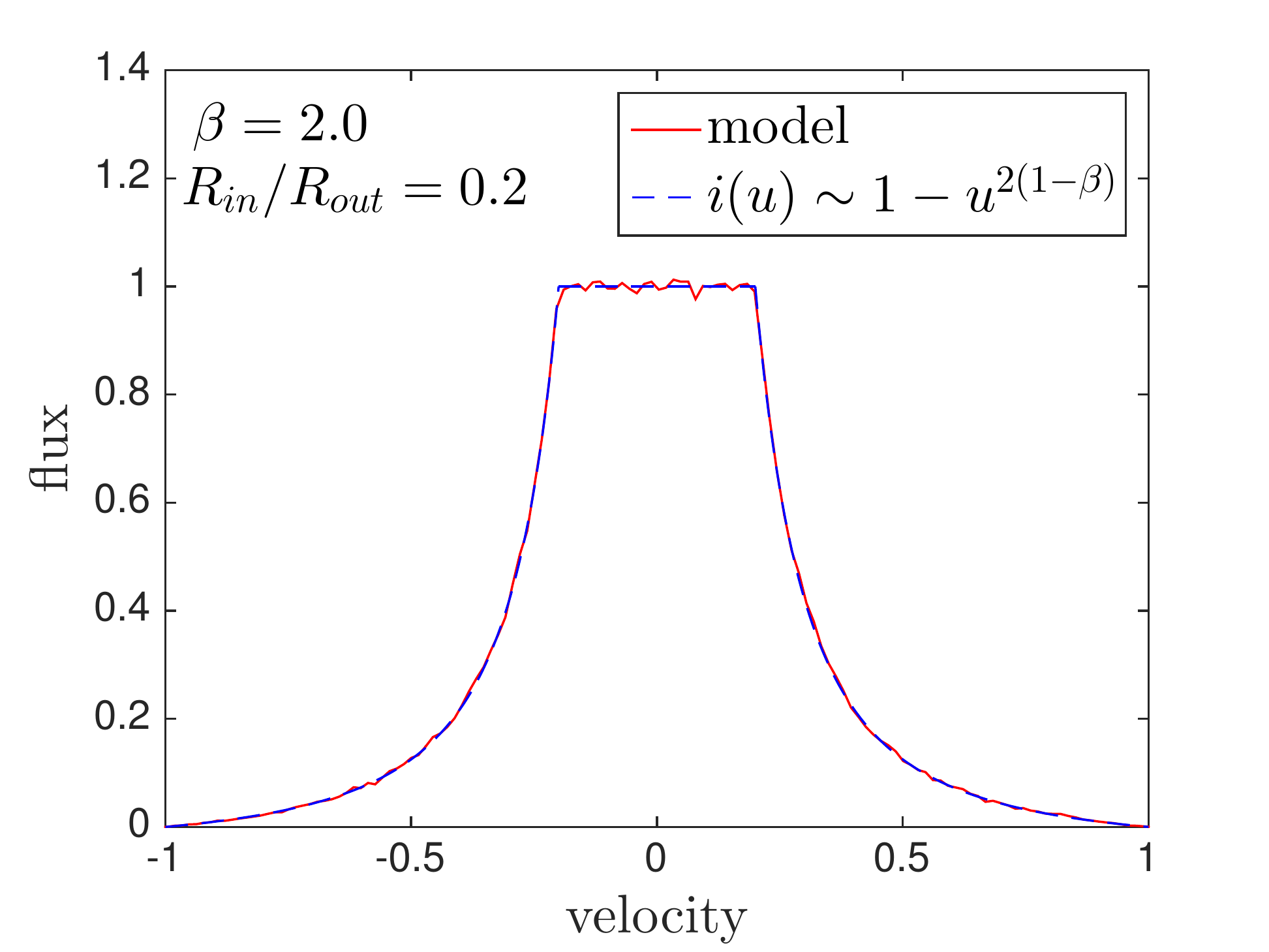}
\\
\includegraphics[trim =10 0 45 15,clip=true,scale=0.34]{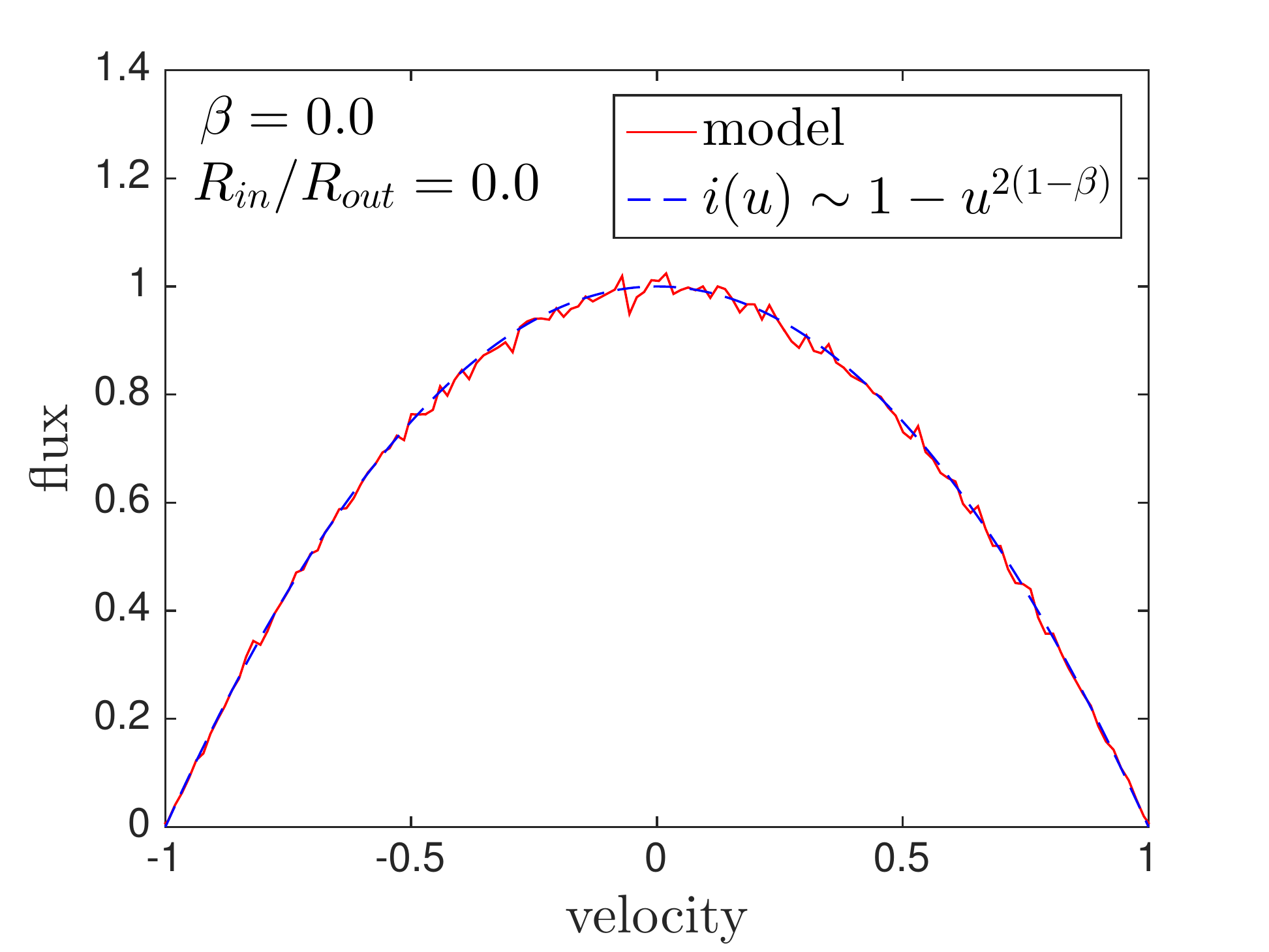}  
\includegraphics[trim =37 0 45 15,clip=true,scale=0.34]{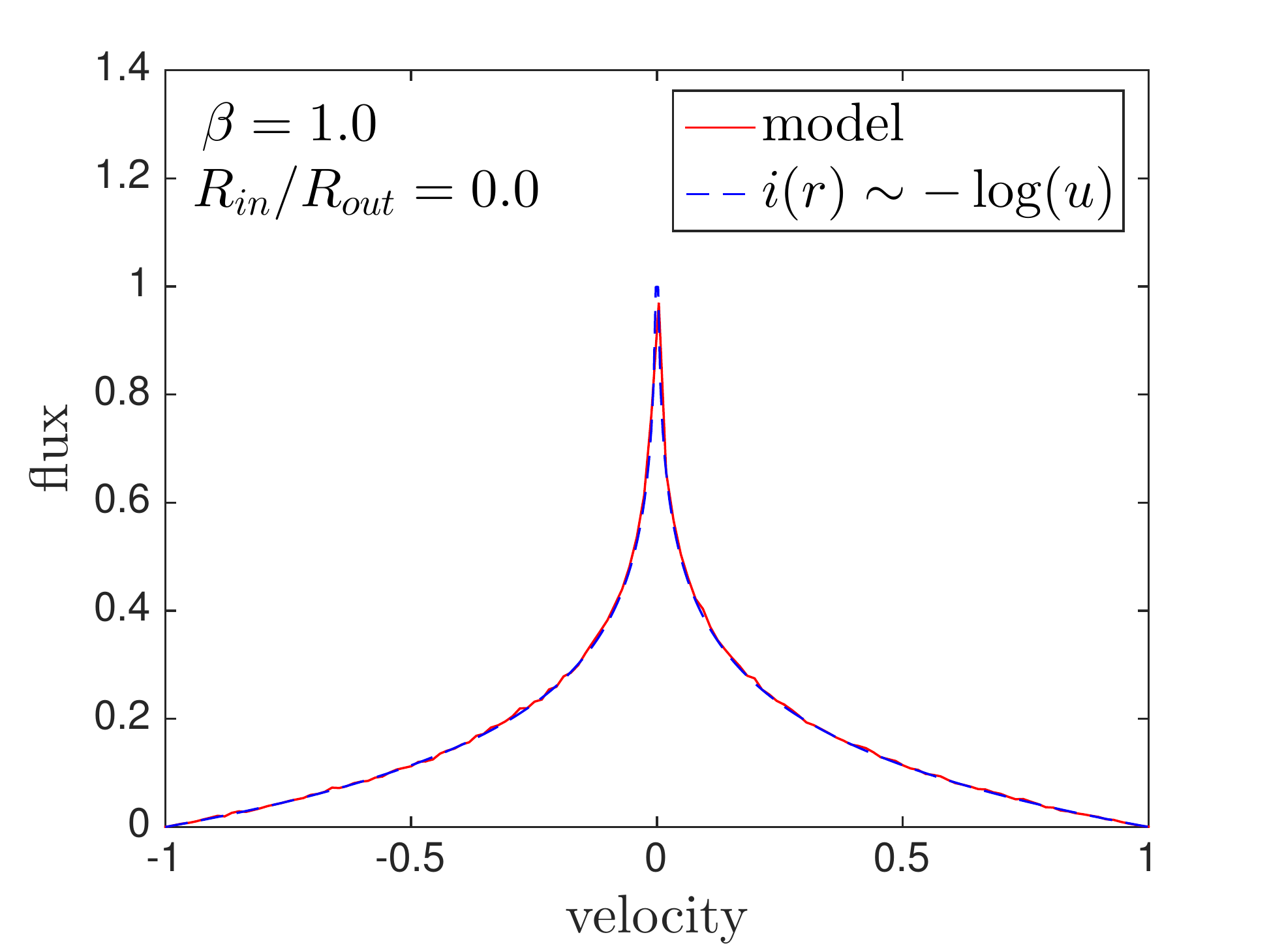} 
\includegraphics[trim =37 0 45 15,clip=true,scale=0.34]{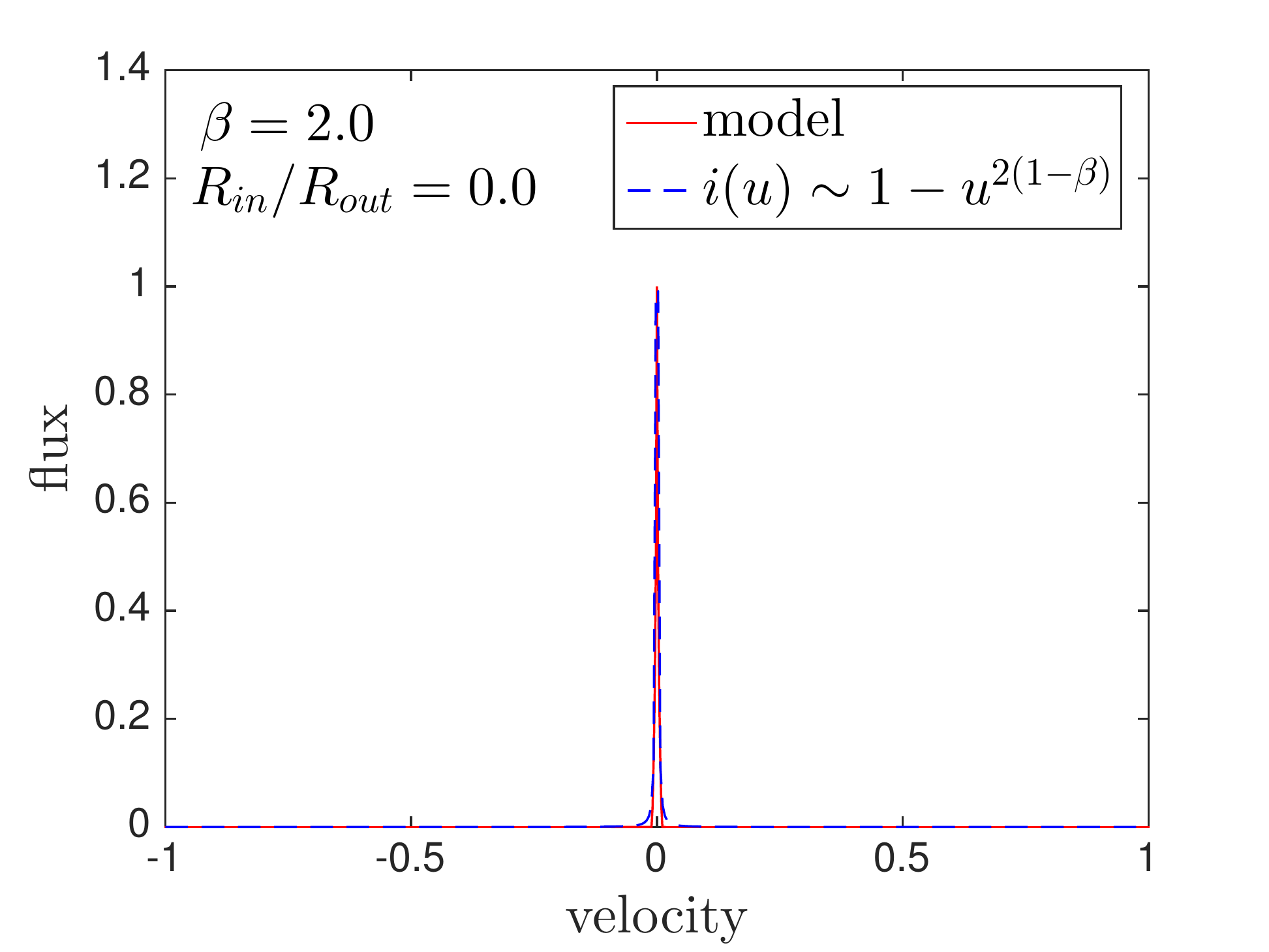}

\caption{Red: benchmark models for optically thin ($\tau =0$) 
line profiles  with fractional velocity $v \propto r$. Left to right: initial emissivity 
profiles $i(r) \propto r^{-2\beta}$ with $\beta=0.0$,1.0 and 
2.0. Cases with $R_{\rm in}/R_{\rm out}=0.2$ are on the top and 
with $R_{\rm in}/R_{\rm out}=0.0$ on the bottom.  The presence of a plateau in the upper plots is due to the finite inner radius (detached shell). Blue: the analytical case 
with $i(u) \sim 1-u^{2(1-\beta)}$ except in the case of $\beta=1$ where 
$i(u) \sim -\log u$. Peak fluxes are scaled to unity.}
\label{fig:analytics}
\end{figure*}

\subsection{The geometry of the ejecta and the grid}
\label{grid}

The supernova ejecta is approximated by a three-dimensional Cartesian 
grid, each cell of which is assumed to have uniform density and 
composition.  The grid is a cube with sides of width $2R_{\rm out}$ and a 
declarable number of divisions.  After the initial emission of energy 
packets, the gas plays no further role in the simulation and thus only 
dust properties are considered.  By default, the dust is coupled to the 
gas (although it may be decoupled) and thus follows the smooth 
distribution described above ($\rho \propto r^{-\beta}$).  The dust 
density in each cell is therefore calculated accordingly and any cell 
whose centre falls outside of the bounds of the supernova ejecta has 
density set to zero.

It is worth noting that if a constant mass-loss rate is required, the 
exponent of the velocity profile and the exponent of the density profile 
are not independent.  A constant mass loss rate implies that $4\pi \rho 
vr^2 \propto k$, where $k$ is a constant, and thus for $v \propto 
r^\alpha$ and $\rho\propto r^{-\beta}$, we require that $\beta-\alpha=2$.  
However, it is possible that the supernova event may have induced a 
mass-flow rate that is not constant with radius and thus both exponents 
may be declared independently.

It is known from SED modelling that clumped environments produce very 
different results to environments assumed to have a smooth distribution of 
dust and gas.  Specifically, clumped models tend to require a higher dust 
mass in order to reproduce a similar level of infrared dust emission 
compared to a smoothly distributed model.  The capacity for modelling a 
clumped dusty medium is therefore included in the code.  The fraction of 
the dust mass that is in clumps is declared ($m_{\rm frac}$) and the total 
volume filling factor of the clumps ($f$) is also specified.  Dust that is 
not located in clumps is distributed according to a smooth radial profile.  
The clumps occupy a single grid cell and their size can therefore be 
varied by altering the number of divisions in the grid.  The clumps are 
distributed stochastically with the probability of a given cell being a 
clump proportional to the smooth density profile (i.e. $p(r) \propto 
r^{-\beta}$).  The density within all clumps is constant and is calculated 
as

\begin{equation}
\rho_{\rm clump}=\frac{M_{\rm clumps}}{V_{\rm clumps}}=\frac{m_{\rm frac}M_{\rm tot}}{\frac{4}{3} f\pi (R_{\rm out}^{3}-R_{\rm in}^{3} )}
\end{equation}

\noindent where $M_{\rm tot}$ is the total dust mass, $M_{\rm clumps}$ is the 
total dust mass in clumps and $V_{\rm clumps}$ is the total volume occupied by 
clumps.  $m_{\rm frac}$ and $f$ are defined as above.

\subsection{The radiative transport mechanism}
\label{transport}

Following emission, a packet must be propagated through the grid until it 
escapes the outer bound of the ejecta at $R_{\rm out}$.  The probability that the 
packet travels a distance $l$ without interacting is $p(l)={\rm e} ^{-n \sigma 
l}={\rm e} ^{-\tau} $ where $n$ is the grain number density, $\sigma$ is the 
cross-section for interaction and $ \tau = n\sigma l$ for constant $n$ and 
$\sigma$ (as in a grid cell).  Noting that the probability that a packet 
will interact within a distance $l$ is $1-{\rm e}^{-\tau}$, we may 
sample from the cumulative probability distribution to give:

\begin{align}
\xi = 1 - {\rm e}^{-\tau} \implies \tau= -\ln (1-\xi)
\end{align}

\noindent where $0<\xi<1$ is a random number sampled from a uniform distribution.
The frequency of the 
photon packet and the mass density of the cell are then used to calculate 
the opacity of that cell. Using the fact that $n\sigma=\kappa\rho$, 
the distance $l$ that the packet travels before its next interaction is 
calculated.  If this value is greater than the distance from its position 
to the edge of the cell then the packet is moved along its current 
trajectory to the cell boundary and the process is repeated.  If the 
distance is less than the distance to the boundary then an event occurs 
and the packet is either scattered or absorbed, with the probability of 
scattering equal to the albedo of the cell

\begin{equation}
	\omega=\frac{\sigma_{\rm sca}}{\sigma_{\rm sca}+\sigma_{\rm abs}}
\end{equation}

If the packet is absorbed then it is simply removed from the simulation as 
discussed above.  If the packet is scattered then a new trajectory is 
sampled from an isotropic distribution in the comoving frame of the dust 
grain and the frequency of the packet is recalculated using Lorentz 
transforms subject to the velocity at the radius of the interaction (see 
Appendix A for further details).  This process is repeated until the 
packet has either escaped the outer boundary of the supernova ejecta or 
has been absorbed. Escaped photon packets are added to frequency bins, weighted by $w$, in 
order to produce an overall emergent line profile.  For our grain radii between 0.35$\mu$m and 3.5$\mu$m, Mie theory predicts grains to be forward scattering ($g=0.7-0.8$).  However, models that incorporate these values with the \citet{Henyey1941} phase function are not significantly different to isotropically scattering grain models.

\subsection{Properties of the Dusty Medium}

Dust of any composition for which optical data are available may be used 
and the relative abundances of the species may be declared by the user.  
A grain size may be specified for each species.  The extinction due to dust is only dependent on 
the cross-sectional area of the grains and not on the overall 
distribution.  It is therefore not possible to determine details of a grain size distribution and only to constrain a single grain size parameter.  The capacity to declare a size distribution is however 
included for the sake of ease of comparison with SED models.  Mie theory 
and optical properties are used to calculate the overall $Q_{\rm abs}(\nu)$ 
and $Q_{\rm sca}(\nu)$ for each species and the derived opacities are summed 
over each species weighted according to their relative abundances.

As will be discussed in Section \ref{params}, the effects of scattering on 
the shapes of line profiles can potentially be quite pronounced and it is 
therefore important to consider the effects of electron scattering as well 
as those of dust scattering.  Electron densities are estimated from the observed luminosity of 
H${\alpha}$ using an average temperature of 10,000~K and assuming that the electron density distribution is coupled to the emissivity distribution. The total optical depth  to electron scattering between $R_{\rm in}$ and $R_{\rm out}$ is then calculated.  Electron scattering is treated in an identical manner to dust 
scattering, with $\tau = \tau_{\rm dust}+\tau_{\rm e}$ in each cell.  If, for a 
given packet, an event occurs, it is first calculated whether this is an 
electron scattering event or a dust event (either scattering or 
absorption) by considering the ratio of the optical depths to each.  If 
the packet is scattered by an electron then the velocity of that electron 
is calculated by considering the bulk velocity at that radius and adding a 
thermal velocity component following the formalism described by 
\citet{Hillier1991}.  The scattering process is then identical to that for 
dust.  If the event is a dust event then the process continues as 
described above.  Including the electron scattering mechanism in models is optional.  

In the majority of cases the electron scattering optical depths are not 
high enough to discernibly affect the overall shape of the profile.  
However, there may be some early epoch cases (the concept is discussed for 
SN 2010jl by \citet{Fransson2014}) where the electron scattering optical 
depths are high enough to have a significant effect on the observed 
profiles.

\begin{figure*}
\includegraphics[trim =20 5 45 15,clip=true,scale=0.5]{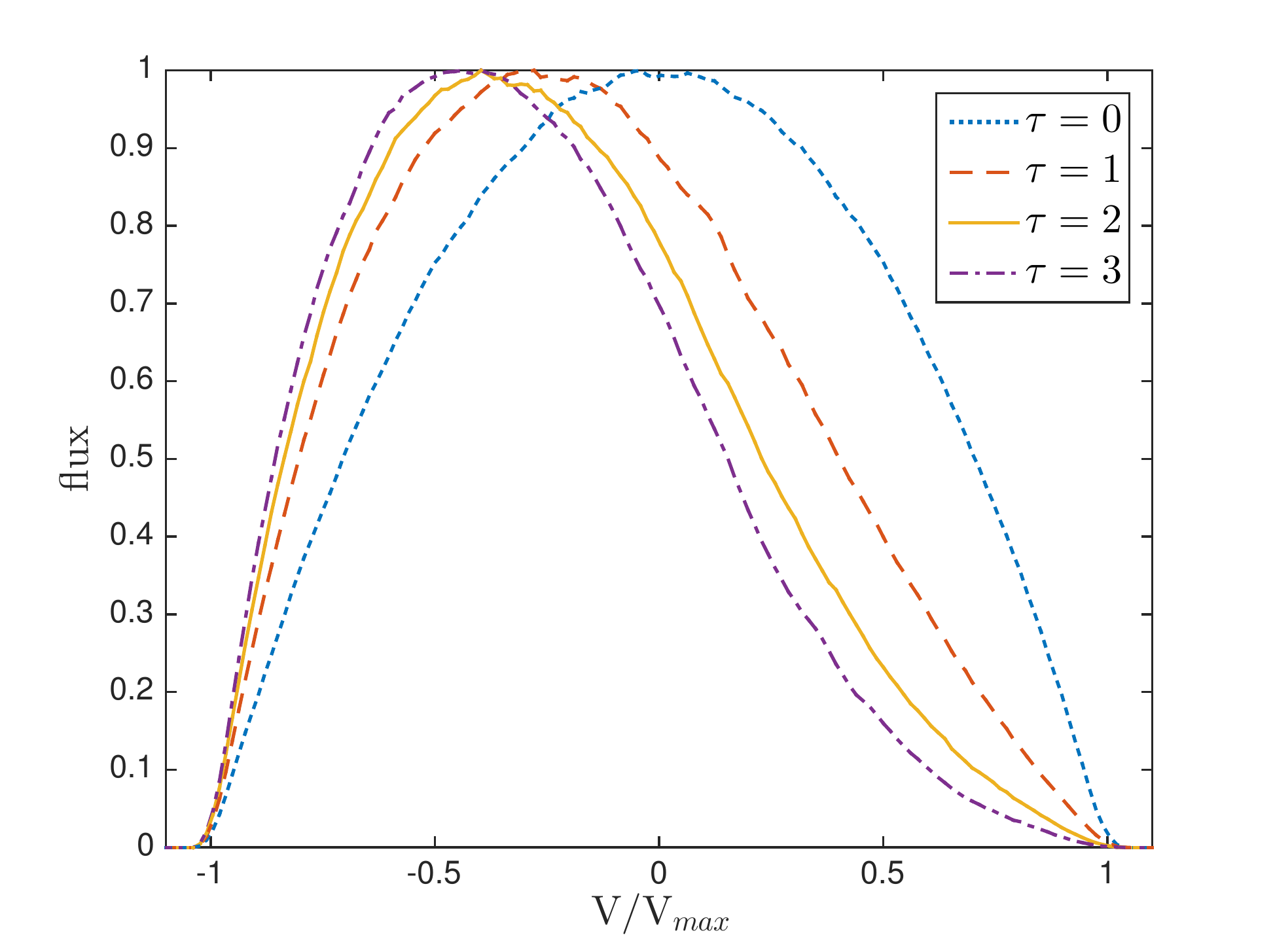} 
\includegraphics[trim =20 5 45 15,clip=true,scale=0.5]{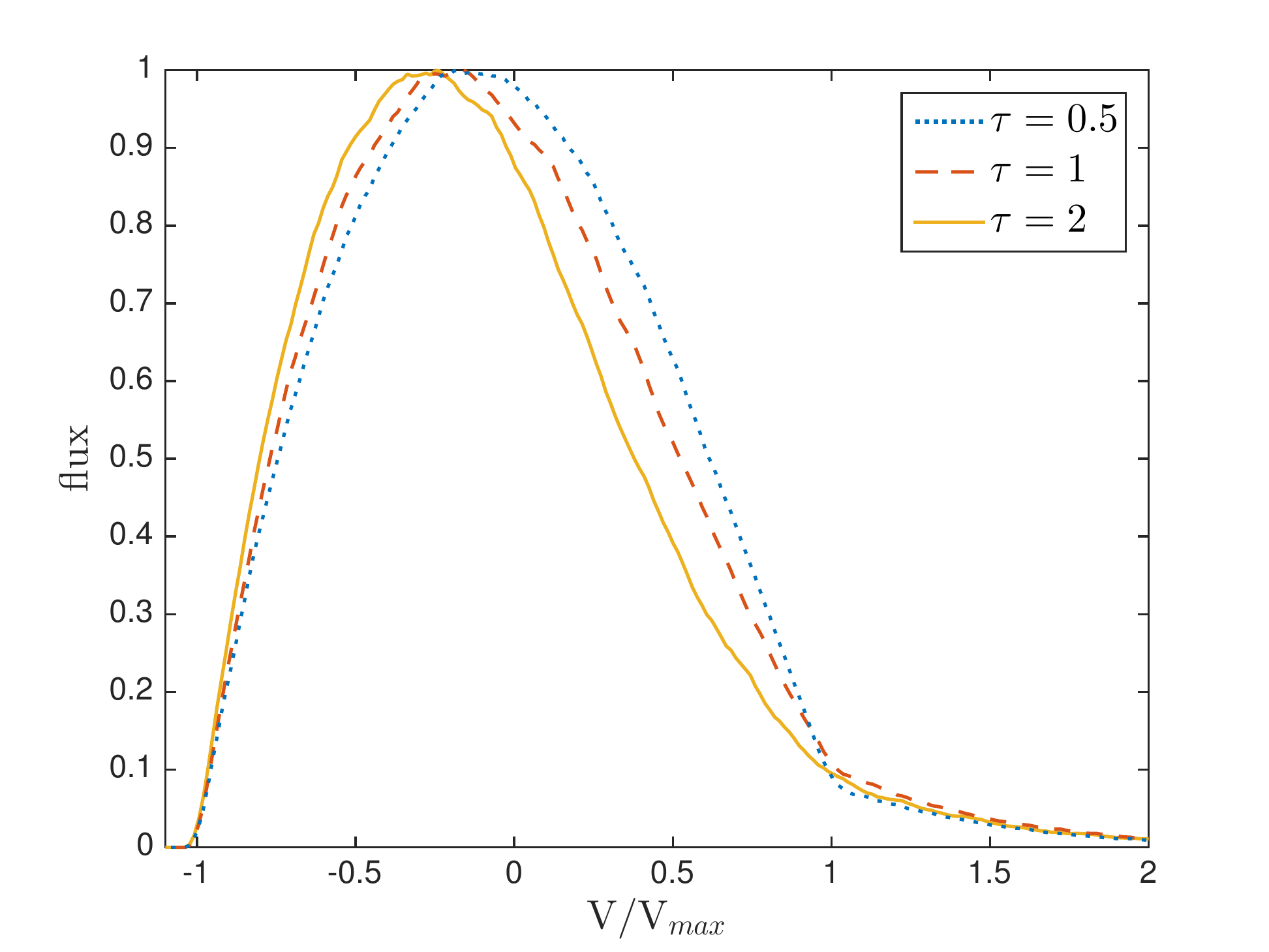}  
\caption{Benchmark models for line profiles  with $v \propto r$, $i(r) \propto$ constant and a filled sphere with $R_{\rm in}/R_{\rm out}=0$.  Pure dust absorption models ($\omega = 0$) are presented in the left-hand plots, whilst partially scattering models are presented on the right ($\omega = 0.6$) as per \citet{Lucy1989} Models II and III. All resulting profiles have been scaled to unit flux at their peaks.}
\label{fig:Lucy}
\end{figure*}

\subsection{Technical details}
{\sc DAMOCLES} is written in {\sc FORTRAN} 95 and parallelized for shared memory machines using {\sc OPENMP}.  It has been developed on and currently runs on a MacBook Pro 11.2 quad core with Intel Core i7 2.8GHz processors and 16GB of memory.  A typical, medium resolution simulation using 125,000 grid cells and 10$^5$ packets takes approximately 15 s to run.  The number of packets transported and the total dust optical depth are the most important factors in determining runtime.  We intend to make {\sc DAMOCLES} available in the public domain after some further development and documentation.

\section{Comparison of {\sc DAMOCLES} models with analytical and previously published results}
\label{params}

There is a general lack of published models in the literature that 
consider dust absorption-affected asymmetric line profiles.  We therefore 
tested the code by comparing the results to optically thin profiles that 
may be derived analytically.  We then tested the absorption and scattering 
components of the code by comparing our results for the case of an 
optically thick medium with those derived by \citet{Lucy1989} in their 
Model II and Model III scenarios.

\subsection{Comparison of {\sc DAMOCLES} models with analytical results}
\label{analytics}

Analytical profiles may be calculated in the dust-free case.  We ran a 
number of models based on the methods of \cite{Gerasimovic1933} 
who derived equations for line profiles emitted from a transparent 
expanding shell.

Describing the fractional expansion velocity of the shell as $v(r) \propto 
r^\alpha$ with $\alpha \neq 0$ such that $v(r)=\frac{V(r)}{V_{\rm max}}$ where 
$V(r)$ and $V_{\rm max}$ represent physical velocities and $v_{\rm max}=1$, the 
energy emitted by the nebula between line of sight velocities $u$ and $u+du$ is 
proportional to

\begin{equation}
\int _\tau i(r) r \sin (\theta) \, r \, {\rm d}\theta \, {\rm d}r
\end{equation}

\noindent where $i(r)$ represents the emission per unit volume at radius 
$r$ and $\theta$ is the angle to the observer's line of sight.  We adopt inner radius $R_{\rm in}=q$  and outer radius $R_{\rm out}=1$ such that $q=R_{\rm in}/R_{\rm out}$.

Setting $i(r) \propto r^{-2\beta}$ (for a recombination or collisionally excited line emitted from 
a medium with an assumed density profile for the emitter $\rho \propto 
r^{-\beta}$) then gives

\begin{equation}
\begin{split}
i(u) \, {\rm d}u &\sim \frac{{\rm d}u}{\alpha u^{\frac{2\beta-3+\alpha}{\alpha}}} \int^{\theta_1}_{\theta_0} \cos^{\frac{2\beta-3}{\alpha}} \theta \sin \theta \, {\rm d}\theta 
\\
&\sim  \frac{{\rm d}u}{u^{\frac{2\beta-3+\alpha}{\alpha}}} \Bigg[\frac{\cos^{\frac{2\beta - 3 + \alpha}{\alpha}} \theta}{2\beta -3 + \alpha}\Bigg]^{\theta_1}_{\theta_0}
\end{split}
\end{equation}

\noindent for $\frac{2\beta-3}{\alpha} \neq -1$ where $i(u) \,{\rm d}u$ is the energy emitted in a volume element and $\theta_0$ and $\theta_1$ are the bounds of this element.  The case 
$\frac{2\beta-3}{\alpha} = -1$ results in a logarithmic relationship.

In the case of a ``filled'' nebula, i.e. one where the inner radius is 
vanishingly small in comparison to the outer radius, we obtain

\begin{equation}
\label{eqn:sides}
	i(u) \, {\rm d}u \sim \pm \frac{{\rm d}u}{(2\beta-3+\alpha) u^{\frac{2\beta-1+\alpha}{\alpha}}} \Big(1-u^{\frac{2\beta-3+\alpha}{\alpha}} \Big)
\end{equation}

If the nebula is not ``filled'', that is to say, the inner radius is some 
fraction of the outer radius and the remnant is a detached shell, the 
above formula becomes valid only from some critical value $u'=q^\alpha$ to  
$u=1$. For $u<u'$, we obtain

\begin{equation}
i(u) \, {\rm d}u \sim \pm \frac{{\rm d}u}{(2\beta-3+\alpha)} \Big( \frac{1}{q^\alpha} - 1 \Big)
\end{equation}

\noindent and therefore the top of the line is flat while the sides are 
sloping.

Crucially, the width of the flat section is determined by $u'=q^\alpha$ or 
simply $u'=q$ in the case where $v \propto r$, whilst the shape of the 
profile outside of the flat top is described by equation \ref{eqn:sides}.

Profiles with a variety of shapes may be derived from these formulae 
depending on the relative values of $\alpha$ and $\beta$.  Here we 
consider three main families of curves:

\begin{enumerate}\parskip3pt

	\item \ \ $\quad i(u)  \sim u^{-\gamma}-1$ \quad ($\alpha>0$, $2\beta-3+\alpha>0$)
	\item \ $\quad i(u)  \sim 1-u^\gamma$ \quad \ \ ($\alpha>0$, $2\beta-3+\alpha<0$)
	\item  $\quad i(u) \sim -\log u$ \quad \ \ ($\alpha>0$, $2\beta-3+\alpha=0$)

\end{enumerate}

\noindent where $\gamma$ is defined as $\gamma= \lvert 
\frac{2\beta-3+\alpha}{\alpha} \rvert$.

Models are presented for each of these cases, both for a filled nebula and 
for a shell structure with $R_{\rm in}/R_{\rm out}=0.2$.  A velocity profile $v 
\propto r$ appropriate for supernova ejecta in the free expansion phase is 
used throughout \citep{Li1992,Xu1992,McCray1996,Baron2005}.  Values of 
$\beta = 0, 1$ and $2$ are adopted.  Fig. \ref{fig:analytics} 
illustrates the excellent agreement between the analytical case and the 
models.  All fluxes are scaled to unity at the peak.

\subsection{Comparison of {\sc DAMOCLES} models with previously published results}
\label{opt_thick_testing}

In addition to the tests for optically thin lines described above, we also 
compared our outputs to those derived by \citet{Lucy1989} in order to 
assess the accuracy of the scattering and absorption aspects of the code.  
We consider two similar cases, equivalent to Models II and III of 
\citet{Lucy1989}. In the first case, dust with zero albedo (pure absorption) is 
uniformly distributed throughout a filled nebula with a velocity profile 
$v \propto r$.  In the second case, the same scenario is considered but in a 
medium of dust with albedo $\omega =0.6$.

\begin{figure*}
\includegraphics[trim =75 40 40 15,clip=true,scale=0.515]{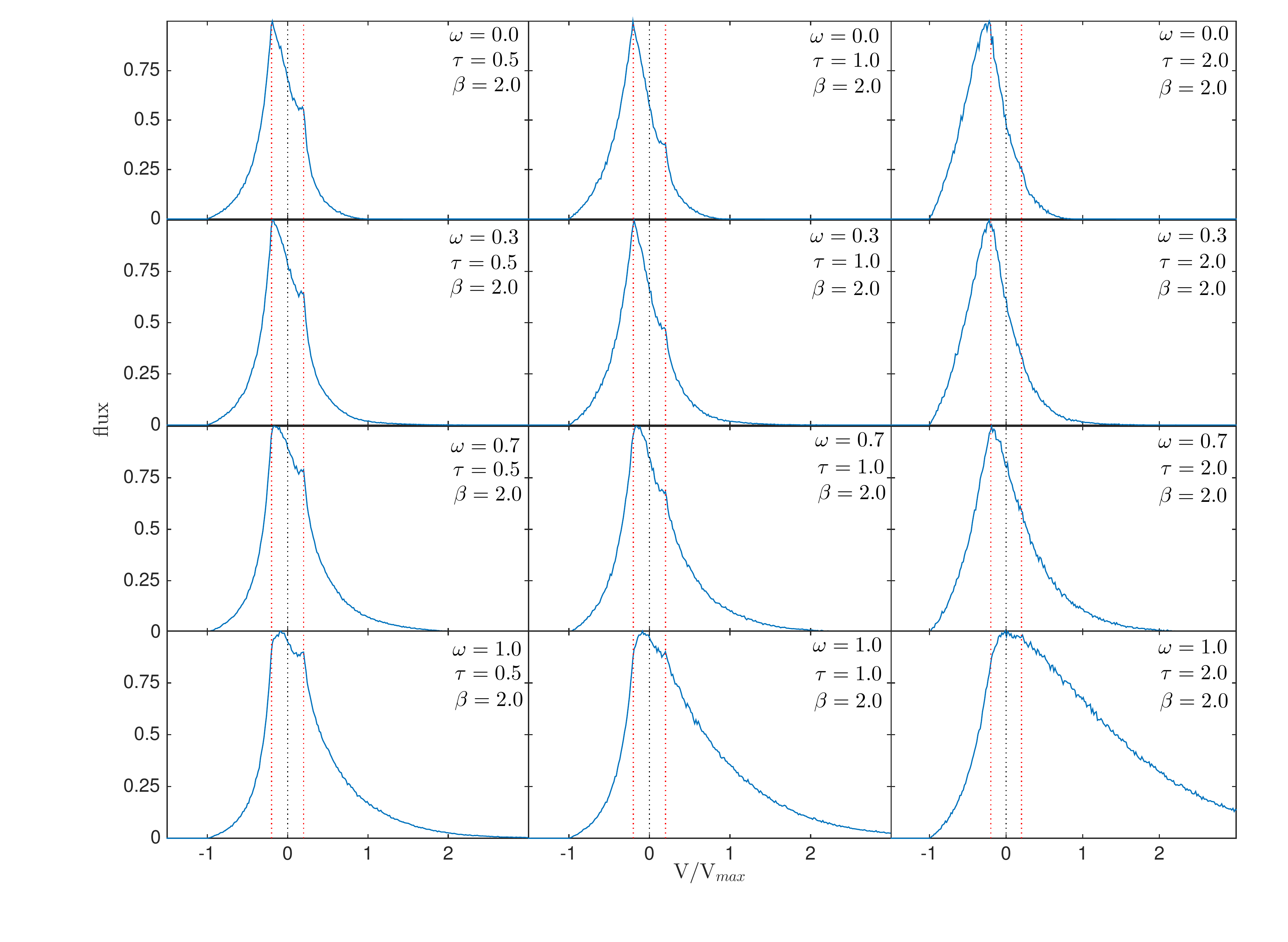} 
\caption{Set of models with $i(r) \propto r^{-4}$ (i.e. $\beta=2.0$), $R_{\rm in}/R_{\rm out}=0.2$, $v(r) \propto r$  
and $v_{\rm max}=1$ illustrating the effects of varying the dust optical depth $\tau$ and albedo $\omega$. 
Peak fluxes are scaled to unity.  The black dotted line marks $v=0$ and the red dotted lines mark $-v_{\rm min}$ and $+v_{\rm min}$.}
\label{wt}
\end{figure*}

In the first case, the profile may once again be derived analytically from 
the basic geometry using the fact that radiation will be attenuated by a 
factor ${\rm e}^{-2\tau_{\nu} v}$ between points with line-of-sight fractional 
velocities $-v$ and $+v$ where $\tau_{\nu}$ is the optical depth at 
frequency $\nu$ from the centre to the outer edge of the ejecta.  The line 
profile is therefore given by

\begin{equation}
\frac{I(v)}{I(-v)} = \exp(-2\tau_{\nu} v)  
\end{equation}

\citet{Lucy1989} presented several examples for both the analytical case of 
the perfect absorber and a Monte Carlo model for grains with $\omega 
=0.6$.  We present the same cases in Fig. \ref{fig:Lucy} and note that 
the resulting profiles exhibit the same features and shape. Of particular 
interest is the scattering wing that appears beyond the maximum velocity 
($v_{\rm max}=1$) on the red side of profiles in the partial 
scatterer case as a result of the packets doing work on the expanding sphere.  
This was noted by \citet{Lucy1989} as a potential diagnostic for the 
presence of dust in the ejecta of a supernova and we will discuss this 
further in Section \ref{ps}.

\subsection{The sensitivity of the variable parameters}
\label{ps}

\begin{figure*}
\includegraphics[trim =80 40 6 15,clip=true,scale=0.515]{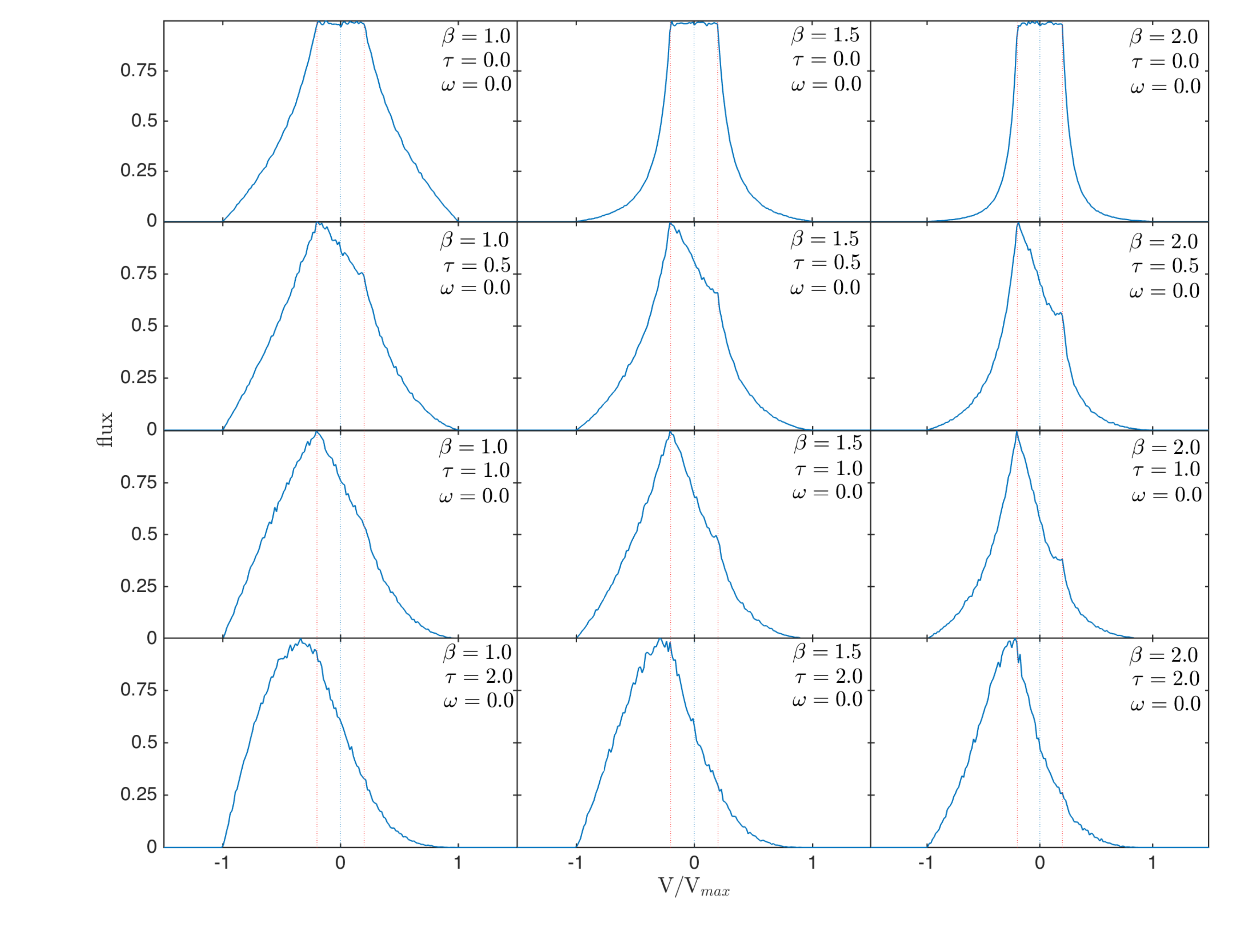} 
\caption{Set of models with $i(r) \propto r^{-2\beta}$ for $\beta=1.0$ \textit{(left)}, $\beta=1.5$ (middle) or $\beta=2.0$ \textit{(right)}, $\omega=0$, 
$R_{\rm in}/R_{\rm out}=0.2$, $v(r) \propto r$ and $v_{\rm max}=1$ illustrating the effects of varying 
the dust optical depth $\tau$.  Peak fluxes are scaled to unity. The black dotted line marks $v=0$ and the red dotted lines mark $-v_{\rm min}$ and $+v_{\rm min}$.}
\label{bt}
\end{figure*}

It is of general interest to establish potential diagnostic signatures in 
the line profiles of supernovae and their remnants in order to trace dust 
formation more effectively.  We here discuss the effects of the main 
parameters of interest, namely:

\begin{itemize}
\item the maximum velocity $V_{\rm max}$
\item the ejecta radius ratio  $R_{\rm in}/R_{\rm out}$
\item the dust optical depth $\tau$
\item the dust albedo $\omega$ 
\item the density profile index $\beta$, where $\rho \propto r^{-\beta}$
\end{itemize}

\subsubsection{The maximum velocity $V_{\rm max}$}

The maximum velocity is defined as the velocity at the outer edges of the 
line emitting region for a given line.  The maximum velocity may vary 
between different spectral lines or doublets due to different locations of 
species having differing ionization thresholds.  Clearly, the larger the 
maximum velocity used the wider the profile becomes.  To some extent 
therefore the steepness of the density profile and the maximum velocity 
can act to counter each other since a steeper density distribution narrows 
the profile (see Section \ref{beta}).  The shape of the wings of the 
profiles, however, generally precludes much degeneracy in this aspect -- the 
overall shape of the line profile can be used to determine the exponent of 
the density distribution to within a relatively small range.

More important is the effect that the maximum velocity has on the overall 
optical depth.  Since the outer radius is calculated directly from the 
maximum velocity (as $R_{\rm out}=V_{\rm max} \times t$ where $V_{\rm max}$ is determined from the blue side of the observed line profile), the overall volume of the ejecta is determined solely by 
this value and the ratio of the inner and outer radii.  The total dust 
optical depth to which the radiation is exposed can therefore be greatly 
affected by even a relatively small change in the maximum velocity for 
fixed values of the other parameters.  Practically, however, the maximum 
velocity can usually be fairly well determined from the observations 
(identified as the point where the flux vanishes on the blue side) and may 
be further constrained through modelling.

\subsubsection{The ejecta radius ratio $R_{\rm in}/R_{\rm out}$}

As already discussed in Section \ref{analytics}, the width of the flat top 
is determined by the ratio of the inner and outer radii, the exponent of 
the velocity profile and the maximum velocity.  We assume that the 
supernova is in free expansion from just a few months after the explosion 
and therefore $r=vt$ such that within the ejecta the velocity profile 
takes the form $v \propto r$ at a fixed time i.e. the supernova expands 
self-similarly \citep{Li1992,Xu1992,Kozma1998b}.  For this case, 
$R_{\rm in}/R_{\rm out}$ is given by

\begin{equation}
\frac{R_{\rm in}}{R_{\rm out}}=\frac{V_{\rm min}}{V_{\rm max}}
\end{equation}

\noindent where it is often possible to constrain $V_{\rm min}$ and $V_{\rm max}$ 
to a relatively narrow range simply from the observed line profile.

The majority of spectral lines emitted from supernovae and supernova 
remnants are expected to have a flat top before dust attenuation effects 
since it is rare for these objects to form a completely filled nebula.  
However, even a very small amount of dust attenuation may result in the 
line profile appearing to be smoothed at its peak.

\begin{figure*}
\centering
\includegraphics[trim=0 0 0 0,clip=true,scale=0.43]{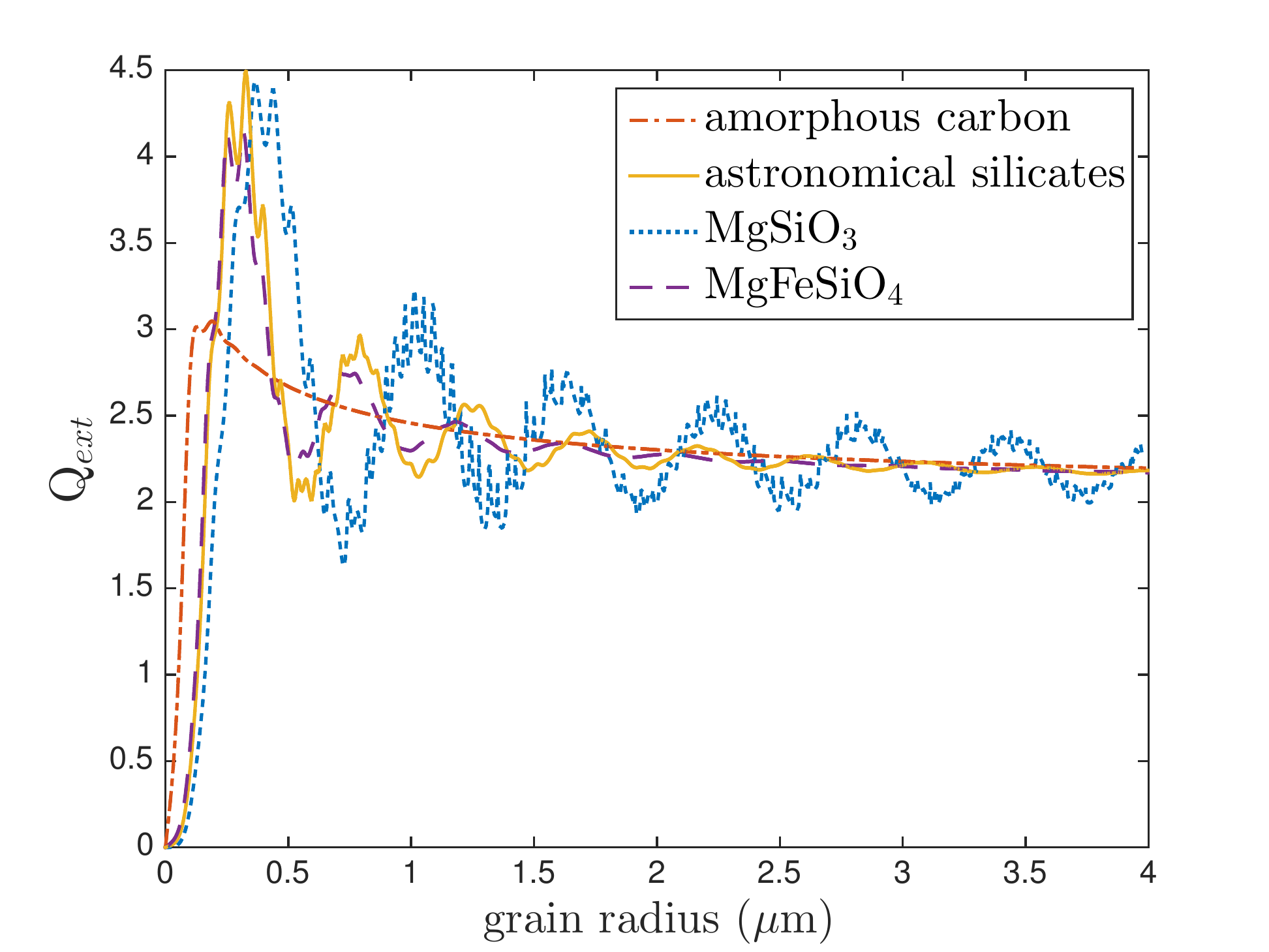}
\includegraphics[trim = 0 0 0 0,clip=true,scale=0.43]{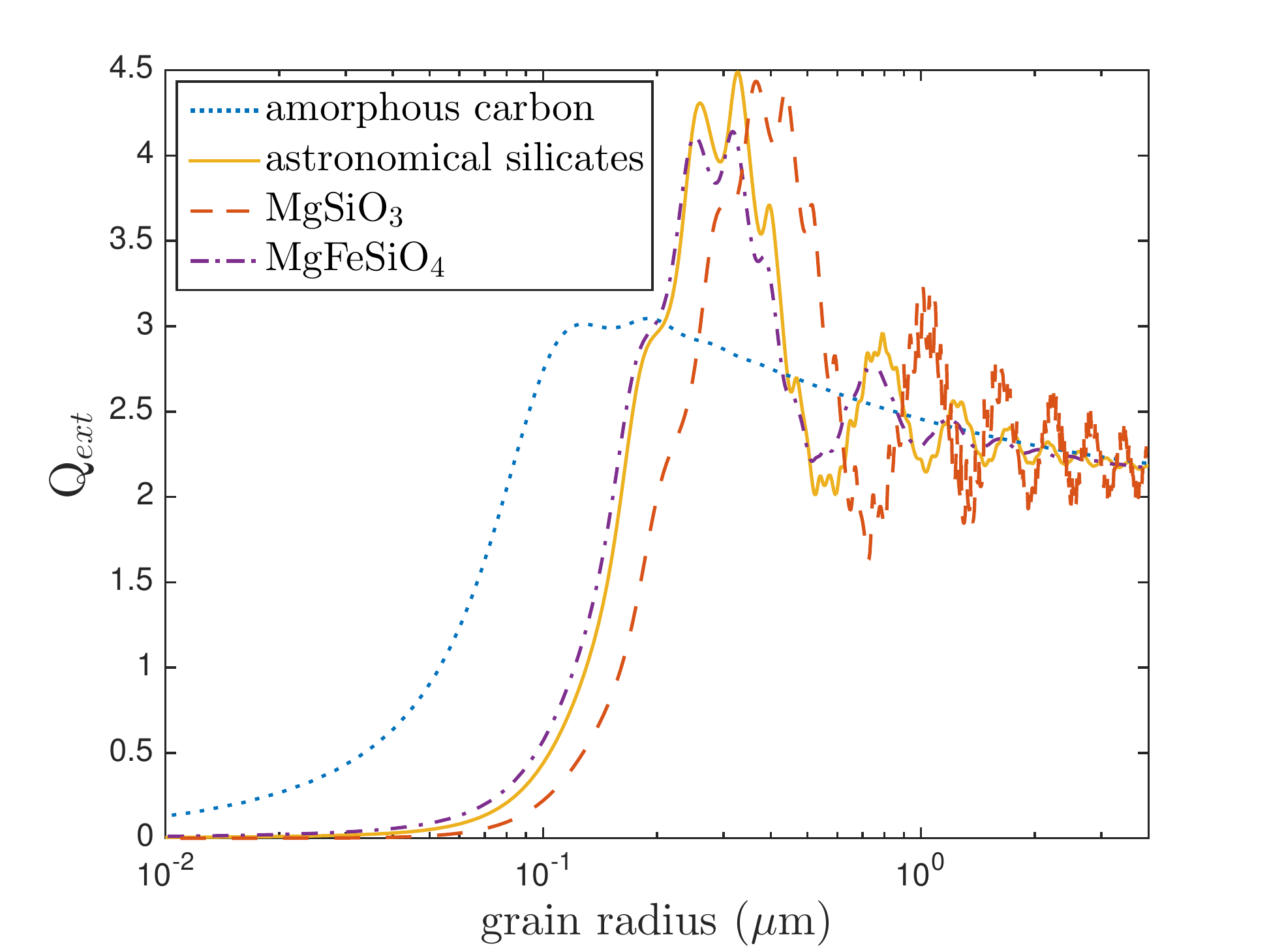}
\\
\hspace{0.001mm}
\includegraphics[trim =0 0 0 0,clip=true,scale=0.43]{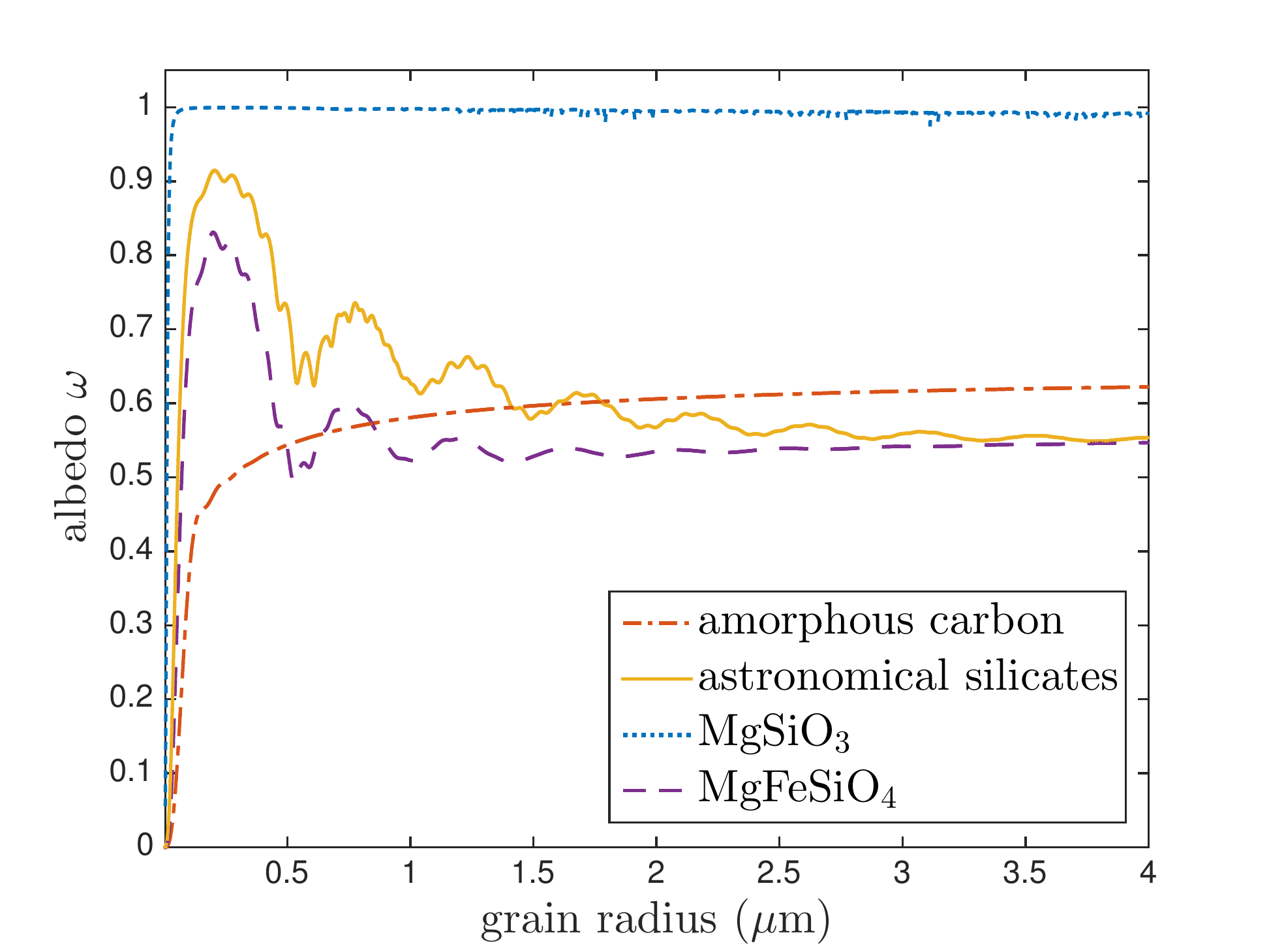}
\includegraphics[trim =0 0 0 0,clip=true,scale=0.43]{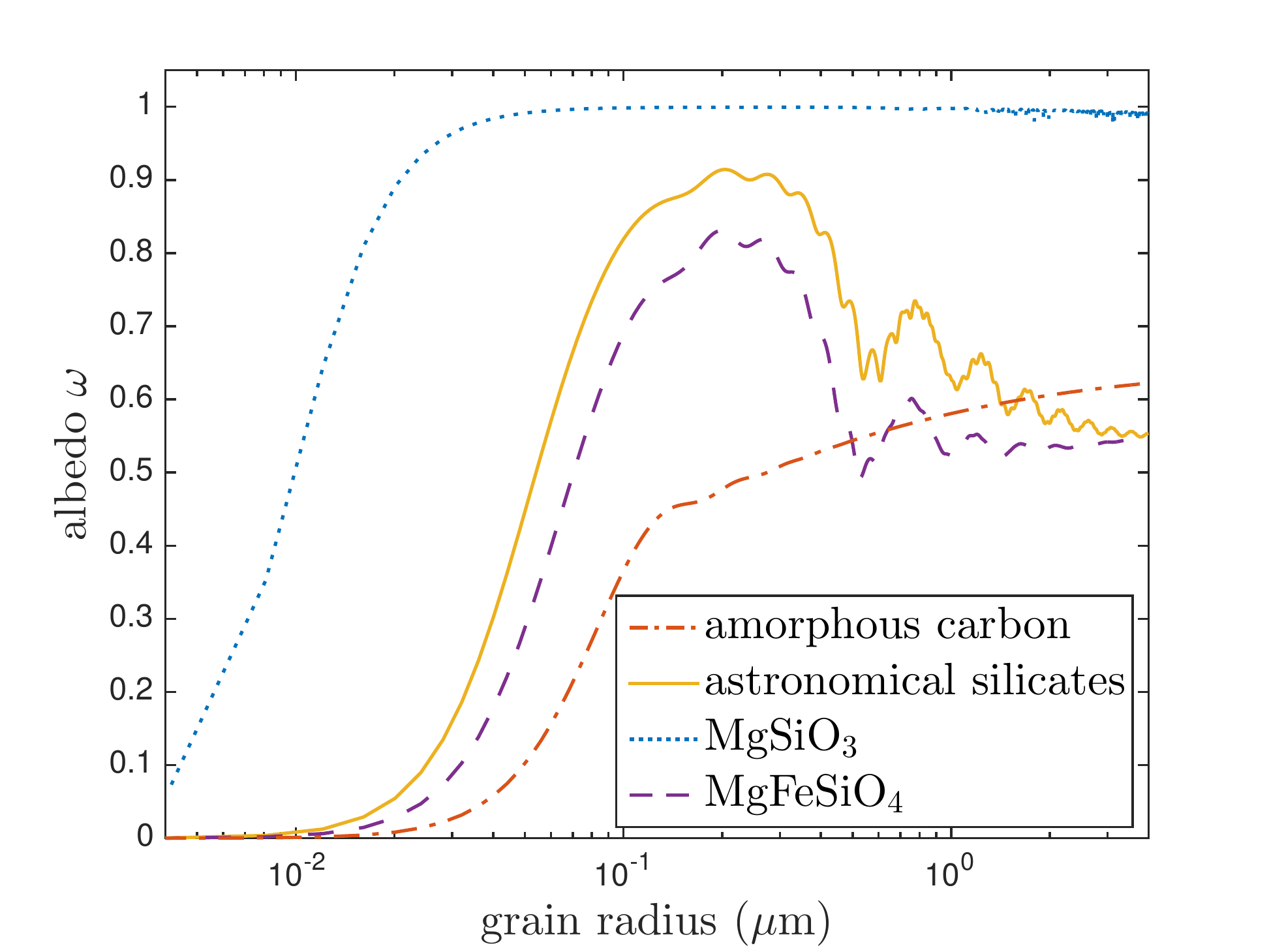}
\caption{The variation of albedo and extinction efficiency ($Q_{\rm ext}$) 
with grain radius at $\lambda$ =  656~nm for \citet{Zubko1996} BE amorphous 
carbon, \citet{Draine1984} astronomical silicate and the
MgSiO$_3$ and MgFeSiO$_4$ samples of \citet{Jager2003} and 
\citet{Dorschner1995}, respectively. A linear grain size scale is 
presented on the left and a log scale on the right.}
\label{albedo_grain}

\end{figure*}

The effects of absorption by dust on a line profile for a filled nebula 
with $R_{\rm in}/R_{\rm out}=0$, as opposed to a detached shell, are shown in 
Fig. \ref{fig:Lucy}. 
All profiles have been scaled to unit flux at their peaks.

\subsubsection{The dust optical depth $\tau$ (detached shell case)}
\label{tau}

As expected, greater attenuation of the original line profile is seen on 
the red side (see Figs \ref{wt} and \ref{bt} ).  The profiles are most 
revealing at lower dust optical depths since the effects of the asymmetric 
absorption can be seen in different sections of the profiles and the 
profiles therefore tend to exhibit more features. The region of the 
profile that is most clearly affected by dust absorption is the 
flat-topped region.  A small amount of absorption in this region results 
in a skewed profile, with a fraction of the flat-topped section removed.  
The peak becomes blueshifted as a result, but only to the original value 
of $-V_{\rm min}$, the minimum velocity corresponding to $R_{\rm in}$. In addition 
to the attenuation in this region, the red wing of the profile is also 
somewhat reduced, and the blue wing somewhat increased relative to their 
original symmetric positions.  The result is a relatively ``jagged'' 
looking profile, often with sharp changes at $\pm V_{\rm min}$.  The profile 
is generally asymmetric, although the degree of absorption in the 
flat-topped region may sometimes make it seem as though the profile is in 
fact symmetric and uniformly blueshifted (see Section \ref{asym} for 
further discussion).  Observationally, these sharp features might become 
smoothed due to insufficient spectral resolution.

At high dust optical depths or when the ratio of the inner and outer radii 
is small, the entire profile is shifted to the blue and the peak moves 
beyond $-V_{\rm min}$ further into the blue.  The profiles also tend to become 
more smooth and featureless.  A set of models showing the effects of 
varying optical depths for different density profiles and dust albedos are 
presented in Figs \ref{wt} and \ref{bt} with $R_{\rm in}/R_{\rm out} = 0.2$.

\subsubsection{The dust albedo $\omega$}
\label{omega}

In the past, there has often been a focus on the effects of absorption by 
dust on the shapes of line profiles and less attention has been paid to 
the potential effects of scattering by dust.  In fact, line profiles can 
be significantly affected by scattering of radiation.  The greater 
attenuation of radiation received from the receding portion of the ejecta 
results in an asymmetry of the line profile whereby the majority of the 
observed emission is located bluewards of the peak.  However, the effects 
of repeated dust scattering events within the ejecta can substantially 
alter the shape of a line profile and potentially can act to counter the 
blueshifted asymmetry.

Not only does repeated scattering of photons increase the number of 
potential opportunities for a given photon to be absorbed but it also 
results in continuous shifting of the frequency of the photon to the red.  
The photon must do work on the expanding shell of dust in order to escape 
and thus many of the photons are reprocessed beyond the theoretical 
maximum velocity on the red side of the profile.  Even in the case of dust 
grains with a relatively low albedo, a surprisingly persistent wing on the 
red side of the profile is seen, generally beyond the maximum theoretical 
velocity of the emitting region. In the case of strong dust scattering and 
high dust optical depths, this can actively result in a shift in the 
overall asymmetry of the profile, with the majority of the emission being 
emitted redwards of the peak. The peak however, remains blueshifted (for 
example the bottom left panel of Fig. \ref{wt}) or central (for example 
the bottom right panel of Fig. \ref{wt}).  For the line profile to 
exhibit this effect requires the dust to be a nearly perfect scatterer;
of the albedos plotted in Fig. \ref{albedo_grain}
only the nearly transparent MgSiO$_3$ sample of \citet{Jager2003}
exhibits such a behaviour.

The combination of relatively low dust optical depths, initially 
flat-topped profiles, greater attenuation on the blue side along with 
increased flux on the red side due to scattering can result in a profile 
that sometimes ends up appearing almost symmetrical, particularly if 
contaminants, such as narrow lines or blending with other broad lines, are 
present or if the resolution of the data is low.  The potential for 
apparently symmetrical profiles that appear to have been uniformly 
blueshifted should be noted (see Figs \ref{wt} and \ref{bt} for 
examples of this).

\subsubsection{Density profile $\rho \propto r^{-\beta}$}
\label{beta}

Whilst the density profile of the dust may have some effect on the 
resulting profiles, it is the initial emissivity profile (dependent on the 
gas density profile) that has the greatest effect on the shape of the line 
profile.  In general, the steeper the emissivity distribution, the 
narrower the line profile becomes.  The sides of the line profile may 
become almost vertical for a very steep distribution since the majority of 
the emission then comes from a very narrow velocity range (see Figure 
\ref{fig:analytics}).

The dependence of the shape of the line profile on the emissivity 
distribution is described analytically in Section \ref{analytics} for the 
case of very optically thin dust.  However, for even fairly low dust 
optical depths, the density profile plays a significant role in 
determining the shape of the line profile where it is affected by dust 
absorption.  As previously discussed, at relatively small optical depths 
for reasonable $R_{\rm in}/R_{\rm out}$, a section of the flat-topped region is 
removed resulting in a peak at $-V_{\rm min}$.  The shape of the profile in 
this region is significantly affected by the density profile.  Shallow 
density profiles (low $\beta$) produce a virtually linear variation in 
flux between $-V_{\rm min}$ and $+V_{\rm min}$ (for example the profiles in the 
left-hand column of Fig. \ref{bt}).  For a fixed dust optical depth, the 
steeper the distribution becomes, the more concave the profile becomes 
between $-V_{\rm min}$ and $+V_{\rm min}$, ultimately resulting in a clear 
shoulder to the profile at $+V_{\rm min}$ (for example the profiles in the 
right-hand column of Fig. \ref{bt}).  For extremely steep density 
distributions this can result in a double peaked profile with a trough to 
the red of $V=0$.  An illustration of the effects on the line profiles of 
varying $\beta$ and $\tau$ is shown in Fig. \ref{bt}.  As previously 
noted, these features may not be apparent in observed line profiles with 
poor spectral resolution.

\subsection{Inferring properties of the dust from the models}

The presence of an extended red wing at large positive velocities in 
combination with increased extinction on the red side at smaller positive 
velocities can allow the values of $\tau$ and $\omega$ to be well 
constrained.  Where this occurs it is possible to translate these values into a 
dust mass and grain size for a given species or combination of 
species using grain optical properties and Mie theory (see Figure 
\ref{albedo_grain}).

For amorphous carbon, the albedo generally increases with grain size.  
The presence and extent of any scattering wing on the red side of the 
observed profile can therefore help to place limits on the grain radius.  
However, the greater the grain radius used the smaller the available 
cross-section for interaction per unit dust mass.  Larger masses of dust 
are therefore required to fit the same degree of absorption if a larger 
grain size is used.  This is in contrast to SED radiative transfer 
modelling where larger grain sizes generally result in less dust being 
required to fit the IR portion of the SED (W15).  These two techniques in 
tandem may therefore provide limits on grain sizes for different species 
or combinations thereof.

It is known that the use of different optical properties may substantially 
alter dust masses derived using SED fitting for a given species of 
specific grain size (e.g. \citet{Owen2015}).  However, the use of 
different sets of grain optical constants in our models seems to have only 
a minor effect on the required dust masses, except for cases where the 
albedo is close to unity (pure scattering grains).

\begin{figure*}
\includegraphics[clip=true,scale=0.34,trim = 0 0 48 10]{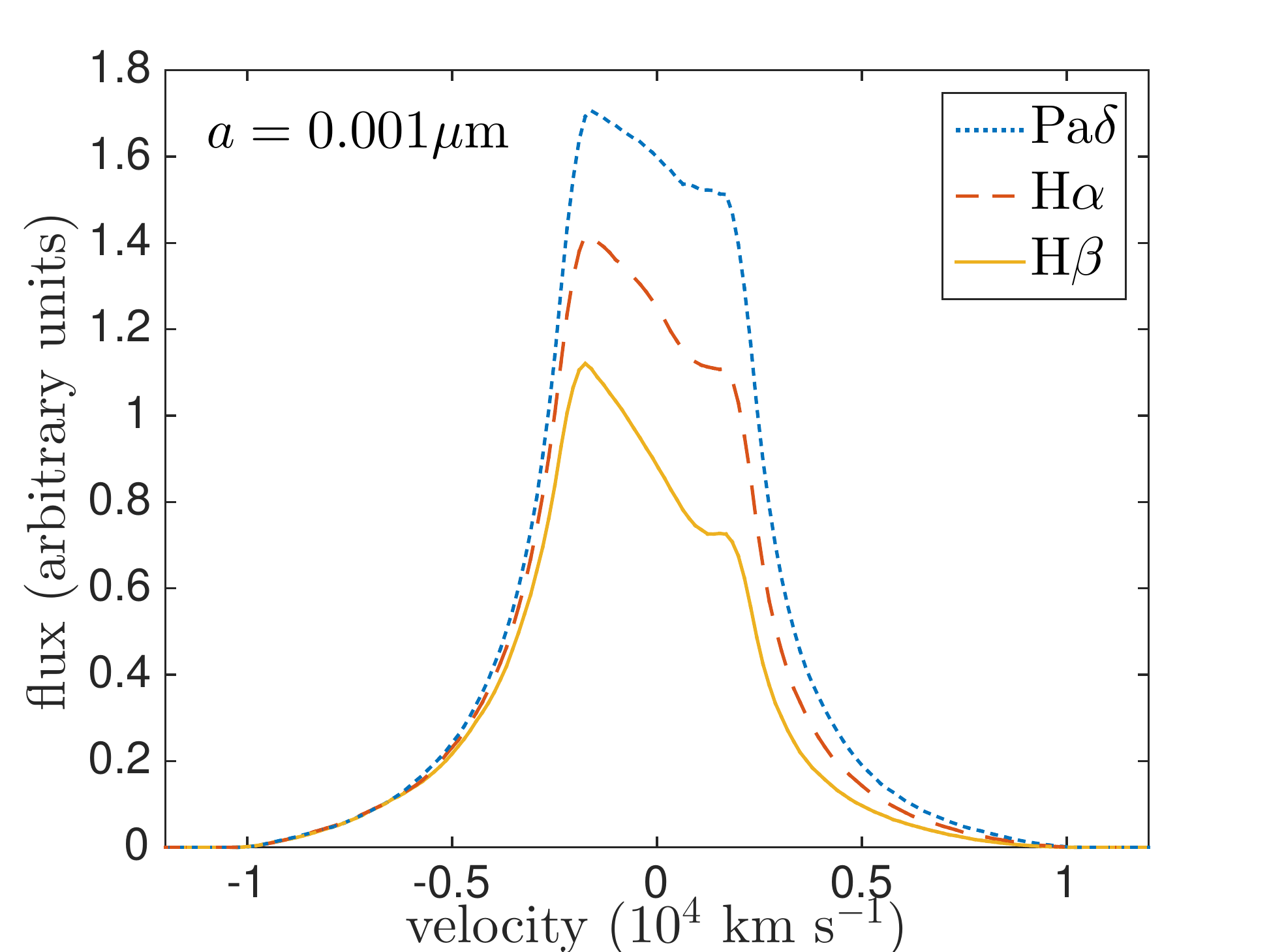}
\includegraphics[trim=38 0 48 10,clip=true,scale=0.34]{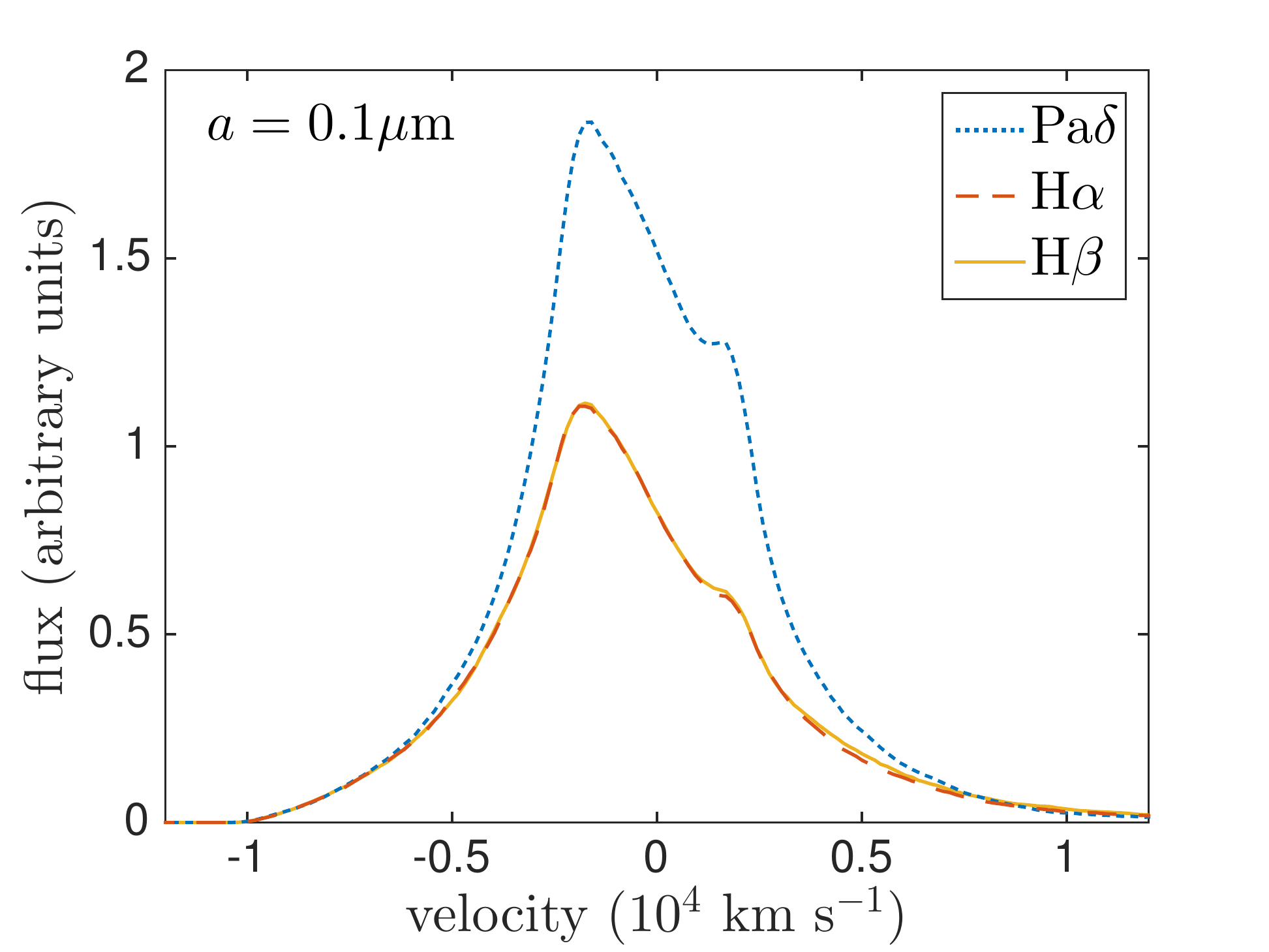}
\includegraphics[trim=38 0 48 10,clip=true,scale=0.34]{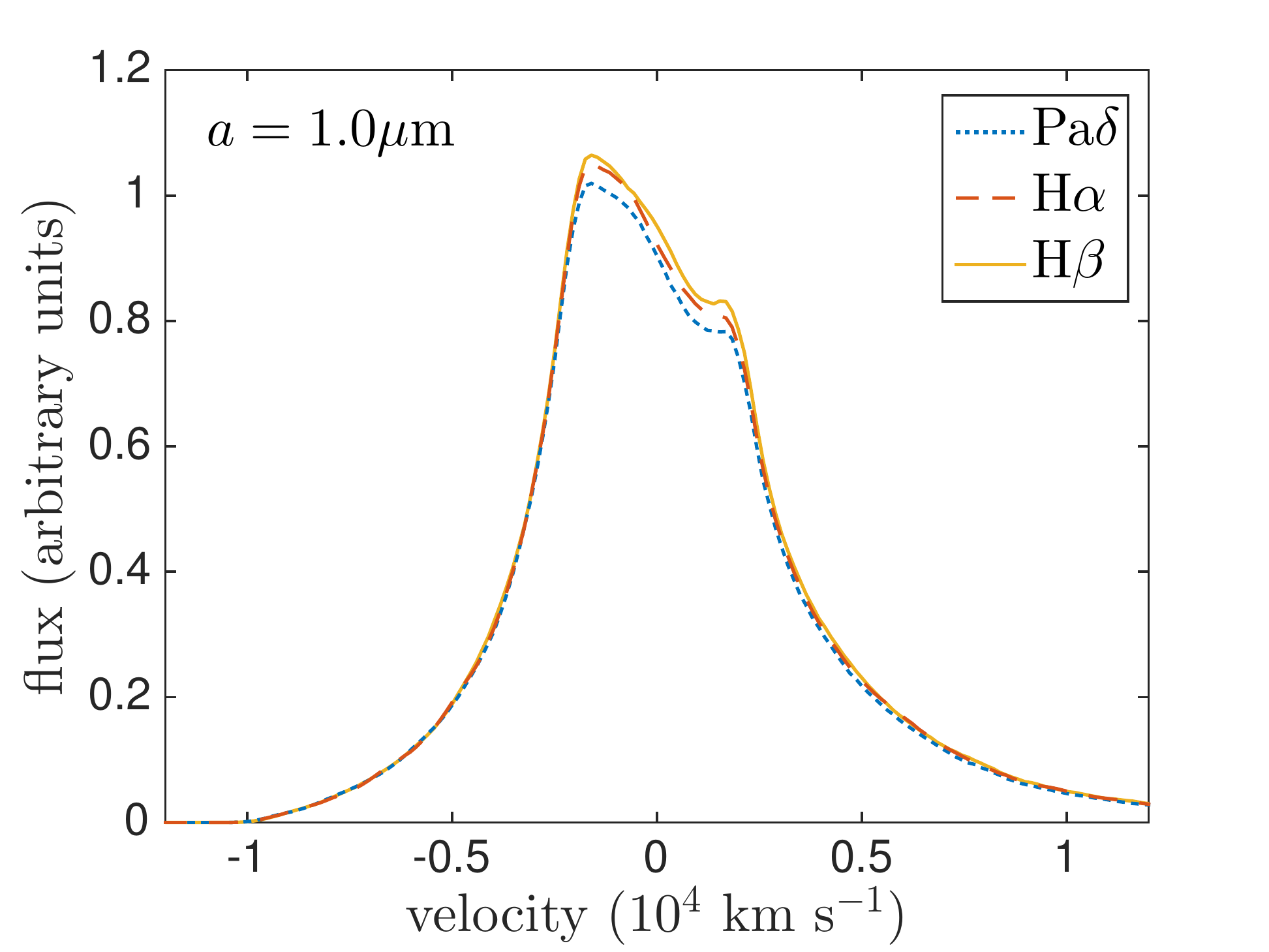}

\vspace{1mm}\includegraphics[trim=0 0 48 10,clip=true,scale=0.34]{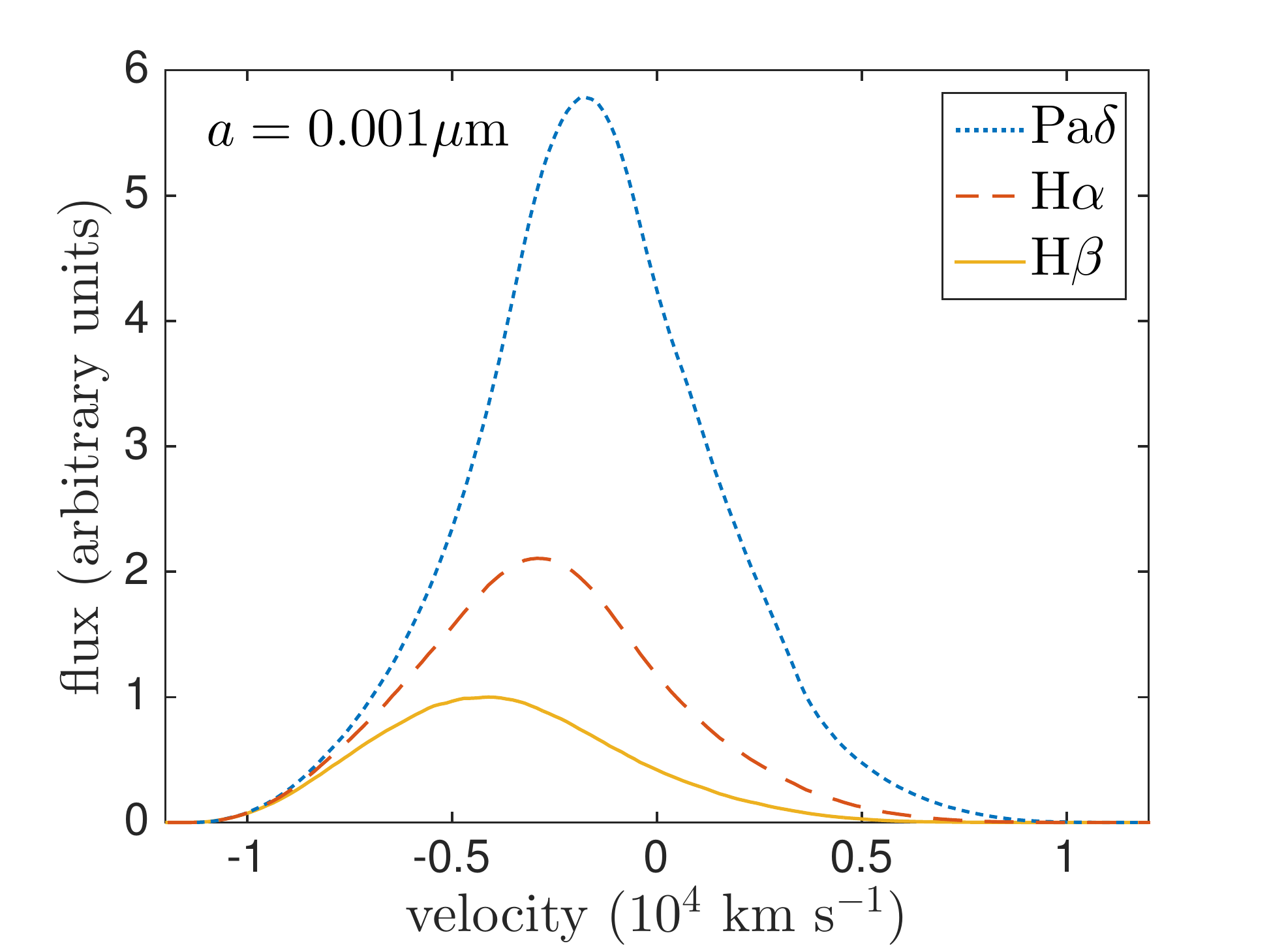}
\includegraphics[trim=38 0 48 10,clip=true,scale=0.34]{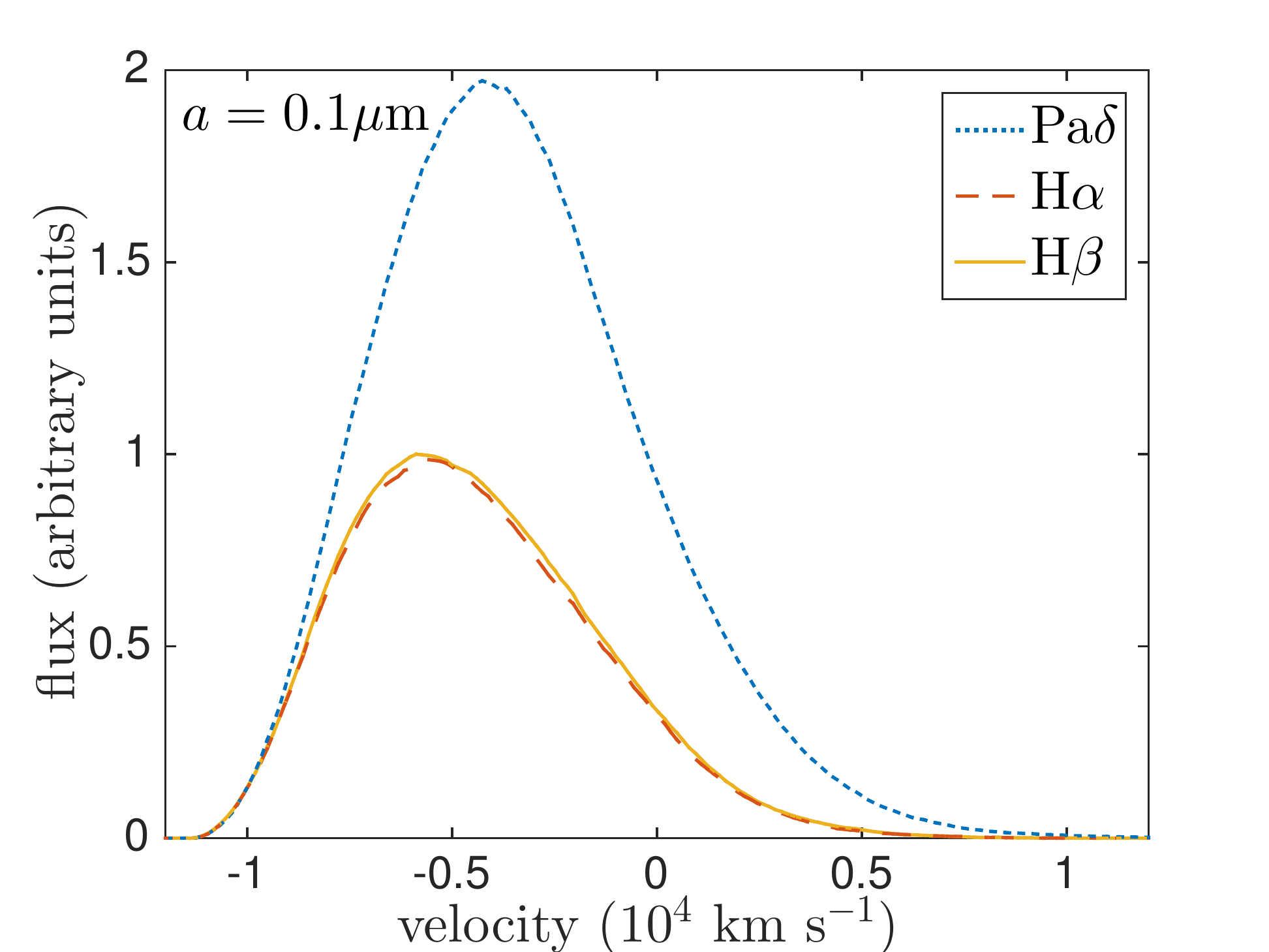}
\includegraphics[trim=38 0 48 10,clip=true,scale=0.34]{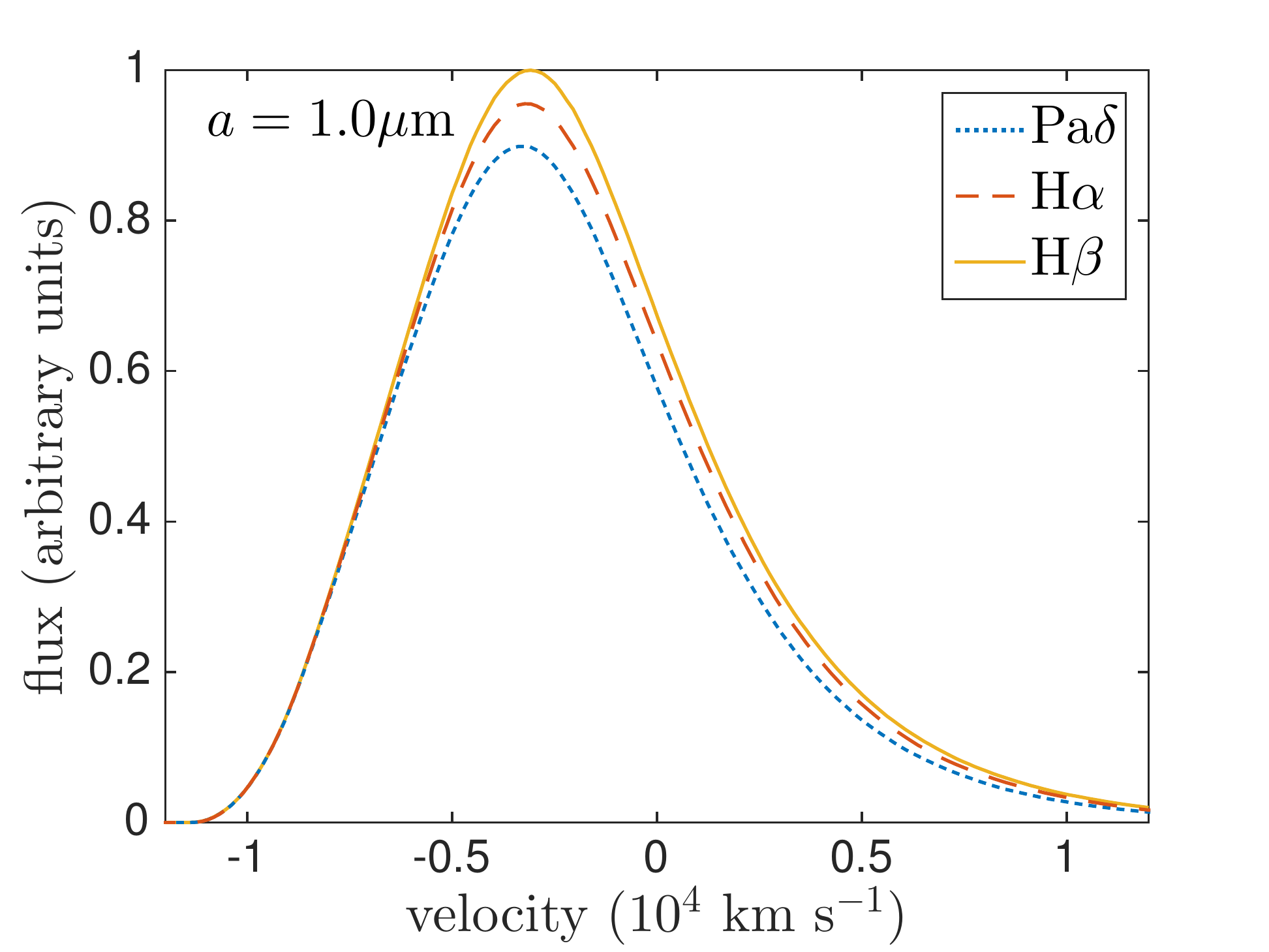}
\caption{Model line profiles for H$\alpha$ (6563\AA\ in \textit{red}), 
H$\beta$ (4861\AA\ in \textit{yellow}) and Pa$\delta$ (10049\AA\ in 
\textit{blue}) for optically thin \textit{(upper)} and  optically thick 
\textit{(lower)} cases respectively. All models adopted density profile 
$\rho(r) \propto r^{-4}$ (i.e. $\beta = 2$), velocity profiles $v(r) 
\propto r$ and radii ratio $R_{\rm in}/R_{\rm out}=0.2$. The grain radii used 
were $a=0.001~\mu$m \textit{(left)}, $a=0.1~\mu$m \textit{(middle)} and 
$a=1.0~\mu$m \textit{(right)}. All the above models used 
\citet{Zubko1996} BE amorphous carbon.}
\label{wav_dep}
\end{figure*}

\begin{figure*}
\includegraphics[trim =0 0 52 0,clip=true,scale=0.33]{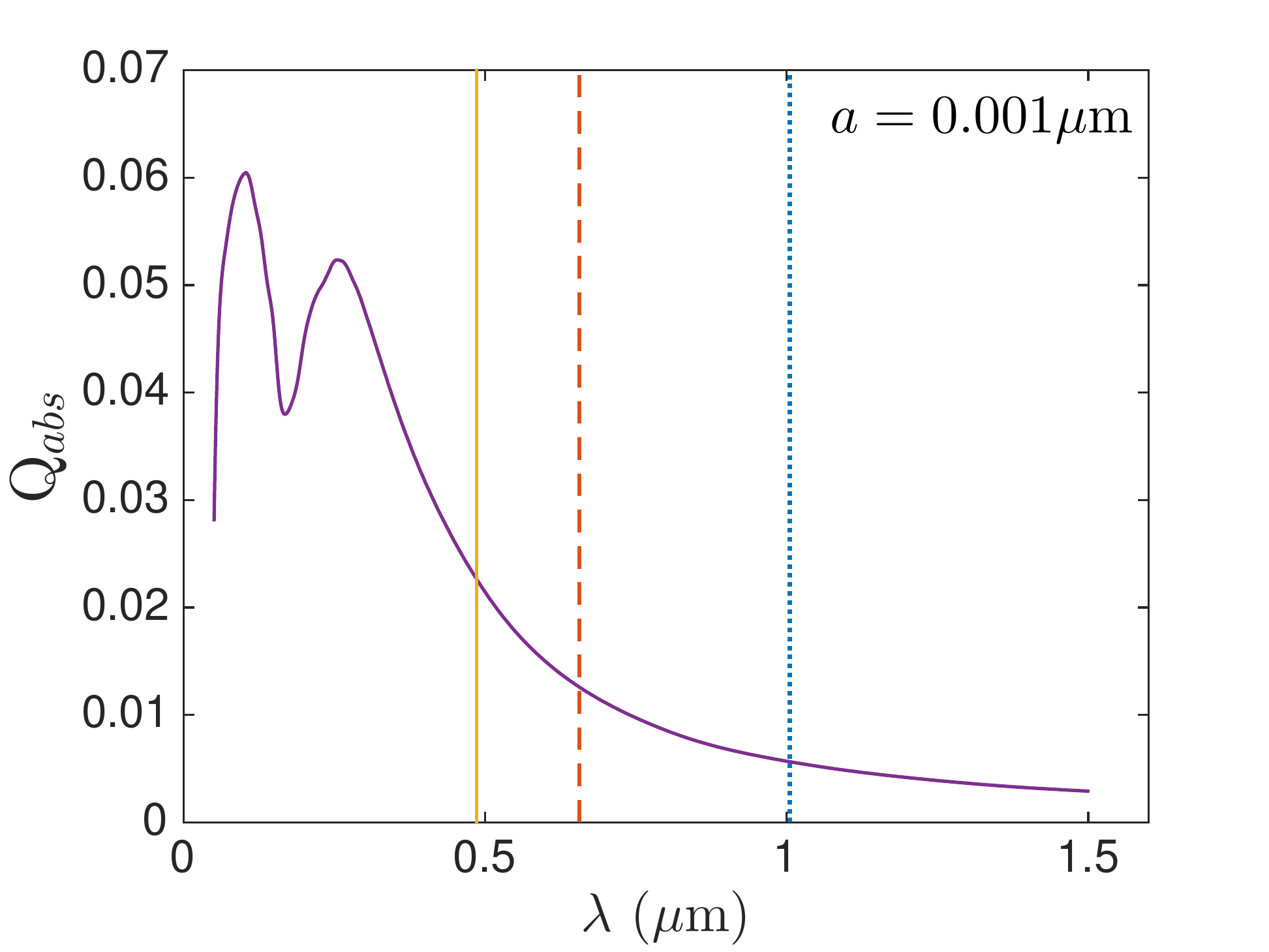} \hspace{0.5mm}
\includegraphics[trim =39 0 52 0,clip=true,scale=0.33]{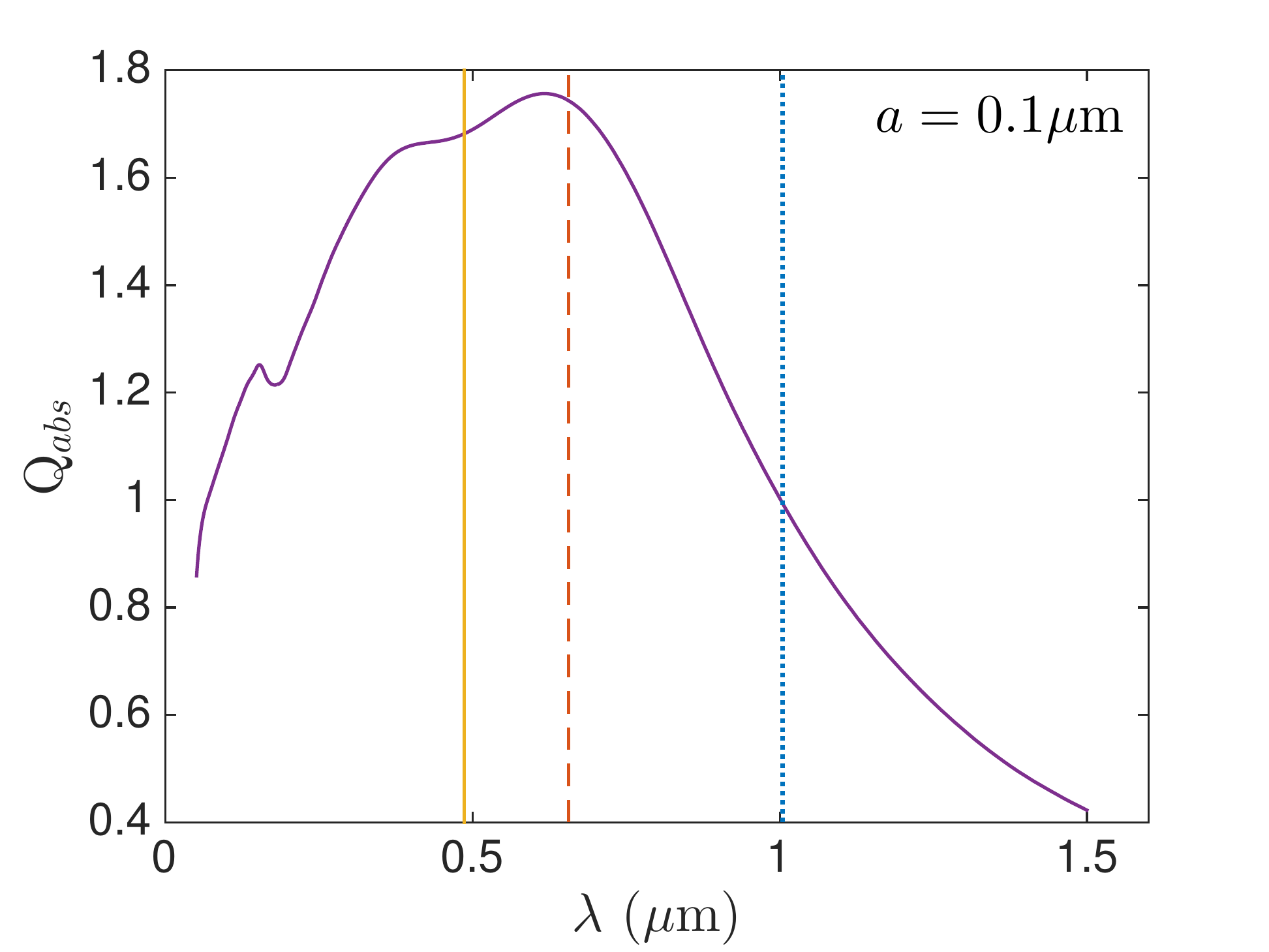} \hspace{0.5mm}
\includegraphics[trim =32 0 52 0,clip=true,scale=0.33]{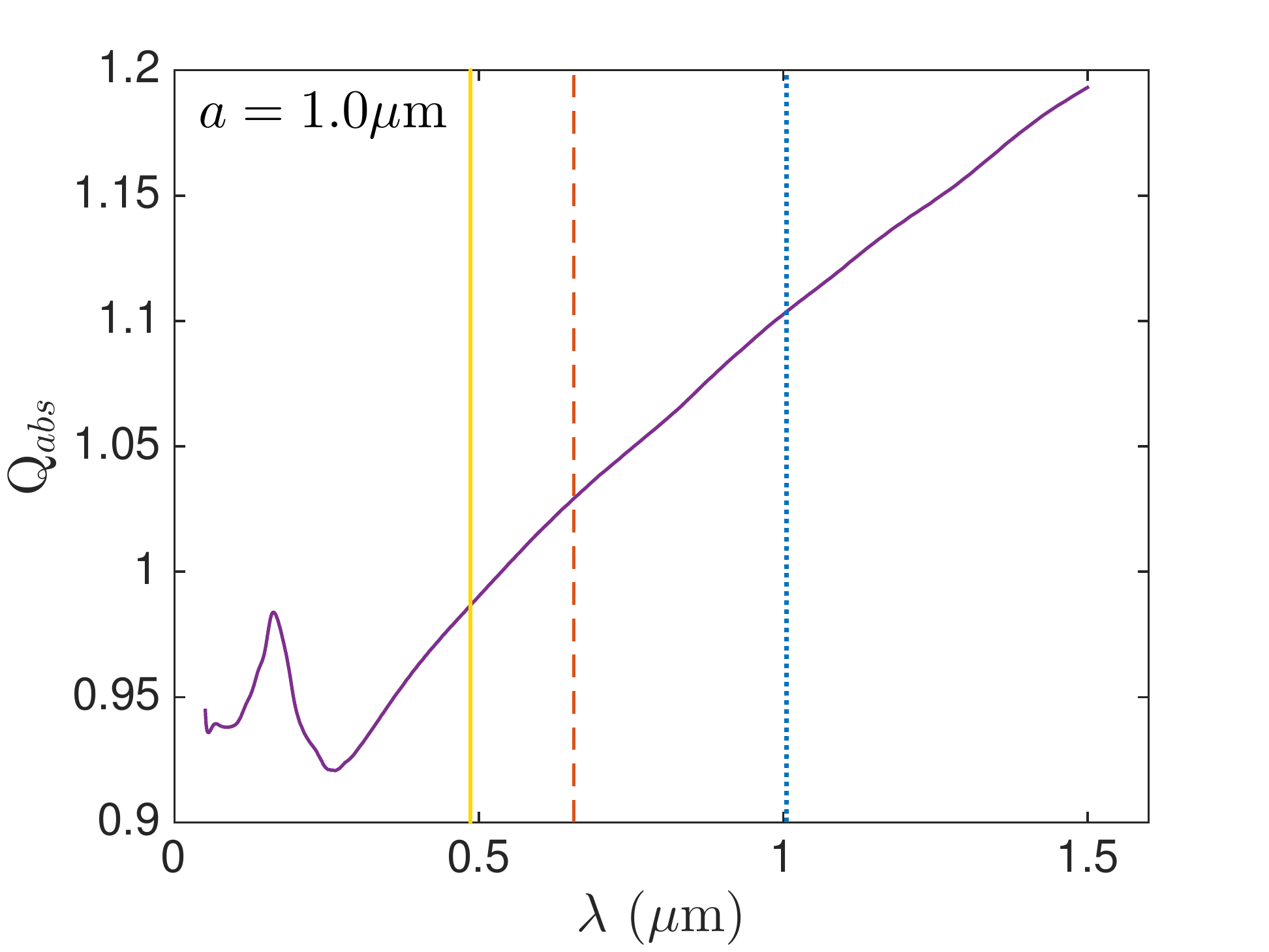}
\caption{The variation of amorphous carbon dust absorption efficiency 
with grain size. The grain radii plotted are $a=0.001~\mu$m 
\textit{(left)}, $a=0.1~\mu$m \textit{(middle)} and $a=1.0~\mu$m \textit{(right)}.  
The vertical lines mark the wavelengths of H$\alpha$ (6563\AA\ in red), 
H$\beta$ (4861\AA\ in yellow) and Pa$\delta$ (10049\AA\ in blue).}
\label{wav_dep2}
\end{figure*}

\subsection{The wavelength dependence of dust absorption}
\label{asym}

The greater the dust optical depth, the more attenuation of the line 
there will be.  As expected, the red side of the profile suffers a greater 
degree of absorption than the blue side.  The resulting asymmetry is 
somewhat more complex than perhaps previously thought.  Dust has 
repeatedly been cited as the agent responsible for the apparent 
blueshifting of supernova line profiles in the manner of the profiles 
presented in Fig. \ref{fig:Lucy}; that is, relatively high optical 
depths result in an overall shift of the entire profile towards the blue.
 The relationship between the blueshifting of the peaks 
of profiles and their wavelength has been discussed by several authors in 
relation to dust formation \citep{Smith2012, Fransson2014, Gall2014}.  
  
In practice, a relatively large dust optical depth is required to actively 
shift the peak of the profile bluewards of its natural $-V_{\rm min}$ position 
(corresponding to the velocity at the inner radius of the shell) unless 
this value is very small in comparison to $V_{\rm max}$ i.e. the profile 
originally had a very narrow flat top.  In many cases, it seems likely that 
the dust may not be optically thick and the blueshifting of the peak of 
the profile is just a result of attenuation in the flat-topped section (close 
to $R_{\rm in}$).  The peak would then tend to be located at $-V_{\rm min}$.

Since dust absorption is wavelength dependent for $2\pi a < \lambda$, one 
might expect the position of the peak line flux to be dependent on the 
wavelength of the line being considered.  We note here that whilst 
variations of the peak velocity of a line as a function of line wavelength 
may occur in cases of high dust optical depths or small $R_{\rm in}/R_{\rm out}$, 
this may not be the case for many supernova lines emitted from ejecta with 
low dust optical depths.  The wavelength-dependence of dust absorption 
instead can result in differing degrees of extinction in the flat-topped 
region of each profile but still leave the peak at its blueshifted 
position of $-V_{\rm min}$.  If this is the case then there would be no reason 
to expect a variation in the position of the peaks of profiles to be 
correlated with the wavelength dependence of dust absorption.  Instead one 
would expect it potentially to trace the location of different ions within 
the ejecta, possibly with different $V_{\rm min}$ values observed for 
different species.

For lines from the same ion, for example the Balmer and Paschen lines of 
H{\sc i}, we might expect to see peaks at the same position but differing 
degrees of absorption. At high spectral resolutions, it might be possible 
to detect differences in the shapes of the line profiles, particularly 
between $-V_{\rm min}$ and $+V_{\rm min}$ where the steepness of the incline 
traces the degree of dust absorption.  This can be seen in Figure 
\ref{wav_dep} where we illustrate the effects of the wavelength dependence 
of dust absorption for three lines, H$\alpha$ (6563\AA), H$\beta$ 
(4861\AA) and Pa$\delta$ (10049\AA).  All lines were modelled using three 
different grain sizes and for both optically thin and thick dust cases.  
We also show the variation of the absorption efficiency with wavelength 
for three different amorphous carbon grain sizes in Fig. \ref{wav_dep2}.

\section{Results for SN~1987A}
\label{results}

\begin{figure}
\centering
\includegraphics[trim =20 0 45 15,clip=true,scale=0.48]{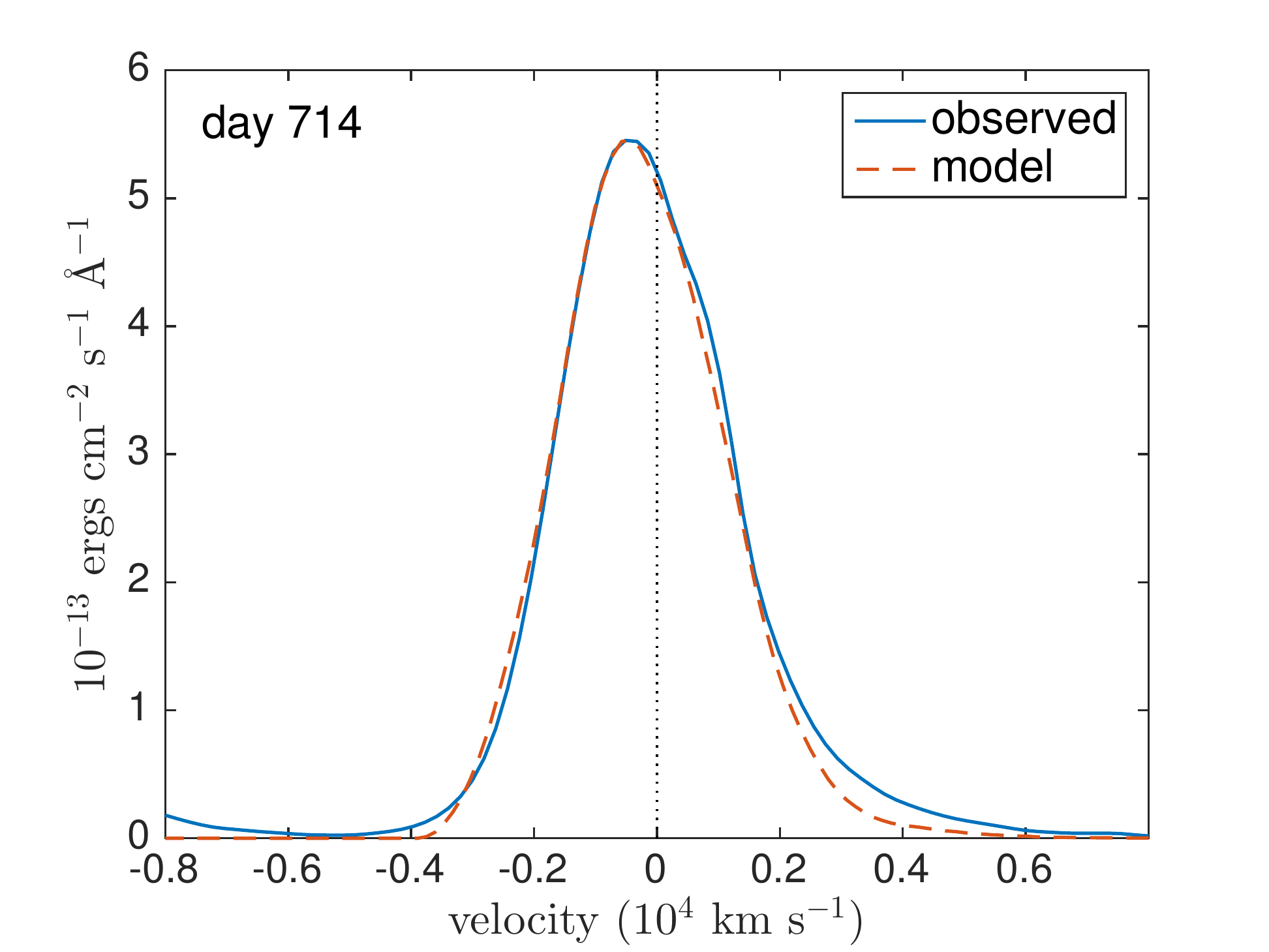}
\caption{Amorphous carbon smooth dust fit to the day 714 H$\alpha$ 
line of SN~1987A using an MRN size distribution,
illustrating the underestimation of the red scattering wing for small 
grain radii.  Model parameters are the same as the smooth dust fit for 
day 714 (Table \ref{smooth1}) except for the 
grain size distribution and dust mass:  $M_{\rm dust}=8.0 \times 10^{-6} 
M_{\odot}$, $a_{\rm min}=0.005~\mu$m, $a_{\rm max}=0.25~\mu$m and $n(a) \propto 
a^{-3.5}$.}
\label{MRN}

\end{figure}

\begin{figure*}
\centering
\includegraphics[trim =0 0 40 0,clip=true,scale=0.25]{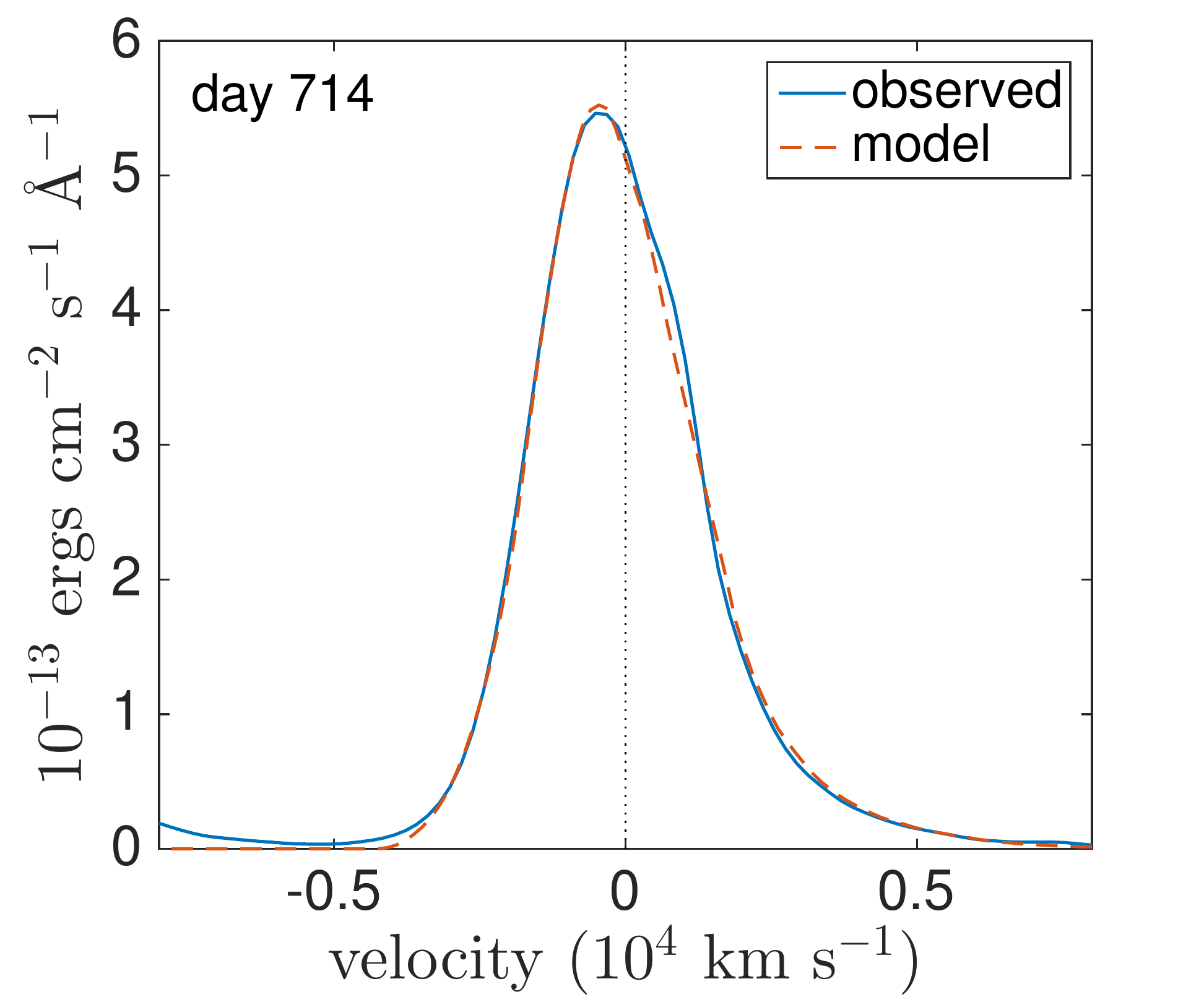}
\includegraphics[trim =35 0 40 0,clip=true,scale=0.25]{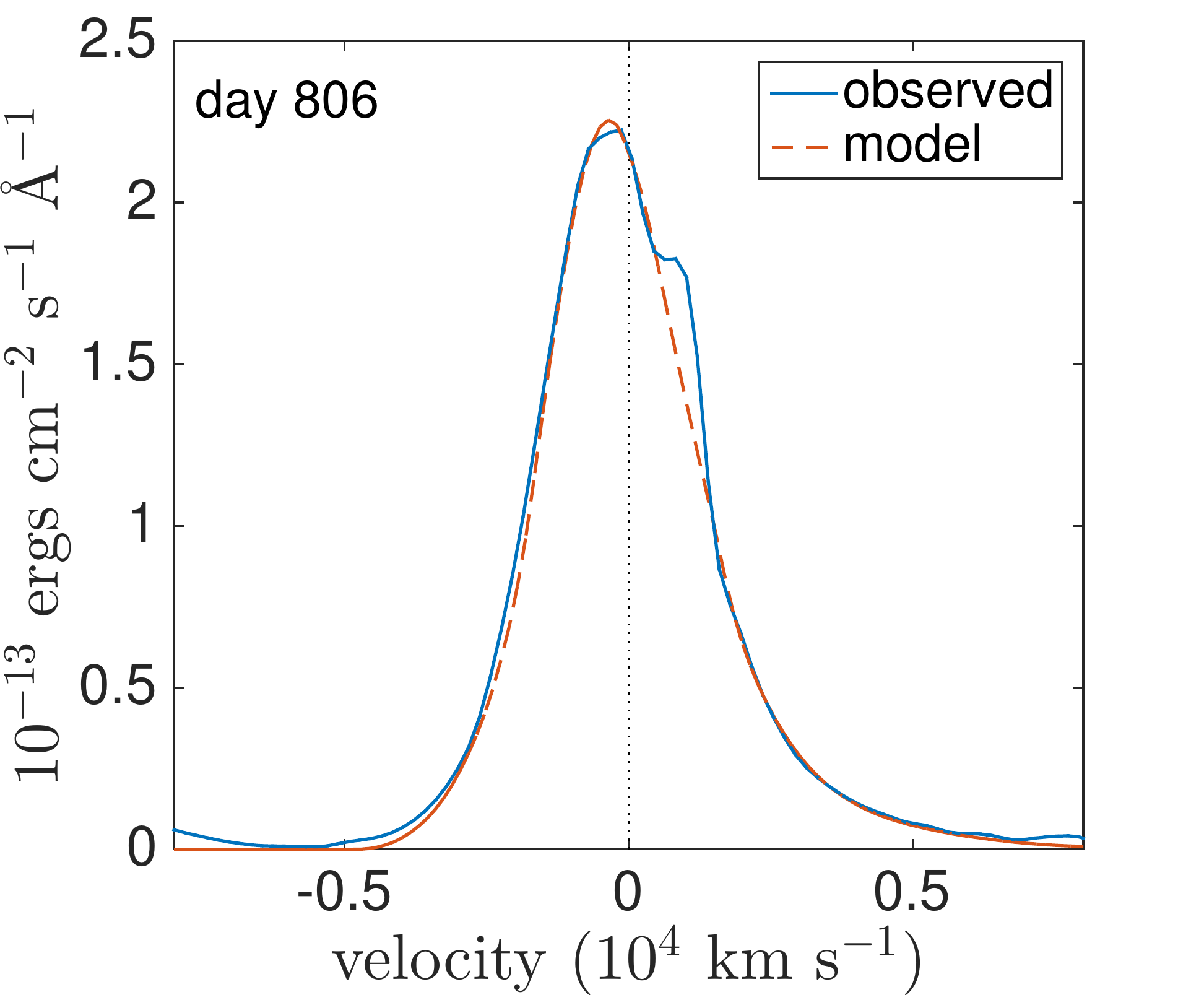}
\includegraphics[trim =47 0 40 0,clip=true,scale=0.25]{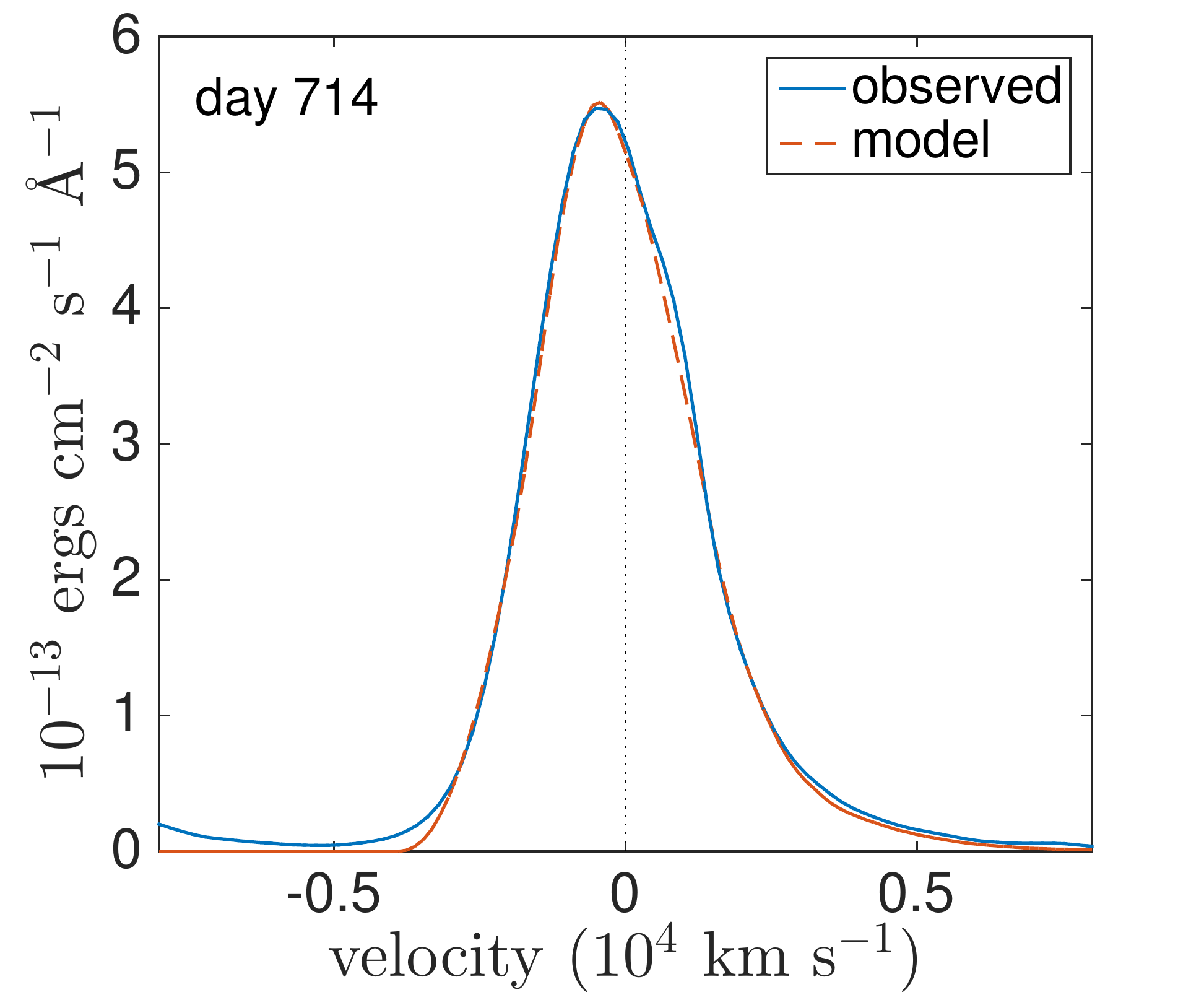}
\includegraphics[trim =35 0 40 0,clip=true,scale=0.25]{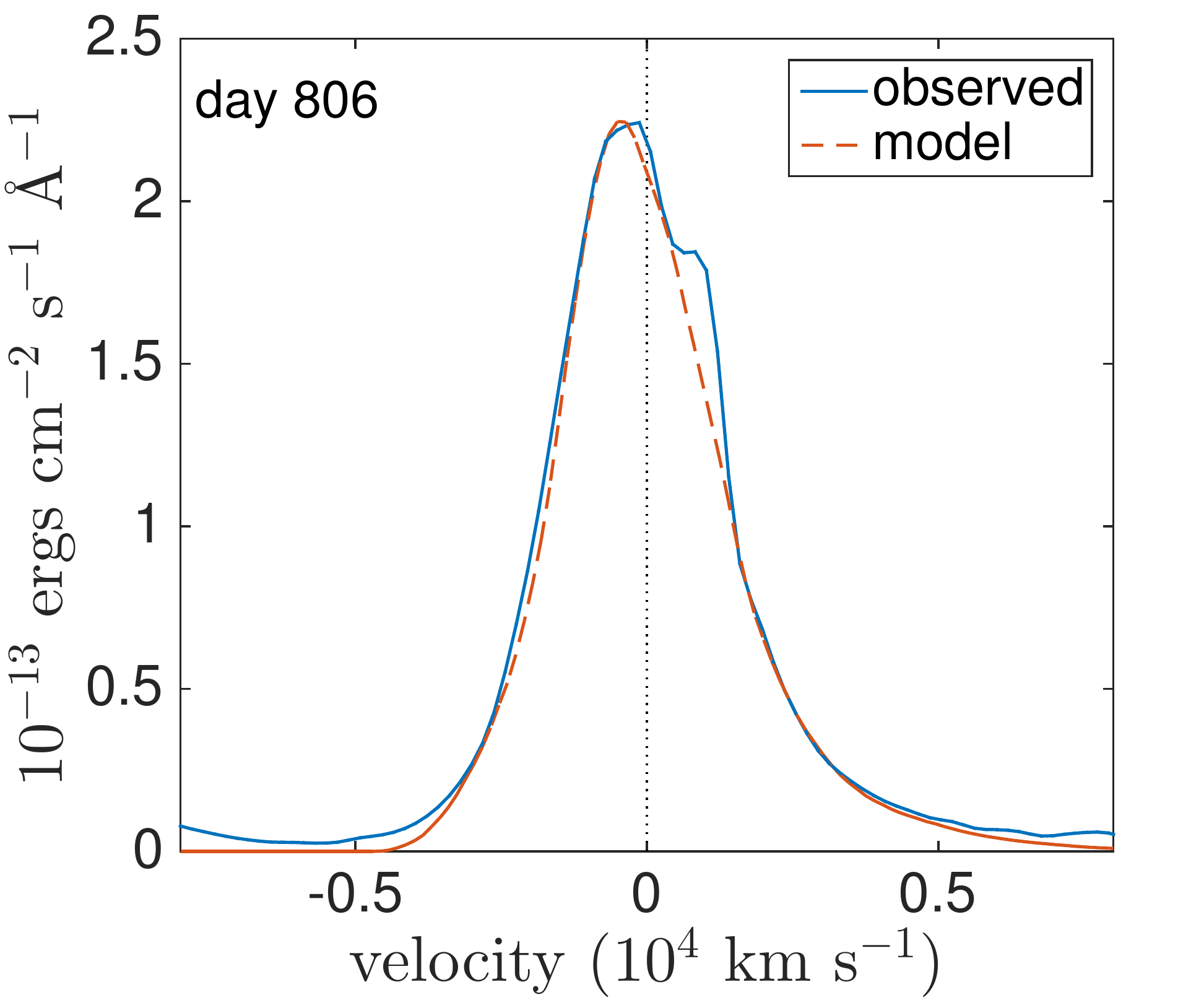}
\caption{Best model fits to the SN~1987A H$\alpha$ line at day 714 
and day 806 for the parameters detailed in Tables \ref{smooth1} and \ref{clumped1}. The two fits on the left are smooth dust models using amorphous carbon grains of radius $a=0.35~\mu$m and the two fits on the right are clumped dust models using amorphous carbon grains of radius $a=0.6~\mu$m.}
\label{Ha}

\end{figure*}

We have modelled the H$\alpha$ line of SN~1987A at days 714, 806, 1862, 
2211, 2875, 3500 and 3604, and the [O~{\sc i}]~$\lambda$6300,6363~\AA\ 
doublet at days 714, 806, 1054 and 1478.  After day 3604 the H$\alpha$ 
profile begins to become dominated by emission from the reverse shock and 
the structure of the emitting region may no longer be approximated by a 
single shell model as we do here \citep{Fransson2013}.  The [O~{\sc 
i}]~$\lambda\lambda$6300,6363~\AA\ doublet becomes too weak to model after day 
1478 (see Fig. \ref{Ha_evol_early}).  We continue to adopt a velocity 
profile $V(r) = \frac{V_{\rm max}}{R_{\rm max}}r$ and treat the variable 
parameters listed at the start of Section \ref{ps}.  Whilst the albedo and 
optical depth are not varied directly, they are altered by adjusting the 
dust mass, $M_{\rm dust}$, and the grain size, $a$, which together determine 
the albedo and optical depth via Mie theory and the optical properties of 
the dust.
\setlength{\tabcolsep}{3pt}
\begin{table}
\caption{Observed luminosities of the H$\alpha$ line and estimated 
electron scattering optical depths from $R_{\rm in}$ to $R_{\rm out}$ for the 
radii detailed in Tables \ref{smooth1} and \ref{clumped1} based on an 
assumed gas temperature of 10,000~K.}
\centering
\begin{tabular}{@{}cccccc@{}}
\hline
& \multicolumn{2}{c}{H$\alpha$} &  \multicolumn{2}{c}{[O~{\sc i}]}  \\
day &  $L_{\rm obs}$ & $L_{\rm undep}$/  &  $L_{\rm obs}$ & $L_{\rm undep}$/   & $\tau_{\rm e}$ \\
& (10$^{37}$ erg s$^{-1}$) &$L_{\rm obs}$& (10$^{37}$ erg s$^{-1}$) & $L_{\rm obs}$& ($10^{-2}$) \\
\hline
714 & 1.36 & 1.65 &0.313&3.57& 1.44  \\
806 & 0.57 & 1.77 &0.0942&3.57& 0.840 \\
1054 &&&0.0242 & 3.23\\
1478 &&& 0.00185&2.70 \\
1862 & 0.0063 & 2.06 &&& 0.159  \\
2211 & 0.0041 & 2.07 &&& 0.0378  \\
2875 & 0.0019 & 2.84 & & &0.0219  \\
3500 & 0.00079 & 3.16 & &&0.0125  \\
3604 & 0.00098 & 3.27 &&&0.0149  \\

\hline
\end{tabular}

\label{tau_e}
\end{table}%
\setlength{\tabcolsep}{8pt}

In all models, the ejecta occupies a shell with inner radius $R_{\rm in}$ and 
outer radius $R_{\rm out}$.  Packets are emitted according to a smooth density 
profile assuming recombination or collisional excitation such that $i(r) 
\propto \rho(r)^2 \propto r^{-2\beta}$.  Initially the dust is considered 
to have a smooth density distribution and is assumed to be coupled to the 
gas so as to follow the same radial profile.  A clumped distribution of 
dust is considered later (see Section \ref{clumped_models}).

Assuming an electron temperature of 10,000~K, we estimated the total electron scattering optical depths between $R_{\rm in}$ and $R_{\rm out}$ based on the 
observed fluxes of the H$\alpha$ recombination line.  A temperature of 10,000~K for the recombining material is 
likely too high at the epochs considered but we adopt it in order
not to underestimate electron scattering optical depths.  The values 
we calculate from the observed H$\alpha$ luminosities are listed in Table 
\ref{tau_e}.  Since the total electron scattering optical depths at these epochs 
are negligibly small we therefore do not include electron scattering in 
the models.

There is rarely a unique set of parameters that provide the best fit to 
the data.  However, the majority of the parameters of interest can be well 
constrained from our modelling by considering different elements of the 
shape of the profile.  In particular, by constructing fits to the data 
using minimum and maximum limits for the grain radius, credible lower and 
upper bounds on the dust mass formed within the ejecta may be derived.  
We present here fits to the data obtained using both small and large 
values of the grain radius $a$ since it is the grain size which has the 
most significant effect on the overall dust mass required to reproduce the 
line profile (see Section \ref{params}).

All of our models are of a dusty medium composed solely of amorphous 
carbon grains. We use the optical constants from the BE sample presented 
by \citet{Zubko1996}.  Although previous SED modelling of SN~1987A  
limited the fraction of silicates present in the dusty ejecta to a maximum 
of 15\% (\citet{Ercolano2007}, W15), the recent work of \citet{Dwek2015} has suggested that a large mass of mostly silicate dust may have formed at early epochs ($\sim$ 615 d).  It is therefore useful to consider the effects 
on our models of using silicate dust.  We discuss this in detail in 
Sections \ref{species} and \ref{dwek}.

For each profile, the maximum velocity is initially identified from the 
data as the point where the emission vanishes on the blue side and is then 
varied throughout the modelling in order to produce the best fit.  The 
equivalent point on the red side is indeterminate from observations due to 
the effects of dust scattering.  We determine the approximate value of 
$V_{\rm min}$ by examining the width of the profile near its peak. Using the features and shapes presented in Figs \ref{wt} and \ref{bt} as a guide, we first examined the observed profile for any obvious points of inflection or abrupt changes in the steepness of the profile.  If these were observed then they were compared to similar changes in theoretical profiles which allowed us to estimate the value of $V_{\rm min}$.  If none were observed, then a model setting $V_{\rm min}$ to be the velocity of the profile peak was considered.  Where neither of these approaches yielded a good model (this was rare) we iterated over a range of values of $V_{\rm min}$ as with other variable parameters such as the dust mass. On the red side the theoretical minimum velocity often 
falls at a similar velocity to the 6583\AA\ line so any dust-induced 
features near this wavelength that would allow a more accurate 
determination of $V_{\rm min}$ can be overwhelmed by the nebular line.  
Having determined the minimum and maximum velocities, the ratio of the 
inner and outer radii of the supernova ejecta can be determined since 
$R_{\rm in}/R_{\rm out}=V_{\rm min}/V_{\rm max}$.  The outer radius is calculated from the 
epoch and the maximum velocity.

The only parameters that  remain to be determined are the exponent of 
the density profile $\beta$, the mean grain radius and the total dust 
mass.  The shape of the blue wing is solely a product of the density 
profile and the dust mass; the height and shape of the red wing is a 
product of these and also of the scattering efficiency of the grains (the 
albedo $\omega$); the extent and shape of the asymmetry in the flat-topped 
portion of the profile is a function of only the total dust optical depth 
determined by the dust mass and the grain radius.  By iterating over these 
three parameters, an excellent fit to the data can usually be 
obtained.

Models are produced in the same manner for the [O~{\sc 
i}]~$\lambda\lambda$6300,6363~\AA\ doublet as for the single H$\alpha$ line, with 
each component of the doublet being modelled independently and the 
resulting profiles added according to a specified ratio.  Although the 
theoretical intrinsic flux ratio is 3.1 for optically thin emission \citep{Storey2000}, the 
actual ratio between the two components can be affected by self-absorption 
\citep{Li1992} and we therefore left it as a free parameter.  The deduced 
doublet ratios are listed in Tables \ref{smooth1}, \ref{clumped1} and 
\ref{clumped2}.

\begin{figure*}
\centering
\includegraphics[trim =0 30 52 0,clip=true,scale=0.3]{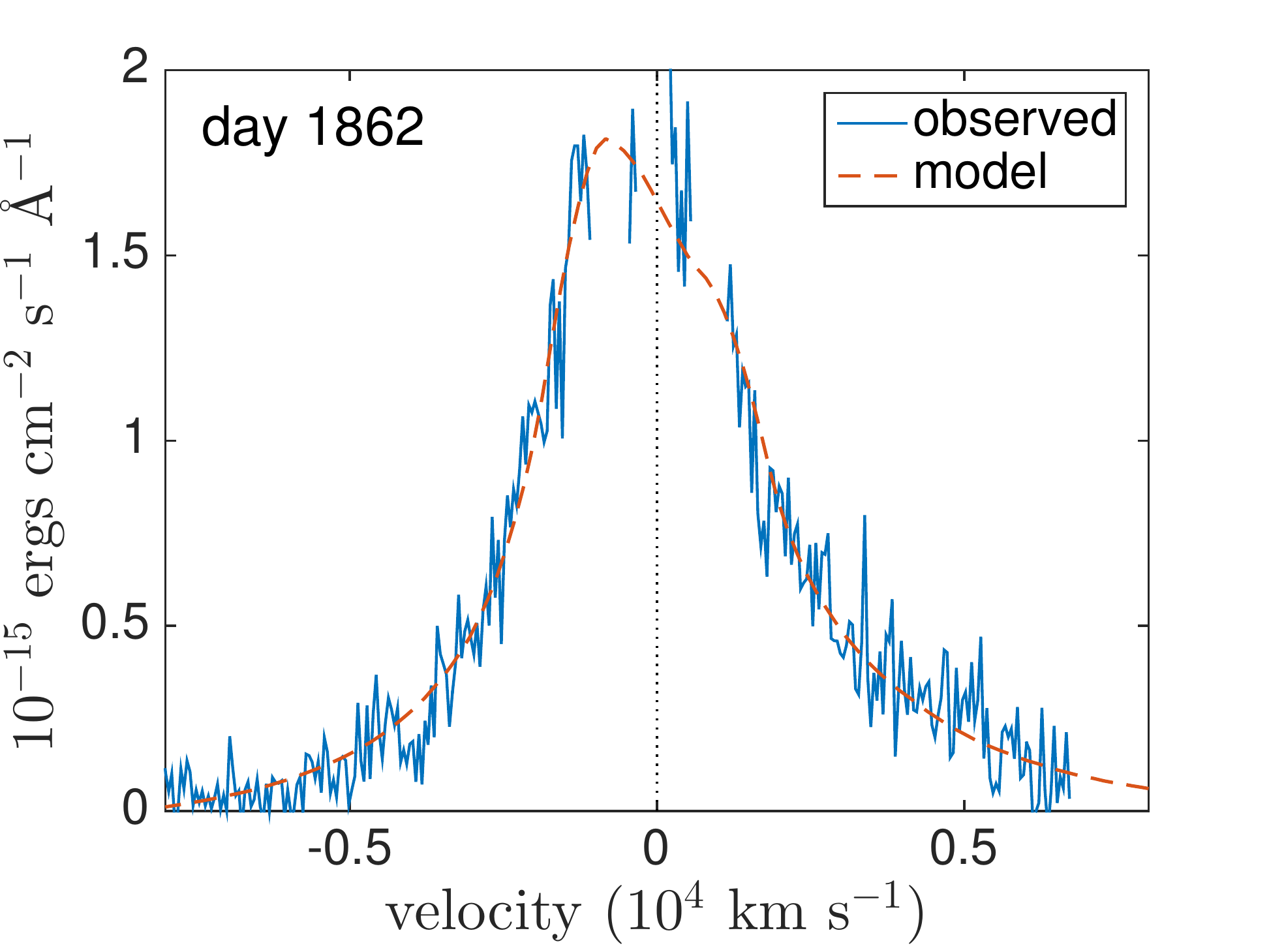}
\hspace{2mm}
\includegraphics[trim =35 30 52 0,clip=true,scale=0.3]{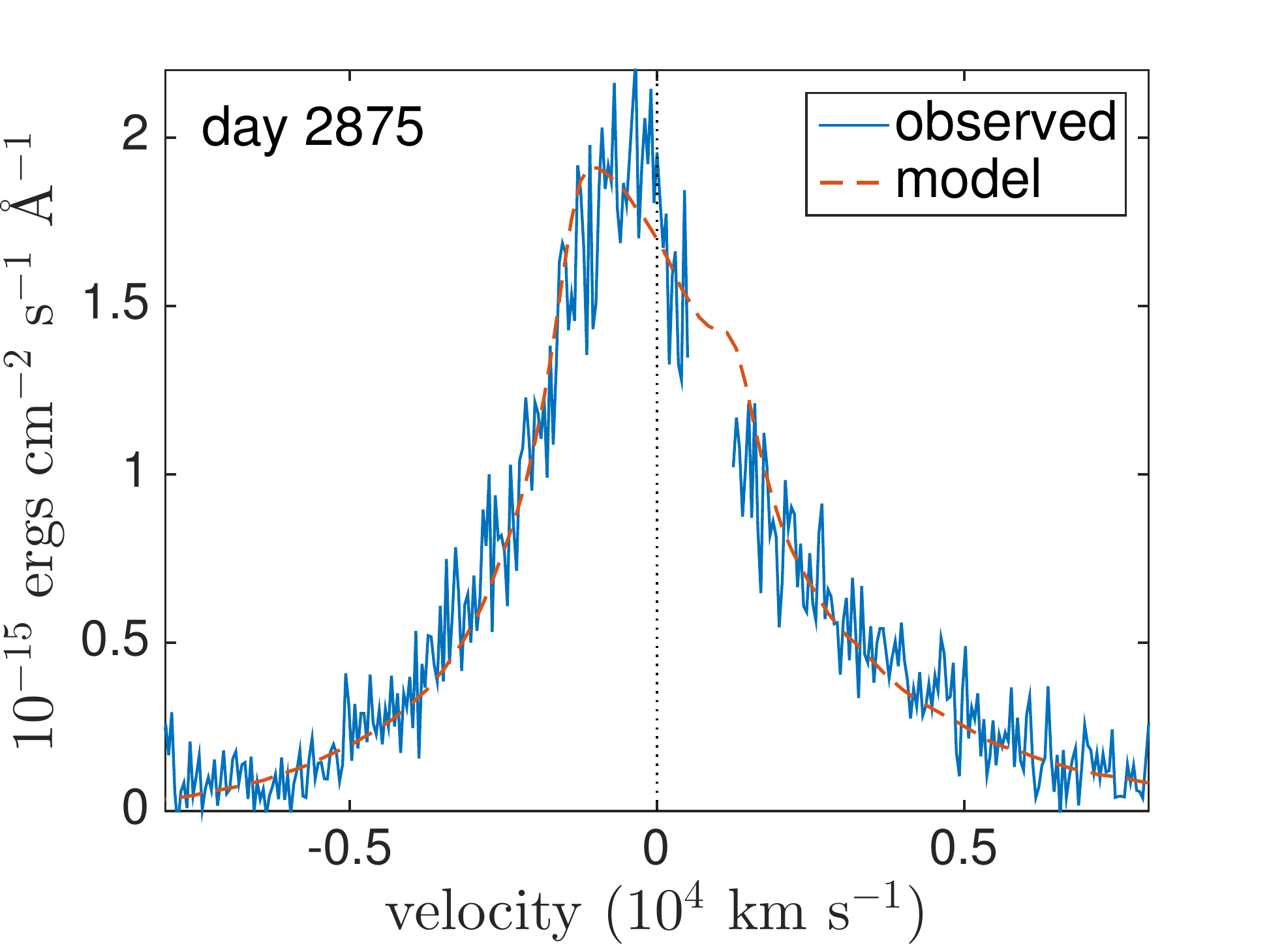}
\hspace{2mm}
\includegraphics[trim =35 30 52 0,clip=true,scale=0.3]{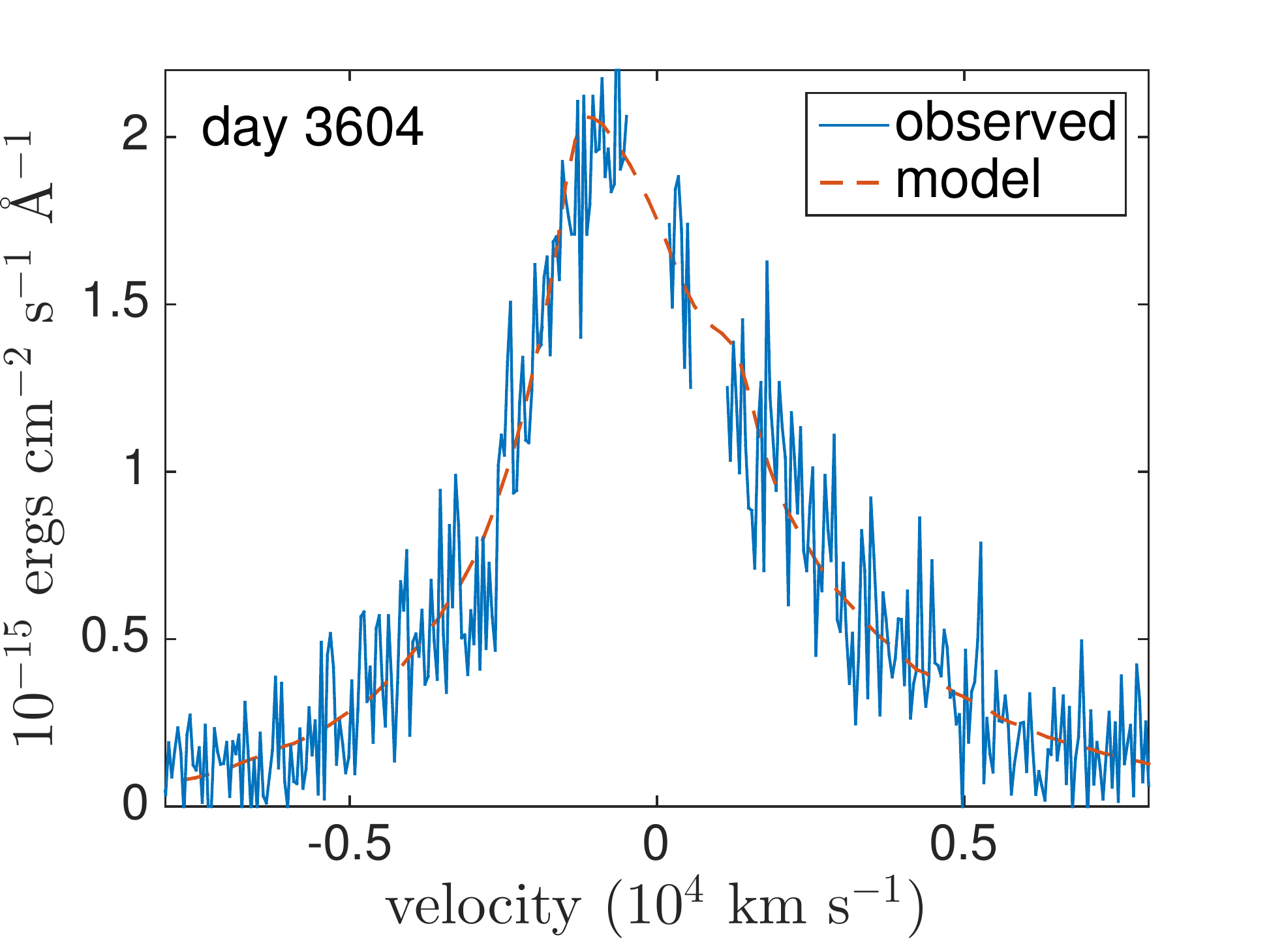}
\\
\includegraphics[trim =0 30 52 20,clip=true,scale=0.3]{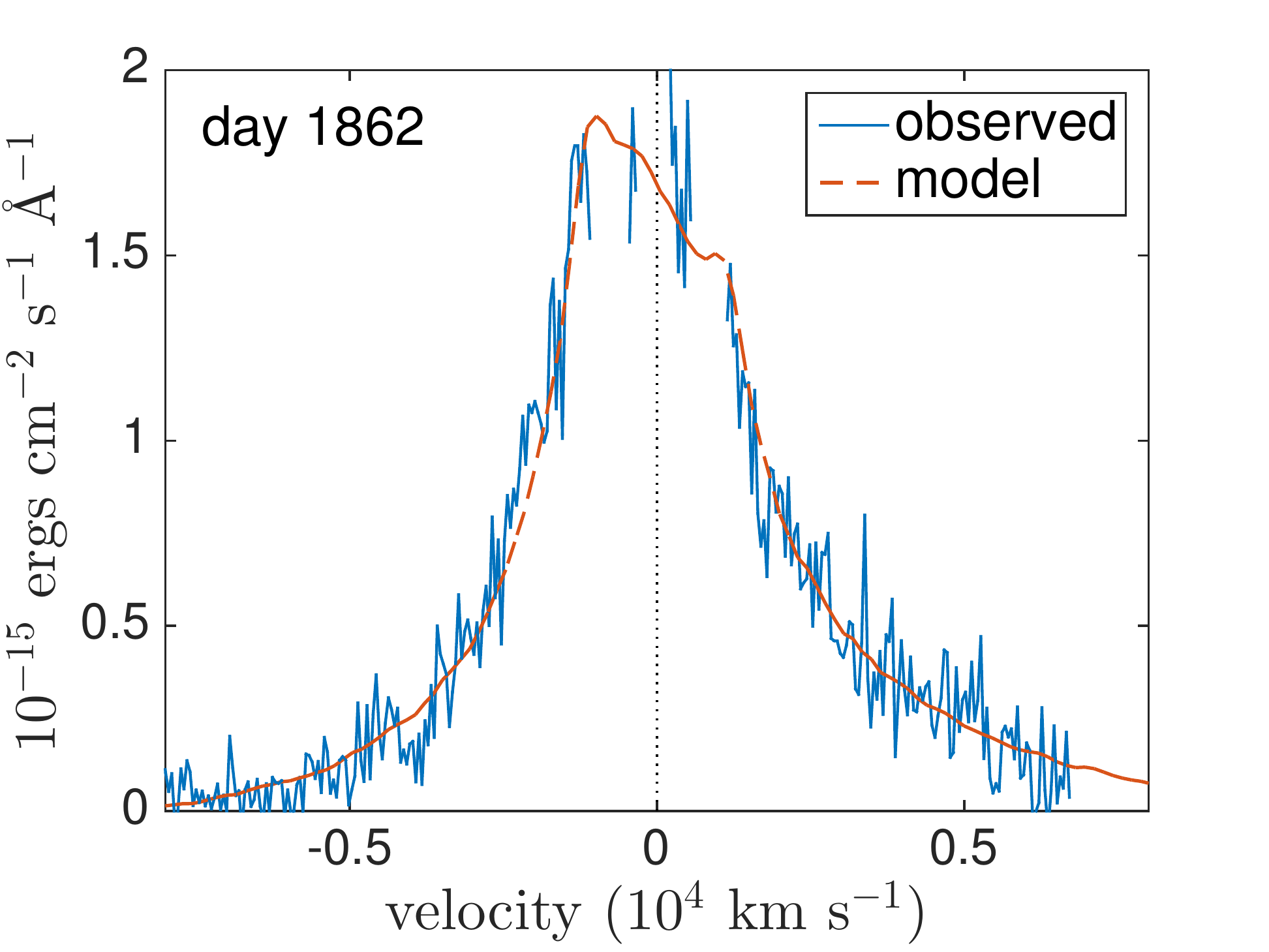}
\hspace{2mm}
\includegraphics[trim =35 30 52 20 ,clip=true,scale=0.3]{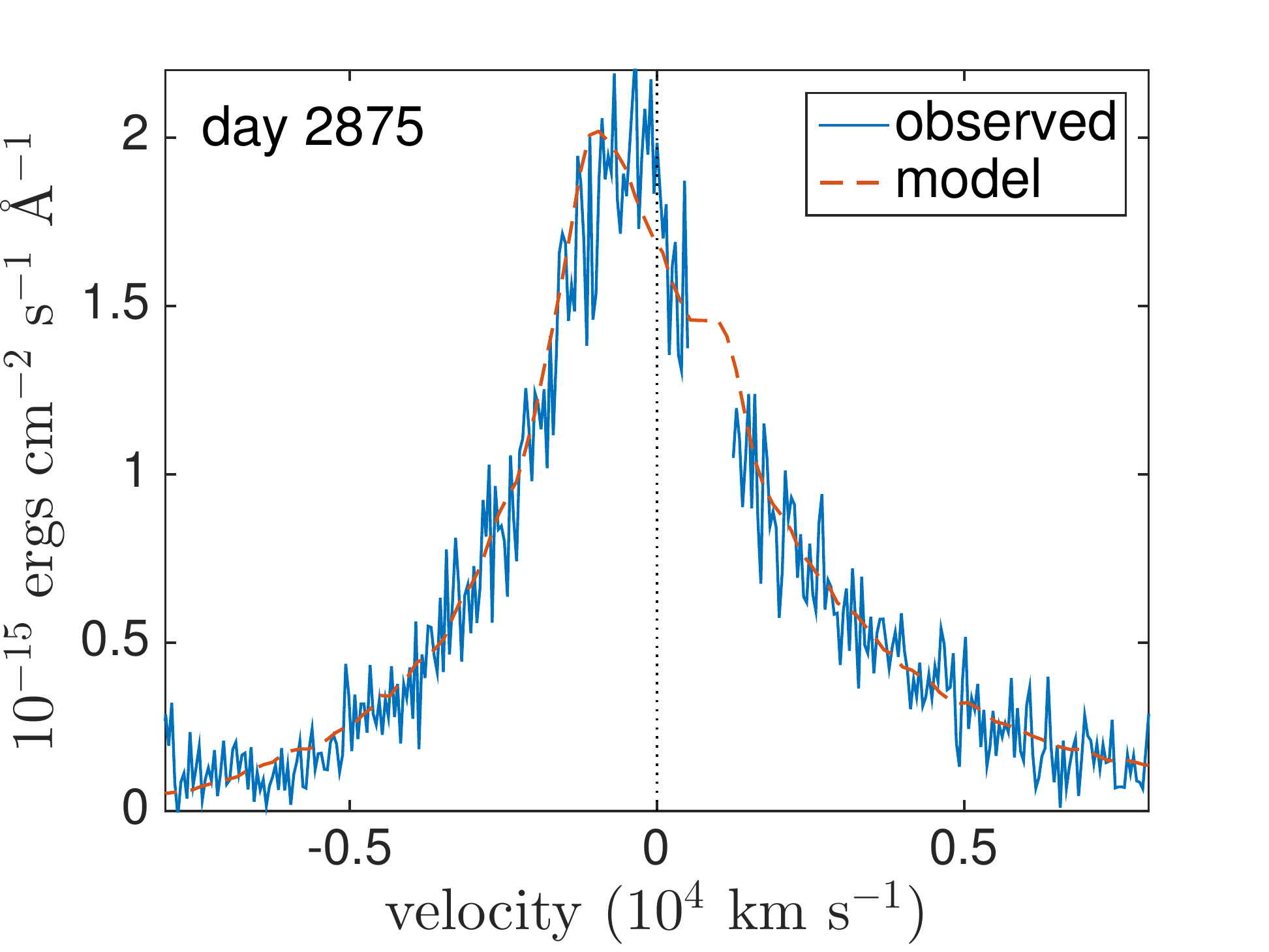}
\hspace{2mm}
\includegraphics[trim =35 30 52 20,clip=true,scale=0.3]{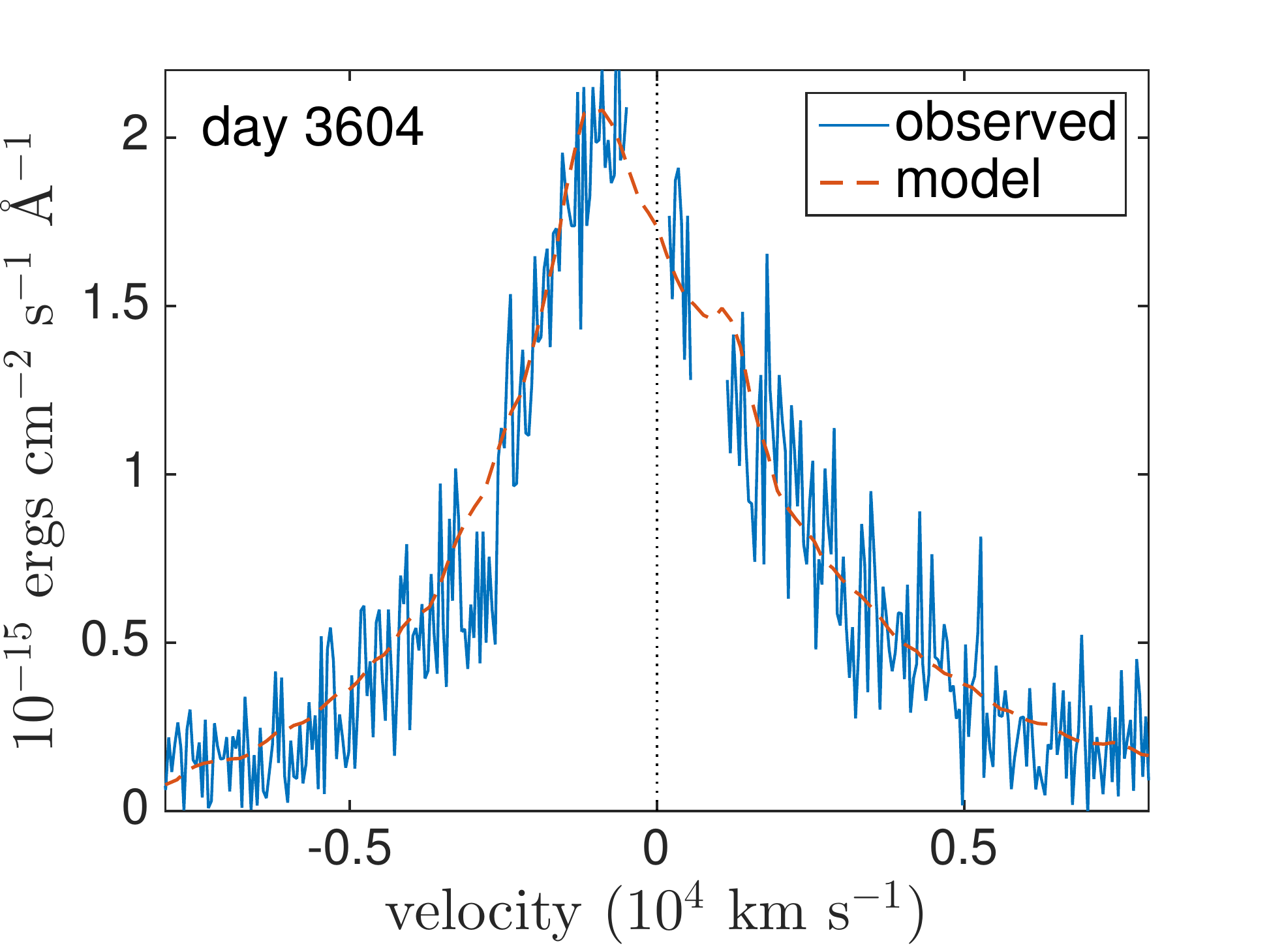}
\\
\includegraphics[trim =0 0 52 20,clip=true,scale=0.3]{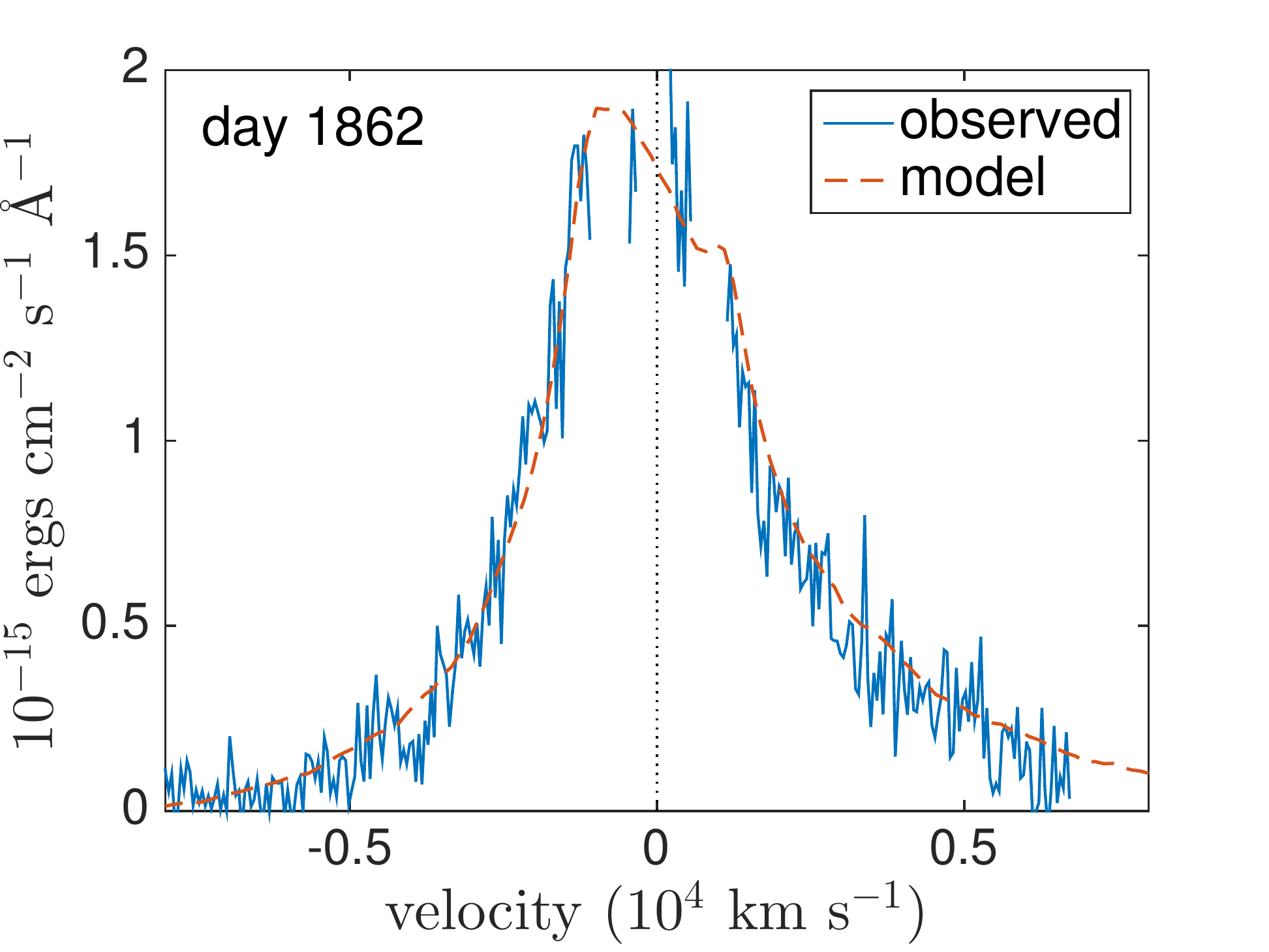}
\hspace{2mm}
\includegraphics[trim =35 0 52 20,clip=true,scale=0.3]{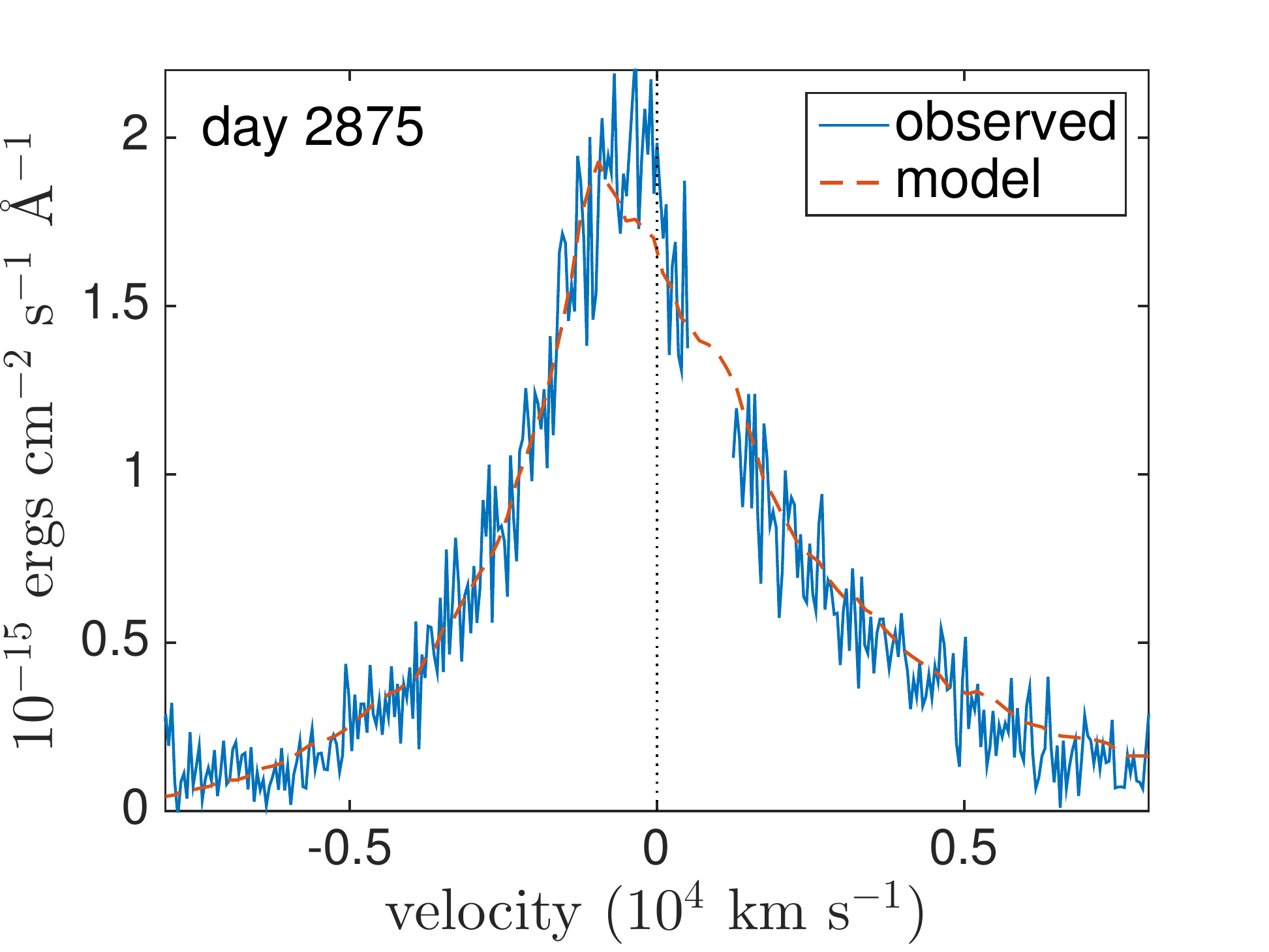}
\hspace{2mm}
\includegraphics[trim =35 0 52 20,clip=true,scale=0.3]{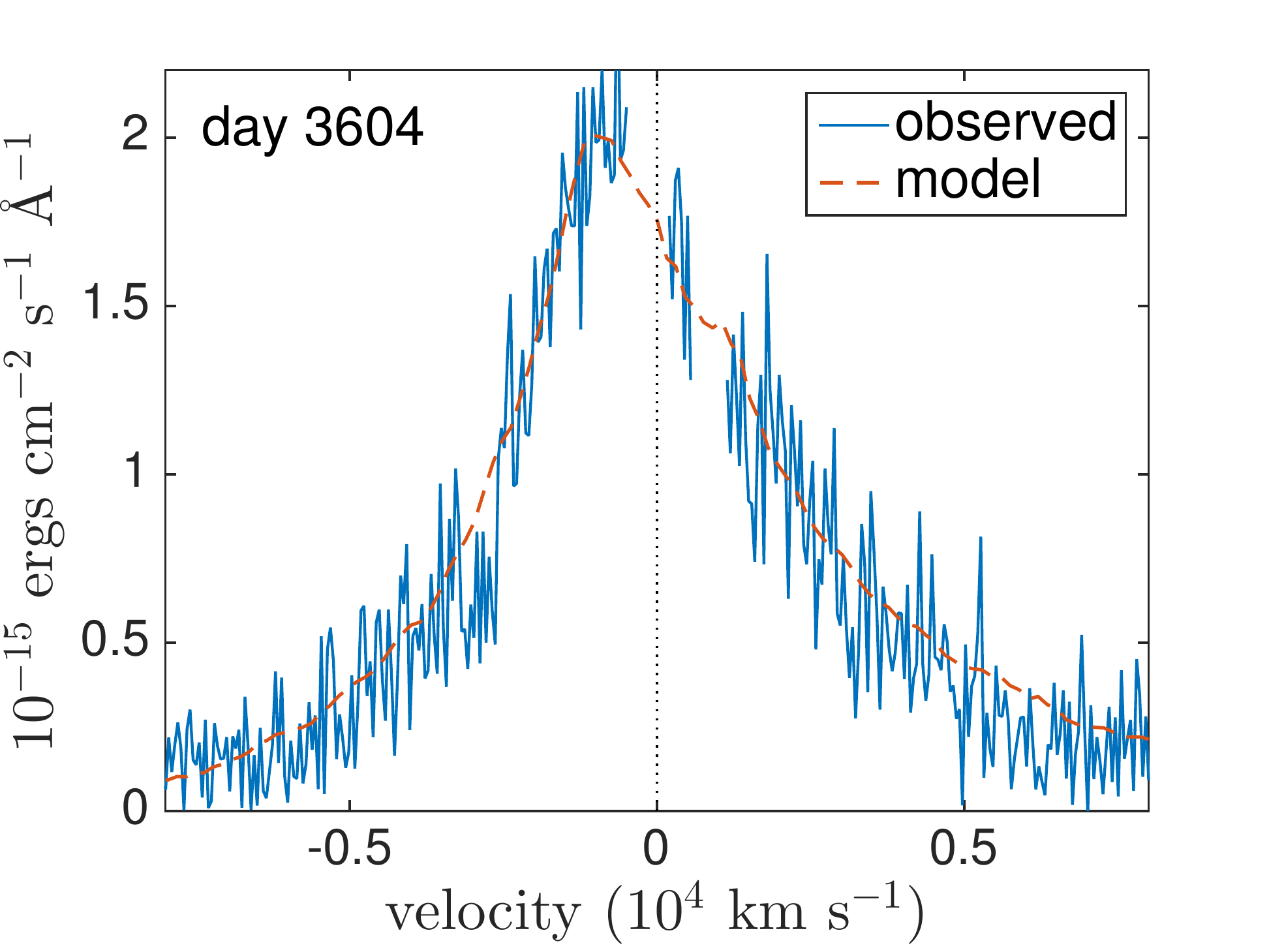}
\caption{Best model fits to the SN~1987A H$\alpha$ line at days 1862, 2875 and 
3604 for the parameters detailed in Tables \ref{smooth1} --  \ref{clumped2}.  On the top row are smooth model fits with amorphous carbon grains of radius $a=0.35~\mu$m.  On the middle and bottom rows are clumped model fits with amorphous carbon grains of radii $a=0.6~\mu$m and $a=3.5~\mu$m, respectively.}
\label{d1862_3604}

\end{figure*}

For all lines, though particularly at very late epochs, even small 
fluctuations in the adopted value of the continuum level can have a 
substantial effect on the fit to the resulting profile.  Since it is not 
feasible to establish the level of the continuum so precisely, the value 
of the continuum has been left as a free parameter that may be adjusted 
(to within sensible margins) in order to allow for the widest possible 
dust mass range to be determined.  We generally find it is necessary to 
assume a continuum level that is slightly lower where the dust mass is 
higher.  The [O~{\sc i}]$\lambda$6300,6363~\AA\ doublets at days 1054 and 
1478 are weak relative to the continuum and are also blended with the 
wings of other lines making it difficult to fit their wings accurately.  
We aim to fit the lines between approximately $-3000$ km s$^{-1}$ and +5000 
km s$^{-1}$ but present a wider velocity range for context (for example 
see Fig. \ref{OI_smooth}).

Fits to the H$\alpha$ line profile at days 2211 and 3500 are omitted for 
the sake of space but are very similar to those of days 1862 to 3604.  
All profiles have been smoothed to approximately the same resolution as 
the observed profiles using a moving-average procedure.  Parameters for 
the models at all epochs including days 2211 and 3500 are detailed in 
Tables \ref{smooth1} to \ref{clumped2}.

\subsection{Smooth Density Models for SN 1987A}
\label{smooth_models}

Even at the earliest epochs there is a substantial wing on the red side of 
the H$\alpha$ line profile that cannot be fitted by scattering from moving 
grains with a low albedo.  The minimum required albedo is approximately 
$\omega \approx 0.5$ implying relatively large grain radii.  As previously 
discussed, the larger the grain size the larger the mass of dust required 
to reproduce the same optical depth.  Fig. \ref{MRN} illustrates the fit 
for the day 714 H$\alpha$ profile for the case where a classic MRN 
\citep{Mathis1977} grain size distribution is adopted, with $a_{\rm min}=0.005 
\mu$m, $a_{\rm max}=0.25~\mu$m and $n(a) \propto a^{-3.5}$.  It can be seen 
clearly that the extended red wing is significantly underestimated.  
Since the albedo of amorphous carbon grains varies significantly with 
grain radius (see Fig. \ref{albedo_grain}) we can establish a strong 
lower bound to the mean dust grain radius, which we estimate to be $a \ge 
0.35~\mu$m.  This is the smallest grain size that is still capable of 
reproducing the red scattering wing at all epochs and we therefore use 
this lower limit value throughout our smooth density modelling.

The inner and outer radii of the ejecta are calculated at each epoch from 
the maximum velocity used, the day number and the specified ratio 
$R_{\rm in}/R_{\rm out}$.  The radii generated are consistent with those used in 
previous models of SN 1987A (\citet{Ercolano2007}, W15) and the 
minimum velocities for both the [O~{\sc i}] and H$\alpha$ line emitting 
regions are relatively consistent with those obtained by \citet{Kozma1998b} 
who estimate that  hydrogen extends into the core to a depth of 
$\lesssim 700$~km~s$^{-1}$ and the oxygen reaches down to $\sim 
400$~km~s$^{-1}$.  They are also consistent with predictions from 3D 
explosion models at the time of shock-breakout that predict the oxygen to 
reach to a depth of $\sim 200$~km~s$^{-1}$ 
\citep{Hammer2010,Wongwathanarat2015}. Figs \ref{Ha} to 
\ref{OI_smooth} show the best fits to the data for days 714 to 3604 
whilst Table \ref{smooth1} details the parameters used.

It can be seen from Tables \ref{smooth1} -- \ref{clumped2} that, in order 
to reproduce the blueshifts seen in the [O~{\sc 
i}]~$\lambda\lambda$6300,6363~\AA\ doublet, considerably larger dust masses are 
required than to fit the H$\alpha$ line at the same epoch.  Although the 
same maximum velocities and therefore outer radii are used in our [O~{\sc 
i}] and H$\alpha$ models, the inner radii for the [O~{\sc i}] models are 
significantly smaller and the density distribution much steeper.  This 
implies that [O~{\sc i}] is concentrated towards the centre of the 
ejecta whereas H$\alpha$ is more diffuse.  This is broadly in 
agreement with 3D explosion dynamics models that suggest that a few hours 
after the explosion the heavier elements will, in comparison to hydrogen, 
be located more centrally in the ejecta with ``bullets" of heavier 
material reaching the outer edges \citep{Hammer2010}.  If dust is forming 
in the inner regions of the ejecta then the majority of the [O~{\sc i}] 
emission must travel through the newly formed dust whereas the more 
diffuse H$\alpha$ emission has a greater chance of escaping unaffected.  
This may explain the difference between the dust masses needed for the 
[O~{\sc i}] and H$\alpha$ models.

\begin{table*}
	\begin{minipage}{180mm}
	\caption{The parameters used for the best fitting 
smooth models of SN~1987A with amorphous carbon grains of radius $a=0.35~\mu$m.  Optical depths are given from $R_{\rm in}$ to $R_{\rm out}$ at $\lambda = 6563$~\AA\ for H$\alpha$ and $\lambda = 6300$~\AA\ for [O~{\sc i}]. Values of $\tau_V$ are very close to the quoted values of $\tau_{H\alpha}$.}
	\label{smooth1}
	\centering
  	\begin{tabular}{@{} cccccccccccc @{}}
    	\hline
 & day & $V_{\rm max}$ & $V_{\rm min}$ & $R_{\rm in}/R_{\rm out}$ & $\beta$ & $M_{\rm dust}$ & $R_{\rm out}$ & $R_{\rm in}$ & [O~{\sc i}] ratio & $\tau_{\lambda}$   \\
	&& (km~s$^{-1} $)& (km~s$^{-1} $) & & & ($M_{\odot}$) & (cm) & (cm) \\
	\hline

[O~{\sc i}]  & 714 & 3250 & 228 & 0.07  & 2.9 & 9.65$\times 10^{-5}$ & 2.00$\times 10^{16}$ & 1.40$\times 10^{15}$ & 2.6 & 3.60  \\ \relax
[O~{\sc i}]  & 806 & 4000 &  240 & 0.06 & 2.4 & 1.50$\times 10^{-4}$ & 2.79$\times 10^{16}$ & 1.67$\times 10^{15}$ & 2.3 & 2.86 \\ \relax
[O~{\sc i}]  & 1054 & 4300 & 215& 0.05  & 2.1 & 2.35$\times 10^{-4}$ &   3.92$\times 10^{16}$ & 1.96$\times 10^{15}$ & 2.7 & 2.23  \\ \relax
[O~{\sc i}]  & 1478 & 4500 & 180 & 0.04  & 1.7 & 2.95$\times 10^{-4}$ &   5.75$\times 10^{16}$ & 2.30$\times 10^{15}$ & 3.0 & 1.30 \\
H$\alpha$ & 714 & 3250 & 813 & 0.25  & 1.2 & 2.10$\times 10^{-5}$ &   2.00$\times 10^{16}$ & 5.01$\times 10^{15}$ & & 0.61\\
H$\alpha$ & 806 & 4000  & 880 & 0.22 & 1.9 & 3.80$\times 10^{-5}$ &   2.79$\times 10^{16}$ & 6.13$\times 10^{15}$ & & 0.59 \\
H$\alpha$ & 1862 & 8500 &  1275 & 0.15  & 1.9 & 5.00$\times 10^{-4}$ &   1.37$\times 10^{17}$ & 2.05$\times 10^{16}$ & & 0.35\\
H$\alpha$ & 2211 & 9000 & 1260& 0.14 & 1.9 & 9.25$\times 10^{-4}$ &   1.72$\times 10^{17}$ & 2.41$\times 10^{16}$ & & 0.42\\
H$\alpha$ & 2875 & 9500 & 1330 & 0.14 & 1.9 & 1.50$\times 10^{-3}$ &   2.36$\times 10^{17}$ & 3.30$\times 10^{16}$ & & 0.36 \\

H$\alpha$ & 3500 & 10000 & 1400 & 0.14 & 1.9 & 3.35$\times 10^{-3}$  & 3.02$\times 10^{17}$ & 4.23$\times 10^{16}$ && 0.49   \\

H$\alpha$ & 3604 & 10250 & 1333 & 0.13 & 1.9 & 4.20$\times 10^{-3}$ &   3.19$\times 10^{17}$ & 4.15$\times 10^{16}$ & & 0.55 \\ 

    \hline
  \end{tabular}

\end{minipage}
\end{table*}

\begin{table*}
	\begin{minipage}{180mm}
	\caption{The parameters used for the best fitting  
clumped models of SN~1987A with amorphous carbon grains of radius $a=0.6~\mu$m. Optical depths are given from $R_{\rm in}$ to $R_{\rm out}$ at $\lambda = 6563$~\AA\ for H$\alpha$ and $\lambda = 6300$~\AA\ for [O~{\sc i}]. Values of $\tau_V$ are very close to the quoted values of $\tau_{H\alpha}$.}
	\label{clumped1}
\centering
  	\begin{tabular}{@{} ccccccccccccc @{}}
    	\hline
 & day & $V_{\rm max}$ & $V_{\rm min}$ & $R_{\rm in}/R_{\rm out}$ & $\beta$ & $M_{\rm dust}$ & $R_{\rm out}$ & $R_{\rm in}$ &  [O~{\sc i}] ratio & $\tau_{\lambda}$    \\
	&& (km~s$^{-1} $) & (km~s$^{-1} $)& & & ($M_{\odot}$) & (cm) & (cm)   \\
	\hline
[O~{\sc i}]  & 714 & 3250 & 228& 0.07 & 2.7 & 2.00$\times 10^{-4}$ & 2.00$\times 10^{16}$ & 1.40$\times 10^{15}$ & 2.3 & 3.84   \\ \relax
[O~{\sc i}]  & 806 & 4000 & 240&0.06 & 2.3 & 4.00$\times 10^{-4}$ & 2.79$\times 10^{16}$ & 1.67$\times 10^{15}$ & 2.0 & 4.02  \\ \relax
[O~{\sc i}]  & 1054 & 4300 & 215&0.05 & 2.3 & 7.50$\times 10^{-4}$ &   3.92$\times 10^{16}$ & 1.96$\times 10^{15}$ & 2.3 & 3.85  \\ \relax
[O~{\sc i}]  & 1478 & 4500 & 180&0.04 & 2.0 & 1.10$\times 10^{-3}$ &   5.75$\times 10^{16}$ & 2.30$\times 10^{15}$ & 2.8 & 2.65  \\
H$\alpha$ & 714 & 3250 & 813&0.25 & 1.4 & 5.50$\times 10^{-5}$ &   2.00$\times 10^{16}$ & 5.01$\times 10^{15}$ & & 0.87  \\
H$\alpha$ & 806 & 4000 & 880&0.22 & 1.8 & 9.00$\times 10^{-5}$ &   2.79$\times 10^{16}$ & 6.13$\times 10^{15}$ & & 0.76 \\
H$\alpha$ & 1862 & 8500 & 1190&0.14 & 1.9 & 1.20$\times 10^{-3}$ &   1.37$\times 10^{17}$ & 1.91$\times 10^{16}$ & & 0.46  \\

H$\alpha$ & 2211 & 9000 & 1260&0.14 & 1.9 & 3.00$\times 10^{-3}$ &   1.72$\times 10^{17}$ & 2.41$\times 10^{16}$ & & 0.73 \\

H$\alpha$ & 2875 & 9500 & 1140&0.12 & 2 & 8.00$\times 10^{-3}$ &   2.36$\times 10^{17}$ & 2.83$\times 10^{16}$ & & 1.05  \\

H$\alpha$ & 3500 & 10000 & 1200&0.12 & 2 & 1.35$\times 10^{-2}$  & 3.02$\times 10^{17}$ & 3.63$\times 10^{16}$ && 1.08   \\

H$\alpha$ & 3604 & 10250 & 1230&0.12 & 2 & 1.70$\times 10^{-2}$ &   3.19$\times 10^{17}$ & 3.83$\times 10^{16}$ & & 1.22 \\ 

    \hline
  \end{tabular}

\end{minipage}
\end{table*}

\begin{table*}
	\begin{minipage}{180mm}
	\caption{The parameters used for the best fitting 
clumped models of SN~1987A with amorphous carbon grains of radius $a=3.5~\mu$m. Optical depths are given from $R_{\rm in}$ to $R_{\rm out}$ at $\lambda = 6563$~\AA\ for H$\alpha$ and $\lambda = 6300$~\AA\ for [O~{\sc i}]. Values of $\tau_V$ are very close to the quoted values of $\tau_{H\alpha}$.}
	\label{clumped2}
\centering
  	\begin{tabular}{@{} cccccccccccccc @{}}
    	\hline
 & day & $V_{\rm max}$ & $V_{\rm min}$ & $R_{\rm in}/R_{\rm out}$ & $\beta$ & $M_{\rm dust}$  & $R_{\rm out}$ & $R_{\rm in}$ & [O~{\sc i}] ratio & $\tau_{\lambda}$ \\
	&& (km~s$^{-1} $) &  (km~s$^{-1} $) & & & ($M_{\odot}$)  & (cm) & (cm)  \\
	\hline
[O~{\sc i}]  & 714 & 3250 &228& 0.07 & 2.9 & 1.50$\times 10^{-3}$ & 2.00$\times 10^{16}$ & 1.40$\times 10^{15}$ & 2.3 & 4.20   \\ \relax
[O~{\sc i}]  & 806 & 4000 &240& 0.06 & 2.3 & 2.70$\times 10^{-3}$ & 2.79$\times 10^{16}$ & 1.67$\times 10^{15}$ & 2.1 & 3.95   \\ \relax
[O~{\sc i}]  & 1054 & 4300 &215& 0.05 & 2.3 & 5.50$\times 10^{-3}$ &   3.92$\times 10^{16}$ & 1.96$\times 10^{15}$ & 2.5 & 4.12  \\ \relax
[O~{\sc i}]  & 1478 & 4500 &180& 0.04 & 1.9 & 8.00$\times 10^{-3}$ &   5.75$\times 10^{16}$ & 2.30$\times 10^{15}$ & 2.8 & 2.81  \\
H$\alpha$ & 1862 & 8500 &1190& 0.14 & 1.9 & 1.00$\times 10^{-2}$  & 1.37$\times 10^{17}$ & 1.91$\times 10^{16}$ && 0.55   \\
H$\alpha$ & 2211 & 9000 &1260& 0.14 & 1.9 & 2.40$\times 10^{-2}$ &   1.72$\times 10^{17}$ & 2.41$\times 10^{16}$ & & 0.85\\
H$\alpha$ & 2875 & 9500 &1140& 0.12 & 2 & 6.00$\times 10^{-2}$  & 2.36$\times 10^{17}$ & 2.83$\times 10^{16}$ && 1.15   \\
H$\alpha$ & 3500 & 10000 &1200& 0.12 & 2 & 1.15$\times 10^{-1}$  & 3.02$\times 10^{17}$ & 3.63$\times 10^{16}$ && 1.34   \\
H$\alpha$ & 3604 & 10250 &1230& 0.12 & 2 & 1.25$\times 10^{-1}$  & 3.19$\times 10^{17}$ & 3.83$\times 10^{16}$ && 1.31   \\ 

    \hline
  \end{tabular}

\end{minipage}
\end{table*}

\subsection{Clumped Dust Models for SN 1987A}
\label{clumped_models}

A number of investigators have presented arguments for the material in the 
ejecta of SN~1987A being clumped \citep{Lucy1991,Li1992,Kozma1998b} and so 
we consider clumped models for the ejecta dust to be more realistic than 
smoothly distributed dust models. It has been shown through the modelling 
of optical-IR SEDs that when dust is assumed to have a clumped 
distribution then the derived dust masses can be significantly larger than 
for the case of dust that is distributed smoothly between the inner and 
outer radii (e.g. \citet{Ercolano2007,Owen2015}). We present two sets of 
fits to the line profile based on the clumped dust modelling of W15, one 
set with a minimum grain size and one set with a maximum grain size.  
Each fit is based on the best fitting smooth model such that the photon 
packets are emitted assuming a smooth radial density profile.  However, 
the dust is no longer coupled to the gas but instead is located entirely 
in clumps of size $R_{\rm out}/25$.  The clumps are distributed stochastically 
between $R_{\rm in}$ and $R_{\rm out}$ with the probability of a given grid cell 
being a clump proportional to $r^{- \beta }$ where $i(r) \propto r^{-2 
\beta}$.  The number of clumps used is determined by the clump filling 
factor $f$ which is kept constant at $f=0.1$.  All properties are fixed 
from the smooth models with the exception of the grain radius, density 
profile exponent ($\beta$) and the total dust mass.

\begin{figure*}
\centering

\includegraphics[trim =0 35 50 0,clip=true,scale=0.24]{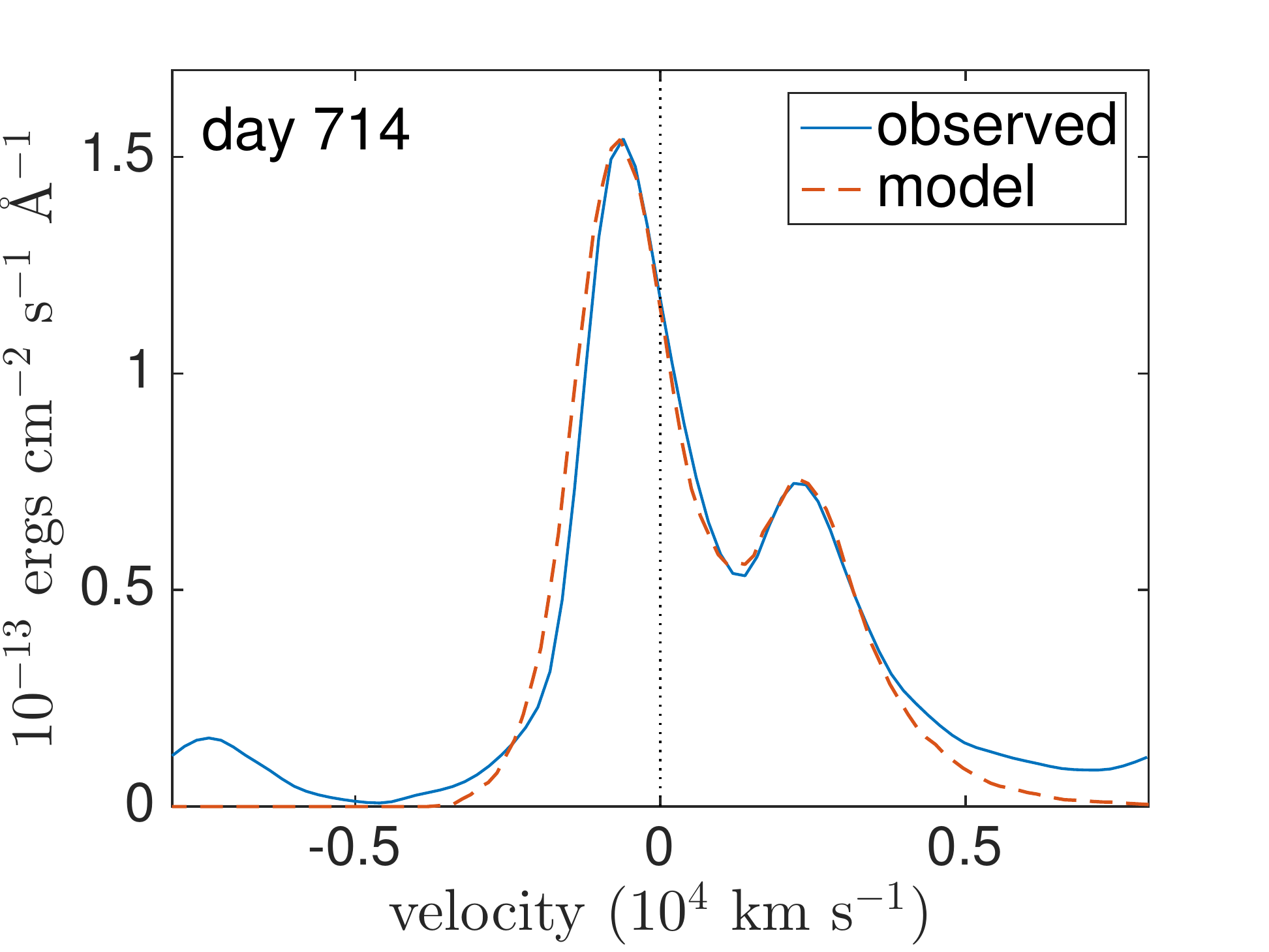}
\includegraphics[trim =0 35 50 0,clip=true,scale=0.24]{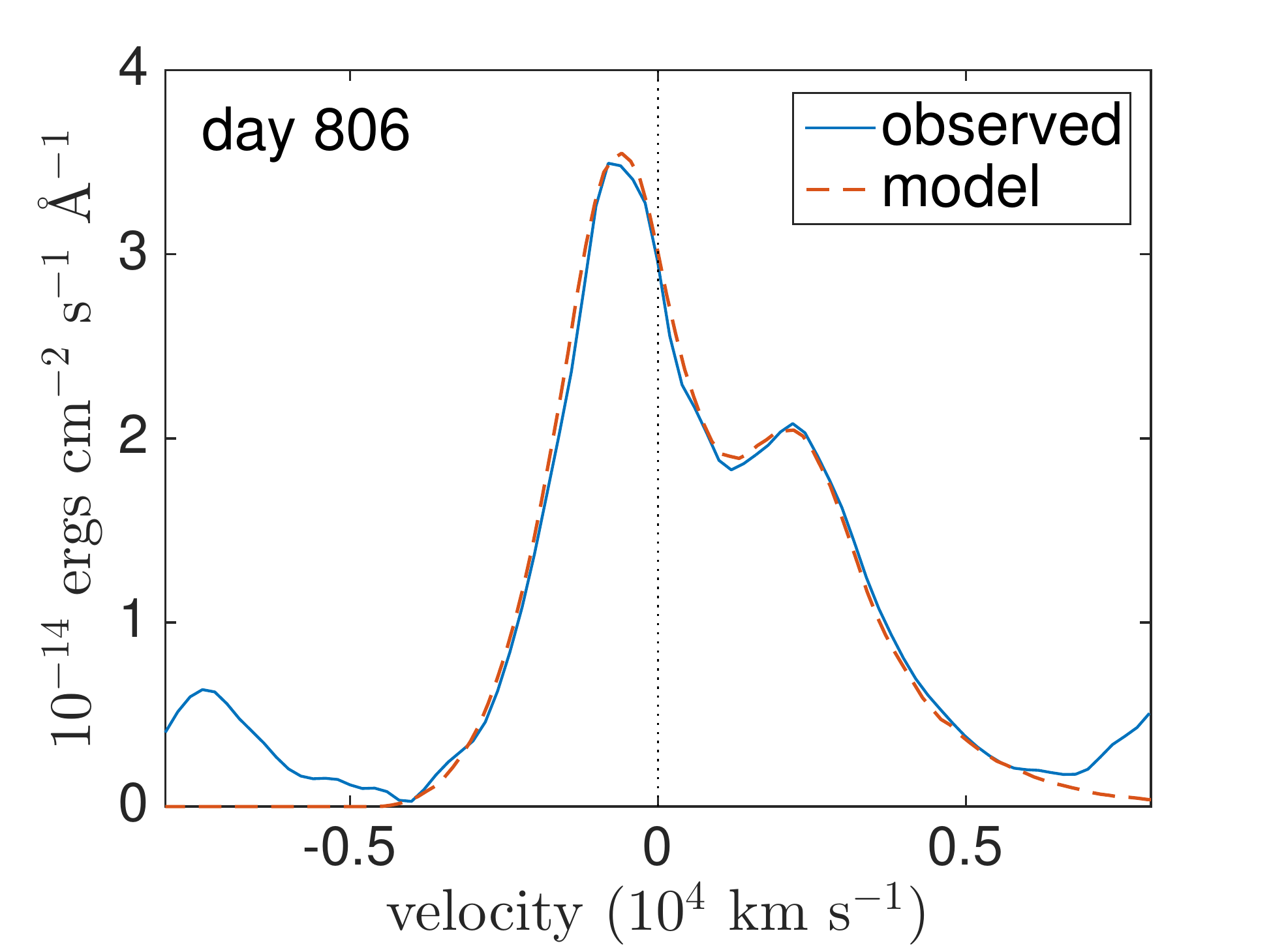}
\includegraphics[trim =0 35 50 0,clip=true,scale=0.24]{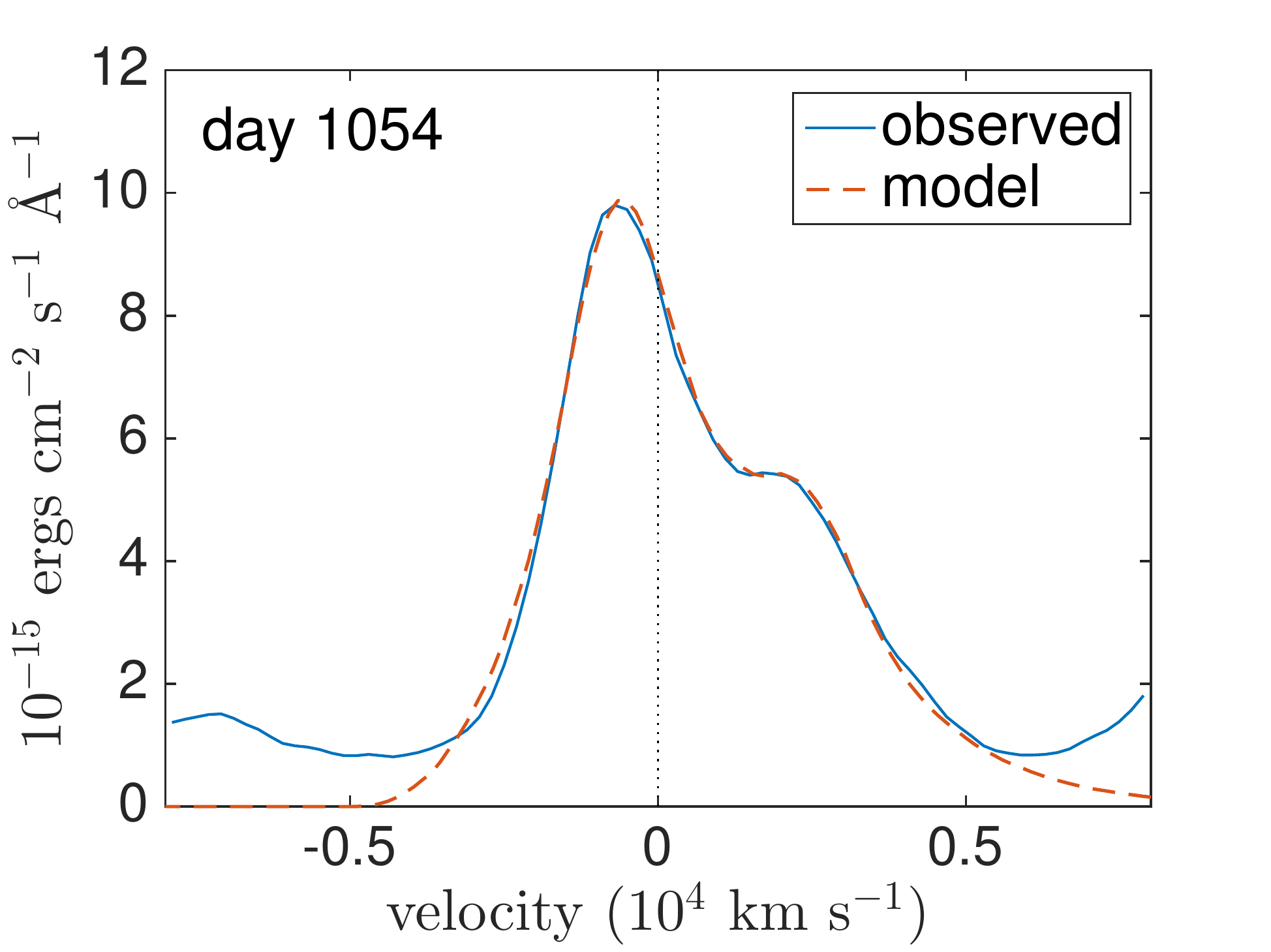}
\includegraphics[trim =0 35 40 0,clip=true,scale=0.24]{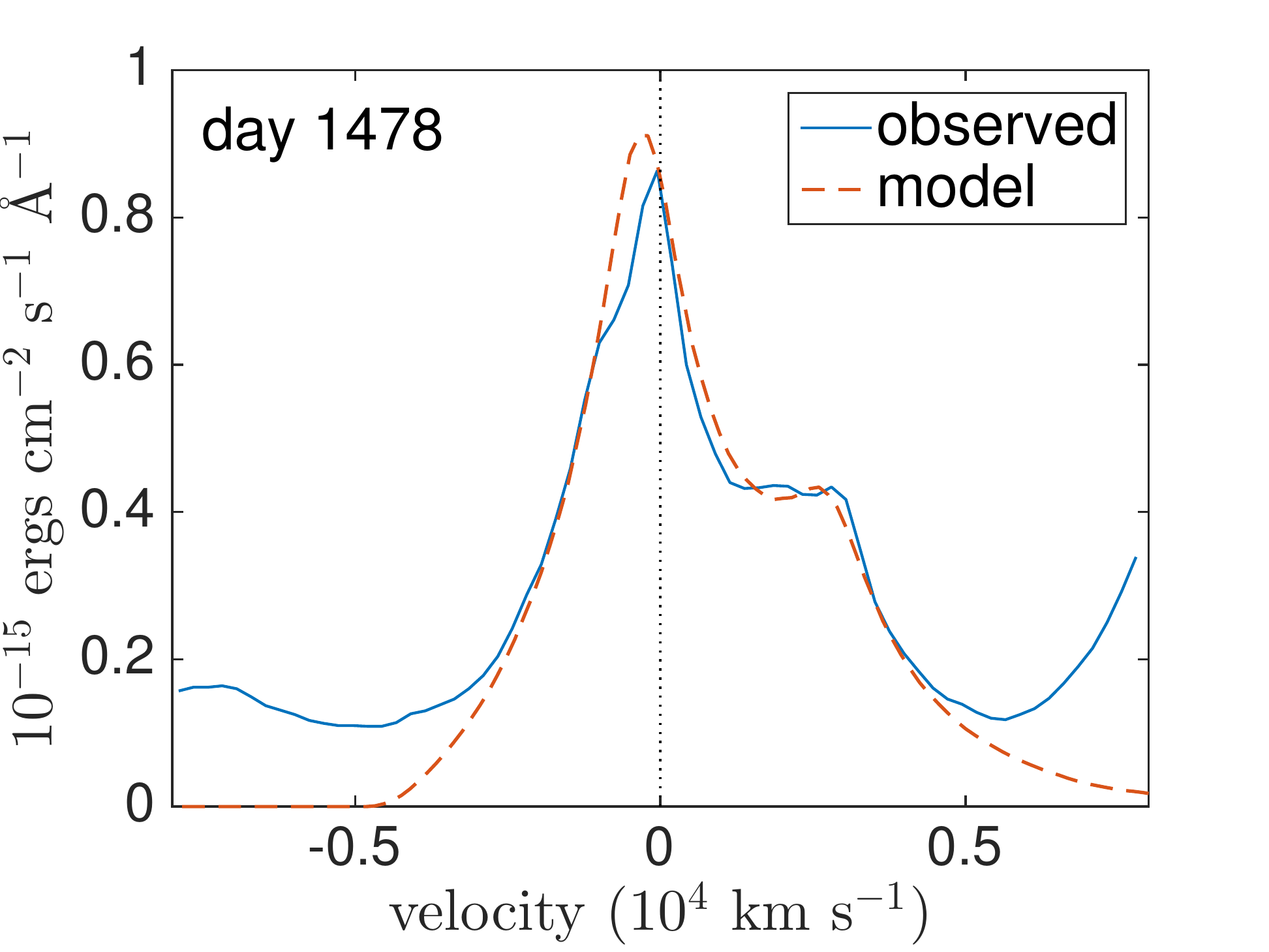}

\includegraphics[trim =0 35 50 -20,clip=true,scale=0.24]{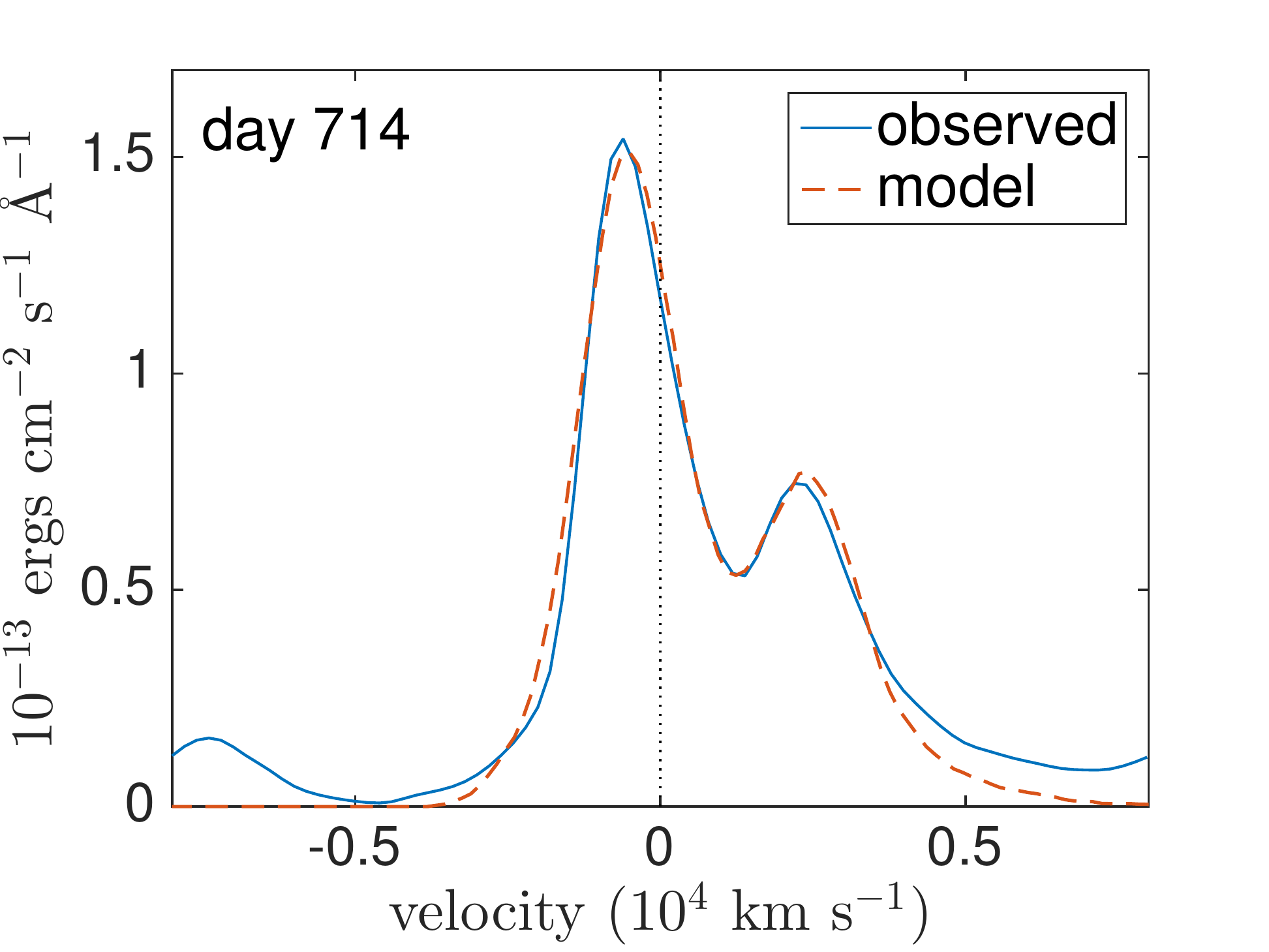}
\includegraphics[trim =0 35 50 -20,clip=true,scale=0.24]{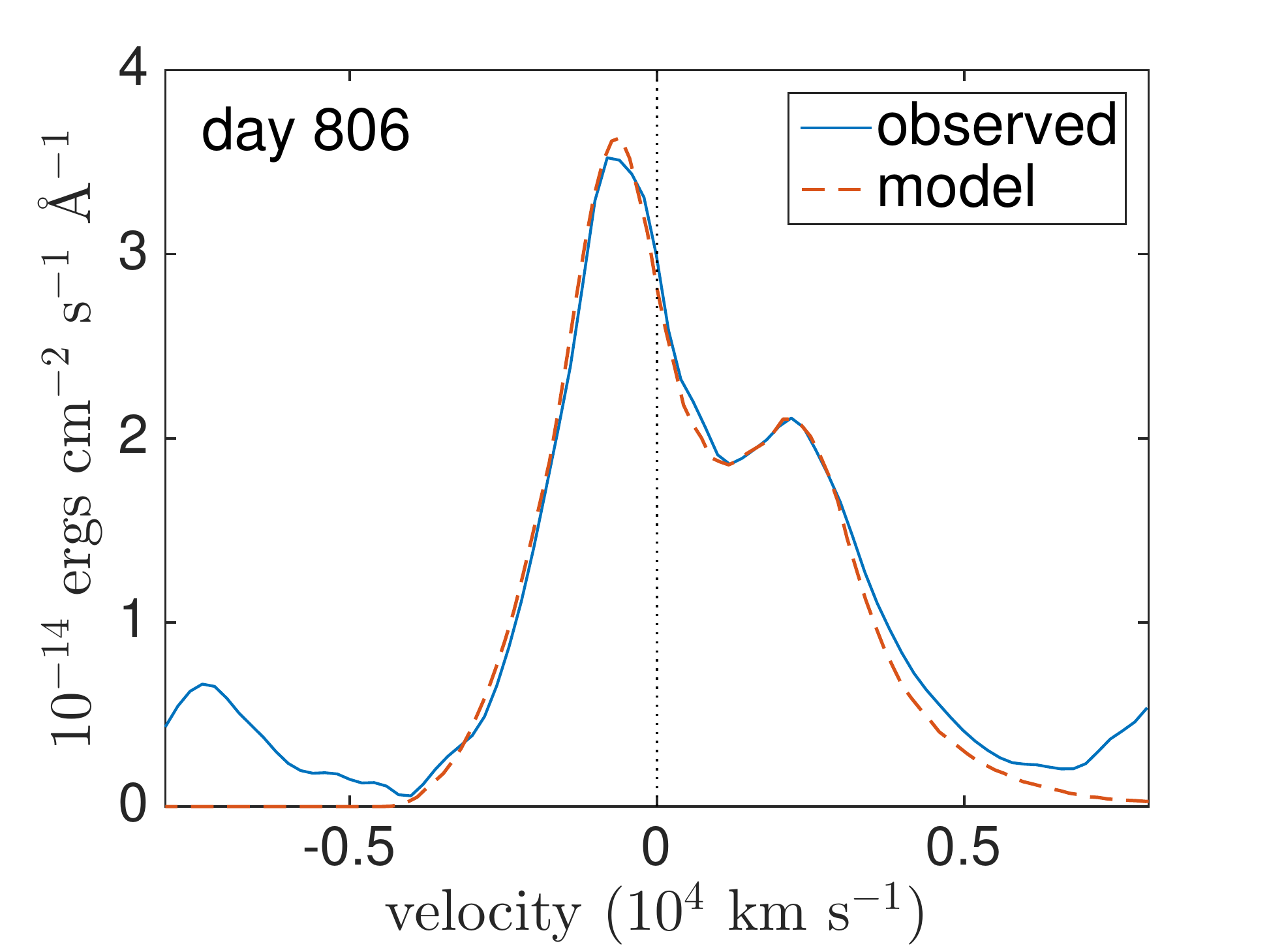}
\includegraphics[trim =0 35 50 -20,clip=true,scale=0.24]{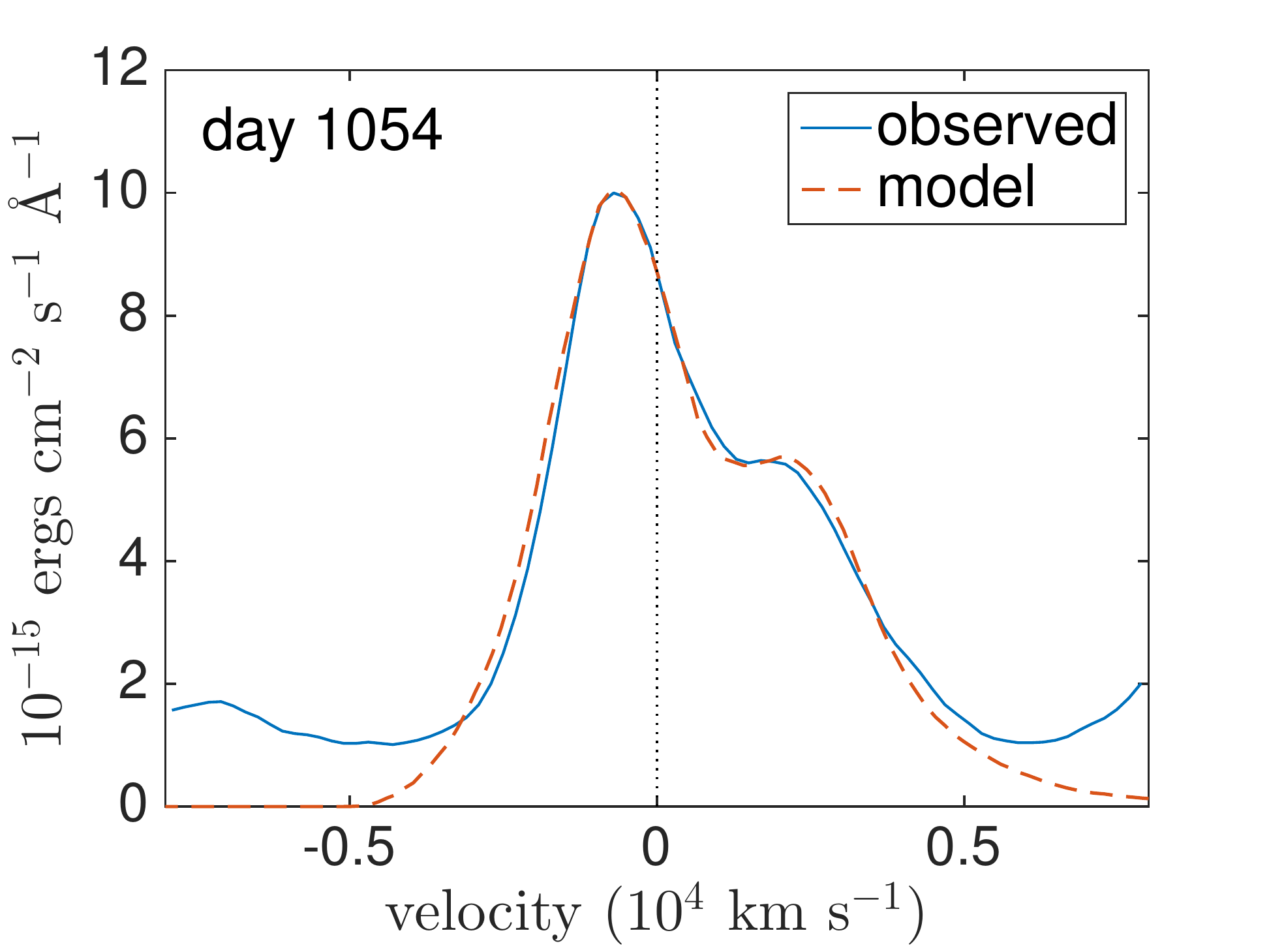}
\includegraphics[trim =0 35 50 -20,clip=true,scale=0.24]{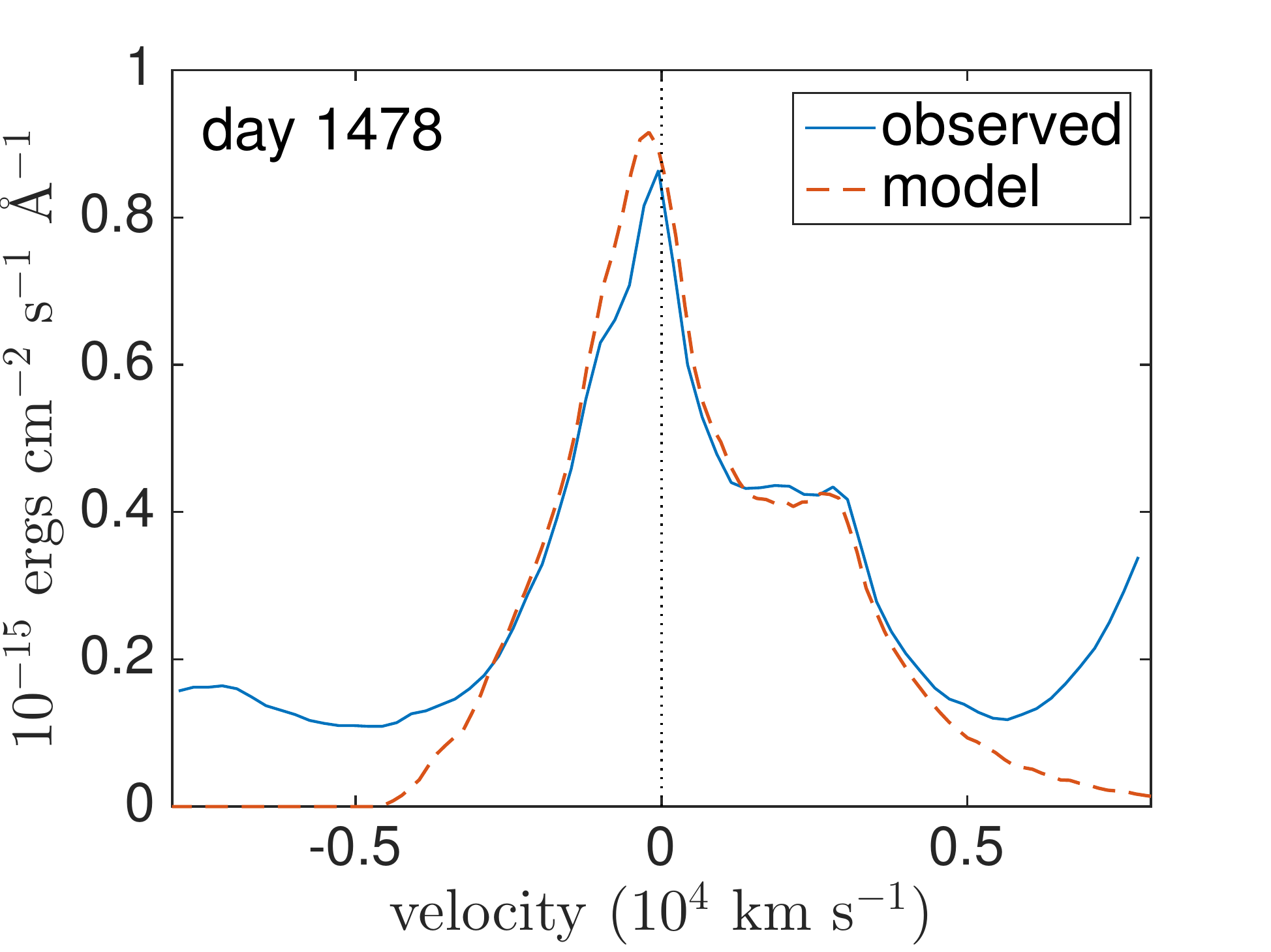}

\includegraphics[trim = 0 0 50 -20,clip=true,scale=0.24]{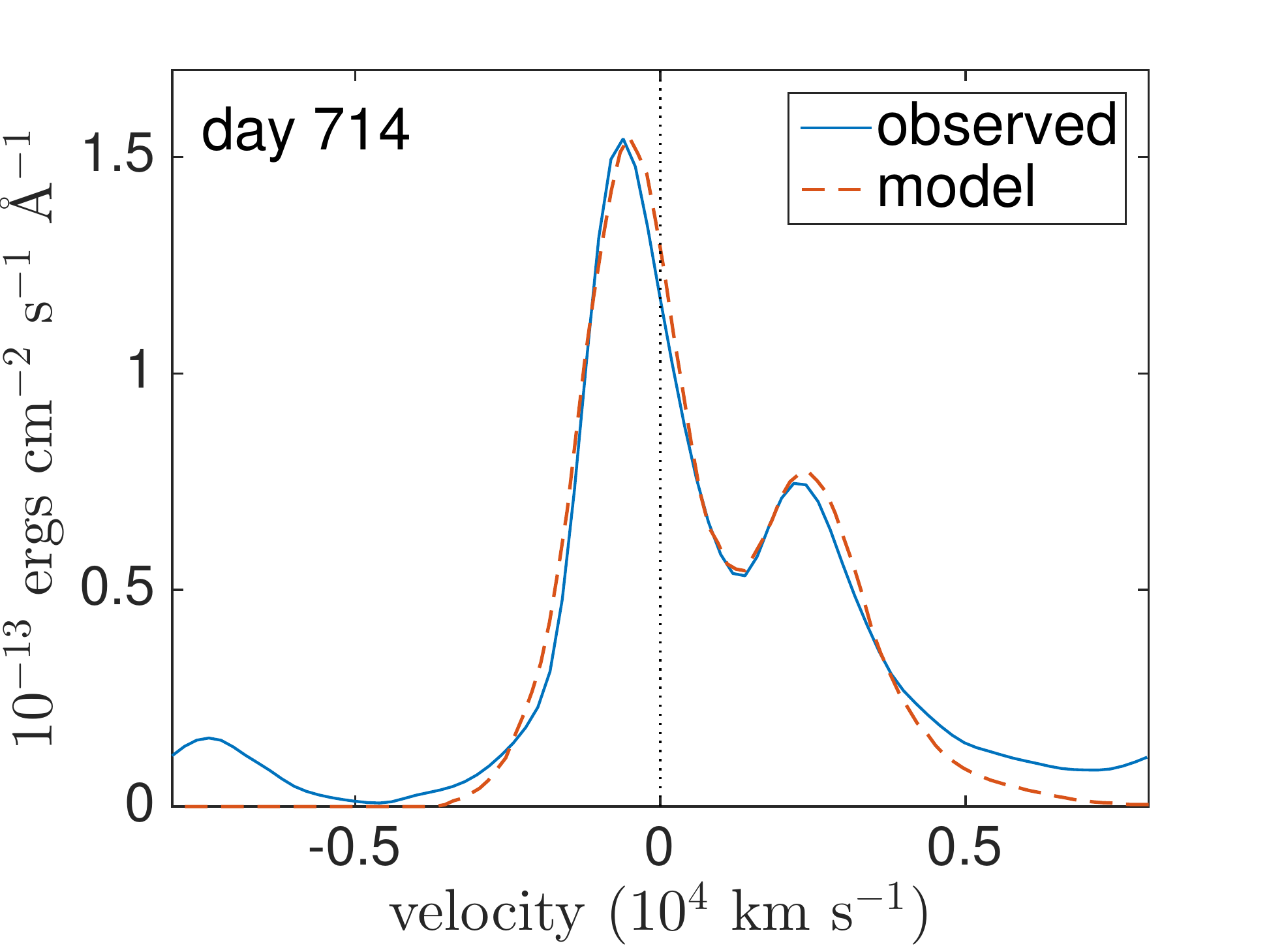}
\includegraphics[trim =0 0 50 -20,clip=true,scale=0.24]{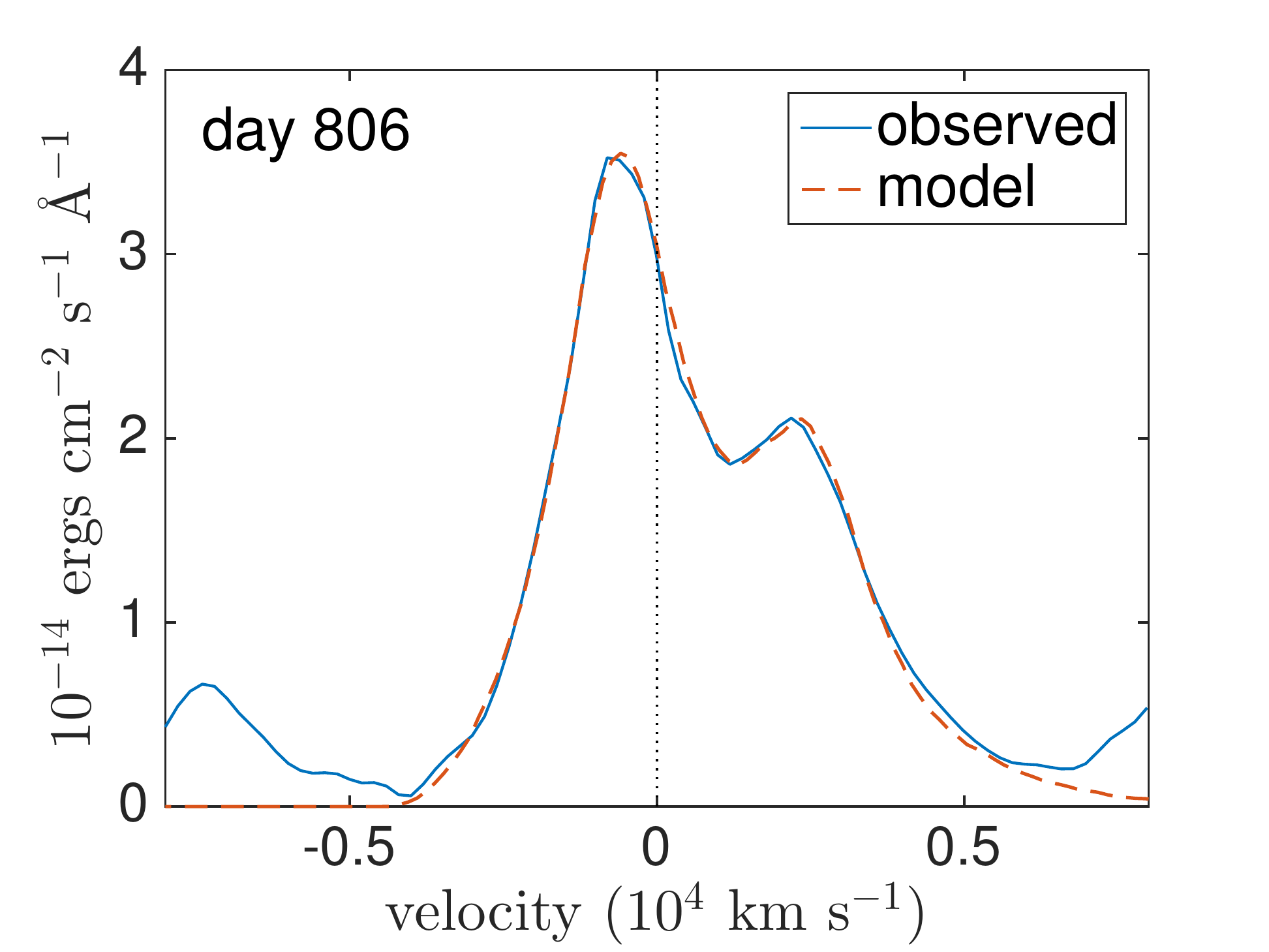}
\includegraphics[trim =0 0 50 -20,clip=true,scale=0.24]{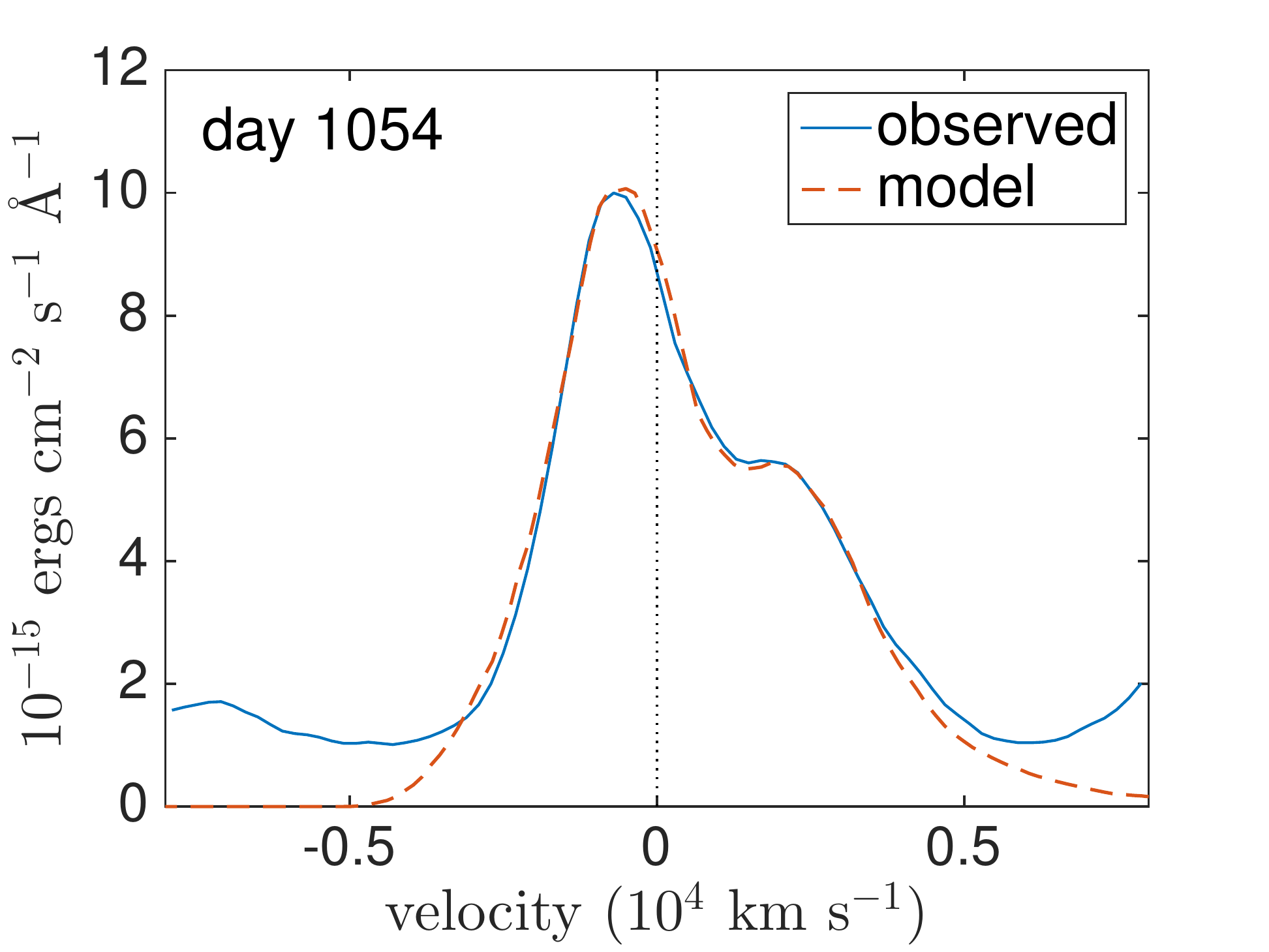}
\includegraphics[trim =0 0 50 -20,clip=true,scale=0.24]{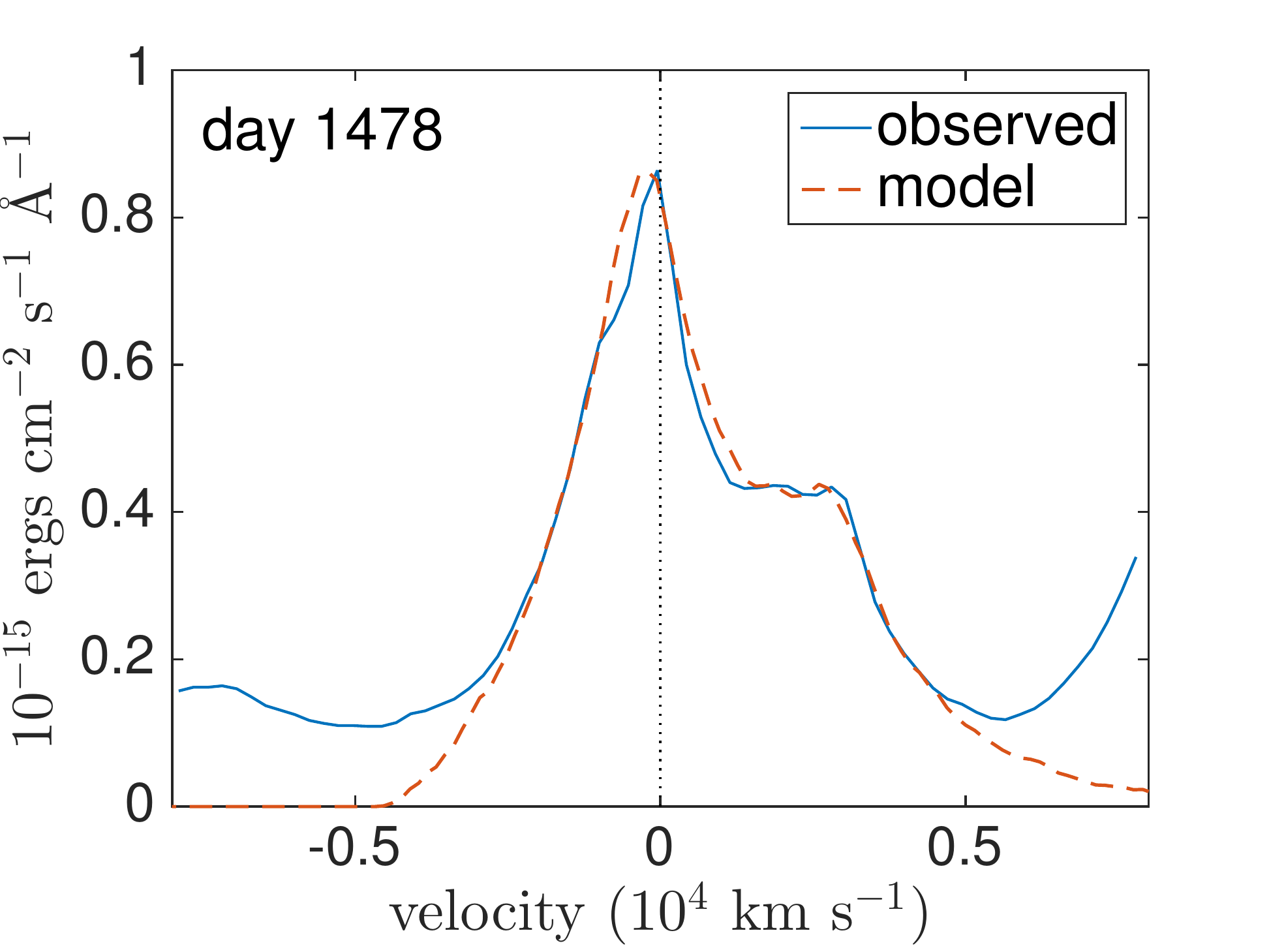}

\caption{Best model fits to the SN~1987A [O~{\sc i}]~$\lambda$6300,6363~\AA\ doublet at days 714, 806, 1054 and 1478 for the parameters detailed in Tables \ref{smooth1} -- \ref{clumped2}.  On the top row are smooth dust fits with amorphous carbon grains of radius $a=0.35 \mu$m.  On the middle and bottom rows are clumped dust fits with amorphous carbon grains of radii $a=0.6 \mu$m and $a=3.5 \mu$m respectively.}
\label{OI_smooth}

\end{figure*}

Models were again constructed using the smallest possible grain radius (a=0.6~$\mu$m in the clumped case) in order to derive minimum dust masses 
for clumped distributions.  By considering the extent of the red 
scattering wing, upper limits to the grain size were also derived with the 
purpose of limiting the maximum dust mass at each epoch.  By steadily 
reducing the grain radius from an initial value of 5~$\mu$m (motivated by 
the maximum possible grain size derived by W15 for their day 8515 model), 
we produced a set of models with a maximum grain radius of $a=3.5~\mu$m.  

The increase in grain size from the smooth case to the clumped case is 
necessary in order to have a slightly larger albedo.  Grains of radius 
$a=0.35~\mu$m do not reproduce the red side of the profiles well for a 
clumped medium.  This is because when the dust is located in clumps the 
radiation is subject to less scattering as well as to less absorption.  
The reduction in scattering appears not to be compensated for by the 
increased dust mass and a larger grain radius is therefore required, 
particularly at day 714.

For all but the H$\alpha$ line at days 714 and 806 a similar fit could be 
obtained with either a grain radius of $a=0.6~\mu$m or $a=3.5~\mu$m (see 
Figs \ref{Ha} -- \ref{OI_smooth}).  However, for H$\alpha$ 
at days 714 and 806 even a small change to the grain radius from 0.6~$\mu$m resulted in a 
significantly poorer fit, either over-estimating or under-estimating the red wing. 
We therefore conclude that the dust mass estimates produced for the 
H$\alpha$ lines at days 714 and 806 for a grain radius of $a=0.6~\mu$m are 
the best H$\alpha$-based estimates of the dust mass at this epoch.

In our subsequent analyses, we adopt the values derived from our clumped 
models.  Details of the parameters used are presented in 
Tables \ref{clumped1} and \ref{clumped2} and the fits are presented in Figures 
\ref{Ha} and \ref{d1862_3604}.

\noindent
\setlength{\tabcolsep}{16pt}
\begin{table*}
\centering
\caption{MSEs illustrating the variation in goodness of fit for the H$\alpha$ line profile for a range of dust masses with other parameters fixed at their best-fitting values for the clumped model with $a=0.6\mu$m as detailed in Table \ref{clumped1}.  The MSE is calculated between $-5000$ and $+7000$~km~s$^{-1}$ for the day 714 H$\alpha$ profile and between $-8000$ and $+8000$~km~s$^{-1}$ for the day 2875 H$\alpha$ profile.  A factor of zero represents the dust-free model.  The best-fitting model is italicized.}
\begin{tabular}{l c c c c c c}
\hline
& \multicolumn{6}{c}{\textit{multiple of best-fitting mass}}\\
\hline
& 0 & 0.1 & 0.5 & \textit{1.0} & 2.0 & 10 \\
\hline
Day 714 MSE ($10^{-13}$~erg~cm$^{-2}$~s$^{-1}$)  &0.167 & 0.133 &0.043 & \textit{0.005} &0.115&1.15\\
Day 2875 MSE ($10^{-15}$~erg~cm$^{-2}$~s$^{-1}$) &0.0791 & 0.0604 & 0.0258 &\textit{0.0182}&0.0563&0.288 \\
\hline
\end{tabular}
\label{MSE_mass}
\end{table*}%

\setlength{\tabcolsep}{20pt}
\begin{table*}
\centering
\caption{Mean square errors illustrating the variation in goodness of fit for the H$\alpha$ line profile for a range of density profiles with other parameters fixed at their best-fitting values for the clumped model with $a=0.6\mu$m as detailed in Table \ref{clumped1}. The MSE is calculated between $-5000$~km~s$^{-1}$ and $+7000$~km~s$^{-1}$ for the day 714 H$\alpha$ profile and between $-8000$~km~s$^{-1}$ and $+8000$~km~s$^{-1}$ for the day 2875 H$\alpha$ profile. The best-fitting model is italicised.}
\begin{tabular}{l c c c c c}
\hline
& \multicolumn{5}{c}{\textit{density profile exponent ($\beta$)}}\\
\hline
& 1.0 & 1.2 & \textit{1.4} & 1.6 & 1.8 \\
\hline
Day 714 MSE ($10^{-13}$~erg~cm$^{-2}$~s$^{-1}$)  & 0.0328 & 0.0117 & \textit{0.005} & 0.0184 & 0.0410\\
\hline
\\
\hline
& 1.6 & 1.8 & \textit{2.0} & 2.2 & 2.4 \\
\hline
Day 2875 MSE ($10^{-15}$~erg~cm$^{-2}$~s$^{-1}$) & 0.0282 & 0.0205 & \textit{0.0182} & 0.0193 & 0.0255 \\
\hline
\end{tabular}

\label{MSE_beta}
\end{table*}%

\subsection{Goodness of fit}

We detailed at the start of Section \ref{results} the process by which parameters were constrained in order to obtain good fits to the data.  These fits were judged both by eye and by minimizing the MSE between the model and the observed data for each line profile.  For those interested in the sensitivity of the fits to various parameters, in Tables \ref{MSE_mass} and \ref{MSE_beta} we detail the mean square error (MSE) for the H$\alpha$ profile at days 714 and 2875 for a range of dust masses and density profile exponents.  All other parameters were kept fixed at their best-fitting values for the clumped models of H$\alpha$ with a grain radius $a=0.6\mu$m as in Table \ref{clumped1}.  The MSE is calculated as
\begin{equation}
\frac{1}{N} \sum_i (f_{{\rm obs},i} - f_{{\rm mod},i})^2
\end{equation}

where $N$ is the number of data points, $f_{{\rm obs},i}$ is the observed flux at the $i^{\rm th}$ data point 
and $f_{{\rm mod},i}$ is the modelled flux at the $i^{\rm th}$ data point. The MSEs were calculated between $-5000$ and $+7000$~km~s$^{-1}$ for the day 714 H$\alpha$ profile and between $-8000$ and $+8000$~km~s$^{-1}$ for the day 2875 H$\alpha$ profile.  Note that the MSEs should only be compared between models for a given observed line profile and not between different line profiles since each observation is associated with a different inherent error.

For day 714, we find that increasing or decreasing the total dust mass by a factor of two with all other parameters fixed causes a substantial increase in the mean square error (by factors of 23 and 8.6 respectively) effectively ruling out these values.  For day 2875 a similar variation is seen but with the MSE varying by factors of 1.4 and 3.0 for each case.  The narrower range of MSEs at day 2875 compared to day 714 is due to a noisier profile which results in a greater allowed range of good fits.   The sensitivity of the goodness of fit to the dust mass and density profile is similar for the other modelled epochs.

\subsection{The effects of clumping}

As in the case of SED radiative transfer models, the dust masses required 
to reproduce the observations in the clumped scenario are considerably 
higher than for the smooth scenario.  The dust masses differ between our 
smooth models for $a=0.35~\mu$m and clumped models for $a=0.6\mu$m by a 
factor of approximately 3.  The dust mass estimates are even larger when
comparing clumped $a=0.6~\mu$m models to clumped $a=3.5~\mu$m models at 
later epochs. This does not take into account the increase in grain radius 
between the two cases however.  This increase accounts for a reasonable 
fraction of this difference. We estimate the effects of clumping alone to 
increase the required dust mass by a factor of approximately 1.5-2.0 from 
the smooth case.

\subsection{More complex models}
\label{complex}

Where blueshifted lines are observed in the spectra of CCSNe it is often the case that the Balmer lines of H{\sc I} are less affected than the [O~{\sc i}] lines \citep{Milisavljevic2012}.  This may be due to a difference in the location or distribution of the emitting elements; if the neutral hydrogen was diffusely distributed throughout the envelope but the oxygen was co-located with the dust in the core and in clumps then this could result in [O~{\sc i}] emission undergoing greater attenuation than H$\alpha$.  This geometry would be in line with previous models of SN~1987A that suggested that the dust-forming regions are likely to include those which are oxygen-rich \citep{Kozma1998a}.  Clearly, any model of dust formation in the ejecta of a CCSN must consistently reproduce all of the line profiles at a given epoch.  The models presented in this paper thus far have coupled the gas and dust distributions for a fixed clump volume filling factor and clump size.  The H$\alpha$ and [O~{\sc i}] models therefore require different dust masses with the [O~{\sc i}] models usually requiring a dust mass $\sim4$ times larger than the H$\alpha$ models. 

We now present a model that reconciles this difference by additionally varying the clump filling factor, clump size and emissivity distribution.  We assume that neutral hydrogen is likely diffuse throughout the ejecta and so maintains a smoothly distributed power-law emissivity distribution between $R_{\rm in}$ and $R_{\rm out}$ for H$\alpha$.  However, we now assume that dust mostly forms in dense regions of high metallicity and so restrict the [O~{\sc i}]$\lambda$6300,6363~\AA\ emission to originate largely from the dusty clumps with only a small fraction emitted from the inter-clump medium.  As previously discussed, the greater the covering factor of the dust the greater the albedo required in order to reproduce the H$\alpha$ red scattering wing. In order to obtain both the strong blueshifting of the [O~{\sc i}] line and the extended red scattering wing observed in H$\alpha$ a small number of dense clumps were required along with a small mass of diffusely distributed highly scattering dust in the inter-clump medium.

\begin{figure}
\centering
\includegraphics[scale=0.45,clip=true]{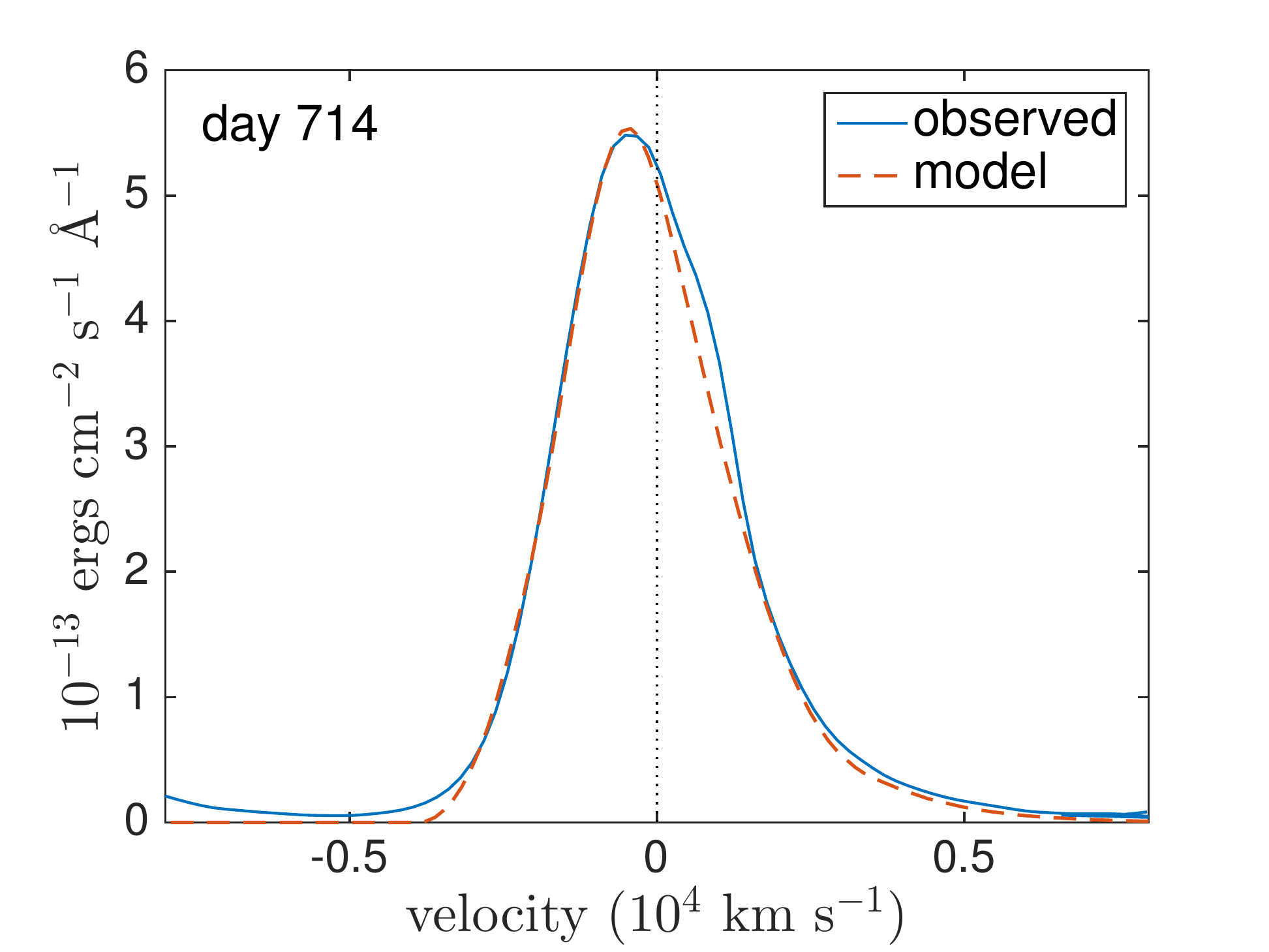}
\includegraphics[scale=0.45, clip=true]{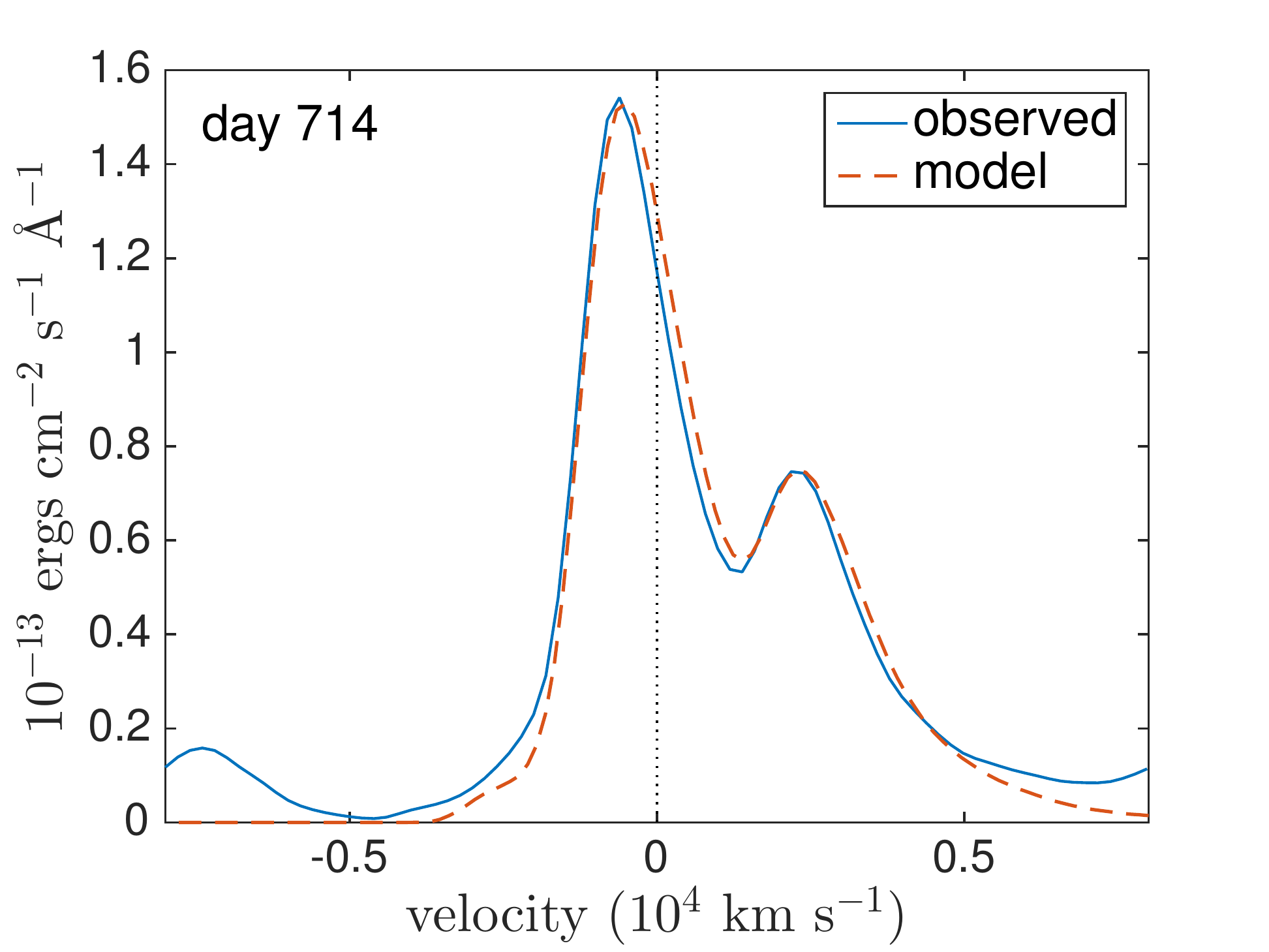}
\caption{Fits to the H$\alpha$ and [O~{\sc i}]$\lambda$6300,6363~\AA\ lines at day 714 using the more complex dust model described in Section \ref{complex} with a dust mass of  $2.3 \times 10^{-4}$M$_{\odot}$.}
\label{HaOImod}
\end{figure}

In order to fit both line profiles simultaneously, we required a very high albedo ($\omega > 0.8$) that demanded the inclusion of some fraction of silicate dust. Amorphous carbon grains alone are incapable of producing this level of scattering for any grain size.  We adopted a grain radius of $a=0.6\mu$m, the same as that used in our initial clumped models and we varied the relative proportions of amorphous carbon and MgSiO$_3$ in order to achieve the necessary albedo.  The adopted grain densities were $\rho_{\rm c}=1.85$~g~cm$^{-3}$ for amorphous carbon grains and $\rho_{\rm s} = 2.71$~g~cm$^{-3}$ for MgSiO$_3$.  The resulting dust model for day 714 used 75\% MgSiO$_3$ and 25\% amorphous carbon by cross-sectional area with a volume filling factor $f_V=0.1$ and a clump size $R_{\rm out}/5$.  90\% of the dust mass was located in clumps with the remaining 10\% distributed smoothly between $R_{\rm in}$ and $R_{\rm out}$ according to a power law $\rho \propto r$. Clumps were distributed stochastically with probability $\propto r^{-8}$ compared to $r^{-2.7}$ in our standard models discussed earlier. Equal numbers of [O~{\sc i}] packets were emitted from each clump. The increased steepness of the density profile is required to compensate for the clumped packet emission relative to the previous smooth distribution.  Since the clumps are distributed stochastically according to the density profile, less flux is emitted from the central regions in a clumped emission model than in a smooth distribution model (since there are gaps between the clumps).  In order to obtain a sufficiently steeply rising line profile, the density profile must therefore be steepened in clumped emission models. The adopted value of $\beta$ does not significantly affect the best-fitting values of the other parameters of interest however. H$\alpha$ was distributed smoothly according to a density power law $\rho(r) \propto r^{-1.3}$.  $R_{\rm out}$ was the same for all components (i.e. clumped dust, diffuse dust, [O~{\sc i}] emission and H$\alpha$ emission) and was calculated using a maximum velocity of 3250~km~s$^{-1}$.  The inner radius was $R_{\rm in} = 0.07 R_{\rm out}$ for all components except the smooth H$\alpha$ emission which was emitted between $R_{\rm in}=0.25R_{\rm out}$ and $R_{\rm out}$.  

The total dust mass used was $M_{\rm dust}=2.3 \times 10^{-4}$~M$_{\odot}$.  This dust mass is very similar to that derived from our original clumped models of [O~{\sc i}] using amorphous carbon grains of radius $a=0.6\mu$m.  The slight increase over our amorphous carbon dust mass of $1.5 \times 10^{-4}$~M$_{\odot}$ is largely due to the higher grain density of MgSiO$_3$.  At this grain radius amorphous carbon and MgSiO$_3$ have similar extinction efficiencies and so the change in species and geometry does not substantially alter the dust mass. We therefore adopt the [O~{\sc i}] dust masses in our further analyses and consider the differences in our derived dust masses between H$\alpha$ and [O~{\sc i}] to be the result of the clumped emission of [O~{\sc i}].

Fits to both the [O~{\sc i}]$\lambda$6300,6363~\AA\ and H$\alpha$ lines for day 714 using these parameters are presented in Fig. \ref{HaOImod}.


\subsection{The effect of a grain size distribution}
\label{gs_distn}

It is important to consider the potential effect on the dust mass of 
modelling a grain size distribution instead of a single grain size.  For a 
grain size distribution the overall extinction cross-section, $C_{\rm ext}$, 
at a given wavelength is

\begin{equation}
 C_{\rm ext}=\int^{a_{\rm max}}_{a_{\rm min}} Q_{\rm ext}(a) n(a) \pi a^2 da 
 \end{equation}

where $Q_{\rm ext}(a)$ is the extinction efficiency for a grain size $a$ and 
$n(a)$ is the number of grains with size $a$. The overall extinction 
efficiency is then

\begin{equation}
 Q_{\rm ext} = \frac{C_{\rm ext}}{ \int^{a_{\rm max}}_{a_{\rm min}} n(a) \pi a^2 da} 
 \end{equation}
 
The scattering cross-section $Q_{\rm sca}$ is similarly calculated.  As a 
result of these calculations, there is rarely a single grain size that has 
the same albedo and extinction efficiency as a size distribution.  
Modelling a size distribution may therefore alter the deduced dust mass.  
Since the models are only sensitive to the overall optical depth and 
albedo, it is not possible to deduce the grain size range or distribution 
and only single grain sizes are investigated (as presented above).

Whilst this apparently limits the scope of the results, it is useful to 
consider the extent to which different grain size distributions would 
alter the derived dust masses.  By considering a number of grain radius 
ranges and adopting a power law distribution with a variable exponent, we 
may gain some insight into the effects of adopting a distribution rather 
than a single size.  As discussed in Section \ref{smooth_models}, for a classical MRN power law ($n(a) \propto 
a^{-3.5}$) with a wide grain radius range ($a_{\rm min} = 0.001~\mu$m to 
$a_{\rm max} = 4.0~\mu$m) the derived albedo is much too small to reproduce 
the required wing seen at early epochs.  We therefore adopt an approach 
whereby, for a number of grain size ranges, we adjust the exponent of the 
distribution until the overall albedo is the same as that seen for the 
best fitting single grain radius for the clumped distributions.  We may 
then approximately calculate the required dust mass as

\begin{equation}
\label{distn_conv}
M_{\rm d}= \frac{M_{\rm s} Q_{\rm ext,s}(a_s)}{a_{\rm s}} \times \frac{\int^{a_{\rm max}}_{a_{\rm min}} n(a) a^3 da}{\int^{a_{\rm max}}_{a_{\rm min}} Q_{\rm ext}(a) n(a) a^2 da}
\end{equation}

where the subscript $s$ represents the single grain size quantities and 
the $d$ subscript represents quantities for the grain size distribution.

We calculate the required dust masses for the clumped H$\alpha$ model on 
day 714 for a selection of distributions with varying $a_{\rm min}$.  These 
are presented in Table \ref{tb_distn}.  It can be seen that in all cases, 
a larger dust mass is required for grain size distributions in order to reproduce the same profile as a 
single grain size.  The conversion factors presented in the table are 
valid for any model with grain size $a=0.6~\mu$m and may therefore also be 
applied to the models for day 806.  We repeated the process for 
$a=3.5~\mu$m but found that, in order to reproduce the required albedo, the 
distribution had to be heavily weighted towards the larger grains and that 
the value of $a_{\rm min}$ had no effect on the required dust mass.  
Increasing the value of $a_{\rm min}$ to larger values ($>2~\mu$m) does not 
have a significant effect either.  This is because both extinction 
efficiency and albedo tend to a constant value with increasing grain 
radius and the adoption of different grain size ranges and distributions 
above a certain threshold results in only insignificant variations in 
these quantities.
\setlength{\tabcolsep}{12pt}
\begin{table}
	\caption{Dust masses for day 714 clumped models of the H$\alpha$ 
line using different grain size distributions and 100\% amorphous carbon. The final column shows the factor of increase over the dust mass for the single size model ($M=7 \times 10^{-5} M_{\odot}$ with $a=0.6~\mu$m) and $p$ is the exponent of the grain size distribution $n(a) \propto a^{-p}$.}
	\label{tb_distn}
	\begin{center}
  	\begin{tabular}{@{} ccccc @{}}
    	\hline
$a_{\rm min}$ & $a_{\rm max}$ & $p$ & $M$ & $M/M_{0.6}$  \\
($\mu$m) & ($\mu$m) & & ($M_{\odot}$) & \\
\hline
0.001 & 4.0 & 2.45 & 1.93 $\times 10^{-4}$ & 2.76 \\
0.01 & 4.0 & 2.45 & 1.93 $\times 10^{-4}$ & 2.76 \\
0.05 & 4.0 & 2.52 & 1.84 $\times 10^{-4}$ & 2.62 \\
0.1 & 4.0 & 2.72 & 1.61 $\times 10^{-4}$ & 2.3\\ 
0.5 & 4.0 & 8.20 & 7.23 $\times 10^{-5}$ & 1.03 \\

    \hline
  \end{tabular}
  \end{center}
\end{table}
\setlength{\tabcolsep}{7pt}

We conclude that if a distribution of grain sizes is indeed present, the 
deduced single size dust masses are likely to under-estimate the true mass 
of newly formed dust.

\subsection{The effect of different grain species}
\label{species}

In our analyses so far, we have mostly focused on amorphous carbon as the 
species of interest.  This was motivated by previously published early epoch optical 
and IR SED analyses that found that the silicate mass fraction must be  limited to $\leq$15\% (\citet{Ercolano2007}, W15).  The recent suggestion by 
\citet{Dwek2015} that large masses of the  glassy silicate MgSiO$_3$ 
may have formed at early epochs is discussed further in the next 
subsection.  The parameters that affect the quantity of dust required by 
our models are the mean albedo and optical depth of the dust.  There could 
be multiple combinations of grain species and sizes that result in a good 
fit to the data.

We can evaluate the required change in dust mass when a medium of 100\% 
silicates is used instead of amorphous carbon. Using the astronomical 
silicate optical constants of \cite{Draine1984}, which are `dirtier' (with 
lower albedos) than the glassy pure MgSiO$_3$ sample of \citet{Jager2003}. 
In a similar manner to the approach detailed in Section \ref{gs_distn}, we 
can calculate the mass of DL silicate that gives a fit equivalent to that 
for a single carbon grain radius.  We consider the albedo for the grain 
radius needed for the best-fitting amorphous carbon model, calculate the 
equivalent grain radius for DL silicate that gives the same albedo and 
then calculate a new dust mass by allowing for the change in the 
extinction cross-section:

\begin{equation}
M_{\rm sil} = M_{\rm amc} \Big( \frac{Q_{\rm amc}}{Q_{\rm sil}} \Big) \Big(\frac{a_{\rm sil}}{a_{\rm amc}}\Big) \Big(\frac{\rho_{\rm sil}}{\rho_{amC}}\Big)
\end{equation}

Because of the nature of the variation of albedo with grain radius for the 
\citet{Draine1984} astronomical silicate (see Fig. \ref{albedo_grain}), 
there is often more than one silicate grain radius that will give rise to 
the same albedo at a given wavelength.  Some of the possibilities and the 
resulting mass conversion factors are given in Table \ref{tb_sil}.  For 
our best fitting amorphous carbon models with $a=0.6~\mu$m (the first two 
entries in Table \ref{tb_sil}), using any fraction of silicates with 
either $a=0.6~\mu$m or $a=3.5~\mu$m would increase the dust mass.  
However, 
for the case of an amorphous carbon grain radius of $a=3.5~\mu$m (the last 
three entries), using silicate dust would reduce the dust mass by a factor 
of about 2 relative to our amorphous carbon values.

\begin{table}
	\caption{Dust mass conversion factors for single size models using  
grains of 100\% Zubko BE amorphous carbon or 100\% Draine \& Lee silicate 
at $\lambda \sim 656$~nm. $f$ is the factor by which the dust mass 
changes on going from amorphous carbon to silicates.}
	\label{tb_sil}
	\begin{center}
  	\begin{tabular}{@{} cccccccc @{}}
    	\hline
	\multicolumn{3}{c}{\textit{carbon}} && \multicolumn{3}{c}{\textit{silicates}} & \\
$a$ &$\omega$ &  $Q_{\rm ext}$ & &$a$&$\omega$ & $Q_{\rm ext}$ & M$_{\rm sil}$/M$_{\rm amc}$ \\
($\mu$m) &&&&($\mu$m)\\
\hline
0.6 & 0.56 & 2.61 & &0.0583 & 0.58 &0.08 & 5.37 \\
0.6 &0.56 & 2.61 & &4.00 & 0.56 & 2.18 & 13.0 \\
 \\
3.5 & 0.62 &2.21 & &0.0641 & 0.64 & 0.10 & 0.65 \\
3.5 & 0.62 &2.21 & &1.020 & 0.63 & 2.15 & 0.49 \\
3.5 & 0.62 & 2.21 & &1.376 & 0.62 & 2.35 & 0.61 \\

    \hline
  \end{tabular}
  \end{center}
\end{table}

\begin{figure*}
\includegraphics[trim =0 28 40 0,clip=true,scale=0.25]{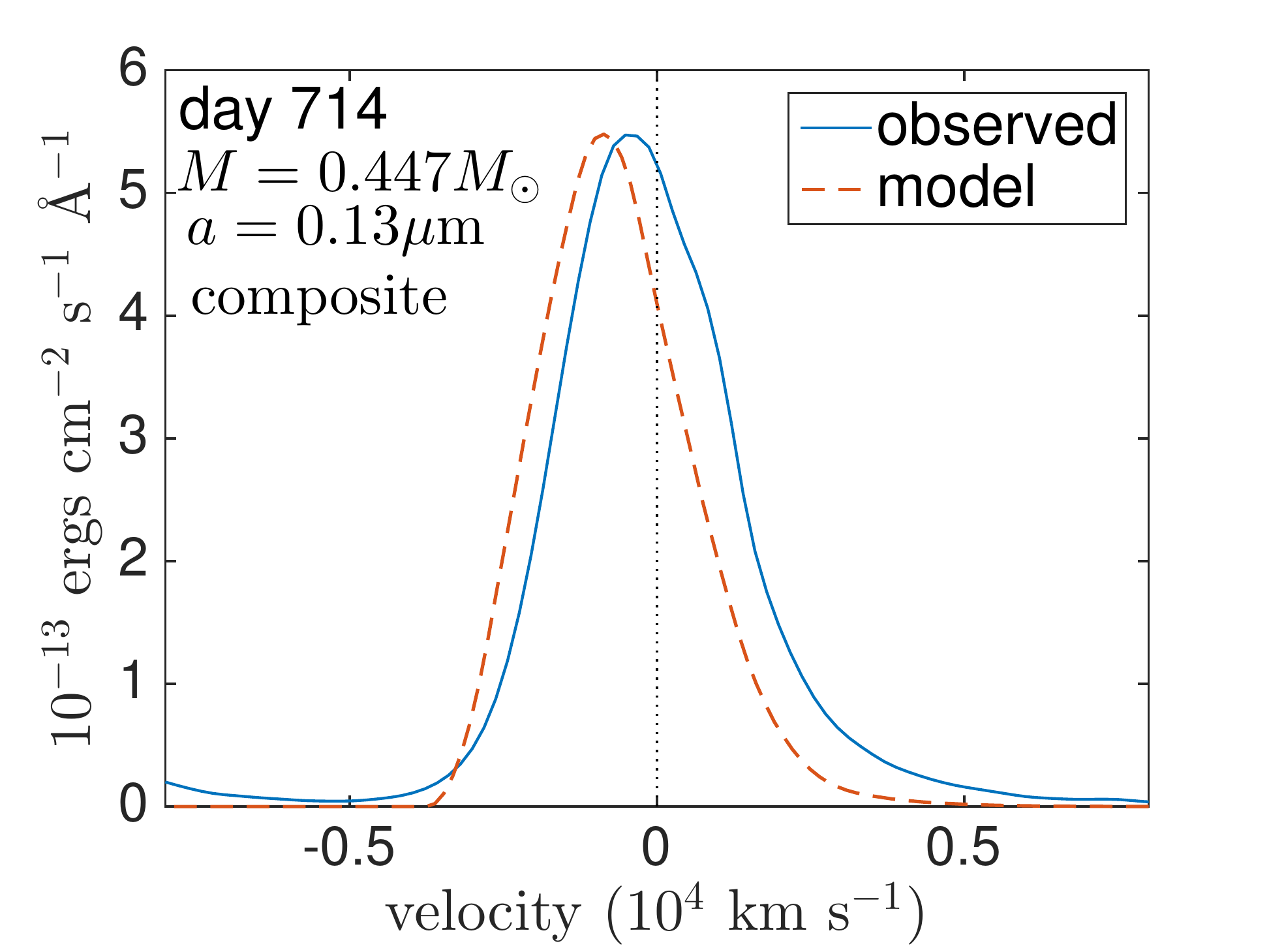}
\includegraphics[trim =29 28 -10 0,clip=true,scale=0.25]{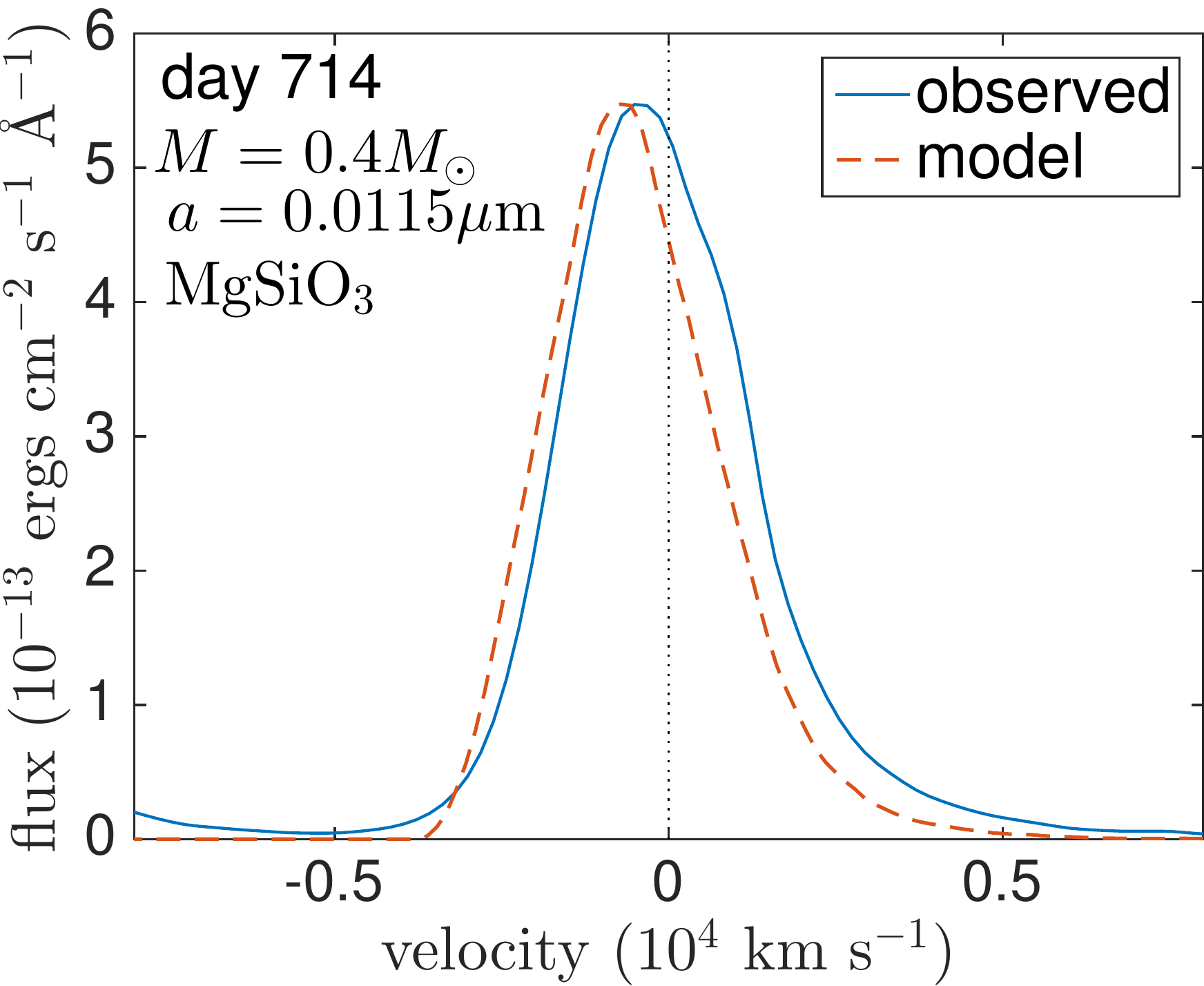}
\includegraphics[trim =29 28 -10 0,clip=true,scale=0.25]{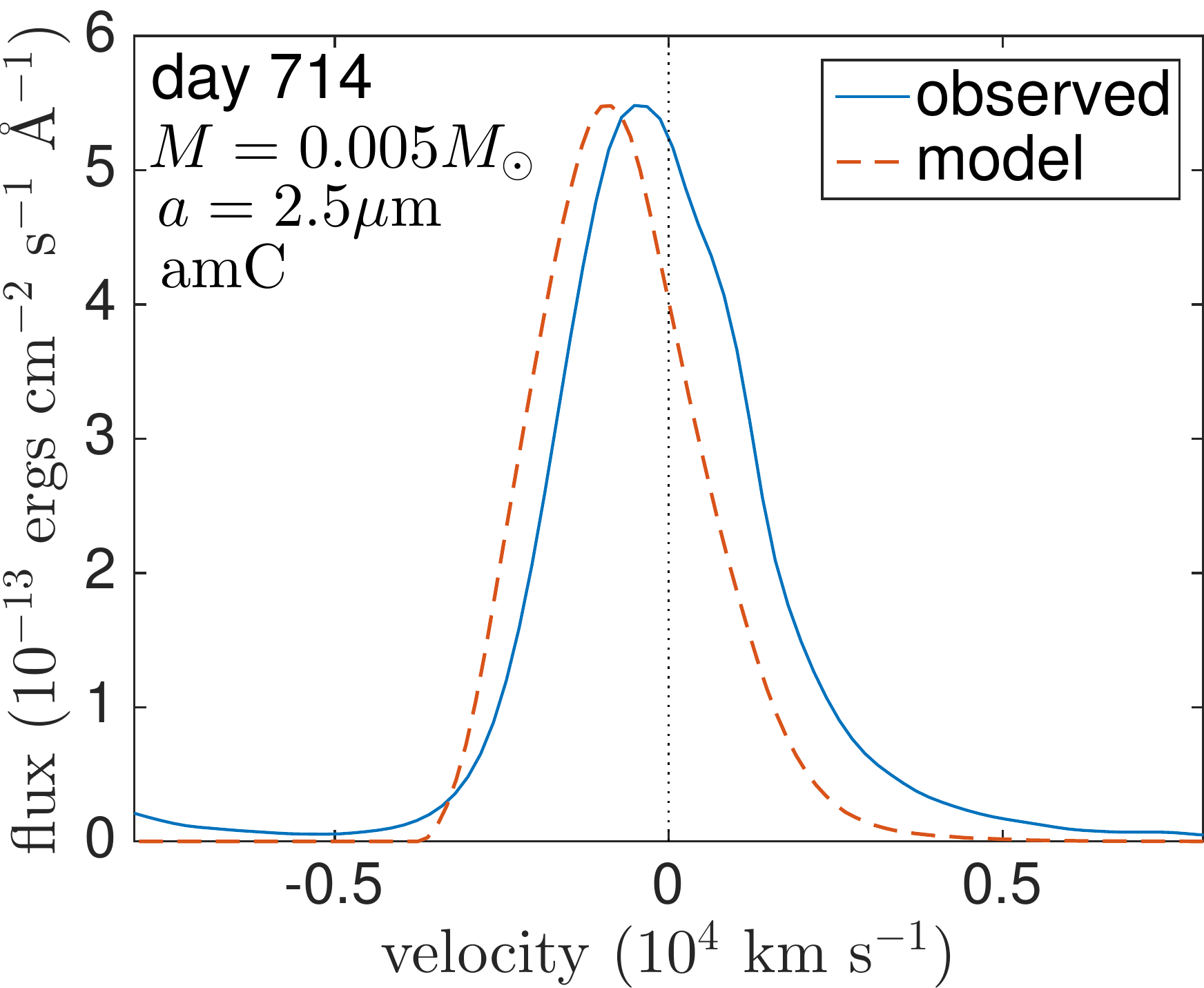}
\includegraphics[trim =29 28 0 0,clip=true,scale=0.25]{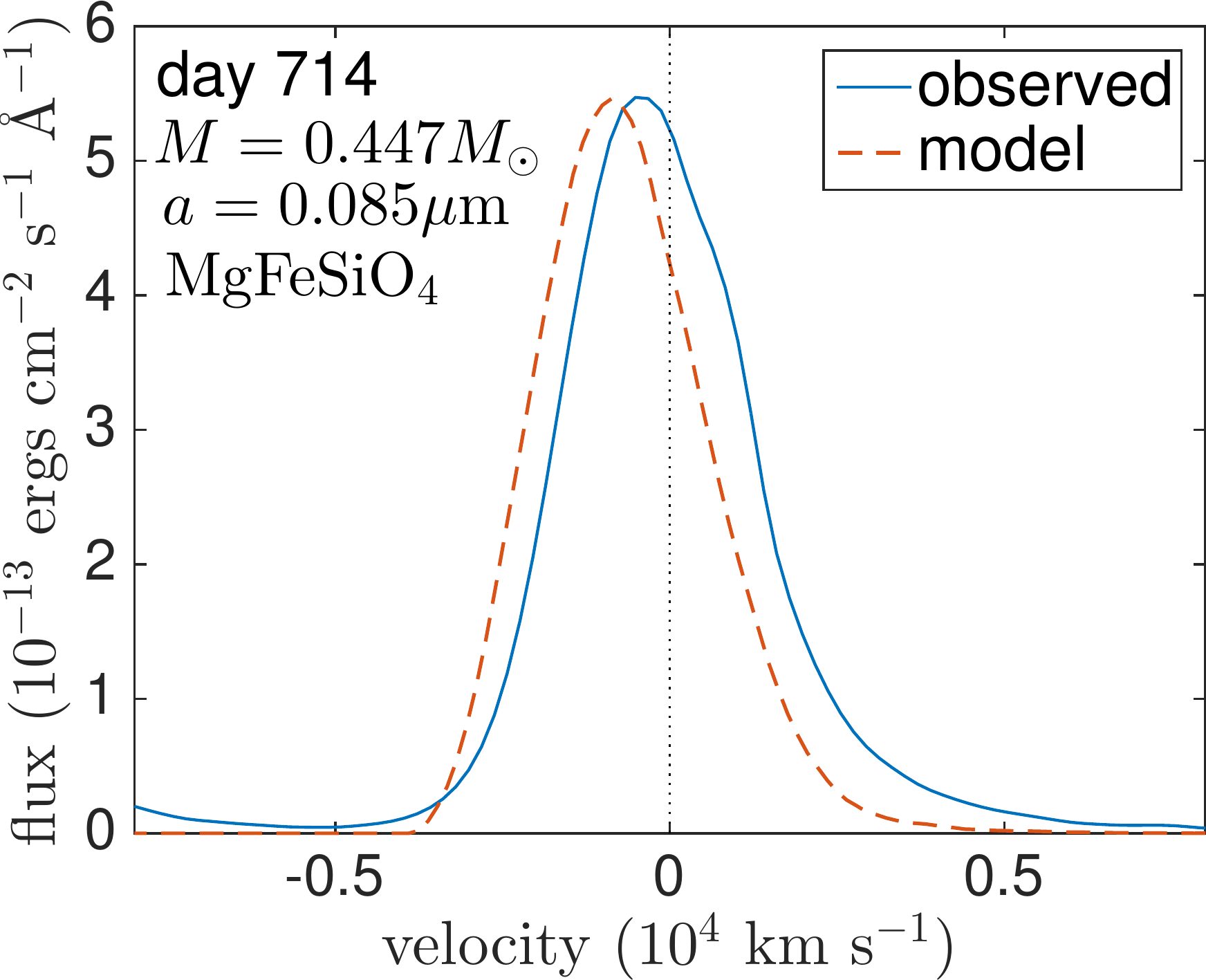}

\includegraphics[trim =0 0 40 -10,clip=true,scale=0.25]{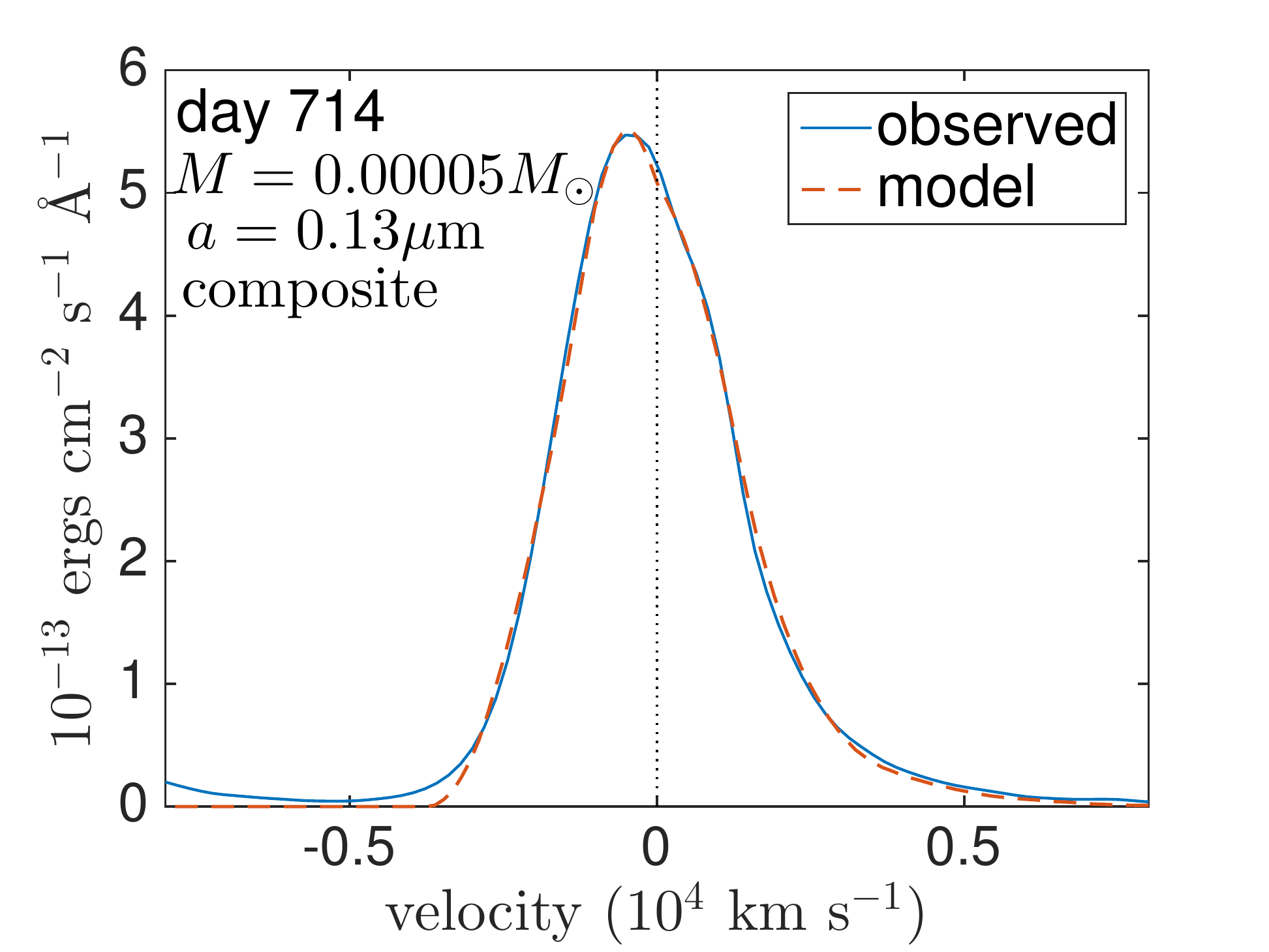}
\includegraphics[trim =29 0 -10 -10,clip=true,scale=0.25]{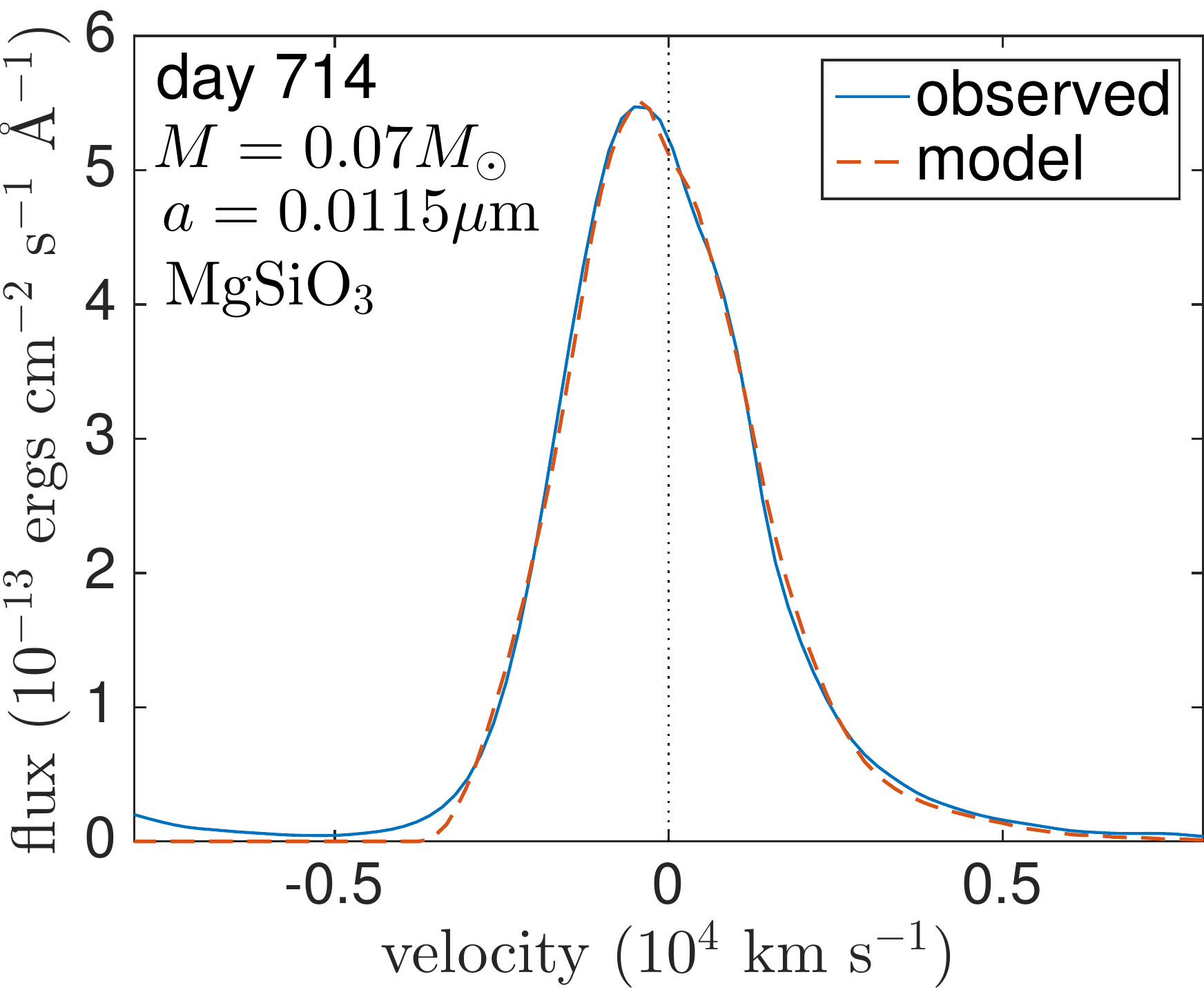}
\includegraphics[trim =29 0 -10 -10,clip=true,scale=0.25]{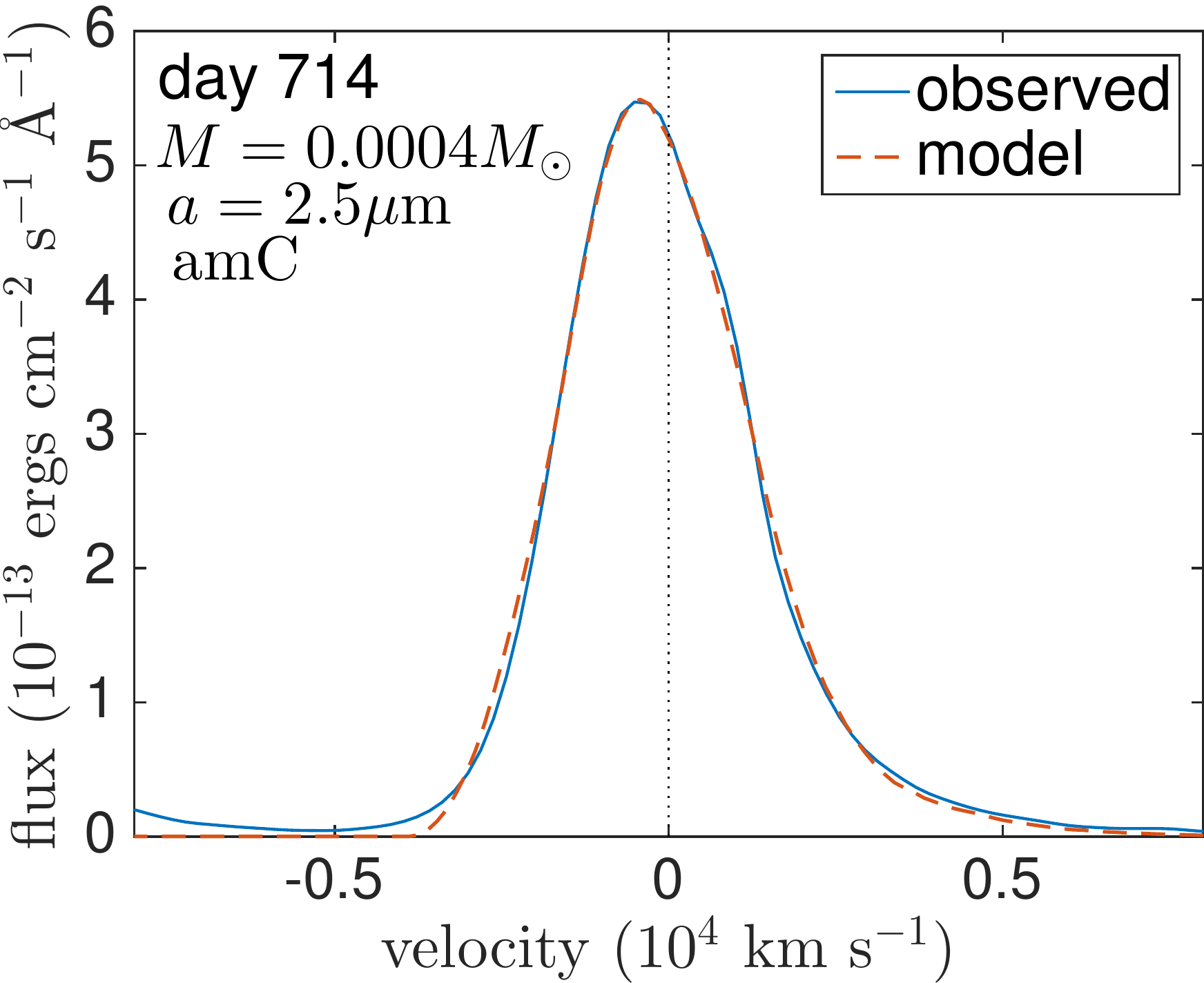}
\includegraphics[trim =29 0 0 -10,clip=true,scale=0.25]{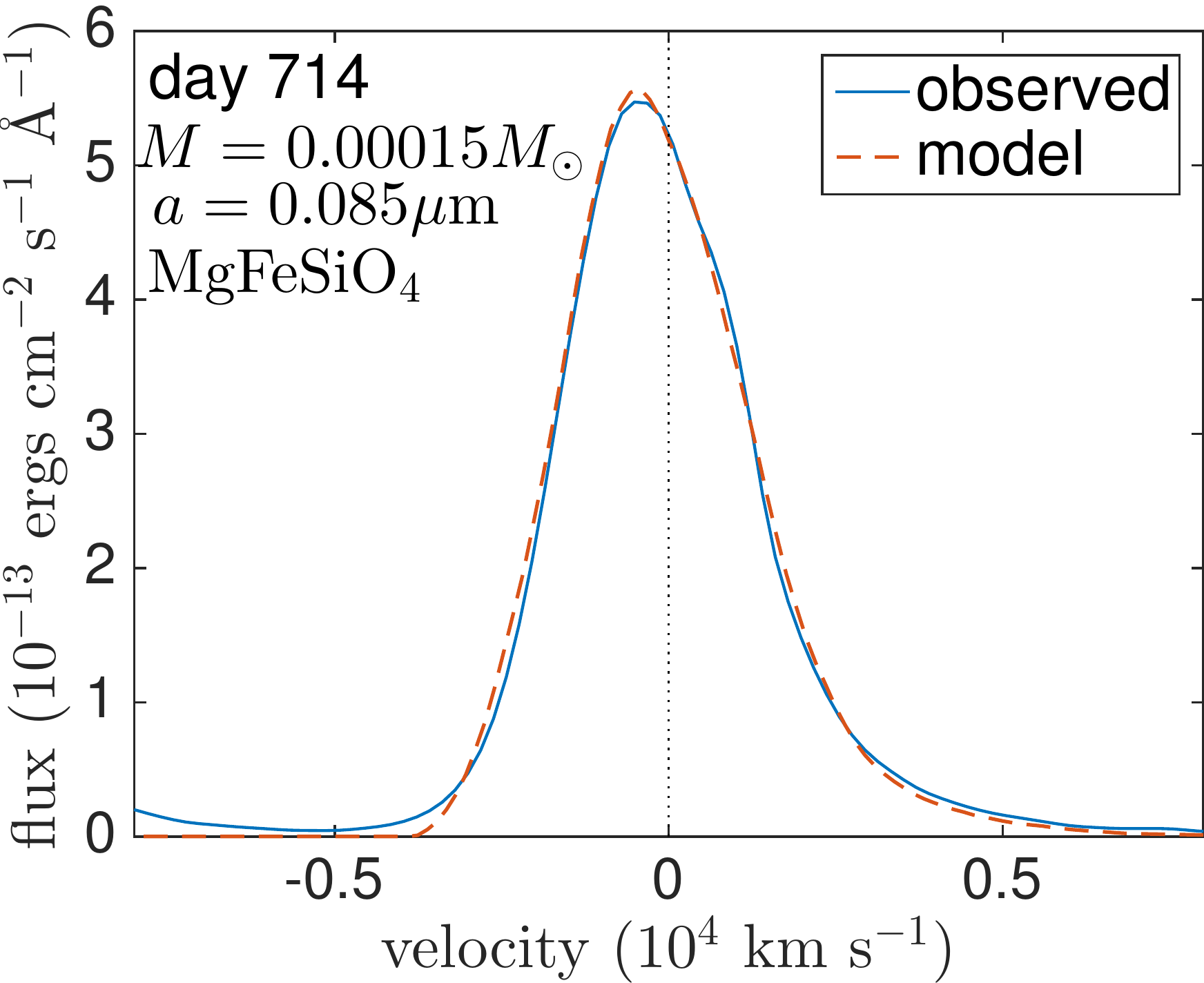}
\caption{H$\alpha$ models using different grain species and dust masses.   
Models for the dust masses presented by \citet{Dwek2015} are on the top and models 
using our minimum required dust masses are on the bottom.  From left to 
right the dust species are composite grains (82\% MgSiO$_3$ and 18\% 
amorphous carbon by volume), pure MgSiO$_3$, pure amorphous carbon and pure 
MgFeSiO$_4$. A density distribution with $\beta=2.3$ was adopted with a 
filling factor $f=0.09$ and an effective clump radius 
$R_{\rm eff}/R_{\rm out}=0.044$.  All other parameters are the same as in Table 
\ref{clumped1}.}
\label{Dwek_models_Ha}
\end{figure*}

\subsection{Modelling large masses of dust at early epochs: comparison 
with the results of \citet{Dwek2015}}
\label{dwek}

In a recent analysis of infrared SED data, DA15 suggested that it may be possible for a large mass (0.4M$_\odot$) of 
MgSiO$_3$ silicate dust to have been present in SN~1987A even at 
relatively early epochs ($t\sim615$ d), since that species has very low 
IR emissivities.  Up to this point, we have constructed models using 
\citet{Zubko1996} BE amorphous carbon dust but in the previous section we 
discussed the effect on derived dust masses of instead using 
\citet{Draine1984} astronomical silicate which has higher optical and IR 
emissivities than the glassy MgSiO$_3$ species considered by DA15. Our 
clumping structure in our models was based on that used by W15.

\begin{figure*}
\includegraphics[trim =0 28 40 0,clip=true,scale=0.24]{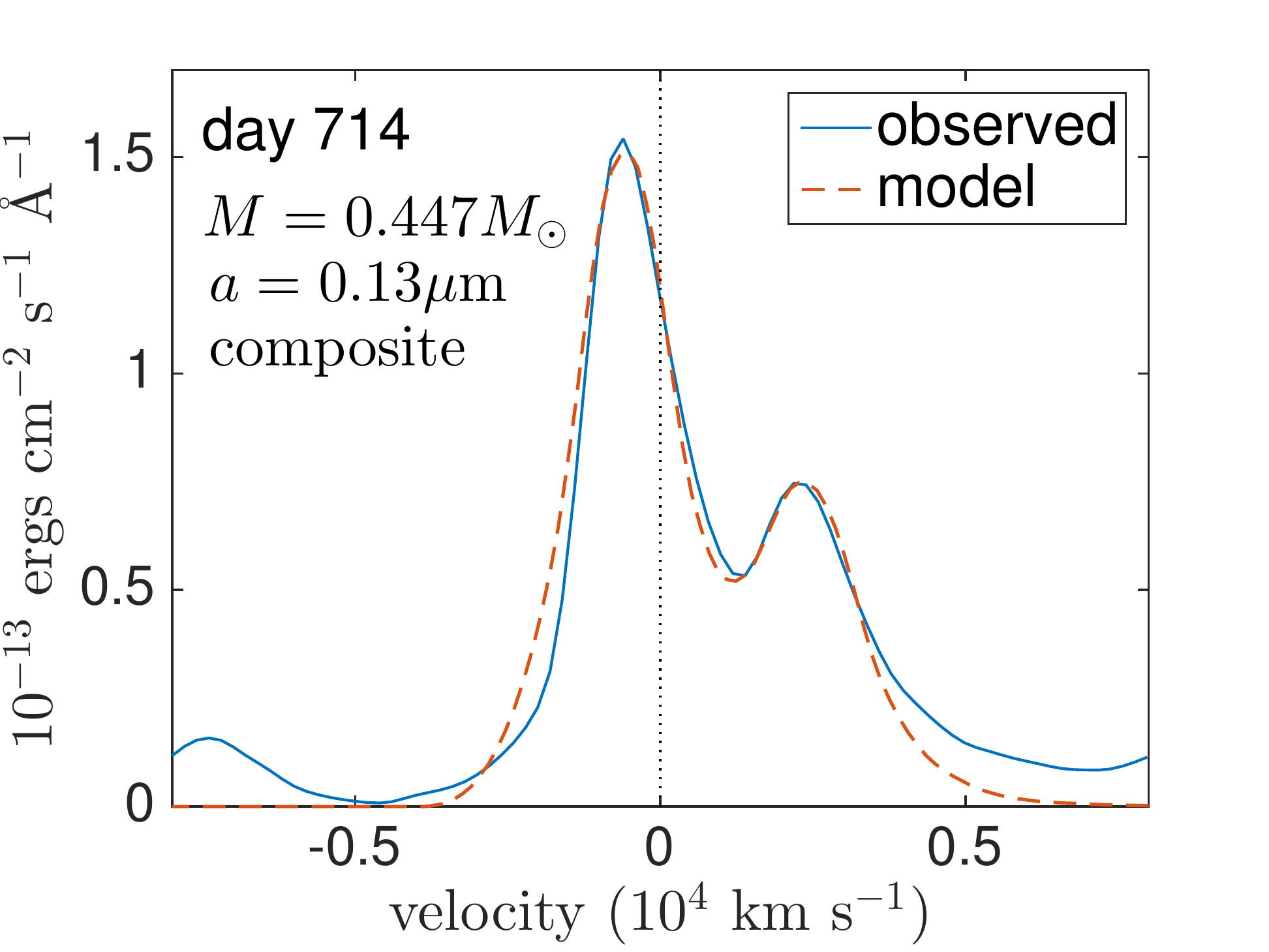}
\includegraphics[trim =29 28 -10 0,clip=true,scale=0.24]{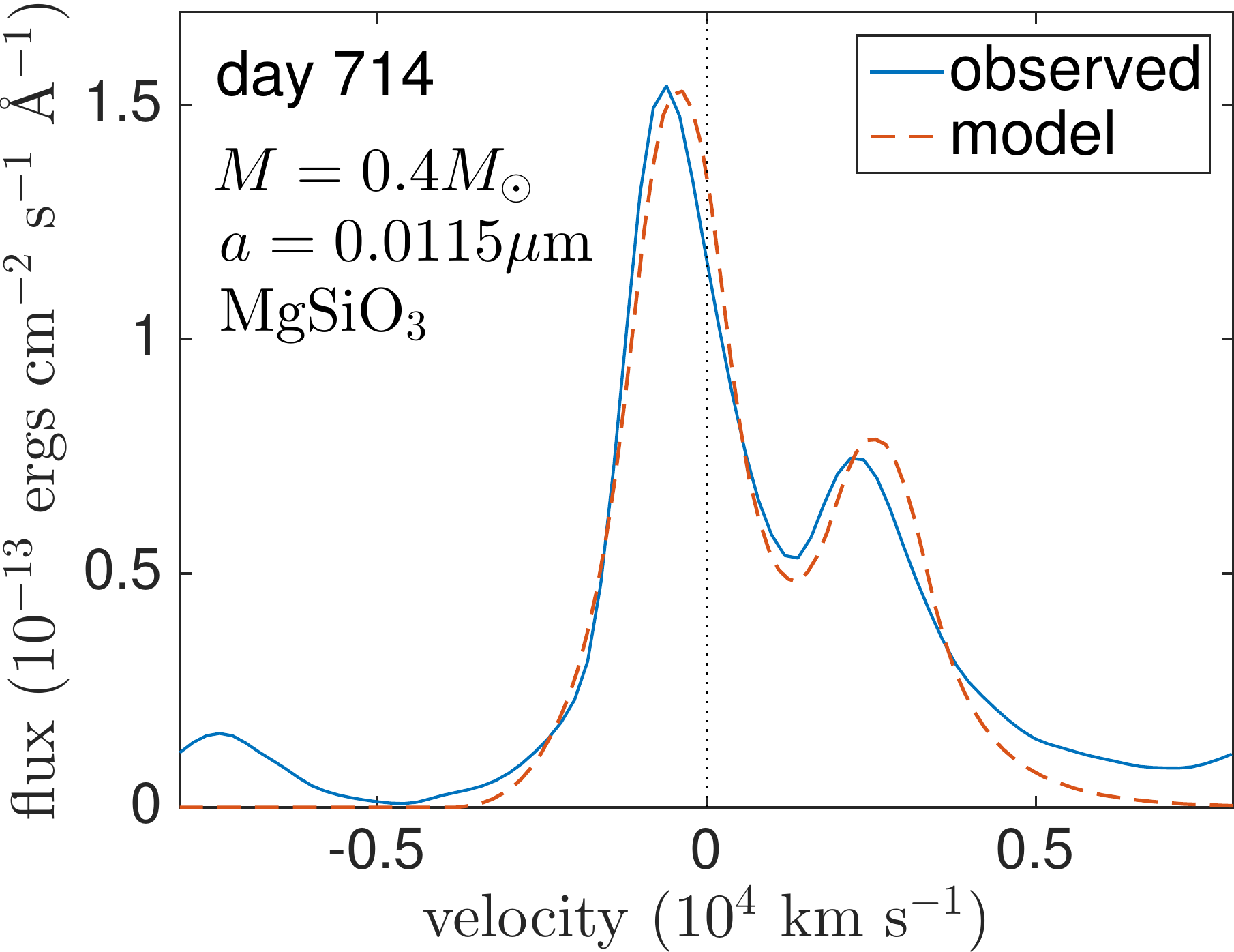}
\includegraphics[trim =29 28 -10 0,clip=true,scale=0.24]{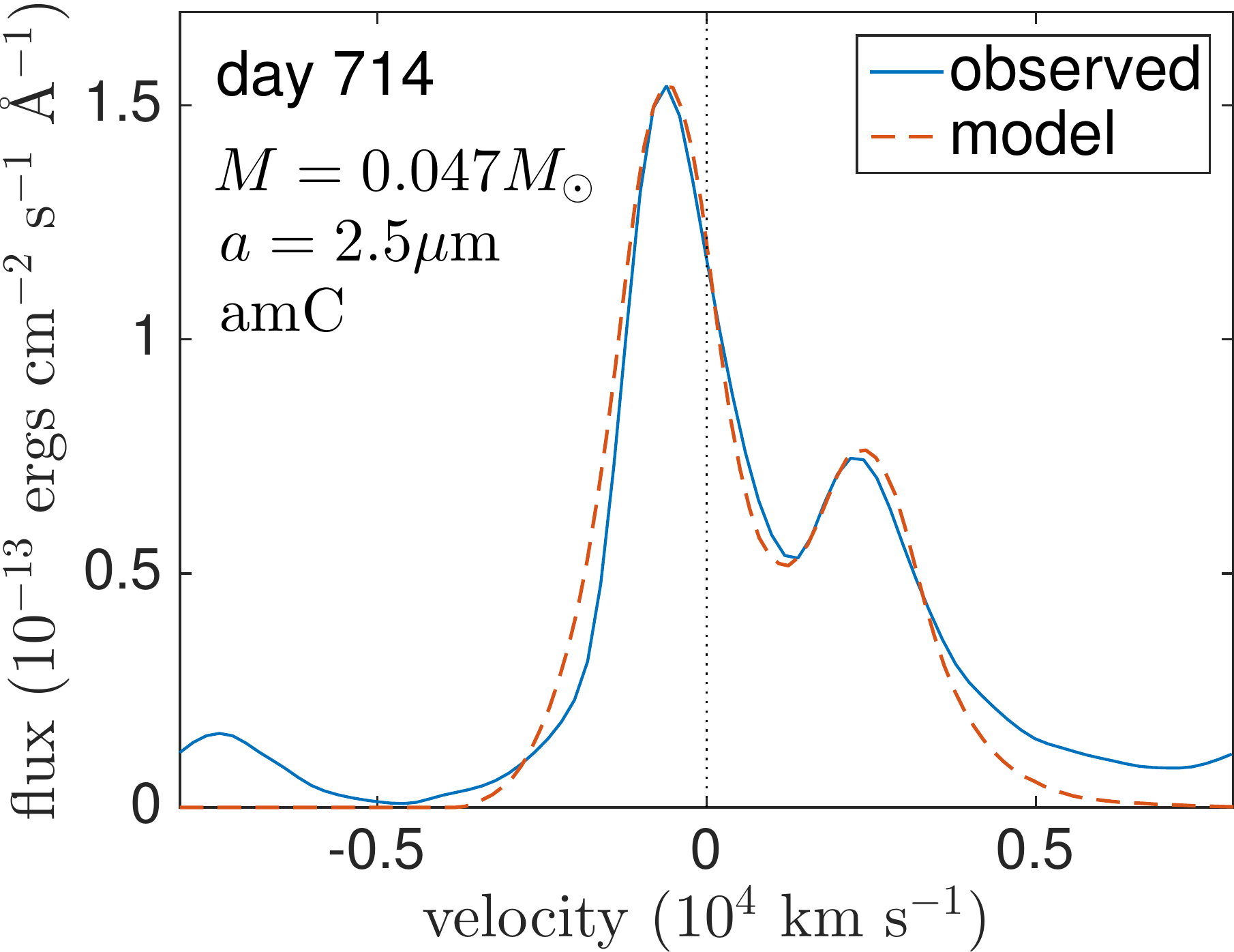}
\includegraphics[trim =29 28 0 0,clip=true,scale=0.24]{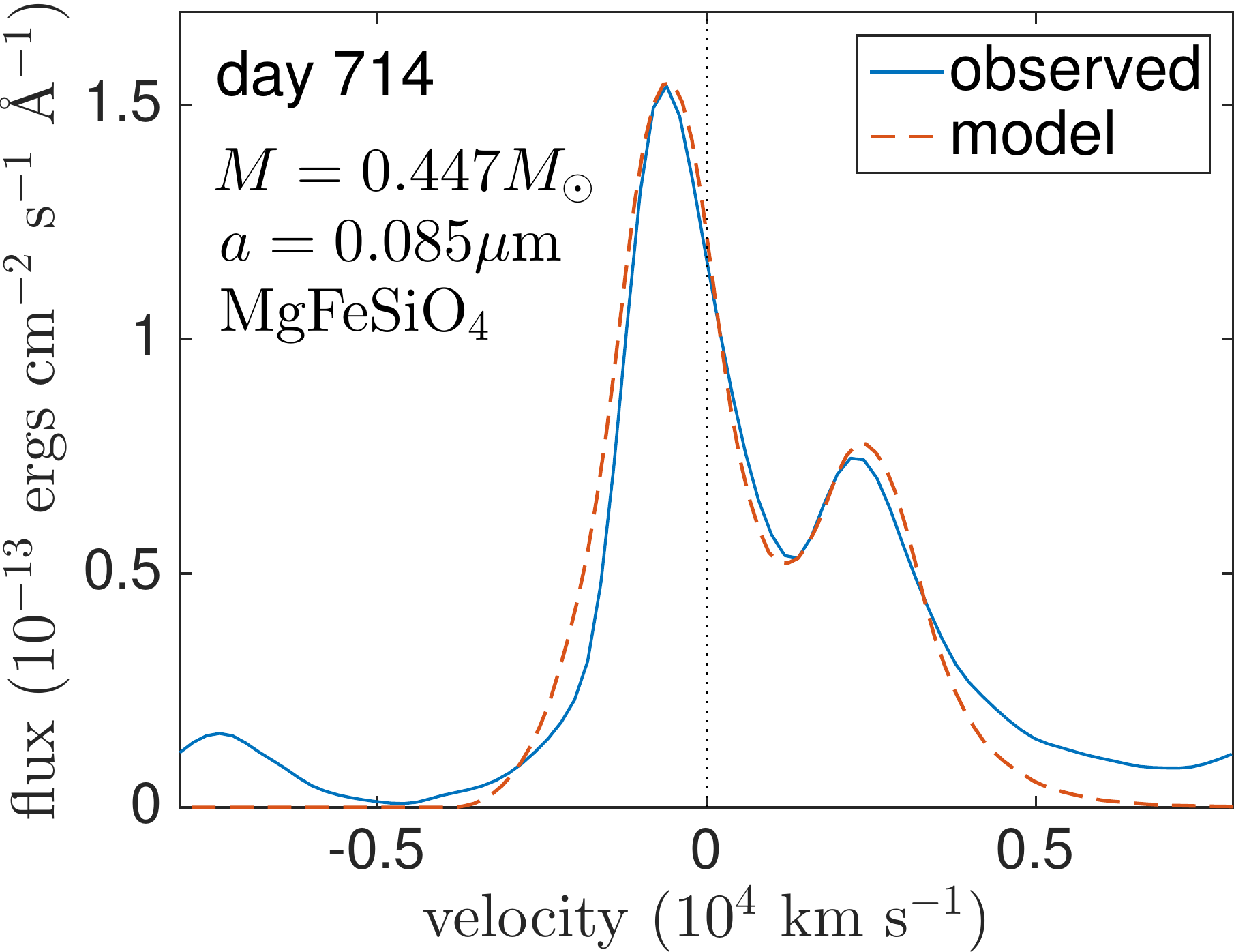}

\includegraphics[trim =0 0 25 -10,clip=true,scale=0.24]{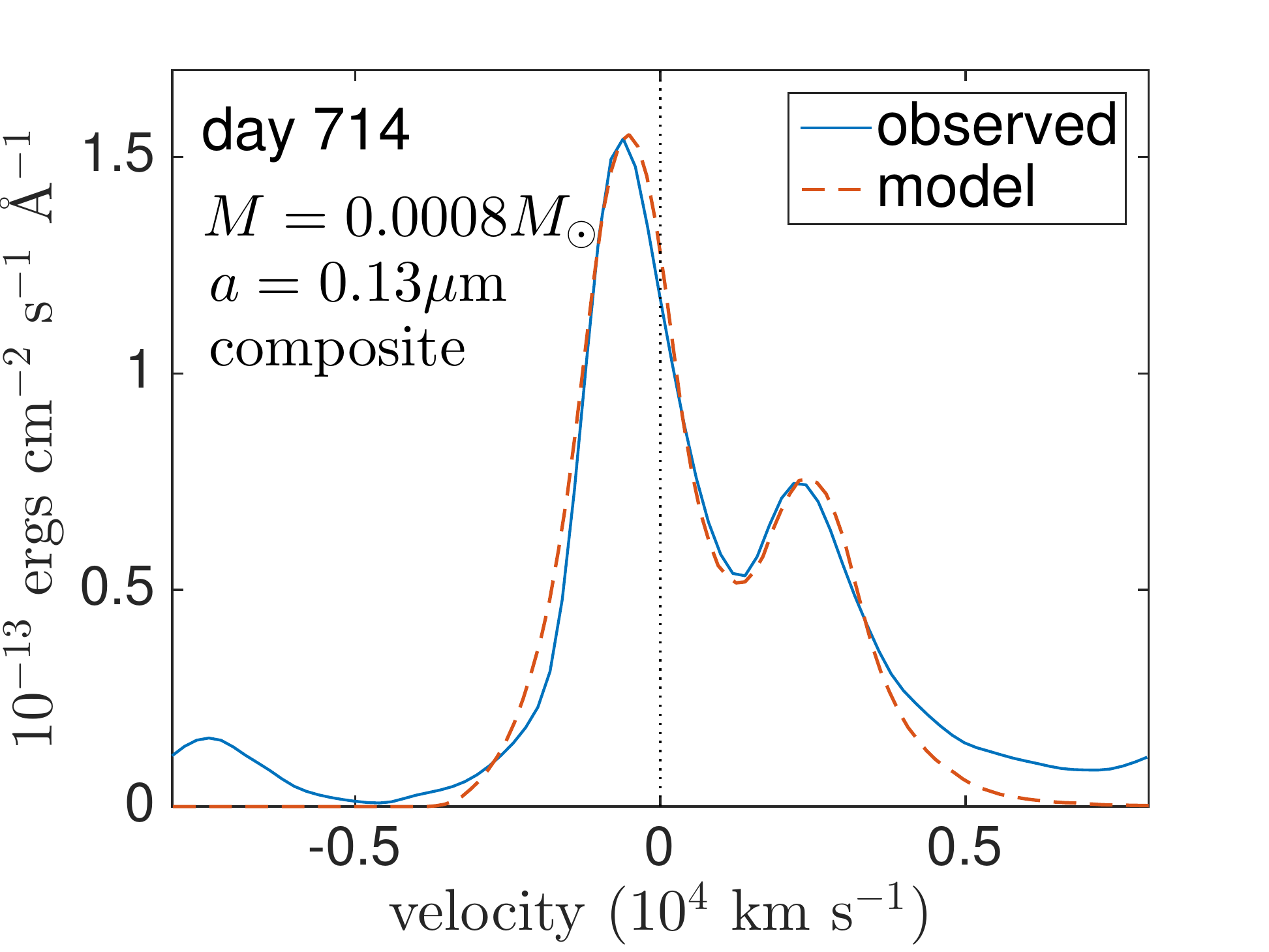}
\includegraphics[trim =29 0 -20 -10,clip=true,scale=0.24]{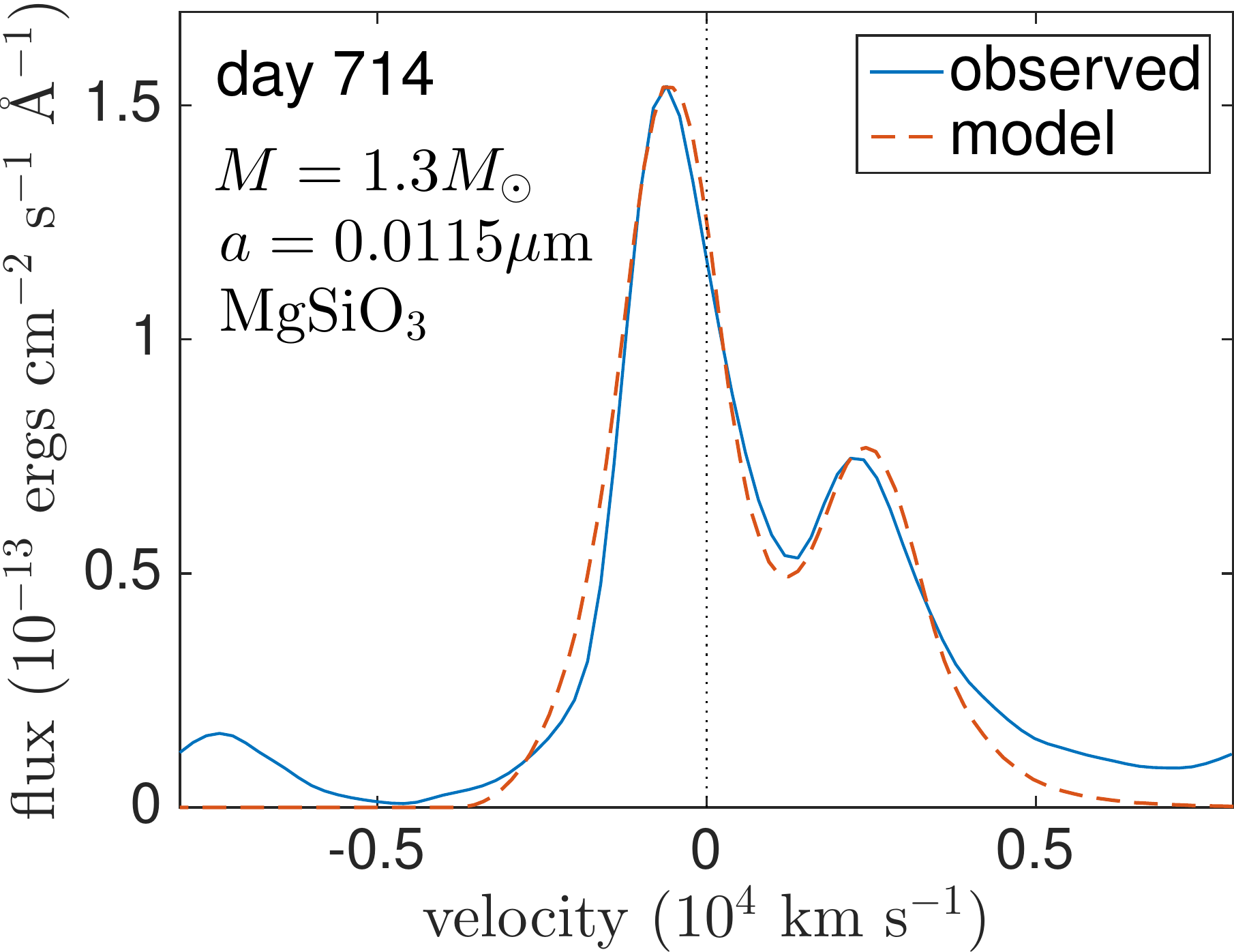}
\includegraphics[trim =29 0 20 -10,clip=true,scale=0.24]{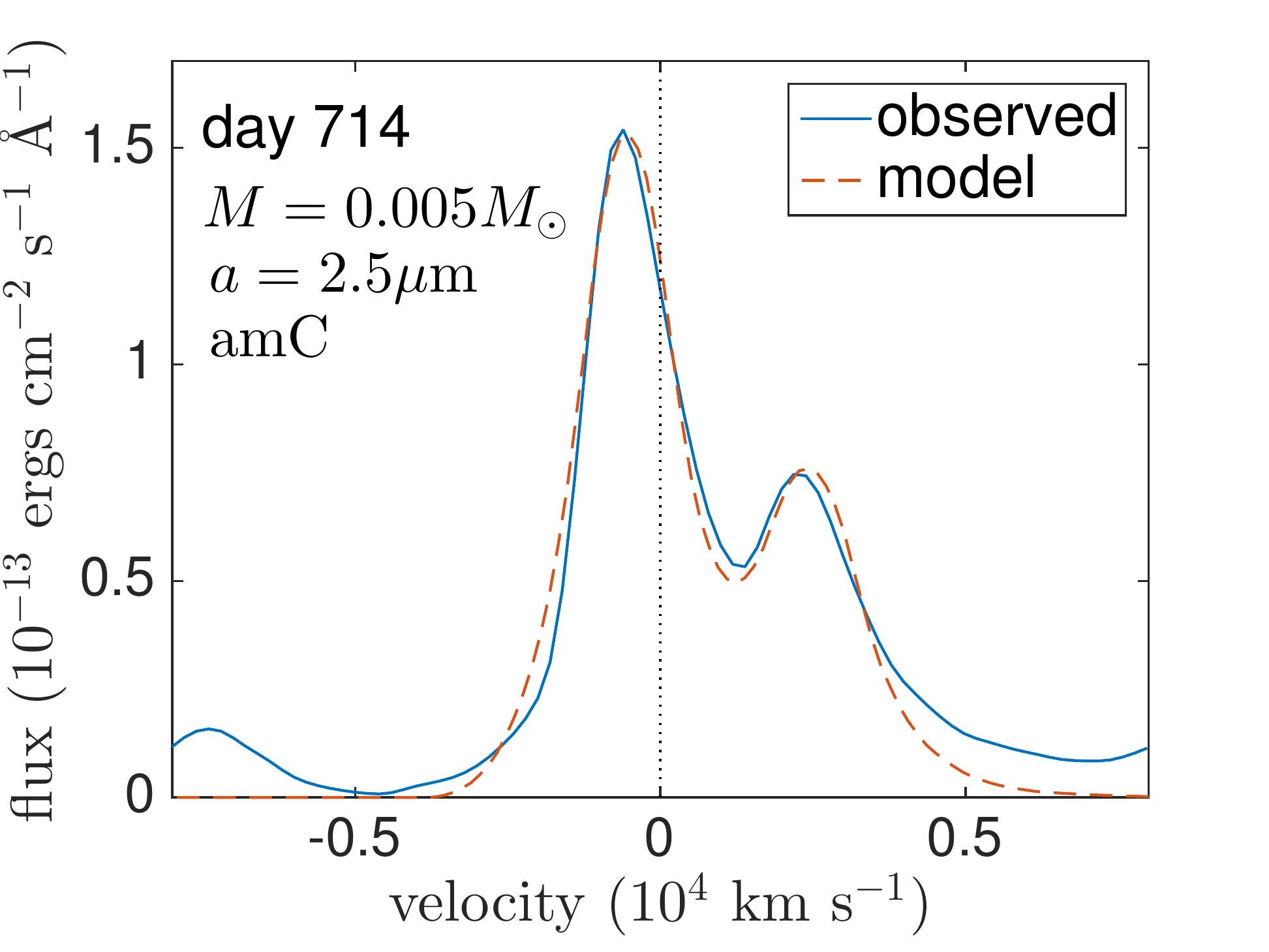}
\includegraphics[trim =35 0 40 -10,clip=true,scale=0.24]{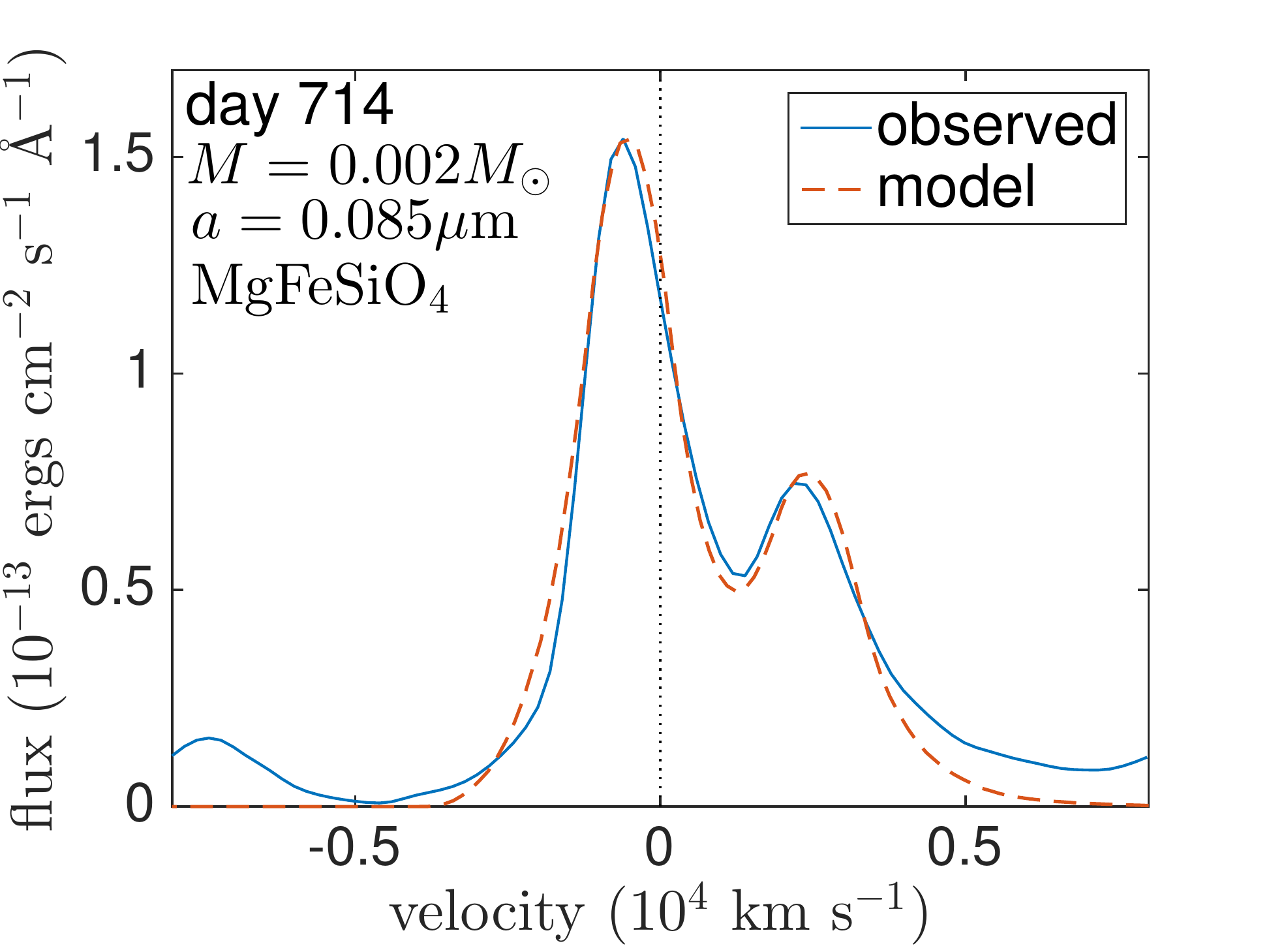}

\caption{[O~{\sc i}]$\lambda$6300,6363~\AA\ models using different grain 
species and dust masses.  Models using the dust masses presented by DA15 
are on the top and models using our minimum required dust masses are on 
the bottom.  From left to right the species are composite grains (82\% 
MgSiO$_3$ and 18\% amorphous carbon by volume), pure MgSiO$_3$, pure 
amorphous carbon and pure MgFeSiO$_4$.  A density distribution with 
$\beta=1.3$ was adopted with a filling factor $f=0.09$ and an effective 
clump radius $R_{\rm eff}/R_{\rm out}=0.044$. The ratio between the 
doublet components was 2.2. All other parameters are the same as in 
Table \ref{clumped1}.}
\label{Dwek_models_OI}
\end{figure*}

We now consider models for day 714 based on the grain types
used by DA15.  We adopt a clumped structure equivalent to the 
preferred model of DA15 who considered 1000 clumps with a filling factor of 
0.09 and a negligible dust mass in the inter-clump medium.  We calculate 
the effective spherical radius of our clumps by equating the volume of our 
cubic clumps to a sphere of radius $R_{\rm eff}$.  Clumps of width 
$R_{\rm out}/14$ generate the desired $R_{\rm eff}/R_{\rm out}=0.044$ equivalent to 
that of DA15.  In our code, using a filling factor of 0.09 then generates 
1034 clumps, similar to the number used by DA15.  We ran a series of models 
(presented in Figs \ref{Dwek_models_Ha} and \ref{Dwek_models_OI}) for 
both the H$\alpha$ and [O~{\sc i}]$\lambda$6300,6363~\AA\ line profiles.  
In each case we modelled the lines using a dust grain mixture as described 
by DA15 such that the medium comprised 18\% amorphous carbon and 82\% 
MgSiO$_3$ by volume.  We adopted the same optical constants as used in 
their work (i.e. \citealt{Jager2003} for MgSiO$_3$ grains and 
\citealt{Zubko1996} for amorphous carbon) and the same grain mass densities as DA15, 
$\rho_s=3.2$~g~cm$^{-3}$ and $\rho_c=1.8$~g~cm$^{-3}$.  In addition to 
modelling their composite grain case, we also considered three single 
species models, using Zubko BE amorphous carbon, MgSiO$_3$, and 
MgFeSiO$_4$ (in the latter two cases the optical constants were taken from 
\citealt{Jager1994} and \citealt{Dorschner1995}). For each species we 
adopted the smallest single grain size that has an albedo of $\omega 
\approx 0.6$. The ejecta 
parameters were as listed in Table \ref{clumped1}, with the exception of 
the density distribution which we took to be $\rho(r) \propto r^{-1.3}$ 
for H$\alpha$ and $\rho(r) \propto r^{-2.3}$ for [O~{\sc i}] in order to 
optimise the best fits.

For each species, two models are presented.  The first adopts the minimum 
possible dust mass that provides a reasonable fit to the observed line 
profiles and the second uses the dust mass derived by DA15 for that 
specific species ($M=0.4~M_{\odot}$ for MgSiO$_3$ and $M=0.047~M_{\odot}$ 
for amorphous carbon giving a total composite dust mass of 
$M=0.447~M_{\odot}$).  We treated MgFeSiO$_4$ as we do the composite 
grains and adopted a dust mass of $M=0.447~M_{\odot}$ for it.  Results 
from the models are presented in Figs \ref{Dwek_models_Ha} and 
\ref{Dwek_models_OI}.

The [O~{\sc i}] models can display similar profiles for substantially 
different dust masses.  This is a result of the relatively high optical 
depths within the clumps themselves. If a clump is optically thick then 
the majority of radiation that hits it will be absorbed and the profile 
becomes insensitive to how much dust is actually contained within the 
clump.  For our [O~{\sc i}] minimum dust mass models, the optical depths 
within a clump over an effective clump radius $R_{\rm eff}$ at 6300\AA\ are 
around $\tau_{\rm clump} \approx 0.4$.  Over the entire nebula optical depths 
are very high and $\sim$72\% of the total flux is absorbed.  Increasing the 
total dust mass therefore has only a small effect on the emergent line 
profile and once $\tau_{\rm clump}>1$ then the line profile remains unchanged 
for increasingly large dust masses.  It is because of this fact that we 
present only the smallest dust mass capable of reproducing the [O~{\sc i}] 
profiles seen in Fig. \ref{Dwek_models_OI}.  The insensitivity of the 
[O~{\sc i}] profiles to dust mass is not the case for the H$\alpha$ 
profile models (where $\tau_{\rm clump} < 0.05$ for all of our models) and the 
H$\alpha$-fit dust masses presented in Fig. \ref{Dwek_models_Ha} 
therefore represent the most sensitive diagnostic of the dust mass for 
each grain type.  All of our models discussed in previous sections have 
significantly smaller clump optical depths ($\tau_{\rm clump}<0.1$), making 
them sensitive to dust mass variations.

For all the [O~{\sc i}] line profile models, except for those using pure 
MgSiO$_3$ or pure Mg$_2$SiO$_4$ dust, the required dust masses are 
significantly less than those proposed by DA15. The [O~{\sc i}] profile 
obtained using DA15's very large MgSiO$_3$ dust mass of 0.4~M$_{\odot}$ 
provides a reasonable fit, but the same dust mass significantly 
overestimates the blueshifting of the H$\alpha$ line 
(Figure~\ref{Dwek_models_Ha}). We can place an upper limit on the 
mass of pure MgSiO$_3$ on day 714 of 0.07~M$_\odot$, as this 
is the highest mass for which a fit to the observed H$\alpha$ profile can 
be obtained (Figure~\ref{Dwek_models_OI}).

Pure MgSiO$_3$ is extremely glassy, with very high albedos in the optical 
for a wide range of grain radii.  At grain radii small enough to reduce 
the albedo to $\omega \approx 0.6$, in order to fit the observed line 
profiles, the extinction efficiency in the optical becomes extremely low 
(see Fig. \ref{albedo_grain}), with large masses of dust therefore 
required in order to produce even a small amount of line absorption. 
However, for a given albedo, the extinction efficiencies increase by large 
factors if either carbon or iron is included in the grain. In the 
composite grain model the amorphous carbon component dominates the overall 
extinction due to its much larger extinction efficiency at small grain 
radii. Similarly, for MgFeSiO$_4$ (or Mg$_{0.5}$Fe$_{0.5}$SiO$_3$) grains
the iron component leads to much larger optical and IR extinction
efficiencies and much lower dust mass upper limits.
If the dust that formed at early epochs contained some fraction of 
elements such as carbon, iron or aluminium, yielding `dirtier' silicate 
grains or composite grains, then fits to the observed blueshifted line 
profiles imply low dust masses. We conclude that for dust masses as large
as 0.07~M$_\odot$ to have been present in SN~1987A's ejecta as early as 
days 600-1000 then the dust would have to have been formed of glassy pure 
magnesium silicates.

In order to be certain that there was no set of parameters for which a dust mass of $M=0.447M_{\odot}$ comprising 82\% MgSiO$_3$ and 18\% amorphous carbon by volume could result in a good fit, a thorough investigation of the variable parameters was performed.  Having fixed the clump size, filling factor, dust mass and composition as per the values detailed above and in DA15, we varied the density profile ($\beta$) and grain radius $a$.  Varying the maximum velocity and the ratio of the inner and outer radii was found to have little effect on the goodness of fit.  The MSE for the H$\alpha$ profile presented in the upper left panel of Fig. \ref{Dwek_models_Ha} was 0.599 (in units of 10$^{-13}$~erg~cm$^{-2}$~s$^{-1}$).  This was improved to 0.246 by increasing the grain radius to $a=0.6\mu$m and the density profile exponent to $\beta=1.5$, which represents the best fit that we could achieve using the values described by DA15 and a dust mass of $M=0.447M_{\odot}$.  However, the overall best fit we obtain for this scenario (see the lower left panel of \ref{Dwek_models_Ha}) used a dust mass of $M=5 \times 10^{-4}M_{\odot}$ giving a MSE=0.0058, substantially improving the fit.

\subsection{Unattenuated line fluxes}

\begin{figure}
\centering
\includegraphics[clip=true,scale=0.43]{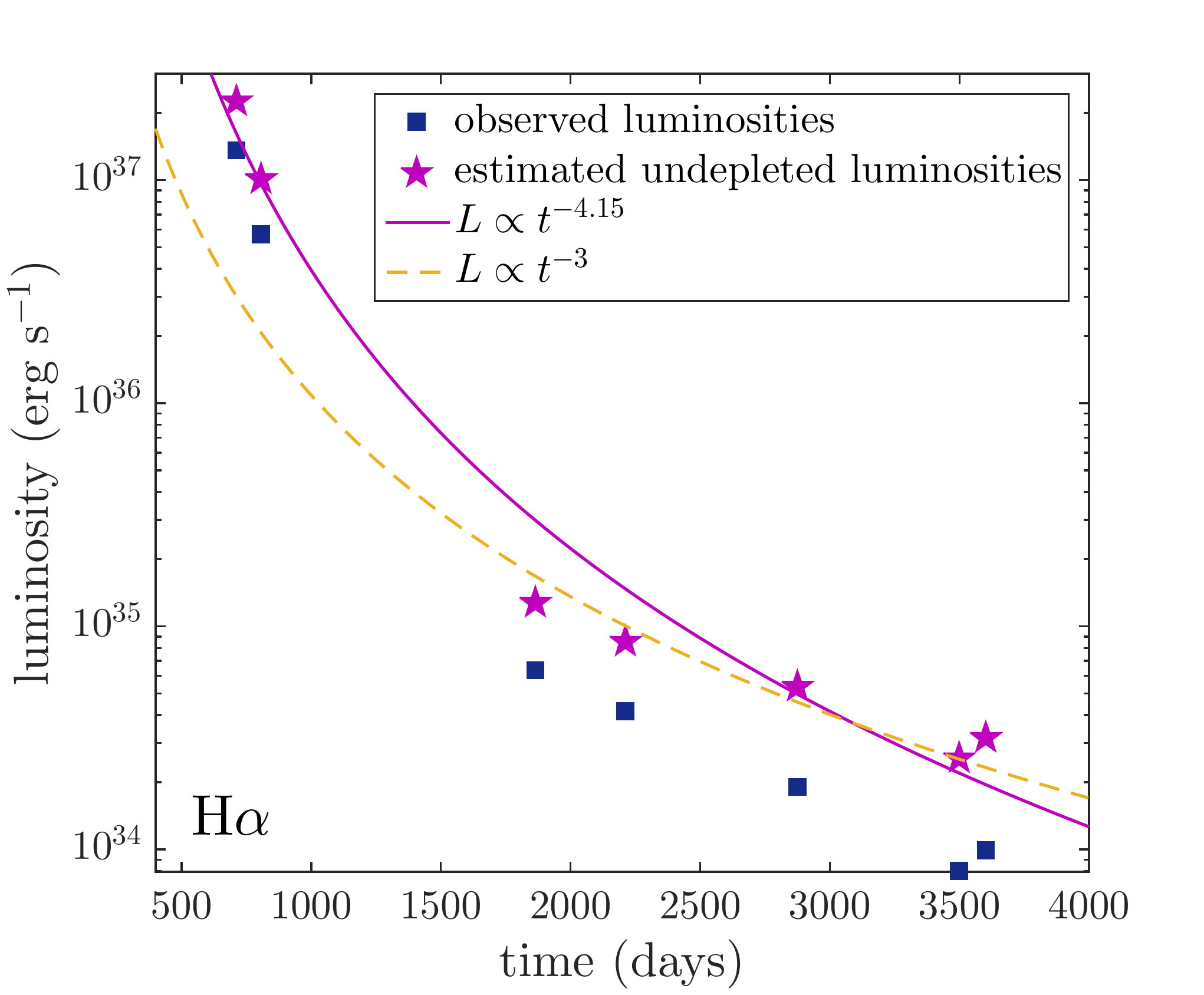}

\vspace{2mm} \includegraphics[clip=true,scale=0.44]{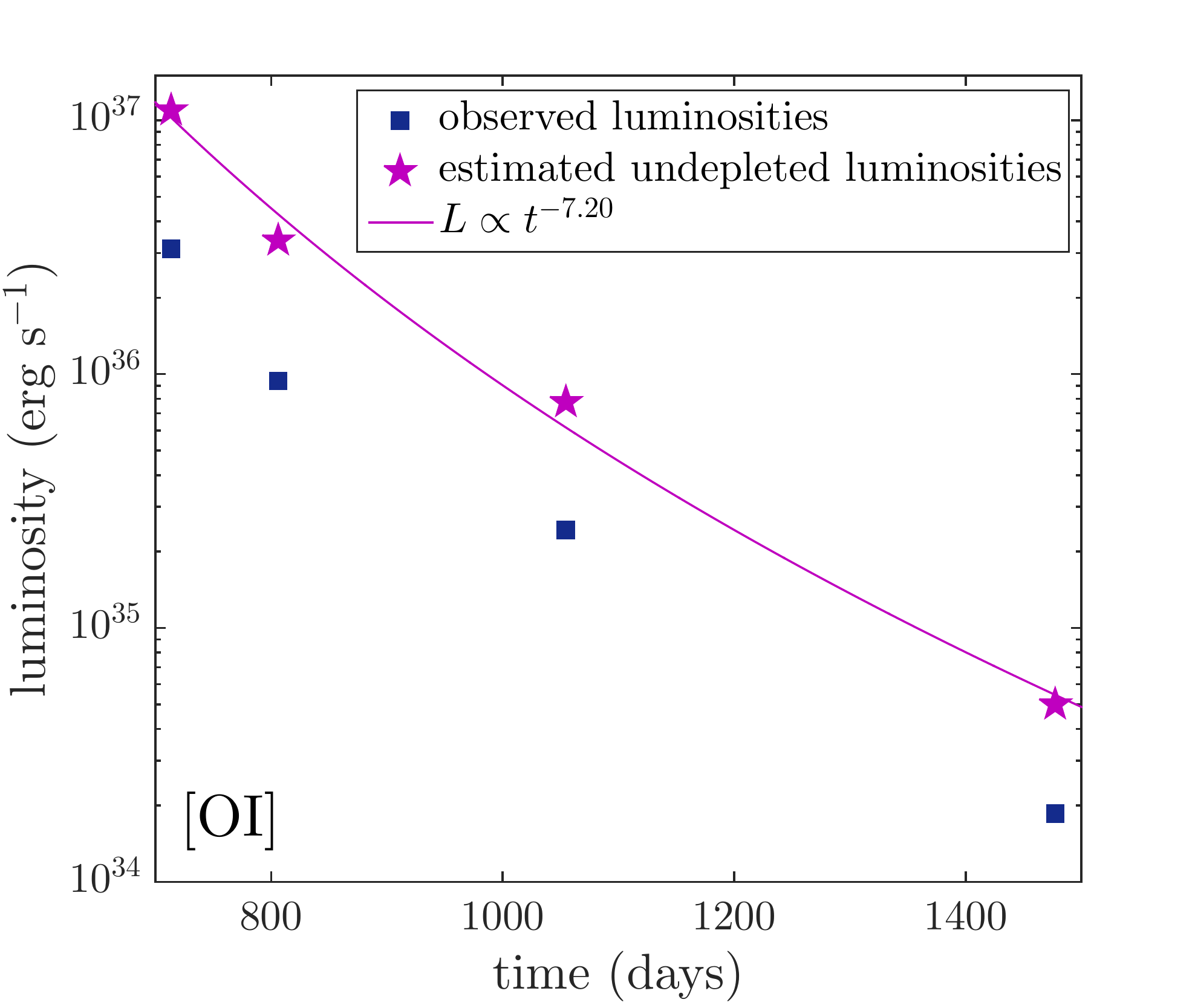}
\caption{Predicted undepleted luminosities for the H$\alpha$ line 
(above) and [O~{\sc i}]$\lambda$6300,6363~\AA\ doublet 
(below) presented with the best power-law fit to the data.}
\label{undep}
\end{figure}

The evolution of the SN~1987A H$\alpha$ and [O~{\sc 
i}]$\lambda$6300,6363~\AA\ line fluxes over time has been discussed 
previously by, for example, \citet{Li1992}, \citet{Xu1992} and 
\citet{Kozma1998b}. We may use our clumped models to predict the 
unattenuated 
emitted line fluxes and consider their evolution through time.  For each 
model, the fraction of the total line energy absorbed by the dust was 
predicted.  We determined the total flux for each observed line profile 
and used the absorbed fraction from our clumped models for $a=3.5\mu$m to 
predict the undepleted flux of the line before attenuation by the dust.  
Gaps in the observed data due to contamination by narrow line emission 
were interpolated over in order to estimate the flux of the broad line 
component. The observed H$\alpha$ luminosities and predicted undepleted 
luminosities are given in Table \ref{tau_e} along with the energy fraction 
absorbed by the dust in each model. No correction has been made for 
interstellar extinction along the sightline to SN~1987A.
There is very little change in 
these values if we adopt the models with $a=0.6~\mu$m instead of 
$a=3.5~\mu$m.  Plots of the observed and undepleted line luminosities are 
given for all modelled epochs of H$\alpha$ and [O~{\sc i}] in Figure 
\ref{undep}.

We also present power-law fits to the time evolution of the unattenuated 
H$\alpha$ and [O~{\sc i}] line fluxes.  For H$\alpha$, we find that 
$L_{H\alpha}(t) \propto t^{-4.15}$ between days 714 and 3604.  We can 
compare this value to the theoretical time dependence of the flux of a 
recombination line based on the dynamics of the ejecta for an 
environment in a Hubble-type flow $r=vt$.  For a frozen-in ionization 
structure, the mean intensity of a recombination or collisionally-excited 
line per unit volume is locally proportional to the product of the 
densities of the recombining species i.e. $J_{H\alpha} \propto n_{\rm e} n_{\rm p} 
\propto n_{\rm e}^2$.  The total luminosity of the line is therefore dependent 
on the volume $V$ as $L_{H\alpha} \propto 1/V $.  Assuming a constant 
maximum expansion velocity, the luminosity should vary with time as 
$L_{H\alpha}(t) \propto t^{-3}$.

This relationship is only true for a constant ionization fraction.  This 
`freeze-out' phase is estimated to have begun at $\sim 800$ d and 
first sets in at lower density high velocity regions, gradually moving 
inwards with time \citep{Danziger1991,Fransson1993}.  Since our modelling 
begins at day 714, the ionization fraction in the inner higher density 
regions is likely still decreasing due to recombination during our first 
two epochs.  This presumably accounts for the slightly steeper 
$L_{H\alpha}(t) \propto t^{-4.15}$ that we find across all epochs.  
\citet{Kozma1998b} estimate that H$\alpha$ emission from the outer regions 
begins to dominate over H$\alpha$ emission from core regions for $t>$ 
900 d. If earlier epochs are ignored, the last five epochs ($t \ge 
1862$ d) plotted in (Fig. \ref{undep}) exhibit a shallower trend that 
is in good agreement with the expected $L_{\rm H\alpha}(t) \propto t^{-3}$ 
evolution.

The [O~{\sc i}]$\lambda$6300,6363~\AA\ doublet exhibits a much 
steeper evolution, $L_{[OI]}(t) \propto t^{-7.2}$, than the H$\alpha$ line 
(Fig. \ref{undep}). These collisionally excited lines are very sensitive 
to the gas temperature, with emissivities that fall to low values for 
temperatures below $\sim$3000~K. The models of \citet{Li1992} and \citet{Kozma1998a} 
predict that the gas temperature in the relevant [O~{\sc i}] emitting 
regions should have fallen below 1000~K after day $\sim$1000.

\section{Discussion}
\label{discuss}

Using Monte Carlo models that consider both the absorbing and scattering 
effects of dust, we have modelled the evolution of the H$\alpha$ and 
[O~{\sc i}]~$\lambda$6300,6363~\AA\ line profiles over time, enabling us 
to place constraints on the evolution of newly formed dust in the ejecta 
of SN 1987A.

As can be seen in Fig. \ref{d1862_3604}, even a small degree of 
asymmetry in observed supernova line profiles can be indicative of dust 
formation within the ejecta.  In addition to this, a line profile that is 
consistently asymmetric through time requires increasingly large dust 
masses to account for a similar degree of blueshifting since the 
expansion of the ejecta would otherwise cause the dust optical depth to 
the edge of the ejecta to be reduced.

In Section \ref{dwek} we compared our results with those of \citet{Dwek2015} and 
concluded that large dust masses can only have been present at early 
epochs if the grains were formed purely of glassy magnesium silicates that 
contained no iron or carbon component and that even for pure magnesium 
silicates no more than 0.07~M$_\odot$ could have been present. We now 
compare our results with those of \citet{Lucy1989} and W15.

\citet{Lucy1989} analysed the [O~{\sc i}]~$\lambda$6300,6363~\AA\ doublet 
for SN~1987A and estimated dust optical depths for a number of epochs. 
They translated these into dust masses for day 775 only. From our smooth 
flow modelling of the [O~{\sc i}] doublets, we obtain $\tau_V \approx 3.60$ at day 
714 and $\tau_V \approx 2.86$ at day 806.  These values are higher 
than the values given by \citet{Lucy1989} who derived $\tau_V=1.19$ at day 
725 and $\tau_V=1.25$ at day 775.  The value of the assumed albedo 
accounts for the majority of this discrepancy.  \citet{Lucy1989} 
considered line profiles before and after dust condensation and concluded 
that any evidence of an extended red scattering wing was unconvincing.  
Accordingly, they adopted a model with perfectly absorbing dust ($\omega = 
0$).  For our amorphous carbon models for the [O~{\sc 
i}]~$\lambda\lambda$6300,6363~\AA\ profile using a grain radius $a=0.35\mu$m, we 
obtain an albedo of approximately $\omega = 0.5$ at $\lambda=6300$ \AA.

\begin{figure*}
\begin{center}
\includegraphics[trim =70 30 85 15,clip=true,scale=0.54]{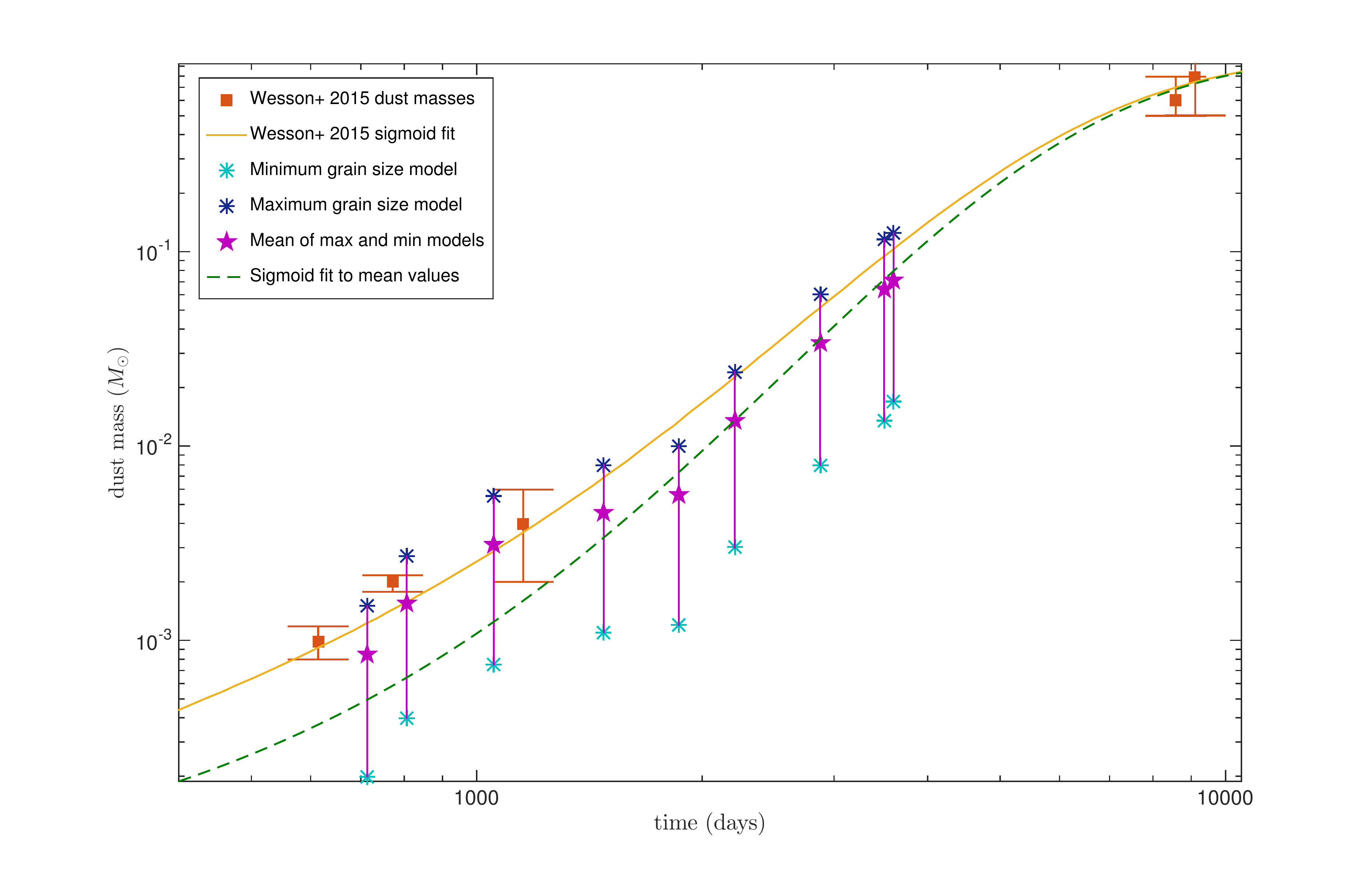}
\caption{Derived dust masses for SN~1987A as a function of epoch. 
\textit{Red squares -} dust masses derived by W15 
from their photometric SED modelling of SN 1987A. \textit{Solid yellow line} - 
W15's sigmoid fit to 
their values. \textit{Dark and light blue asterisks -} maximum 
($a=3.5~\mu$m) and 
minimum ($a=0.6~\mu$m) dust masses respectively for the [O~{\sc i}] models 
for $t \le 1478$ days and for the H$\alpha$ models for $t \ge 1862$ days. 
\textit{Purple 
stars -} predicted dust masses calculated as the mean of the maximum and 
minimum dust masses.
\textit{Dashed green line -} sigmoid fit 
to our predicted dust masses.}
\label{Mdust}
\end{center}
\end{figure*}

The dust masses derived by \citet{Lucy1989} at day 775 (e.g. $M_{\rm dust}=4.4 
\times 10^{-6} M_{\odot}$ for amorphous carbon) are  
different to those obtained from our smooth dust modelling of the [O~{\sc 
i}]~$\lambda\lambda$6300,6363~\AA\ doublet at day 806 ($M_{\rm dust}=1.5 \times 
10^{-4} M_{\odot}$ for amorphous carbon).  There are three main reasons 
for the discrepancy.  First, the albedo is significantly larger in our 
modelling as already discussed.  A larger dust mass is therefore required 
to produce the same amount of absorption.  Secondly, to match the extended 
red wing our required grain radius is considerably larger than the small 
grains ($a < 0.1\mu$m) adopted by \citet{Lucy1989}. Larger grain radii 
reduce the total cross-section of interaction and so a greater dust mass 
must be present to compensate for this. Finally, the adopted maximum 
velocity (4000~km~s$^{-1}$) in our model is larger than the value adopted 
by \citet[1870~km~s$^{-1}$]{Lucy1989}.  The larger value of $V_{\rm max}$ 
increases the total volume of the ejecta significantly and therefore 
significantly more dust is required to produce the same optical depth.

\citet{Lucy1989} also noted that the dust optical depth increased rapidly 
after day 580 and that the rate of increase of the dust optical depth 
appeared to slow between day 670 and day 775, the latest day that they 
considered.  Our results, for both clumped and smooth models, suggest that 
the dust optical depth actually drops between day 714 and day 806 before 
starting to increase again at later epochs.  This is consistent with the 
results of \citet{Lucy1989} where the slowing rate of increase of dust 
optical depth could be consistent with a turning point subsequent to day 
775.

We can also compare our dust masses with the mass estimates derived from 
SED-fitting by W15 (see Fig. \ref{Mdust}).  W15 used a sigmoid fit to 
their dust mass evolution, of the form

\begin{equation}
M_d(t)=a{\rm e}^{b{\rm e}^{ct}},
\end{equation}
 
\noindent where $a=1.0M_{\odot}$ (representing the limiting dust mass), 
$b=-8.53$ and $c=-0.0004$.  Both their dust masses and this sigmoid fit 
are shown in Fig. \ref{Mdust}.  It exhibits an initial period of slow 
growth in mass followed by an intermediate period of accelerating growth 
followed by another slowing until a plateau is ultimately reached.  In 
this sense it may be representative of the process of dust 
formation whereby initial conditions appropriate for grain growth 
gradually develop until optimal conditions are reached at an intermediate 
epoch when grain growth is at its fastest before conditions once again 
deteriorate and the rate slows again (as discussed by W15).  Performing a 
least-squares regression to this function using just our own derived 
clumped dust masses, we obtain a sigmoid fit with coefficients 
$a=1.0M_{\odot}$, $b=-10.0$ and $c=-0.0004$.  These values are 
remarkably similar to those derived by W15.  This sigmoid fit is also 
plotted in Fig. \ref{Mdust}.

We find that at all epochs the dust masses derived by W15 are entirely 
within the dust mass ranges determined by our models.

Our sigmoid fit to the mean of the maximum and 
minimum dust masses does not take into account any systematic effects of 
grain growth.  At earlier epochs, whilst grains are 
still small relative to later epochs, the lower bound to the dust mass 
estimates may be more representative than the upper end; the reverse would 
be true at later epochs.
This is in contrast to the sigmoid fit of W15, whose fits to their early 
epoch SEDs used an MRN distribution with grain radii between 0.005 
and 0.25~$\mu$m, whilst their fits to their last two epochs required grain 
radii between 3.005 and 3.25~$\mu$m. The dust masses used for their 
sigmoid fit thus accounted for the effects of grain growth between the earlier 
and later epochs. As mentioned, we could not fit the extended red wings of 
the profiles at early epochs using an MRN distribution.  W15 found that at 
their earlier epochs they could not obtain SED fits with grain radii as 
large as $\sim 1.0~\mu$m. However, they did not consider radii in between 
these size ranges, such as the grains with $a \approx 0.6~\mu$m that we 
require at earlier epochs.  For SED modelling it is generally the case that 
the larger the grain size used, the less dust is required to produce the 
same level of flux.  This may account for the differences between W15's 
earlier epoch dust masses and our own minimum dust mass estimates at 
similar epochs.  The models of W15 used 15\% silicate dust, in contrast to 
our models which used 100\% amorphous carbon dust.  This could also 
contribute to the differences at early epochs, as could the use of 
different sets of optical constants -- we used the BE amorphous carbon 
optical constants of \citet{Zubko1996} whereas W15 used AC constants from 
\citet{Hanner1988}.  W15 found that in order to fit early epoch SEDs epochs 
(e.g. day 615) with Zubko ACH2 constants, smaller inner and outer ejecta 
radii were needed, with half as much dust ($5.0 \times 10^{-4}M_{\odot}$) 
compared to the Hanner AC results.

W15 derived a maximum possible grain size at late epochs, concluding that 
the grains could not be larger than $\sim 5~\mu$m by day 8515. This is 
consistent with the maximum grain radii that we derive at our latest 
epochs.  We find that grain radii most likely cannot have exceeded $\sim 
3.5~\mu$m at day 3604 - the dust mass that we obtain using this grain 
radius is similar to the value predicted by W15's sigmoid fit at that 
epoch.

The relationship between ejecta dust grain radii and post-explosion 
time is important for understanding the likelihood of dust surviving the 
passage of a reverse shock propagating back through the ejecta. By the 
time the effects of a reverse shock begin to appear in the line profiles 
(around day 5000), our models imply that the grains could already be as 
large as several microns in radius and are likely to be larger than $\sim 
0.6~\mu$m. Grains as large as this are more likely to survive destruction 
by sputtering in supernova reverse shocks and in interstellar shocks 
\citep{Silvia2010, Silvia2012, Slavin2015}.
It has been suggested that very large grains (radii up to 4.2$~\mu$m) 
formed in the ejecta of SN 2010jl within a few hundred days after the 
explosion \citep{Gall2014}. The grain radii that W15 and ourselves obtain 
for SN~1987A at very late epochs are nearly as large as found by 
\citet{Gall2014} for SN~2010jl, with both results suggesting that grains 
large enough to survive the destructive force of a reverse shock have 
formed by a few hundred days post-explosion. 

The dust masses obtained from our modelling of SN~1987A's line profiles 
support the conclusion of W15 that even after $\sim$3000 d the dust 
mass was still only a fraction of its current value. This contrasts with 
the results of \citet{Sarangi2015} whose grain chemistry models predict 
that ejecta dust masses should plateau by around 5 yr after the 
explosion. Our results show that SN~1987A's dust mass had reached of 
the order of $0.1M_{\odot}$ by day 3604.  Since its present dust mass is 
several times larger than this (\citealt{Matsuura2015}, W15), a 
substantial fraction of the current dust mass must have condensed after 
this epoch, in agreement with the conclusions of W15.

Ideally, our models would cover the entire evolution of SN 1987A's 
H$\alpha$ line profiles up to the present day.  However, the excitation of 
gas in the outer edges of the ejecta by the reverse shock after $\sim$ day 
5000 results in significant broad and asymmetric emission that 
dominates the original line profile \citep{Fransson2013}.  In addition to 
this, the narrow lines from the equatorial ring start to become so 
strong relative to the declining broad H$\alpha$ profile that, 
post-removal, not enough of the broad profile remained to be 
able to reliably infer information from the profile structure. These 
factors may be 
common to some other CCSNe that have interactions with surrounding 
circumstellar material. Care should also be taken to ensure that any 
observed late-time line profiles being modelled are not in fact the 
product of a light echo reflecting the spectrum from near maximum light. 
Nonetheless, detailed line modelling of asymmetric line profiles has 
proved effective in determining dust masses in the ejecta of SN~1987A at 
multiple epochs during the first ten years after outburst. The method 
clearly has wider application to other supernovae.

\section{Conclusions}

We have investigated the effects of scattering and absorption by ejecta 
dust on supernova line profile shapes and the different characteristic 
features that may be produced.  In particular, attention is drawn to the 
fact that a classical blueshifted peak and asymmetric profile with most 
flux on the blue side is not the only profile type that can signify the 
presence of dust. In the case of strong dust scattering, line profiles can 
have the majority of their flux on the red side. Even with just some dust 
scattering, profiles can often exhibit an extended red scattering wing, 
although care should be taken to ascertain that this cannot be accounted 
for by electron scattering (electron scattering optical depths should 
usually only be significant at very early epochs, $<$ 200 d). The line 
peak should always lie on the blue side, with a line peak velocity that 
will often correspond to the minimum velocity at the inner edge of the 
ejecta shell. If not obscured by narrow circumstellar [N~{\sc ii}] 
6584~\AA\ emission, a pronounced shoulder or corner may be present on the 
red side of the profile, also corresponding to the minimum velocity at the 
inner edge of the ejecta shell.

We have modelled the H$\alpha$ and [O~{\sc i}]~$\lambda$6300,6363~\AA\ 
line profiles from SN~1987A over a range of epochs and have obtained dust 
masses of the order of $0.1M_{\odot}$ by day 3604.  We derive a sigmoid 
fit to our dust mass data that predicts a current dust mass of 
0.68$M_{\odot}$, in line with current SED-based dust mass estimates for 
SN~1987A.  We find that large grains are necessary in order to reproduce both the extended red scattering wings and the asymmetry seen in 
several of the lines and that grains larger than $0.6~\mu$m have formed by 
day 714, while by day 3604 grain radii of $\sim 3.5~\mu$m are needed. We 
find from fits to the H$\alpha$ profile that dust masses cannot have 
exceeded a few$\times10^{-3}$~M$_\odot$ on day 714 for all the grain types 
investigated, apart from glassy pure magnesium silicate grains, for which 
up to 0.07~M$_\odot$ can be fitted.

The observed red-blue line asymmetries persist right through to day 3604 
and beyond -- if no further dust had formed after day $\sim$800 then the 
expansion of the ejecta dust shell would cause dust optical depths 
to drop rapidly with time thereafter, leading to the disappearance of 
red-blue asymmetries. Just to maintain the observed degree of red-blue 
asymmetry seen at the earlier epochs therefore requires that dust must 
have continued to form beyond those epochs.

\section*{Acknowledgements}

AB would like to thank Dr Jeremy Yates and Dr Patrick Owen for discussions and advice during 
the development of the {\sc DAMOCLES} code.  We thank Dr Raylee Stathakis and Dr Mark Phillips for 
providing us with the AAT and CTIO spectra of SN~1987A respectively.  We also thank the anonymous referee for  helpful comments and suggestions.  AB's work has been 
supported by a Science and Technology Facilities Council Research 
Studentship.  This work is based on data acquired through the Australian Astronomical Observatory, on data obtained from the ESO Science Archive Facility and on observations made with the NASA/ESA Hubble Space Telescope, obtained from the data archive at the Space Telescope Science Institute. STScI is operated by the Association of Universities for Research in Astronomy, Inc. under NASA contract NAS 5-26555.  This work also uses services or data provided by the NOAO Science Archive. NOAO is operated by the Association of Universities for Research in Astronomy (AURA), Inc. under a cooperative agreement with the National Science Foundation.

\bibliography{SN1987A_paper}{}

\begin{thebibliography}{}
\makeatletter
\relax
\def\mn@urlcharsother{\let\do\@makeother \do\$\do\&\do\#\do\^\do\_\do\%\do\~}
\def\mn@doi{\begingroup\mn@urlcharsother \@ifnextchar [ {\mn@doi@}
  {\mn@doi@[]}}
\def\mn@doi@[#1]#2{\def\@tempa{#1}\ifx\@tempa\@empty \href
  {http://dx.doi.org/#2} {doi:#2}\else \href {http://dx.doi.org/#2} {#1}\fi
  \endgroup}
\def\mn@eprint#1#2{\mn@eprint@#1:#2::\@nil}
\def\mn@eprint@arXiv#1{\href {http://arxiv.org/abs/#1} {{\tt arXiv:#1}}}
\def\mn@eprint@dblp#1{\href {http://dblp.uni-trier.de/rec/bibtex/#1.xml}
  {dblp:#1}}
\def\mn@eprint@#1:#2:#3:#4\@nil{\def\@tempa {#1}\def\@tempb {#2}\def\@tempc
  {#3}\ifx \@tempc \@empty \let \@tempc \@tempb \let \@tempb \@tempa \fi \ifx
  \@tempb \@empty \def\@tempb {arXiv}\fi \@ifundefined
  {mn@eprint@\@tempb}{\@tempb:\@tempc}{\expandafter \expandafter \csname
  mn@eprint@\@tempb\endcsname \expandafter{\@tempc}}}

\bibitem[\protect\citeauthoryear{Andrews et~al.,}{Andrews
  et~al.}{2010}]{Andrews2010}
Andrews J.~E.,  et~al., 2010, \mn@doi [\apj] {10.1088/0004-637X/715/1/541},
  715, 541

\bibitem[\protect\citeauthoryear{Barlow et~al.,}{Barlow
  et~al.}{2010}]{Barlow2010}
Barlow M.~J.,  et~al., 2010, \mn@doi [\aap] {10.1051/0004-6361/201014585}, 518,
  L138

\bibitem[\protect\citeauthoryear{{Baron}, {Nugent}, {Branch}  \&
  {Hauschildt}}{{Baron} et~al.}{2005}]{Baron2005}
{Baron} E.,  {Nugent} P.~E.,  {Branch} D.,   {Hauschildt} P.~H.,  2005, in
  {Turatto} M.,  {Benetti} S.,  {Zampieri} L.,   {Shea} W.,  eds,  ASP Conf.
  Ser. Vol. 342, Supernovae as Cosmological Lighthouses. p.~351 (\mn@eprint {}
  {astro-ph/0409659})

\bibitem[\protect\citeauthoryear{Bertoldi, Carilli, Cox, Fan, Strauss, Beelen,
  Omont  \& Zylka}{Bertoldi et~al.}{2003}]{Bertoldi2003}
Bertoldi F.,  Carilli C.~L.,  Cox P.,  Fan X.,  Strauss M.~a.,  Beelen A.,
  Omont A.,   Zylka R.,  2003, \mn@doi [\aap] {10.1051/0004-6361:20030710},
  406, L55

\bibitem[\protect\citeauthoryear{Bouchet, Danziger  \& Lucy}{Bouchet
  et~al.}{1991}]{Bouchet1991}
Bouchet P.,  Danziger I.~J.,   Lucy L.~B.,  1991, \mn@doi [\aj]
  {10.1086/115939}, 102, 1135

\bibitem[\protect\citeauthoryear{Chugai, Chevalier, Kirshner  \&
  Challis}{Chugai et~al.}{1997}]{Chugai1997}
Chugai N.~N.,  Chevalier R.~a.,  Kirshner R.~P.,   Challis P.~M.,  1997,
  \mn@doi [\apj] {10.1086/304253}, 483, 925

\bibitem[\protect\citeauthoryear{{Danziger}, {Bouchet}, {Gouiffes}  \&
  {Lucy}}{{Danziger} et~al.}{1991}]{Danziger1991}
{Danziger} I.~J.,  {Bouchet} P.,  {Gouiffes} C.,   {Lucy} L.~B.,  1991, in
  {Danziger} I.~J.,  {Kjaer} K.,  eds,  Vol. 37, European Southern Observatory
  Conference and Workshop Proceedings. p.~217

\bibitem[\protect\citeauthoryear{Dorschner, Begemann, Henning, Jaeger  \&
  Mutschke}{Dorschner et~al.}{1995}]{Dorschner1995}
Dorschner J.,  Begemann B.,  Henning T.,  Jaeger C.,   Mutschke H.,  1995,
  \aap, 300, 503

\bibitem[\protect\citeauthoryear{Draine \& Lee}{Draine \&
  Lee}{1984}]{Draine1984}
Draine B.~T.,  Lee H.~M.,  1984, \mn@doi [\apj] {10.1086/162480}, 285, 89

\bibitem[\protect\citeauthoryear{{Dwek} \& {Arendt}}{{Dwek} \&
  {Arendt}}{2015}]{Dwek2015}
{Dwek} E.,  {Arendt} R.~G.,  2015, \mn@doi [\apj] {10.1088/0004-637X/810/1/75},
  \href {http://adsabs.harvard.edu/abs/2015ApJ...810...75D} {810, 75}

\bibitem[\protect\citeauthoryear{Dwek, Galliano  \& Jones}{Dwek
  et~al.}{2007}]{Dwek2007}
Dwek E.,  Galliano F.,   Jones A.~P.,  2007, \mn@doi [\apj] {10.1086/518430},
  662, 927

\bibitem[\protect\citeauthoryear{Ercolano, Barlow  \& Sugerman}{Ercolano
  et~al.}{2007}]{Ercolano2007}
Ercolano B.,  Barlow M.~J.,   Sugerman B. E.~K.,  2007, \mn@doi [\mnras]
  {10.1111/j.1365-2966.2006.11336.x}, 375, 753

\bibitem[\protect\citeauthoryear{Fabbri et~al.,}{Fabbri
  et~al.}{2011}]{Fabbri2011}
Fabbri J.,  et~al., 2011, \mn@doi [\mnras] {10.1111/j.1365-2966.2011.19577.x},
  418, 1285

\bibitem[\protect\citeauthoryear{Fransson \& Kozma}{Fransson \&
  Kozma}{1993}]{Fransson1993}
Fransson C.,  Kozma C.,  1993, \mn@doi [\apj] {10.1086/186822}, 408, L25

\bibitem[\protect\citeauthoryear{Fransson et~al.,}{Fransson
  et~al.}{2013}]{Fransson2013}
Fransson C.,  et~al., 2013, \mn@doi [\apj] {10.1088/0004-637X/768/1/88}, 768,
  88

\bibitem[\protect\citeauthoryear{Fransson et~al.,}{Fransson
  et~al.}{2014}]{Fransson2014}
Fransson C.,  et~al., 2014, \mn@doi [\apj] {10.1088/0004-637X/797/2/118}, 797,
  118

\bibitem[\protect\citeauthoryear{Gall et~al.,}{Gall et~al.}{2014}]{Gall2014}
Gall C.,  et~al., 2014, \mn@doi [Nature] {10.1038/nature13558}, 511, 326

\bibitem[\protect\citeauthoryear{Gerasimovic}{Gerasimovic}{1933}]{Gerasimovic1933}
Gerasimovic B.,  1933, Z. Astrophys., 7, 335

\bibitem[\protect\citeauthoryear{Gomez et~al.,}{Gomez et~al.}{2012}]{Gomez2012}
Gomez H.~L.,  et~al., 2012, \mn@doi [\mnras]
  {10.1111/j.1365-2966.2011.20272.x}, 420, 3557

\bibitem[\protect\citeauthoryear{Gr\"{o}ningsson, Fransson, Lundqvist, Nymark,
  Lundqvist, Chevalier, Leibundgut  \& Spyromilio}{Gr\"{o}ningsson
  et~al.}{2006}]{Groeningsson2006}
Gr\"{o}ningsson P.,  Fransson C.,  Lundqvist P.,  Nymark T.,  Lundqvist N.,
  Chevalier R.,  Leibundgut B.,   Spyromilio J.,  2006, \mn@doi [\aap]
  {10.1051/0004-6361:20065325}, 5325, 11

\bibitem[\protect\citeauthoryear{Gr\"{o}ningsson et~al.,}{Gr\"{o}ningsson
  et~al.}{2007}]{Groeningsson2007}
Gr\"{o}ningsson P.,  et~al., 2007, \mn@doi [\aap]
  {10.1051/0004-6361:200810551}, 491, 19

\bibitem[\protect\citeauthoryear{Gr\"{o}ningsson et~al.,}{Gr\"{o}ningsson
  et~al.}{2008}]{Groningsson2008}
Gr\"{o}ningsson P.,  et~al., 2008, \mn@doi [\aap] {10.1051/0004-6361:20077604},
  479, 761

\bibitem[\protect\citeauthoryear{Hammer, Janka  \& M\"{u}ller}{Hammer
  et~al.}{2010}]{Hammer2010}
Hammer N.~J.,  Janka H.-T.,   M\"{u}ller E.,  2010, \mn@doi [\apj]
  {10.1088/0004-637X/714/2/1371}, 714, 1371

\bibitem[\protect\citeauthoryear{Hanner}{Hanner}{1988}]{Hanner1988}
Hanner M.~S.,  1988, NASA Conf. Publ., 3004, NASA, Washington DC

\bibitem[\protect\citeauthoryear{Hanuschik, Spyromilio, Stathakis, Kimeswenger,
  Gochermann, Seidensticker  \& Meurer}{Hanuschik et~al.}{1993}]{Hanuschik1993}
Hanuschik R.~W.,  Spyromilio J.,  Stathakis R.,  Kimeswenger S.,  Gochermann
  J.,  Seidensticker K.~J.,   Meurer G.,  1993, \mnras, 261, 909

\bibitem[\protect\citeauthoryear{{Henyey} \& {Greenstein}}{{Henyey} \&
  {Greenstein}}{1941}]{Henyey1941}
{Henyey} L.~G.,  {Greenstein} J.~L.,  1941, \mn@doi [\apj] {10.1086/144246},
  \href {http://adsabs.harvard.edu/abs/1941ApJ....93...70H} {93, 70}

\bibitem[\protect\citeauthoryear{Hillier}{Hillier}{1991}]{Hillier1991}
Hillier D.~J.,  1991, \aap, 247, 455

\bibitem[\protect\citeauthoryear{Hoyle \& Wickramasinghe}{Hoyle \&
  Wickramasinghe}{1970}]{Hoyle1970}
Hoyle F.,  Wickramasinghe N.~C.,  1970, \mn@doi [Nature] {10.1038/226062a0},
  226, 62

\bibitem[\protect\citeauthoryear{Indebetouw et~al.,}{Indebetouw
  et~al.}{2014}]{Indebetouw2014}
Indebetouw R.,  et~al., 2014, \mn@doi [\apj] {10.1088/2041-8205/782/1/L2}, 782,
  L2

\bibitem[\protect\citeauthoryear{J\"{a}ger, Mutschke, Begemann, Dorschner  \&
  Henning}{J\"{a}ger et~al.}{1994}]{Jager1994}
J\"{a}ger C.,  Mutschke H.,  Begemann B.,  Dorschner J.,   Henning T.,  1994,
  \aap, 292, 641

\bibitem[\protect\citeauthoryear{J\"{a}ger, Dorshner, Mutschke, Posch  \&
  Henning}{J\"{a}ger et~al.}{2003}]{Jager2003}
J\"{a}ger C.,  Dorshner J.,  Mutschke H.,  Posch T.,   Henning T.,  2003,
  \mn@doi [\aap] {10.1051/0004-6361:20030916}, 408, 193

\bibitem[\protect\citeauthoryear{Jerkstrand, Fransson, Maguire, Smartt, Ergon
  \& Spyromilio}{Jerkstrand et~al.}{2012}]{Jerkstrand2012}
Jerkstrand A.,  Fransson C.,  Maguire K.,  Smartt S.,  Ergon M.,   Spyromilio
  J.,  2012, \mn@doi [\aap] {10.1051/0004-6361/201219528}, 546, A28

\bibitem[\protect\citeauthoryear{Kotak et~al.,}{Kotak et~al.}{2009}]{Kotak2009}
Kotak R.,  et~al., 2009, \mn@doi [\apj] {10.1088/0004-637X/704/1/306}, 704, 306

\bibitem[\protect\citeauthoryear{{Kozasa}, {Hasegawa}  \& {Nomoto}}{{Kozasa}
  et~al.}{1991}]{Kozasa1991}
{Kozasa} T.,  {Hasegawa} H.,   {Nomoto} K.,  1991, \aap, \href
  {http://adsabs.harvard.edu/abs/1991A%26A...249..474K} {249, 474}

\bibitem[\protect\citeauthoryear{Kozma \& Fransson}{Kozma \&
  Fransson}{1998a}]{Kozma1998a}
Kozma C.,  Fransson C.,  1998a, \mn@doi [\apj] {10.1086/305409}, 496, 946

\bibitem[\protect\citeauthoryear{Kozma \& Fransson}{Kozma \&
  Fransson}{1998b}]{Kozma1998b}
Kozma C.,  Fransson C.,  1998b, \mn@doi [\apj] {10.1086/305452}, 497, 431

\bibitem[\protect\citeauthoryear{Li \& McCray}{Li \& McCray}{1992}]{Li1992}
Li H.,  McCray R.,  1992, \mn@doi [\apj] {10.1086/171082}, 387, 309

\bibitem[\protect\citeauthoryear{Lucy}{Lucy}{2005}]{Lucy2005c}
Lucy L.,  2005, \mn@doi [\aap] {10.1051/0004-6361}, 429, 19

\bibitem[\protect\citeauthoryear{{Lucy}, {Danziger}, {Gouiffes}  \&
  {Bouchet}}{{Lucy} et~al.}{1989}]{Lucy1989}
{Lucy} L.~B.,  {Danziger} I.~J.,  {Gouiffes} C.,   {Bouchet} P.,  1989, in
  {Tenorio-Tagle} G.,  {Moles} M.,   {Melnick} J.,  eds,  Lecture Notes in
  Physics, Berlin Springer Verlag Vol. 350, IAU Colloq. 120: Structure and
  Dynamics of the Interstellar Medium. p.~164, \mn@doi{10.1007/BFb0114861}

\bibitem[\protect\citeauthoryear{Lucy, Danziger, Gouiffes  \& Bouchet}{Lucy
  et~al.}{1991}]{Lucy1991}
Lucy L.,  Danziger I.~J.,  Gouiffes C.,   Bouchet P.,  1991, in Woosley S.~E.,
  ed., Supernovae. Springer-Verlag, New York, p.~82

\bibitem[\protect\citeauthoryear{Maeda, Mazzali, Deng, Nomoto, Yoshii, Tomita
  \& Kobayashi}{Maeda et~al.}{2003}]{Maeda2003}
Maeda K.,  Mazzali P.~a.,  Deng J.,  Nomoto K.,  Yoshii Y.,  Tomita H.,
  Kobayashi Y.,  2003, \mn@doi [\apj] {10.1086/376591}, 593, 931

\bibitem[\protect\citeauthoryear{Mathis, Rumpl  \& Nordsieck}{Mathis
  et~al.}{1977}]{Mathis1977}
Mathis J.~S.,  Rumpl W.,   Nordsieck K.~H.,  1977, \mn@doi [\apj]
  {10.1086/155591}, 217, 425

\bibitem[\protect\citeauthoryear{Matsuura et~al.,}{Matsuura
  et~al.}{2011}]{Matsuura2011}
Matsuura M.,  et~al., 2011, \mn@doi [Science] {10.1126/science.1205983}, 333,
  1258

\bibitem[\protect\citeauthoryear{Matsuura et~al.,}{Matsuura
  et~al.}{2015}]{Matsuura2015}
Matsuura M.,  et~al., 2015, \mn@doi [\apj] {10.1088/0004-637X/800/1/50}, 800,
  50

\bibitem[\protect\citeauthoryear{Mauerhan \& Smith}{Mauerhan \&
  Smith}{2012}]{Mauerhan2012}
Mauerhan J.,  Smith N.,  2012, \mn@doi [\mnras]
  {10.1111/j.1365-2966.2012.21325.x}, 424, 2659

\bibitem[\protect\citeauthoryear{McCray}{McCray}{1996}]{McCray1996}
McCray R.,  1996, in Kuhn T.~S.,  ed., IAU Colloq. 145: Supernovae and
  Supernova Remnants. p.~223

\bibitem[\protect\citeauthoryear{Meikle et~al.,}{Meikle
  et~al.}{2007}]{Meikle2007}
Meikle W. P.~S.,  et~al., 2007, \mn@doi [\apj] {10.1086/519733}, 665, 608

\bibitem[\protect\citeauthoryear{Milisavljevic, Fesen, Chevalier, Kirshner,
  Challis  \& Turatto}{Milisavljevic et~al.}{2012}]{Milisavljevic2012}
Milisavljevic D.,  Fesen R.~A.,  Chevalier R.~A.,  Kirshner R.~P.,  Challis P.,
    Turatto M.,  2012, \mn@doi [\apj] {10.1088/0004-637X/751/1/25}, 751, 25

\bibitem[\protect\citeauthoryear{Morgan \& Edmunds}{Morgan \&
  Edmunds}{2003}]{Morgan2003}
Morgan H.~L.,  Edmunds M.~G.,  2003, \mn@doi [\mnras]
  {10.1046/j.1365-8711.2003.06681.x}, 343, 427

\bibitem[\protect\citeauthoryear{Omont, Cox, Bertoldi, McMahon, Carilli  \&
  Isaak}{Omont et~al.}{2001}]{Omont2001}
Omont A.,  Cox P.,  Bertoldi F.,  McMahon R.~G.,  Carilli C.,   Isaak K.~G.,
  2001, \mn@doi [\aap] {10.1051/0004-6361:20010721}, 374, 371

\bibitem[\protect\citeauthoryear{Owen \& Barlow}{Owen \&
  Barlow}{2015}]{Owen2015}
Owen P.~J.,  Barlow M.~J.,  2015, \mn@doi [\apj] {10.1088/0004-637X/801/2/141},
  801, 141

\bibitem[\protect\citeauthoryear{{Phillips}, {Hamuy}, {Heathcote}, {Suntzeff}
  \& {Kirhakos}}{{Phillips} et~al.}{1990}]{Phillips1990}
{Phillips} M.~M.,  {Hamuy} M.,  {Heathcote} S.~R.,  {Suntzeff} N.~B.,
  {Kirhakos} S.,  1990, \mn@doi [\aj] {10.1086/115402}, \href
  {http://adsabs.harvard.edu/abs/1990AJ.....99.1133P} {99, 1133}

\bibitem[\protect\citeauthoryear{Roche, Aitken, Smith  \& James}{Roche
  et~al.}{1989}]{Roche1989}
Roche P.~F.,  Aitken D.~K.,  Smith C.~H.,   James S.~D.,  1989, \mn@doi
  [Nature] {10.1038/337533a0}, 337, 533

\bibitem[\protect\citeauthoryear{Sarangi \& Cherchneff}{Sarangi \&
  Cherchneff}{2015}]{Sarangi2015}
Sarangi A.,  Cherchneff I.,  2015, \mn@doi [\aap]
  {10.1051/0004-6361/201424969}, 575, A95

\bibitem[\protect\citeauthoryear{Silvia, Smith  \& {Michael Shull}}{Silvia
  et~al.}{2010}]{Silvia2010}
Silvia D.~W.,  Smith B.~D.,   {Michael Shull} J.,  2010, \mn@doi [\apj]
  {10.1088/0004-637X/715/2/1575}, 715, 1575

\bibitem[\protect\citeauthoryear{Silvia, Smith  \& Shull}{Silvia
  et~al.}{2012}]{Silvia2012}
Silvia D.~W.,  Smith B.~D.,   Shull J.~M.,  2012, \mn@doi [\apj]
  {10.1088/0004-637X/748/1/12}, 748, 12

\bibitem[\protect\citeauthoryear{Slavin, Dwek  \& Jones}{Slavin
  et~al.}{2015}]{Slavin2015}
Slavin J.~D.,  Dwek E.,   Jones A.~P.,  2015, \mn@doi [\apj]
  {10.1088/0004-637X/803/1/7}, 803, 7

\bibitem[\protect\citeauthoryear{Smith, Silverman, Filippenko, Cooper,
  Matheson, Bian, Weiner  \& Comerford}{Smith et~al.}{2012}]{Smith2012}
Smith N.,  Silverman J.~M.,  Filippenko A.~V.,  Cooper M.~C.,  Matheson T.,
  Bian F.,  Weiner B.~J.,   Comerford J.~M.,  2012, \mn@doi [\aj]
  {10.1088/0004-6256/143/1/17}, 17, 6

\bibitem[\protect\citeauthoryear{Spyromilio, Stathakis, Cannon, Waterman, Couch
   \& Dopita}{Spyromilio et~al.}{1991}]{Spyromilio1991}
Spyromilio J.,  Stathakis R.,  Cannon R.,  Waterman L.,  Couch W.,   Dopita M.,
   1991, \mnras, 248, 465

\bibitem[\protect\citeauthoryear{Spyromilio, Stathakis  \& Meurer}{Spyromilio
  et~al.}{1993}]{Spyromilio1993}
Spyromilio J.,  Stathakis R.~a.,   Meurer G.~R.,  1993, \mnras, 263, 530

\bibitem[\protect\citeauthoryear{Storey \& Zeippen}{Storey \&
  Zeippen}{2000}]{Storey2000}
Storey P.~J.,  Zeippen C.~J.,  2000, \mn@doi [\mnras]
  {10.1046/j.1365-8711.2000.03184.x}, 312, 813

\bibitem[\protect\citeauthoryear{Sugerman et~al.,}{Sugerman
  et~al.}{2006}]{Sugerman2006}
Sugerman B. E.~K.,  et~al., 2006, \mn@doi [Science] {10.1126/science.1128131},
  313, 196

\bibitem[\protect\citeauthoryear{Suntzeff, Phillips, Depoy, Elias  \&
  Walker}{Suntzeff et~al.}{1991}]{Suntzeff1991}
Suntzeff N.~B.,  Phillips M.~M.,  Depoy D.~L.,  Elias J.~H.,   Walker A.~R.,
  1991, \mn@doi [\aj] {10.1086/115938}, 102, 1118

\bibitem[\protect\citeauthoryear{Todini \& Ferrara}{Todini \&
  Ferrara}{2001}]{Todini2001}
Todini P.,  Ferrara A.,  2001, \mn@doi [\mnras]
  {10.1046/j.1365-8711.2001.04486.x}, 325, 726

\bibitem[\protect\citeauthoryear{Tziamtzis, Lundqvist, Groningsson  \&
  Nasoudi-Shoar}{Tziamtzis et~al.}{2010}]{Tziamtzis2010}
Tziamtzis a.,  Lundqvist P.,  Groningsson P.,   Nasoudi-Shoar S.,  2010, \aap,
  35, 15

\bibitem[\protect\citeauthoryear{Wang et~al.,}{Wang et~al.}{1996}]{Wang1996}
Wang L.,  et~al., 1996, \mn@doi [\apj] {10.1086/177570}, 466, 998

\bibitem[\protect\citeauthoryear{Watson, Christensen, Knudsen, Richard,
  Gallazzi  \& Michałowski}{Watson et~al.}{2015}]{Watson2015}
Watson D.,  Christensen L.,  Knudsen K.~K.,  Richard J.,  Gallazzi A.,
  Michałowski M.~J.,  2015, \mn@doi [Nature] {10.1038/nature14164}, 519, 327

\bibitem[\protect\citeauthoryear{Wesson, Barlow, Matsuura  \& Ercolano}{Wesson
  et~al.}{2015}]{Wesson2015}
Wesson R.,  Barlow M.~J.,  Matsuura M.,   Ercolano B.,  2015, \mnras, 446, 2089
  (W15)

\bibitem[\protect\citeauthoryear{Wongwathanarat, M\"{u}ller  \&
  Janka}{Wongwathanarat et~al.}{2015}]{Wongwathanarat2015}
Wongwathanarat A.,  M\"{u}ller E.,   Janka H.-T.,  2015, \mn@doi [\aap]
  {10.1051/0004-6361/201425025}, 577, A48

\bibitem[\protect\citeauthoryear{Wooden, Rank, Bregman, Witteborn, Tielens,
  Cohen, Pinto  \& Axelrod}{Wooden et~al.}{1993}]{Wooden1993}
Wooden D.~H.,  Rank D.~M.,  Bregman J.~D.,  Witteborn F.~C.,  Tielens A. G.
  G.~M.,  Cohen M.,  Pinto P.~A.,   Axelrod T.~S.,  1993, \mn@doi [\apjs]
  {10.1086/191830}, 88, 477

\bibitem[\protect\citeauthoryear{Xu, McCray, Oliva  \& Randich}{Xu
  et~al.}{1992}]{Xu1992}
Xu Y.,  McCray R.,  Oliva E.,   Randich S.,  1992, \mn@doi [\apj]
  {10.1086/171003}, 386, 181

\bibitem[\protect\citeauthoryear{Zubko, Mennella, Colangeli  \&
  Bussoletti}{Zubko et~al.}{1996}]{Zubko1996}
Zubko V.~G.,  Mennella V.,  Colangeli L.,   Bussoletti E.,  1996, \mnras, 282,
  L1321

\makeatother
\end{thebibliography}
\bibliographystyle{mnras}

\appendix

\section[]{The Lorentz transform formalism}

Since the outflow velocities in supernovae are high, the photon packets 
are subject to Doppler shifting upon emission and at each scattering event.  
When the packet is initially emitted, it has a frequency and a trajectory 
in the rest frame of the emitter. Both of these must be transformed to the 
observer's frame in order for the packet to be propagated through the 
grid.  The new direction and frequency in the observer's frame may be 
simply found by transforming the momentum four-vector $\boldsymbol{P}$ which is 
defined as

\begin{equation}
\boldsymbol{P}=
\begin{pmatrix}
	E \\
	p_x \\
	p_y \\
	p_z \\
	\end{pmatrix} =
	\begin{pmatrix}
	h \nu \\
	h \nu x \\
	h \nu y \\
	h \nu z \\
	\end{pmatrix}
\end{equation}

\noindent We may then derive $\boldsymbol{P'}$, the momentum 4-vector in the 
observer's frame using the relation

\begin{equation}
	\boldsymbol{P'}=\Lambda \boldsymbol{P}	
\end{equation}

\noindent where 

\[
	{\Lambda}=
	 \begin{pmatrix} 
	  \gamma & -\gamma \beta_x & -\gamma \beta_y & -\gamma \beta_z \\
	 -\gamma \beta_x & 1+(\gamma-1)\frac{\beta_x^2}{\beta^2} & (\gamma-1)\frac{\beta_x \beta_y}{\beta^2} & (\gamma-1)\frac{\beta_x \beta_z}{\beta^2} \\
	 -\gamma \beta_y  & (\gamma-1)\frac{\beta_y \beta_x}{\beta^2} & 1+(\gamma-1)\frac{\beta_y^2}{\beta^2} & (\gamma-1)\frac{\beta_y \beta_z}{\beta^2} \\
	 -\gamma \beta_z & (\gamma-1)\frac{\beta_z \beta_x}{\beta^2} & (\gamma-1)\frac{\beta_z \beta_y}{\beta^2} & 1+(\gamma-1)\frac{\beta_z^2}{\beta^2} \\
	 \end{pmatrix}
\]

 \noindent and $\boldsymbol{\beta}=\frac{{\bf{v}}}{c}=(\beta_x,\beta_y,\beta_z)$,   $\beta=\lvert \boldsymbol{\beta} \rvert$ and $\gamma = \frac{1}{\sqrt{1-\beta^2}}$.

In practice, the velocities considered are low enough that it is 
unnecessary to consider terms of the order of $O(\frac{v^2}{c^2})$ and thus 
${\Lambda}$ may be reduced to

\begin{equation}
	{\Lambda}=
	 \begin{pmatrix} 
	 1 & - \beta_x & - \beta_y & - \beta_z \\
	- \beta_x & 1 & 0 & 0 \\
	- \beta_y  & 0 & 1 & 0\\
	- \beta_z & 0 & 0 & 1 \\
	 \end{pmatrix}
	 \\
\end{equation}

\noindent The new direction of travel and frequency in the observer's 
frame are therefore given by  
\begin{equation}
\nu'=\nu(1-x\beta_x-y\beta_y-z\beta_z) \\
\end{equation}
\[
x'=\frac{\nu}{\nu'}(x-\beta_x) 
\]
\[
y'=\frac{\nu}{\nu'}(x-\beta_y) 
\]
\[
z'=\frac{\nu}{\nu'}(x-\beta_z) 
\]

For each scattering event, the packet must be transformed both into and 
out of the comoving frame. The reverse transform is applied by using the 
inverse Lorentz matrix $\Lambda^{-1}$ which is obtained by reversing the 
sign of $\boldsymbol{v}$.  Positive $\boldsymbol{v}$ is defined for frames moving away 
from each other and thus $\boldsymbol{v}$ is defined to be negative in the 
direction of the observer.

\bsp

\label{lastpage}

\end{document}